PhD thesis dissertation

# Chemical vapor deposition of hexagonal boron nitride and its use in electronic devices

Author: Ms. Fei Hui

Director: Dr. Mario Lanza

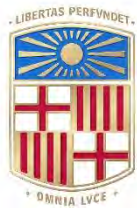

UNIVERSITAT DE BARCELONA

# Chemical vapor deposition of hexagonal boron nitride and its use in electronic devices

Memoria presentada para optar al grado de doctor por la

Universidad de Barcelona

Programa de doctorado en Nanociencias

Author: Ms. Fei Hui

Director: Dr. Mario Lanza

Tutor: Dr. Albert Cirera

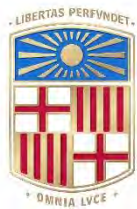

UNIVERSITAT DE BARCELONA

*To my family,*

# Acknowledgements

First of all, I would like to thank my PhD director Prof. Mario Lanza for supervising my work and helping me to meet this enormous challenge. He taught me a lot of knowledge and shared his own research experience, he spent time with me in the lab doing experiments and gave me suggestions on writing the scientific reports, and together we practice my presentation skill. Thanks to his guidance, I feel I have made big progress on both research and communication skills. I really appreciate all the group members in Prof. Lanza's group, we work together happily and have an enjoyable time.

I also send special thanks to the directors of the research groups in which I worked during my stays abroad, Prof. Jing Kong and Mildred Dresselhaus from the Massachusetts Institute of Technology (USA), and Prof. Andrea Ferrari from University of Cambridge (UK). In their groups, I learned different syntheses processes of 2D materials. All their group members deserve my acknowledgment for technical advice. During my visiting time, we worked together and had abundant discussions, which have been very useful and inspired my future research. I also would like to thank all the coauthors of my publications, especially those from Stanford University, IBM company, and Chinese Academy of Sciences.

Finally, I feel very grateful with all institutions that, with their financial support, made this work possible, especially the Royal Society of Chemistry for giving me the researcher mobility award to travel to UK. Moreover, I want to thank to the National Science Foundation of China, and the Young 1000 program for funding my research. I also would like to thank Soochow University for supporting me to attend the international conferences in Spain and in USA. I specially appreciate the support received from the Institute of Functional Nano and Soft Materials of Soochow University, for allowing me using all the characterization techniques as much as I needed and for free.

Finally, I also would like to show my gratitude to my family and friends, thank you all very much for your encouragement and support.

# Index

# Abstract

Dielectrics are insulating materials used in many different electronic devices (e.g. capacitors, transistors, varistors), and play an important role in all of them. In fact, the dielectric is probably the most critical element in most devices, as it is exposed to electrical fields that can degrade its performance. Silicon dioxide ($SiO_2$) has been traditionally the most widely used dielectric material in the industry; however, the scaling down of the devices required a reduction of the $SiO_2$ thickness, which provoked a dramatic increase of the leakage current. This not only results in an increase of the power consumption, but also on the failure of the devices and circuits. Current advanced electronic devices use dielectric materials with a high dielectric constant (e.g. $HfO_2$, $Al_2O_3$ and $TiO_2$) so that their thickness doesn't need to be reduced so much, and high leakage currents are avoided. However, these materials show several intrinsic problems (e.g. large density of native defects, crystallization at high temperatures), and also a bad interaction with adjacent materials (e.g. rough interface with silicon, high diffusivity to polysilicon gate, difficulty to be deposited on two dimensional [2D] materials). Therefore, the race for finding a suitable dielectric material for current and future electronic devices is still open.

In this context 2D materials have become a serious option, not only thanks to their advanced properties, but also to the development of scalable synthesis methods. Graphene has been the most explored 2D material for electronic devices, and it has been used as channel in transistors, and as electrode in capacitors and memristors (among others). However, graphene has no band gap, and therefore it cannot be used as dielectric. $MoS_2$ and other 2D transition metal dichalcogenides (TMDs) are semiconducting 2D materials that can provide more versatility in electronic devices (i.e.

they can increase the current ON/OFF ratio in transistors because the density of carriers can be tuned via electrical field), but their small band gap difficult their use dielectric.

In this PhD thesis I have investigated the use of monolayer and multilayer hexagonal boron nitride ($h$-BN) as dielectric for electronic devices, as it is a 2D material with a band gap of ~5.9 eV. My work has mainly focused on the synthesis of the $h$-BN using chemical vapor deposition, the study of its intrinsic morphological and electrical properties at the nanoscale, and its performance as dielectric in different electronic devices, such as capacitors and memristors. Overall, our experiments indicate that $h$-BN is a very reliable dielectric material, and that it can be successfully used in capacitors and memristors. Its performance depends on several parameters, like the substrate on which it is grown, the growth temperature, the growth time, the vacuum pressure, and even the adjacent electrodes deposited. Moreover, $h$-BN shows additional performances never observed in traditional dielectrics, such as volatile resistive switching, which may also open the door for new applications.

# Abstract in official language

Los dieléctricos son materiales aislantes utilizados en muchos dispositivos electrónicos (por ejemplo condensadores, transistores, varistores), en los que juegan un papel muy importante. En realidad, el dieléctrico es probablemente la parte más crítica en la gran mayoría de dispositivos electrónicos, ya que casi siempre está expuesto a campos eléctricos que pueden degradar sus prestaciones. El dióxido de silicio ($SiO_2$) ha sido el material aislante tradicionalmente utilizado en la industria; sin embargo la miniaturización de los dispositivos requirió una reducción del grosor de los dieléctricos $SiO_2$, lo que provocó un incremento dramático de la corriente de fugas. Esto no sólo produce un aumento de la energía consumida, sino que también puede provocar el fallo del dispositivo entero, e incluso del circuito donde se ha implementado. Actualmente los dispositivos electrónicos más avanzados utilizan materiales aislantes con una constante dieléctrica alta (por ejemplo $HfO_2$, $Al_2O_3$ y $TiO_2$), y así no es necesario reducir tanto su grosor, lo que mantiene una baja corriente de fugas. Sin embargo, estos materiales muestran muchos problemas intrínsecos (por ejemplo grandes cantidades de defectos nativos, cristalización a altas temperaturas), y también una mala interacción con materiales adyacentes (por ejemplo una interfaz rugosa con el sustrato de silicio, una alta difusividad hacia el electrodo de puerta si este está hecho de polisilicio, y gran dificultad para ser depositado sobre materiales bidimensionales). Por lo tanto, la carrera para encontrar un material dieléctrico ideal para dispositivos electrónicos sigue abierta.

En este contexto, los materiales bidimensionales se han convertido en una seria opción, no sólo por sus excelentes propiedades, sino también gracias al desarrollo de nuevos métodos de síntesis escalables. El grafeno ha sido el material bidimensional más estudiado para dispositivos electrónicos, y ha sido utilizado como canal conductor en transistores, y como electrodo en condensadores y memristores (entre otros). Sin embargo, el grafeno no tiene una banda energética prohibida, con lo cual no pude ser usado como dieléctrico. $MoS_2$ y otros materiales derivados son bidimensionales semiconductores que pueden aportar una mayor versatilidad al ser usados en dispositivos electrónicos (por ejemplo pueden incrementar el ratio de corriente ON/OFF en transistores porque la densidad de portadores puede ser controlada aplicando una tensión externa), pero su banda de energías prohibidas es muy pequeña, lo que dificulta su uso como dieléctrico.

En esta tesis doctoral he investigado el uso de nitruro de boro hexagonal (*h*-BN) monocapa y multicapa como material dieléctrico en dispositivos electrónicos, ya su banda de energías prohibidas es de ~5.9 eV. Mi trabajo se ha focalizado en la síntesis de *h*-BN mediante el método *chemical vapor deposition*, el estudio de sus propiedades morfológicas y eléctricas a escala nanométrica, y el análisis de sus prestaciones como dieléctrico en diferentes dispositivos electrónicos (condensadores y memristores). Nuestros experimentos indican que *h*-BN es un material dieléctrico muy fiable, y que es apto para su uso en dispositivos. Sus prestaciones dependen de diferentes parámetros, como el sustrato en el que ha sido crecido, la temperatura y el tiempo de crecimiento, el nivel de vacío y presión, e incluso los materiales usados como electrodos adyacentes. Además, *h*-BN muestra propiedades adicionales nunca observadas en dieléctricos tradicionales, como modulación volátil de la resistividad, lo que podría extender su uso a nuevas aplicaciones.

# Chapter 1:

# Dissertation Summary

## 1.1. Introduction

Thin dielectric films are key elements in a wide range of electronic devices, as they can generate (for example) the capacitance effects required to form the conductive channel in field effect transistors (FETs), or the resistive switching (RS) phenomenon required to induce two logic (resistive) states in non-volatile memories (NVMs) [1-2]. With the scaling down of electronic devices, the traditionally used $SiO_2$ dielectrics became too thin to withstand the electrical fields applied (that problem appeared first in FETs in the early 2000's, around the 45 nm technological node) [3], which threatened the reliability of the devices due to prohibitive leakage currents and dielectric breakdown (BD). The solution adopted by the industry was to replace the ultra thin $SiO_2$ films by thicker high-k dielectric stacks (like $HfO_2$, $Al_2O_3$ and $TiO_2$); the thicker nature of the high-k blocked the leakage current, and its higher dielectric constant produced a similar capacitance effect than the thinner $SiO_2$, which is required to make the devices work. However, the introduction of high-k dielectrics in the semiconductors industry generated several new problems, such as high density of native defects, interaction with the polysilicon gate and Si substrate, severe inhomogeneities, polycrystallization at the temperatures required during the manufacturing process of the devices (>400 °C), and high scattering at the channel region [4-5]. Furthermore, high-k dielectrics interact very badly with other advanced and very promising materials for future electronic devices, such as two dimensional (2D) materials [6-7]. Therefore, research on alternative dielectrics for high performance electronic devices is necessary.

One excellent candidate material for becoming the dielectric of future electronic devices is 2D hexagonal boron nitride (h-BN). 2D *h*-BN is an insulating material from the family of graphene that exhibits several excellent physical [8], chemical [9], electrical [10], mechanical [11], thermal [12], magnetic [13], and optical [14] properties, and it has been already used in several electrical devices, such as: FETs [15], capacitors [16], sensors [17], and memristors [18-20] (among others). However, the most relevant dielectric behaviors, such as tunneling current [21-23], polycrystallization [24-25], charge trapping and de-trapping [26-27], stress induced leakage current (SILC) [28], dielectric strength [29], soft/hard BD [30], and RS [30-34], have never been analyzed in depth in *h*-BN. Moreover, these behaviors will strongly depend on the method used to synthesize the *h*-BN stacks [35], as different methods produce *h*-BN with different sizes, morphologies and densities of defects. This PhD thesis presents a complete and deep study about the synthesis of scalable *h*-BN stacks, and analyzes its performance as dielectric in electronic devices.

## 1.2. Main Contribution of this PhD thesis

### 1.2.1. Objectives of this PhD thesis

The main goal of this PhD thesis is to provide useful knowledge that clarifies if *h*-BN can be reliably used as dielectric in electronic devices. This task can be divided in three objectives. The first one is to develop a scalable method to synthesize *h*-BN that leads to high quality and low amount of defects. The second one is to characterize the intrinsic properties of the materials synthesized, such as thickness, surface roughness, density of defects, Raman signature, and percentage of B and N atoms. And the third is to study its performance as dielectric by applying electrical stresses, both at the nanoscale and at the device level. Another indirect goal of this thesis is to structure the knowledge available until now about the use of *h*-BN as dielectric. This is important because the first studies in this field didn't distinguish between *h*-BN stacks synthesized using different methods, and also because we have detected some literature that (in our opinion) reported irrelevant, unsupported and/or wrong claims. For this reason, this PhD thesis includes, not only four research articles (in the format of letters and full papers), but also two extensive review articles.

### *1.2.2. Key findings*

In the first part of this thesis I present a deep literature review about the synthesis, characterization methods and performance of *h*-BN as dielectric in electronic devices. **Article 1** is a review paper in which I analyze more than 179 references. We make critical comments related to the different performances shown by *h*-BN grown by different methods. Scalability of the synthesis process and its suitability for industrial applications is one of the main criteria when classifying the knowledge available on the use of *h*-BN as dielectric.

In the second part I describe the synthesis of multilayer *h*-BN on different metallic substrates (Pt, Cu, and Fe) using chemical vapor deposition (CVD) approach, and in all cases I check its performance as dielectric in real devices. In **Article 2** and **Article 3** I grew the *h*-BN on Pt substrates. On this specific substrate we observe that the *h*-BN shows important thickness inhomogeneities depending on the metallic grain on which it is grown. While this is an undesired effect, nanoscale electrical characterization via conductive atomic force microscopy (CAFM) reveals that the electrical properties of the *h*-BN stacks (i.e. tunneling current) within one metallic grain are very homogeneous, much more than in high-k dielectric films (e.g. $HfO_2$, $TiO_2$). In **Article 4** I studied *h*-BN grown by CVD on Cu substrates, and I characterized the entire BD process depending on the thickness. I found out that monolayer *h*-BN is extremely resistant to changes in the morphology after the BD. In **Article 5** I grew the *h*-BN on Fe substrates, and fabricated a matrix of memristors that show both volatile and non-volatile RS with low device-to-device variability.

Given the promising performance of *h*-BN as RS medium, in **Article 6** I made an extensive literature review about the use of this and other 2D materials in memristors. This article analyzes more than 364 references, and contains 12 tables comparing the structure, size, current window, endurance, retention, operation voltages, speed, power consumption, transparency and flexibility of the 2D materials based memristors. I also discuss the status and challenges to solve in this direction.

*1.2.3. Thesis Outline*

This thesis is divided in five chapters: Chapter 1 presents the dissertation summary, which introduces the most relevant aspects of this thesis. Chapter 2 provides a technical introduction about the structure and synthesis of *h*-BN, as well as its reliability as dielectric. Chapter 2 features **Article 1**. Chapter 3 describes in depth the synthesis of multilayer *h*-BN stacks by CVD approach, and analyzes the properties of the *h*-BN stacks at the nanoscale (via CAFM) and at the device level (via probe station). Chapter 3 has three sections, one dedicated to *h*-BN grown on Pt, another one for *h*-BN grown on Cu, and another one for *h*-BN grown on Fe. In each study the experimental results have been also compared with theoretical simulations. Chapter 3 features **Articles 2, 3** and **4**. Chapter 4 presents a deep revision of the use of 2D materials as dielectric in memristors, summarizing the best performances and discussing the main challenges. Finally, <u>Chapter 5</u> summarizes the main results of this thesis, conclusions and perspectives.

## 1.3. List of publications

The list of articles shown below only includes the publications which shall be considered for evaluation of this PhD dissertation, although during my PhD I have published many other research articles. A reproduction of each publication can be accessed by the indicated information below. A complete list of the author's publication (updated on May 4th 2018) is included in the scientific curriculum vitae (Appendix A).

Article 1     **Fei Hui**, Chengbin Pan, Yuanyuan Shi, Yanfeng Ji, Enric Grustan-Gutierrez, Mario Lanza, On the use of two dimensional hexagonal boron nitride as dielectric, **Microelectronic Engineering**, 163, 119-133 (**2016**).

*\* Contribution: Deep literature revision, classifying the knowledge, and writing the main parts of the manuscript.*

Article 2     **Fei Hui**, Wenjing Fang, Wei Sun Leong, Tewa Kpulun, Haozhe Wang, Hui Ying Yang, Marco A. Villena, Gary Harris, Jing Kong, Mario Lanza, Electrical homogeneity of large-area chemical vapor deposited multilayer hexagonal boron nitride sheets, **ACS Applied Materials & Interfaces**, 9, 39895-39900 (**2017**).

*\* Contribution: Growth of the h-BN using CVD furnace, characterization of the material and devices, evaluation of the results and writing the main parts of the manuscript.*

Article 3   **<u>Fei Hui</u>**, Xianhu Liang, Wenjing Fang, Wei Sun Leong, Haozhe Wang, Hui Ying Yang, Marco A. Villena, Jing Kong, Mario Lanza, Uniformity of multilayer hexagonal boron nitride dielectric stacks grown by chemical vapor deposition on platinum and copper substrates, Proceedings of the **IEEE-IPFA conference** (**2018**) - *Accepted.*

*\* Contribution: Characterizing the electrical homogeneity of h-BN stacks grown on Pt, evaluation of the results and writing the manuscript.*

Article 4   Lanlan Jiang*, Yuanyuan Shi*, **<u>Fei Hui</u>***, Kechao Tang, Qian Wu, Chengbin Pan, Xu Jing, Hasan Uppal, Felix Palumbo, Guangyuan Lu, Tianru Wu, Haomin Wang, Marco A. Villena, Xiaoming Xie, Paul C. McIntyre, Mario Lanza, Dielectric breakdown in chemical vapor deposited hexagonal boron nitride, **ACS Applied Materials & Interfaces**, 9, 39758-39770 (**2017**).

*\* Contribution: Carrying out the conductive AFM characterization of as-grown h-BN on Cu/Ni and analyzing the results.*

Article 5   **<u>Fei Hui</u>**, Marco A. Villena, Wenjing Fang, Ang-Yu Lu, Jing Kong, Yuanyuan Shi, Xu Jing, Kaichen Zhu, Mario Lanza, Synthesis of large-area multilayer hexagonal boron nitride sheets on iron substrates and its use in resistive switching devices, **2D Materials** (**2018**) - *Minor revision.*

*\* Contribution: Growth of the h-BN by CVD and characterization using SEM, Raman Spectroscopy, cross sectional TEM and probe station. Writing the manuscript.*

Article 6    **<u>Fei Hui</u>**, Enric Grustan-Gutierrez, Shibing Long, Qi Liu, Anna K. Ott, Andrea C. Ferrari, Mario Lanza, Graphene and related materials for resistive random access memories, **Advanced Electronic Materials**, 1600195 (**2017**).

*\* Contribution: Literature research, preparation of the figures and the tables, classifying the knowledge, and writing the manuscript.*

These articles have been developed in collaboration with Massachusetts Institute of Technology (MIT, USA), Stanford University (USA), University of Cambridge (UK) and Soochow University (China). To do this work, I travelled one year to Massachusetts Institute of Technology and half a year to University of Cambridge, where I worked in the groups lead by Prof. Jing Kong and Prof. Andrea Ferrari (respectively). For my stay at the University of Cambridge I won the Royal Society of Chemistry mobility award for PhD students.

*Disclaimer: All these 6 research articles will be entirely reproduced (embedded at the suitable sections or sub-sections) with the permission from the publishers, which are the owners of the copyright. Note that sections, equations and references numbering within the reproduced research articles follow the ones of the published version, not the ones of this thesis. This statement applies to all articles presented in this Thesis dissertation.*

# Chapter 2:

# On the use of 2D layered *h*-BN as dielectric

The aim of this chapter is to describe the status of 2D layered *h*-BN (previous to this PhD thesis). This chapter is divided in three sections. Section 2.1 describes the structure and most remarkable properties of *h*-BN reported until that date. Section 2.2 describes the different synthesis methods, and it discusses their advantages and challenges. Section 2.3 summarizes the most relevant investigations that used *h*-BN as dielectric, and highlights the most remarkable performances. The reliability of *h*-BN as dielectric and the entire BD process will be also described.

## 2.1. Properties of 2D layered *h*-BN

Boron nitride exists in multiple crystalline forms that differ in the arrangement of the B and N atoms, such as cubic boron nitride (*c*-BN), wurtzite boron nitride (*w*-BN) and hexagonal boron nitride (*h*-BN) [36]. Among them, *h*-BN is a typical $sp^2$-hybridized 2D insulator, which is analogous to graphene in terms of their hexagonal lattice structure (see Figure 2.1). *h*-BN is formed by B and N atoms interacting by covalent bonds in plane, forming a hexagonal lattice (i.e. each B atom bonds with three N, and each N atom bonds with three B) [37]. The lattice constant (distance between atoms) in this hexagonal network is 0.25 nm, and it shows a mismatch of only 1.8% with that of graphene [38]. An ideal *h*-BN sheet should be a continuous B and N lattice without any missing bond, and perfectly attached to the substrate. Figures 2.1a and 2.1b show a schematic and an experimental demonstration of the lattice structure of monolayer *h*-BN.

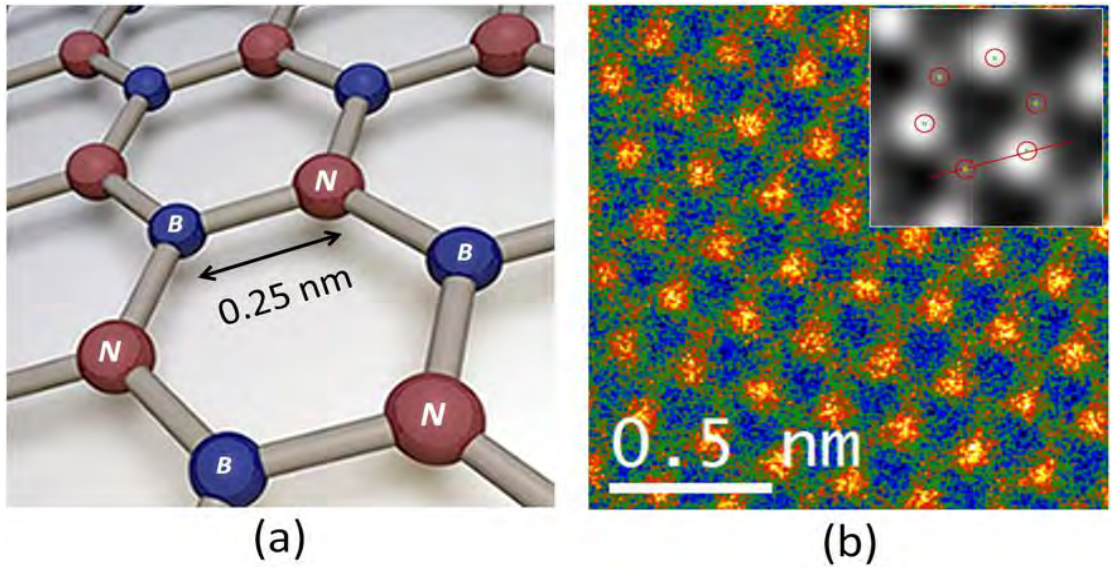

**Figure 2.1**. (a) Schematic of the hexagonal lattice of monolayer *h*-BN. (b) High angle annular dark field scanning transmission electron microscopy image proving the chemical composition of *h*-BN with sub-atomic resolution. Reproduced with permission from Ref. [37], copyright InTech 2013.

*h*-BN holds several extraordinary properties that may be useful for the fabrication of electronic devices. For example, by using an exact numerical solution of the phonon Boltzmann transport equation, Ref. [12] calculated that the thermal conductivity of single layer *h*-BN can be higher than 600 Wm$^{-1}$K$^{-1}$ at room temperature, which is one of the highest values in non-carbon-based materials. Ref. [11] measured the mechanical properties of *h*-BN films by nanoindentation, and reported that the elastic modulus of *h*-BN is in the range of 200-500 N/m. And at the same time, *h*-BN is flexible and can be used to fabricate foldable devices [39]. Ref. [11] demonstrated that thin (<10 nm) *h*-BN stacks are highly transparent and can transmit over 99% of the light with wavelengths in the range of 250-900 nm. Furthermore, it has been reported that *h*-BN stacks are chemically very stable up to 1500℃ in air [9], which allowed their use as anti-oxidation coating [9].

However, recent reports demonstrated that local defects can have a dramatic negative effect on the performances of *h*-BN. For example, it has been proved that local defects (i.e. lattice distortions) in *h*-BN sheets can serve as focus for local oxidation due to the presence of dangling bonds, where oxygen can easily bond [40]. This promotes oxygen migration towards the underlying (protected) substrate, which degrades the material below (and also the *h*-BN). Therefore, lattice defects can remarkably shorten the lifetime of *h*-BN, and it is expected that they also contribute negatively to the performance of *h*-BN based electronic devices. In fact, the presence of defects in a dielectric is something in most of cases unwanted from a reliability point of view, independently that it is a *h*-BN stack or a $SiO_2$/high-k film.

## 2.2. Synthesis of 2D layered *h*-BN

The first synthesis of ultra thin 2D *h*-BN stacks was achieved in 2005 via mechanical exfoliation of an *h*-BN crystal [41]. Taking a piece of scotch tape and repeatedly peeling single crystal or bulk *h*-BN materials, atomically thin *h*-BN nanosheets can be easily recognized on the tape using an optical microscope (see Figure 2.2a). This method is based on a mechanical stripping process, i.e. break the weak van der Waals forces between each two adjacent layers, and doesn't involve the introduction of any type of chemical, nor other alien species. Therefore, the obtained *h*-BN nanosheets retain their original (nearly perfect) crystal structure. However, the size of exfoliated *h*-BN flakes becomes smaller and smaller with the number of peelings, resulting in small lateral size of (in the best cases) few micrometers. Furthermore, exfoliating *h*-BN with a specific number of layers is not doable, and very severe thickness inhomogeneities are always present (see Figure 2.2b) [42], which is not acceptable for the industry. The need of (expensive) human labor is also an important drawback.

Liquid phase exfoliation (LPE) is a scalable synthesis process that consists on applying mechanical stresses to an $h$-BN crystal by sonication [42-43], with the assistance of organic reagents. After sonication, centrifugation to remove large size $h$-BN particles is needed. This method allows exfoliating large amounts of $h$-BN flakes simultaneously, and it is scalable (i.e. suitable for the industry). The $h$-BN produced by this method is presented in the form of flakes suspended in a liquid, and the most important properties defining the quality of LPE $h$-BN are the flakes density, their size and thickness. These properties can be tuned by using different sonication parameters (energy, time), centrifugation parameters (revolutions per minute, time), and type of organic reagents. The integration in the devices is normally done by spin coating one/few drops of the solution containing the $h$-BN flakes on the desired substrate and drying it. This methodology is fast and cheap, but it normally produces thick (>50 nm) films containing a rough network of flakes with folds and random orientations [44]. Therefore, LPE $h$-BN may be suitable for some very specific applications (e.g. coatings), but not for others (e.g. ultra thin gate dielectric in FETs).

It's known that large-area $h$-BN sheets can be grown by physical vapor deposition (PVD) methods. Among them, magnetron sputtering has been successfully used to produce monolayer $h$-BN sheets on a 100 nm Ru/$\alpha$-Al$_2$O$_3$ substrate [45]. The sheets produced by this method show decent layered structure with low density of defects, as confirmed by cross sectional transmission electron microscopy (TEM, see Figure 2.2c), which results in a dielectric strength comparable to that of exfoliated sheets. Molecular beam epitaxy (MBE) also allows growing atomically thin $h$-BN stacks on non-crystal substrates [46], as confirmed by a characteristic Raman signature and by the observation of wrinkles. However, Ref. [46] didn't provide cross sectional TEM images, meaning that one cannot be completely sure if the material has a truly layered structure. In any case, the requirements for the sophisticated equipment are still hindering the use of PVD methods to grow *h*-BN. In the case of magnetron sputtering, it requires an ultrahigh-vacuum system (base pressure of $2 \times 10^{-10}$ torr) and under the atmosphere of high-purity Ar/N$_2$ gas mixtures. In the case of MBE the problem is that both high base pressure ($1.0 \times 10^{-10}$ mbar) and high temperature (1850 ℃ ) conditions are needed to form the *h*-BN films.

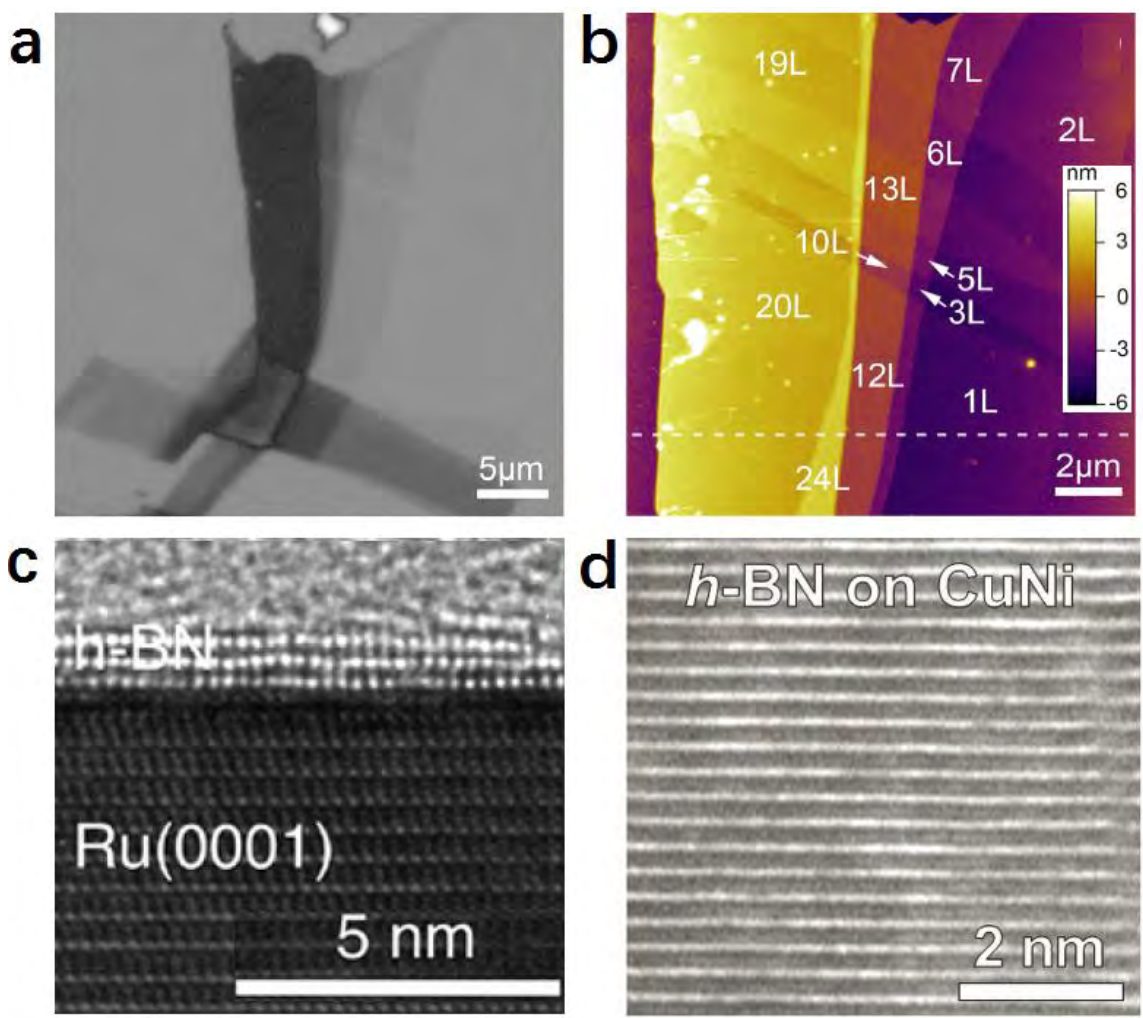

**Figure 2.2**. Characterization of multilayer *h*-BN stacks fabricated following different methodologies. (a) Optical and (b) AFM images of a mechanical exfoliated *h*-BN nanosheets; thickness fluctuations can be seen. Reproduced with permission from Ref. [10], copyright American Chemical Society 2014. Cross-section TEM image of (c) trilayer h-BN grown by magnetron sputtering and (d) CVD-grown multilayer *h*-BN. (c) and (d) are reproduced with permission from Refs. [45] and [18], copyright American Chemical Society 2012, and Wiley-VCH 2017.

When picking up a method to grow scalable *h*-BN with controllable thickness, CVD is the approach that has produced the best results until now [18]. Figure 2.2d shows a cross sectional TEM image of multilayer *h*-BN grown by CVD approach, in which nearly perfect layered structure can be observed. The precursor (borazine, ammonium borane) and carrier gas ($H_2$, Ar) are delivered to the catalytic metallic substrates (such as Cu, Ni, Pt or Fe) under high temperature. The thickness and quality of the *h*-BN stacks can be controlled by tuning the parameters of the CVD process: *i)* carrier gas flow, ii) pressure, *iii)* temperature, *iv)* growth and cooling times, *v)* type of precursor, *vi)* type of substrate, *vii)* type of CVD furnace (cold walled, rapid thermal, plasma assisted). The size of the *h*-BN is only limited by the size of the substrates, which at the same time is limited by the size of the tube furnace. To the best of our knowledge, the largest 2D materials ever produced by this method are 76.2 cm × 76.2 cm, as shown in Ref. [47].

When using CVD, as well as when using PVD, the main problem is the high temperatures used during the growth (> 900 ℃), which make impossible its direct growth on patterned samples (i.e. the high temperatures would destroy any device patterned due to severe diffusion). For this reason, CVD always uses metallic foils as substrate. Ref. [48] reported the growth of *h*-BN on Fe-coated wafers by using a cold walled CVD system (which is 10 times more expensive than a normal one), but no discussion on metal de-wetting and diffusion into the wafer is available. Also, no devices have been reported using this method. In order to avoid this problem, the *h*-BN is grown by CVD on an independent substrate (so far it has been only synthesized on metals) and later it has been transferred to any arbitrary substrate using different methods (e.g. wet transfer [49], dry transfer [50], imprint techniques [51], electrostatic transfer [52], among others). This can be seen as an advantage but also as a problem, as the transfer process may produce cracks, wrinkles and contamination. Interestingly, when fabricating prototype devices aimed to just test the performance of *h*-BN as dielectric, the substrate used during the *h*-BN growth (metallic foil) can be also used as bottom electrode. This strategy does not require the use of a transfer process.

## 2.3. Use of 2D layered *h*-BN as dielectric

When a thin dielectric is placed between two electrodes under polarization the electrical field can generate defects in its microstructure, mainly imperfect bonding (i.e. broken bonds between the atoms that form the dielectric or atoms that penetrate from adjacent electrodes) [53]. If the stress is enough aggressive (high voltage or long time), the density of defects increases prohibitively until forming one/few effective percolation path across the dielectric, leading to the complete loss of the insulating properties, and the circulation of very high currents across it (namely BD) [54]. The currents increase the local temperature, which further increases the current in a self-accelerated manner. Therefore, the BD process strongly depends on several properties of the dielectric material, such as: *i)* density of native bulk defects, *ii)* number of defects at the interfaces with the electrodes, *iii)* chemical stability, *iv)* mechanical stability, and *v)* thermal conductivity [55]. As an example, the BD process in high-k dielectrics show slightly differences compared to $SiO_2$, as the density of defects in high-k materials is much higher, which accentuates charge trapping and de-trapping, producing the observation of random telegraph noise (RTN) signals [56].

2D layered *h*-BN is an insulating material with an energy band gap of ~5.9 eV [8] (measured in exfoliated samples) and a dielectric constant between 2 and 4 [57] (measured in CVD grown samples). Therefore, *h*-BN may be suitable for being used as dielectric in electronic devices. In fact, it is expected that *h*-BN shows an excellent performance as dielectric, because it holds many wanted properties that are relevant during the degradation and BD of a dielectric, such as chemical stability, high mechanical stability, and high thermal conductivity (see also section 2.1). In this context, the strong covalent bonds of *h*-BN may slow down the speed for defect generation, the isolation between planes may difficult the propagation of the defects during the electrical stress, and the high thermal conductivity may avoid the formation of hot spots (which also slows down the BD process). Furthermore, *h*-BN adheres to the adjacent electrodes by van der Waals forces, minimizing the formation of interface defects. In this regard, *h*-BN is very promising for the interaction with graphene and $MoS_2$, two materials that form very bad interfaces with $SiO_2$ and high-k dielectrics [6-7].

Initially, 2D layered *h*-BN was used as substrate to enhance the carrier mobility of graphene FETs [15]. Due to the atomically flat surface of 2D layered *h*-BN (which is free of dangling bonds), the mobility of the carriers in the graphene channel was improved one order of magnitude compared to FETs using traditional $SiO_2$ as substrate. It's also known that the use of *h*-BN in high-mobility graphene devices can enhance the energy gap in multi-terminal measurements of fractional quantum Hall effect [58-59]. Before the starting date of this thesis very few works reported the use of *h*-BN as dielectric. Despite *h*-BN has been used as dielectric in FETs [60-61], most of the reports focus on the properties of channels and/or transistors, not on the performance of the *h*-BN dielectric itself, i.e. direct tunneling current, trap-assisted tunneling, tunneling current homogenity, current across defects, RTN, SILC, and soft/hard BD. So far, the studies on *h*-BN as dielectric have been performed in terms of nanoscale homogeneity and variability, reliability and dielectric breakdown, mainly via CAFM. For example, Ref. [62] studied mechanical exfoliated *h*-BN with different thicknesses using CAFM and reported that the tunneling current across it is extremely homogeneous (see Figure 2.3a). Ref. [62] showed that increasing the thickness of *h*-BN in one layer reduces the current in a factor 50 (see Figure 2.3b), and Ref. [62] claimed that in monolayer, bilayer and trilayer *h*-BN the tunneling current flows by Direct tunneling at low fields and by Fowler Nordheim tunneling at high fields, and that in thicker *h*-BN stacks the current always flows by Fowler Nordheim tunneling [62]. Ref. [63] suggested that exfoliated stacks of *h*-BN reach the BD layer-by-layer, producing dramatic physical removal of material after the BD of each layer (see Figure 2.3c). In that article the authors observed the formation of a hole with increasing depth after sequences of current vs. voltage (I-V) curves.

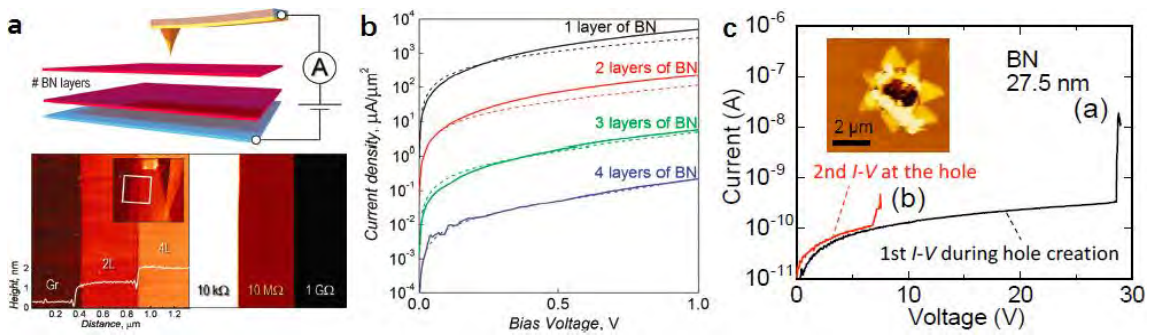

**Figure 2.3.** (a) Schematic of a multilayer *h*-BN stack characterized by CAFM, and topographic (bottom left) and current (bottom right) maps. (b) Both experimental (solid lines) and fitting (dashed lines) I-V curves in log scale for graphite/BN/graphite devices with different thickness of BN insulating layer. Reproduced with permission from Ref. [62], copyright American Chemical Society 2012. (c) I-V curves for a fresh *h*-BN flake (black) and the remaining layers (red) inside the hole. Insert: AFM image of a BD spot. Reproduced with permission from Ref. [63], copyright IEEE 2016.

After that, *h*-BN produced by CVD approach started to be studied as well. In parallel with the development of this thesis, Ref. [64] observed that the tunneling current across CVD-grown *h*-BN shows multilayer insulating islands (Figure 2.4a), which correlate with multilayer areas in the scanning electron microscopy (SEM) images. The CAFM was also able to detect wrinkles in CVD-grown *h*-BN, which manifested as insulating long and straight lines. Also in that work it was demonstrated that monolayer (0.33 nm) $h$-BN resists the electrical stresses much better than six times thicker $HfO_2$ (2 nm) [64]. However, the knowledge available about the electrical properties of CVD-grown $h$-BN stacks before this thesis was very limited. In **Article 1** a deep analysis about the use of $h$-BN as dielectric is presented.

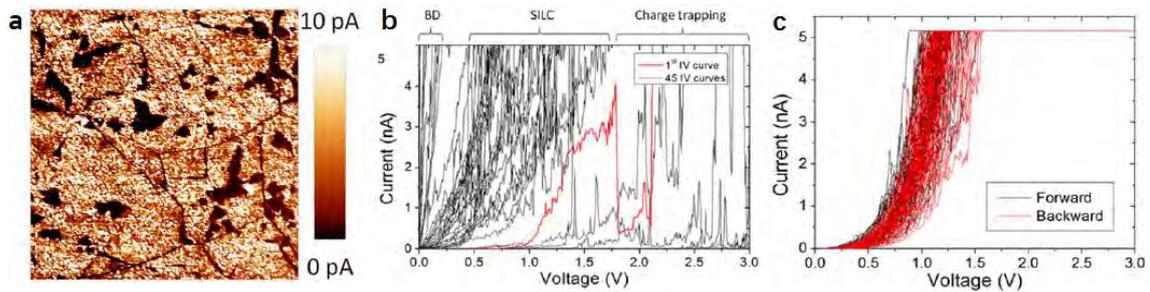

**Figure 2.4.** (a) 10 μm × 10 μm CAFM current map collected on the BN/Cu stack by applying 1V. Sequence of I-V curves collected on a single spot on BN/Cu stack (b) and 2 nm thick $HfO_2$ layer (c). Reproduced with permission from Ref. [64], copyright AIP Publishing LLC 2016.

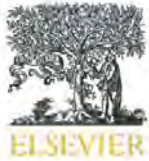
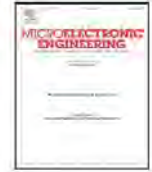
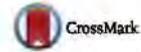

Review article

# On the use of two dimensional hexagonal boron nitride as dielectric

Fei Hui, Chengbin Pan, Yuanyuan Shi, Yanfeng Ji, Enric Grustan-Gutierrez, Mario Lanza *

*Institute of Functional Nano & Soft Materials, Collaborative Innovation Center for Nano Science and Technology, Soochow University, 199 Ren-Ai Road, Suzhou 215123, China*

A B S T R A C T

Recent advances in materials science allowed the incorporation of advanced two dimensional (2D) materials in electronic devices. For example, field effect transistors (FETs) using graphene channels have shown unprecedented carrier mobility at room temperature, which is further complemented by its intrinsic flexibility, transparency, chemical stability and even thermal heat dissipation. Other 2D materials such as transition metal dichalcogenides (TMDs) can provide additional functionalities to the devices, such as band gap induced high ON/OFF ratios in FETs. Interestingly, these 2D metallic (graphene) and 2D semiconducting materials (2D/TMDs) have been mainly implemented in devices using traditional three dimensional (3D) insulators, such as HfO₂, Al₂O₃ and SiO₂, which may not be the best solution given the complex and defective interface bonding. For this reason recently 2D insulators have been started to be used as dielectric in different electronic devices, showing interesting phenomena. A 2D insulator differs from traditional 3D insulators in that it holds a layered structure, in which the bonding in plane is covalent while the plane-to-plane interaction is governed by van der Waals interactions. This genuine structure has been demonstrated to remarkably alter some reliability phenomena like, for example, the entire dielectric breakdown process. In this review, we analyze the performance of 2D layered dielectrics, focusing on hexagonal boron nitride. Different synthesis methods, electrical characterization, reliability and variability analyses, as well as dielectric breakdown process are discussed. Moreover, it should be highlighted that, in many device applications (like capacitors or resistive switching memories), 2D dielectrics may not require the annoying transfer process usually required for graphene and 2D/TMDs, which further facilitates its introduction in the industry.

© 2016 Elsevier B.V. All rights reserved.

## Contents

## 1. Introduction

Dielectrics are key elements in a wide range of electronic devices essential for our society, as they are necessary to provide some of the required functionalities. For example, the omnipresent field effect transistors (FETs), the device that has provoked an unprecedented social progress in human history and that only in 2016 is expected to raise a market of billions of US dollars [1], requires a dielectric to generate the capacitance effect necessary to modulate the conductivity at the channel region [2]. Another example is the wide family of resistive switching non-volatile memories (NVM), which induce a reversible dielectric breakdown in a dielectric to modulate its conductivity, leading to two logic states that can be used to store information [3]. Therefore, a continuous revision of the status and prospects of dielectrics in microelectronics is necessary.

* Corresponding author.
  *E-mail address:* mlanza@suda.edu.cn (M. Lanza).

Traditionally, the dielectric used in most electronic devices was silicon dioxide ($SiO_2$, see Fig. 1a [4]), as it shows an excellent performance as insulator, good compatibility with silicon, low cost and easy fabrication [5]. Around 2005, in the 65 nm technological node, the aggressive scaling down of the transistors made that $SiO_2$ dielectrics became too thin to withstand the electrical fields to which they were subjected [6], raising reliability concerns due to prohibitive leakage currents. For this reason, in successive nodes FETs incorporated high-k materials as gate dielectric (Fig. 1b), as they can provide similar capacitance effect using a larger thickness, which remarkably decreases the leakage current [7–10]. In 2008, in the 45 nm node, the traditional $SiO_2$ gate insulator started to be replaced by thin films of high-k materials, such as $HfO_2$ or $Al_2O_3$ [11–13]. Nevertheless, replacing $SiO_2$, which only has one disadvantage (its low dielectric constant), by high-k dielectrics was a painful process as, in principle, these materials only have the advantage of reducing the leakage current. Among all problems related to the introduction of high-k materials in microelectronic devices are [14–16]: i) high density of native defects, which in FETs can produce instabilities on the threshold voltage; ii) chemical interaction with the polysilicon gate; iii) interaction with the silicon substrate, which produces the apparition of interfacial $SiO_2$ layers, leading to an increase of the equivalent oxide thickness (EOT); iv) phonon scattering at the channel region, which degrades the carriers' mobility; and v) polycrystallization, which increases the inhomogeneity in the material, leading to hot spots and to an overall larger variability. Despite the initial concerns, the industry has been able to adapt to this change, for example by replacing the polysilicon gates by metallic ones, and building controlled high-k/$SiO_2$ superstructures that show low degrees of scattering [6]. It should be noted that replacing materials from the building blocks of the FETs and other electronic devices is in fact a raising parallel methodology to fit the every time more exigent technology requirements. Even the omnipresent silicon core bulk of different devices has been in many cases replaced by other semiconductors (called III-V) with higher carriers' mobility [17–18]. In this context, the apparition of novel two dimensional (2D) materials with superior properties and their application in microelectronics suggests now a similar technological transition (Fig. 1c) [19–21], in which device parameters and materials interfaces will need to be adjusted to build up prototypes with realistic probabilities of being commercialized. Therefore, here arise the question: which is the most suitable dielectric for 2D devices?

The use of advanced 2D materials in electronic devices can be interesting to solve some fundamental limitations, as well as to provide additional thermal, mechanical and optical capabilities (among others). For example, the use of graphene is especially attractive in microelectronic circuits due to its extremely high room-temperature carriers' mobility [22–23] and high saturation velocity [24]. During the explosion of graphene research in 2009–2010 IBM reported FETs with mobilities up to 400 $V^{-1} s^{-1}$ and cut-off frequencies of $f_T = 26$ GHz [25] using standard $SiO_2$ substrates and high-k gate dielectrics, and this value was optimized up to 155 GHz using diamond-like carbon substrates [26]. These values are higher than those of the best silicon based metal-oxide-semiconductor FETs (MOSFETs) with similar gate lengths [27–28]. Using 2D semiconductors from the family of transition metal dichalcogenides (2D/TMDs), including as $MoS_2$ [21,29], $WS_2$ [30], $TiS_2$, $TaS_2$, $MoSe_2$ and $WSe_2$ [31], high carriers' mobility can be almost maintained while magnificently enhancing the current on/off ratio. B. Radisavljevic et al. [21] reported a room-temperature single-layer $MoS_2$ transistor with a mobility of at least 200 $cm^2 V^{-1} s^{-1}$ and on/off ratio exceeding $10^8$. Although later Fuhrer and co-workers demonstrated that this value was overestimated [32–33], there seems to be a consensus that at 240 Kelvin carrier mobilities of 63 $cm^2/V$ s can be achieved [34]. Moreover, the atomically thinness of 2D materials provides an excellent electrostatic control over the channel in FETs [35], especially in short-channel transistors. Quantum confinement, substrate independence and the possibility of building heterostructures are also among their most attractive properties applied to electronic devices [35]. From a mechanical point of view, it has been demonstrated that $MoS_2$ can be combined with graphene to build up high mechanical flexibility, optical transmittance (~74%), and current on/off ratios (>$10^4$) with an average field effect mobility of ~4.7 $cm^2 V^{-1} s^{-1}$ [34]. For these reasons, the introduction of 2D materials in electronic devices has become a global trend, not only in the academia but also in the industry; for example, among all graphene patents registered until June 2014, around 28% have been devoted to build graphene electronic circuits [36], the 1 billion Euro Graphene Flagship Project of the European Community is focused on transferring graphene and related materials from lab to industry, and companies like Samsung and IBM have been very active in 2D materials research.

Interestingly, most of the 2D electronic devices reported until today have used three dimensional (3D) dielectrics like $SiO_2$ and high-k materials ($HfO_2$ and $Al_2O_3$) to enable the different functionalities [25–27,37–38]. Here the concept of 3D refers to the fact that the bonding in these dielectrics is the same in all directions (vertical and horizontal); therefore, despite 3D dielectrics can be atomically thin, they still hold a 3D structure. Despite the successful performances observed in many devices, the combination of 2D metallic and semiconducting materials with 3D dielectrics is complex, as they do not form covalent bonding. For graphene/high-k stacks it is widely known that interface functionalization is necessary for achieving uniform adhesion, being $NO_2$ [39], metal seed layer [40], organic seed layers [41], and ozone ($O_3$) [39–41] the most common strategies. In the case of 2D/TMDs, publications displaying uniform high-k gate dielectrics in FETs with $MoS_2$ channels [42] suggested that high-k nucleation on the surface of 2D/TMDs may be different than on graphene. This hypothesis was later discarded by R. M. Wallace et al. who demonstrated that no covalent bonding is possible between $HfO_2$ and $MoS_2$ [43]. For that reason, some groups used an interfacial layer between high-k and 2D/TMDs to ensure decent adhesion [44]. However, this methodology can generate other electronic concerns, such as charged impurities and dopants that increase the roughness of the 2D material, leading to larger Coulomb scattering that reduces the charge transport and produces hysteretic effects [45–46]. Therefore, finding new dielectrics compatible with graphene and 2D/TMDs is a major requirement for the development of 2D electronic devices.

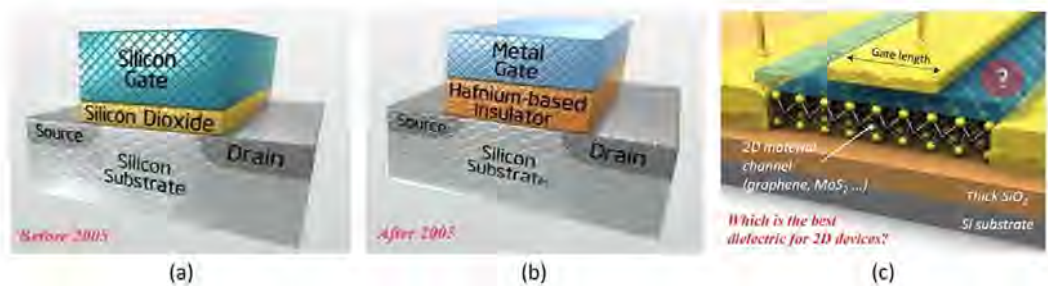

Fig. 1. Evolution of dielectrics in field effect transistors. (a) and (b) show scientific illustrations of the structure of two MOSFET transistors using polysilicon/$SiO_2$/Si and metal/high-k/Si structures, respectively. These architectures correspond to those use in the 65 and 45 nm technological nodes, respectively. (c) Schematic of a field effect transistor using a two dimensional material as channel. Modified and reprinted with permission from [4,21].
(Copyright from Nature Publishing Group 2011.)

In this context, 2D layered dielectrics could be an excellent solution due to their demonstrated excellent compatibility with graphene and other 2D materials [47]. We use the term 2D layered insulator to refer to those insulators made by stacked planes in which the bonding is covalent, while the plane-to-plane interaction is by van der Waals attraction. Unfortunately, the use of 2D insulators as dielectric stack is still at its embryonic stage, and the amount of literature available is scarce, being hexagonal boron nitride (*h*-BN) the only material that has been considerably studied. In brief, *h*-BN is an $sp^2$-hybridized 2D insulator analogue to graphene, in which boron and nitrogen atoms occupy the A and B sublattices of the hexagonal Bernal structure (Fig. 2) [48]. Due to the similar lattice constants of graphene and *h*-BN, which only have a mismatch of 1.7% [49], these two materials have shown a very good interaction and interesting potential for device fabrication. It is the aim of this review paper to compile the knowledge available about the use of *h*-BN as dielectric stack for microelectronic devices.

## 2. Boron nitride production

Most of the methods currently available to produce atomically thin *h*-BN films have been inherit from the research of other 2D materials, especially from graphene. The recent *Science and technology roadmap for graphene, related two-dimensional crystals, and hybrid systems* [36], describes >10 different methods to produce graphene and other 2D materials, including mechanical exfoliation (repeated peeling), liquid phase exfoliation, photoexfoliation, anodic bonding, molecular beam epitaxy, atomic layer epitaxy, heat-driven conversion of carbon, chemical synthesis, growth on silicon carbide, growth on metals by precipitation, physical vapor deposition (PVD) and chemical vapor deposition (CVD). Despite not all these methods can be extrapolated to *h*-BN research, some of them are modifiable to obtain this 2D layered insulator. When using *h*-BN as dielectric, the most exploited synthesis methodologies have been mechanical exfoliation, physical vapor deposition and chemical vapor deposition; for this reason, here we will focus on these three methods (see Fig. 3). Additional information about other methodologies to produce *h*-BN can be found in other fabrication-oriented reviews [50–51].

### 2.1. Exfoliation: mechanical, liquid phase and chemical

As it happened for graphene [52], the first reports on two dimensional *h*-BN used mechanical exfoliation (repeated peeling) on single crystal *h*-BN using Scotch tape [53–55], leading to nanosheets with a lateral size of hundreds of nanometers. Thanks to the layered structure of the boron nitride crystal and the weak Van der Waals' force between the layers, the *h*-BN thin films yielded by this method normally have perfect crystal structure [56]. Therefore, some intrinsic properties, such as the bending modulus [57], luminousness [58], and carrier mobility [59] could be protected to the greatest extent. The most successful source of single crystal *h*-BN has been found by the group of Y. Kubota et al. [60] whose *h*-BN has been used in a wide range of electronic devices all around the world [61–62]. Nevertheless, despite this methodology provides the highest quality *h*-BN ever reported [63–64], its use is still reduced to research purposes, as it is not a scalable technique due to the small size and thickness fluctuations of the nanosheets. Besides, atomically thin nanosheets of *h*-BN can be also obtained by exfoliating powdered raw material instead of single crystalline *h*-BN [46], or even by rubbing bulk crystals against another clean solid surface [65]. In order to relatively increase the scalability of this method, mechanical peeling by ball milling was presented by L. Li et al. [64] as a good way to exfoliate thousands of *h*-BN particles simultaneously. However, the undesirable impurities and the damage to the original crystal structure could limit its development. With the aim of enhancing the scalability and at the same time guarantee the purity of the material, liquid phase exfoliation seems to be the most suitable method due to its fast, cheaper, and energy-saving properties. Generally, this method has three steps: *i*) sonication for dispersing and exfoliating the *h*-BN powder, *ii*) centrifugation for removing the large size *h*-BN particles, and *iii*) drying the supernatant to obtain pure *h*-BN thin films. Additionally, one chemical exfoliation method was carried out by X. Li et al. [66] using the assistance of molten hydroxides at low temperatures. Nevertheless, wet exfoliation methods are not compatible with the microelectronics

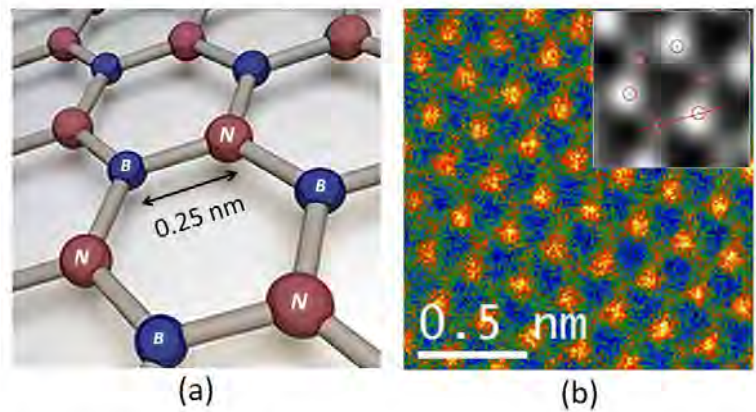

**Fig. 2.** (a) Schematic of the hexagonal lattice of boron nitride. (b) High angle annular dark field scanning transmission electron microscopy (HAADF-STEM) image proving the chemical composition of *h*-BN with sub-atomic resolution.
(Modified and reprinted with permission from [48]. Copyright from InTech 2013.)

# Article 1

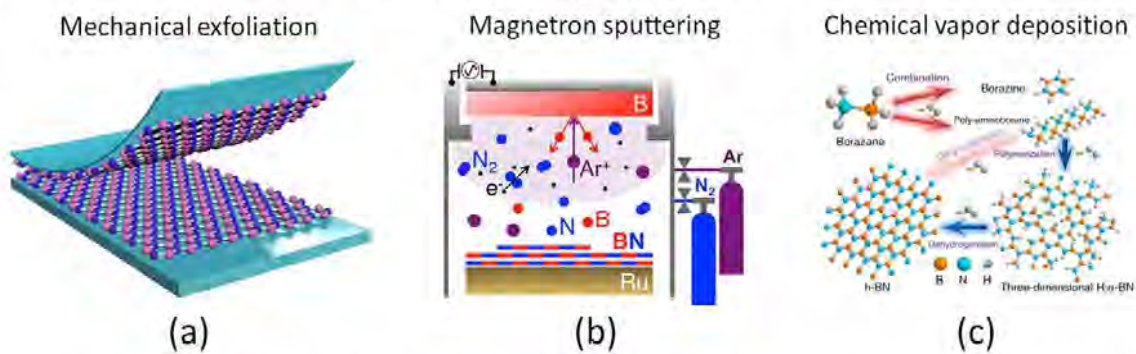

**Fig. 3.** Schematic of various methods for producing the atomically thin *h*-BN films. Modified and reprinted with permission from [1,73,93]. (Copyrights from American Chemical Society 2013, and Nature Publishing Group, 2015.)

industry, and they may be more suitable for other technologies, such as printed electronics [67].

## 2.2. Physical vapor deposition

Physical vapor deposition techniques can be also used to grow thin boron nitride layers. Within PVD family, magnetron sputtering has been the most widely used P. Sutter et al. [68] successfully synthesized monocrystalline and high-quality wafer-scale boron nitride films by magnetron sputtering technique on a 100 nm Ru/α-Al$_2$O$_3$ substrate. During the preparation, a reactive chamber with a boron target (2″ dia., 99.5%, K.J. Lesker) was firstly pumped to $2 \times 10^{-10}$ Torr, and then filled with high-purity Ar/N$_2$ gas mixtures (20% N$_2$) in a total pressure of $10^{-2}$ Torr. The reactive radio frequency power of the sputtering process was tuned to 10 W. After that, the *h*-BN films were grown at high-temperature by reactive deposition. It is also noteworthy that the crystal form of the thin *h*-BN films strongly depends on the substrate bias, indicating that the crystal form is controllable during sputtering process. For instance, J. M. Caicedo et al. [69] found that only *h*-BN can exist when the voltage is less than −150 V d.c. biased and −90 V r.f. biased. However, with the enhancement of the negative bias, *c*-BN also showed up as a part of the boron nitride films. Furthermore, during this fabrication process, there is almost no requirement for the substrate, making it suitable for the industry because no etching and transfer step got involved. In addition, this catalyst-free method avoids replicating the metal grain boundaries into the product, which may efficiently decrease the possibility of current leakage [70]. More interestingly, using this method some specific atoms can be included into the *h*-BN nanofilm on purpose for other particular applications. For instance, M. Goto et al. [71] successfully synthesized hexagonal boron nitride thin films doped with Cu element by co-sputtering deposition procedure on a stainless steel pad. Furthermore, by adjusting the sputter discharge conditions, the domain size of the *h*-BN crystals deposited on the substrate could be well controlled. Although PVD techniques are already a very good way to produce high quality boron nitride nanosheets in a large scale, the requirement for the sophisticated equipment still restricts its development.

More recently, molecular beam epitaxy (MBE) has been proposed as a promising way to produce atomically thin *h*-BN stacks with high-crystalline quality. The main benefit of this technique is that MBE allows direct growth on non-catalytic substrates. For instance, S. Nakhaie et al. [72] successfully synthesized *h*-BN thin films on polycrystalline Ni foils (Alfa Aesar, 99.994% pure, 100 μm thick) through MBE method. In their experiment, the Ni foil with a size of 1 cm² was firstly cleaned and then its surface was modified by ultra high vacuum annealing and Ar sputtering processes. The anneal process was taken place at 1000 °C for 30 min and the sputtering step was carried out under an accelerating voltage of 2 kV for 20 min. The pressure and temperature set in the working chamber for sputtering process was $10^{-4}$ mbar and 600 °C respectively. After the preparation, the main growth process was executed on the substrate with the temperature ranging from 730 to 835 °C for 3–5 h under a pressure of $1.1 \times 10^{-5}$ mbar. Elemental B was generated at 1850 °C with the assistance of a high-temperature effusion cell. Meanwhile, the N-species were activated by an RF plasma source working at 350 W with 0.2 sccm of N$_2$ flow. During this period the isolated *h*-BN flakes connect to each other to form a continuous *h*-BN thin film. Additionally, MBE also has been used in the field of 2D heterostructures fabrication by in situ growing hexagonal boron nitride accurately on transition metal dichalcogenides, which allow tuning the desired band alignments of this novel structure [73].

## 2.3. Chemical vapor deposition

In order to provide an accessible methodology to grow *h*-BN, chemical vapor deposition family techniques have been extensively explored for its ability to control the lateral size, number of layers, as well as the crystalline structure precisely. Different precursors can be used for the synthesis of *h*-BN nano-films by CVD. Normally, they are based on two independent substances containing boron and nitrogen atoms respectively, such as BCl$_3$/NH$_3$ [74–75], BF$_3$/NH$_3$ [76–77], B$_2$H$_6$/NH$_3$ [78–79], and B$_{10}$H$_{14}$/NH$_3$ [80]. However, using two kinds of precursors complicates the experiment procedure and equipment assembly. Moreover, in order to guarantee adequate and equal reactants during the chemical reaction for *h*-BN growth, it's very critical to control the ratio of these two precursors. For this reason, using K. K. Kim et al. [70] used borazine, which contains both boron and nitrogen atoms, to synthesize polycrystalline *h*-BN film with the assistance of Cu foil. Since borazine is in liquid state at room temperature, a bubbler was installed with hydrogen or nitrogen as carrier gas. Prior the growth process, a pre-annealing step at 1000 °C for 30 min under 10 sccm hydrogen atmosphere was necessary to remove possible copper oxides, increase the copper grain size, as well as smoothening the surface. After that, the material synthesis was carried out at 750 °C with 1–3 sccm of borazine and 2000 sccm of hydrogen for around 5–30 min. Finally, the *h*-BN/Cu stack was post-annealed at 1000 °C for 1 h to improve the crystallinity of the *h*-BN thin films. Although borazine has already been able to replace two kinds of independent precursors for the growth of *h*-BN, its liquid state is inevitably setting barriers for its wide utilization. For this reason, a solid precursor called ammonia borane was used by L. Song et al. [81] because of its stability and accessibility. As it happens for the borazine, when using ammonia borane pre-annealing is still an unavoidable step. The main difference is that instead of installing a bubbler system, only a heating belt is needed to decompose the ammonia borane into hydrogen, borazine and aminoborane. Then these products were pushed into the reactive region by H$_2$/Ar gas flow at the temperature of 1000 °C for 30–60 min. Later, the furnace was quickly cooled down and 2–5 layers thick h-BN was obtained. Another valuable addition for this solid precursor is that no high toxicants are generated during the reaction process.

Besides the influence of different precursors, the catalyst substrate also has a very important impact on the growth of h-BN. Until now, many kinds of materials have served as substrate for CVD growth of h-BN, including Ni [82–85], Cu [86–92], Pt [93], SiO₂ [94] and Al₂O₃ [95]. Because Ni and Cu are cheap, accessible and show good catalytic effect, they have become the most widespread substrate materials for CVD-growth of h-BN. Moreover, the relatively small lattice mismatch between h-BN and the 111 faces of Ni or Cu greatly improves the thickness uniformity of boron nitride thin films [80]. Y. M. Shi et al. [82] successfully synthesized hexagonal boron nitride thin films through CVD method on polycrystalline Ni films under ambient pressure. In their experiment, the flat h-BN film grew continuously on the entire Ni foil and the lateral size of the boron nitride film reach around 20 μm, which is actually only limited by the grain size of the Ni foil. In addition, the thickness of the yielded h-BN thin films can be tuned from around 5 nm to around 50 nm by varying the growth parameters. Furthermore, except the post-annealing process that needs great heat (1000 °C) to increase the product's quality, the growth temperature can be dramatically reduced to 400 °C, which greatly reduces the energy consumption. If during the growth process, the temperature is increased up 700 °C under high vacuum condition, a strong reduction atmosphere upon the first boron nitride layer surface would be created to inhibit further decomposition of borazine, easing the production of monolayer h-BN thin films. On the other hand, Y. H. Lee et al. [83] found that the Ni crystal orientations would also greatly affect the CVD growth process of h-BN. It was determined that the h-BN films always grow faster on the surface of Ni(100)-like crystals than on that of Ni(111)-like facets.

Cu foil was also used as the catalyst substrate by K. K. Kim et al. [86] to synthesize monolayer h-BN at low pressure. As when using Ni foils, the morphology of the copper surface is also a crucial element determining the quality of the h-BN film, as it influences the nucleus density and their locations during their formation process. For example, in unpolished copper foils the nucleus tend to form along the copper rolling lines or on surface hillocks related to impurities. On the contrary, in polished foils the nucleation is distributed more homogeneously, and the h-BN islands normally have larger average size. Later, this conclusion was further confirmed by Kang Hyuck Lee [87], who observed that the impurity particles and the amount of allotropes (c-BN) strongly depend on whether the Cu foil was treated by a chemical polishing and thermal annealing. Additionally, the crystallinity of the CVD-grown h-BN is also very closely related to this preparation step [88]. As a goal to further reduce the impurities and improve the flatness of the boron nitride thin films, a filter system was introduced at the entrance of the growth region to effectively block the h-BN nanoparticles [89].

Although both Ni and Cu foils are suitable to be the catalyst substrate for the growth of h-BN thin films, there are still some differences between these two kinds of metal catalysts. First of all, the chemical bonding strength is different at the interface of h-BN/Ni and h-BN/Cu [94]. In most cases it is the strength of the transition metal 3d–h-BN π orbital hybridization that plays a leading role in the chemical bonding strength at the interface. However, compared to weak interfacial reciprocity on h-BN/Cu(111), the one at the Ni(111)/h-BN interface is much stronger, indicating that less wrinkles and hillocks would be created on the Ni substrate. Besides, catalytic speed also varies when using different catalyst. For instance, S. Chatterjee et al. [80] synthesized thin h-BN thin films using the Cu and Ni foils under exactly same conditions by CVD method. As expected, the thickness of the h-BN film on the Ni and Cu foils was 2 nm and 2–15 nm Furthermore, after the thermal pre-annealing process for increasing the Cu or Ni grain size, the grooves in the Ni foil are normally much deeper than the ones in the Cu foil, resulting in a bad quality of the h-BN grown on the grain boundaries. As a result, many authors preferred to use h-BN film yielded on Cu substrate for the application of the electronic devices [70,83–92].

In order to combine the advantages of both Cu and Ni catalyst substrate, recently, a specially designed Cu-Ni alloy was utilized by Xiaoming Xie's group [95] for CVD synthesis of large-scale single crystal h-BN. Surprisingly, during the growth process, the nucleation density greatly reduced and the grain size of the h-BN single crystal spectacularly increased up to 7500 μm², which is almost two orders of magnitude larger than the one reported by other groups before [86,88]. More interestingly, it has been clearly proved by a series of SEM images that the introduction of Ni atoms will greatly decrease the nanoparticles deposited on the grain boundaries of the h-BN films and the surface of the catalyst substrate. Furthermore, the inserted Ni could also help to enhance the poly-aminoborane decomposition by promoting desorption process and the formation of Ni—B and Ni—N bonds [93].

Aiming to get rid of the damage and contamination during the etch and transfer process for electronic devices fabrication, recently, Roland Y. J. Tay et al. [92] successfully synthesized few to multilayer h-BN on SiO₂ wafers directly without the assistance of the metal catalyst through CVD method. During their growth procedure, the temperature in the reaction region was increased to 1000 °C with 500 sccm of Ar and 20 sccm of H₂ at a total pressure of 1.1 Torr. In this condition, due to dehydrogenation, the borazine inside will convert into (BNH_y)₃ molecules which is able to deposit at any substrate. Moreover, as the whole growth process can be carried out under ambient pressure, no self-limiting phenomenon needed to be taken into consideration, allowing to control the film thickness by simply tuning the growth time. Nevertheless, the grain size of the films grown on this substrate is still too small (~25 nm), which may introduce large amounts of defects at the grain boundaries. Therefore, despite this is an interesting approach, the quality of the h-BN on insulating surfaces, as that of graphene, needs to be highly improved to represent a feasible solution.

## 3. Device fabrication

First of all, we would like to emphasize that the use of any 2D material in microelectronic devices usually brings associated many concerns about the compatibility of fabrication processes. One of the most common criticisms is the scalability of the prototypes. While it is true that mechanical exfoliation and electron beam lithography are techniques that don't show realistic potential for mass production of electronic devices, other scalable techniques can be also used. As mentioned above, CVD process allows the fabrication of wafer scale 2D materials with decent quality, and on these large sheets the devices can be easily patterned using conventional photolithography.

From the point of view of device fabrication, there are two processes characteristic of 2D materials that should be highlighted. The first one is the need of 2D material transfer. This is usually the step that scares most device engineers, and strongly difficult the fabrication of competitive devices using 2D materials. The main reasons are: i) the complexity of the whole process (specially the lift off step), ii) the easy formation of cracks related to the mechanical strains derived from the transfer, and iii) the difficulty to remove impurities from the polymer scaffold. Generally, there are two groups of transfer methods, namely dry and wet transfer. Normally, the dry transfer technique is applied to exfoliated h-BN thin films, while wet transfer has been typically more used in CVD-grown 2D materials.

In the procedure of the dry transfer process [46], the h-BN flake is firstly exfoliated from a h-BN crystal by scotch tape. Then a polydimethylsiloxane (PDMS) stamp can be used as the medium for transferring the h-BN nanosheet, by pressing it with and quickly peeling it off from the h-BN/scotch tape. At last, the h-BN layer attached to the PDMS was able to transfer to any target substrate with the assistance of a micromanipulator. The advantage of dry transfer technique its lower polymer contamination compared to wet etching transfer. On the other hand, wet etching transfer has also drawn lots of attention to transfer large area 2D sheets, which are usually grown by CVD, which greatly extends its potential for wafer-scale fabrication of electronic devices. Inspired

124      *F. Hui et al. / Microelectronic Engineering 163 (2016) 119–133*

from the advances made in graphene research [96] the transfer of *h*-BN has been rapidly achieved [82]. After CVD synthesis, polymethylmethacrylate (PMMA) can be spin-coated as a thin (~100 nm) supporting layer. Then the underlying catalytic substrate on which the *h*-BN was grown can be etched using a wide range of solutions, and the PMMA/*h*-BN block can be released. The next step consists of picking up the PMMA/*h*-BN block with the desired substrate, followed by removal of the polymer scaffold using an acetone bath (other polymer media should be removed with the suitable etchant).

Nevertheless, it should be highlighted that the transfer process was initially developed to provide an insulating substrate for graphene sheets. While the graphene-on-metals is not very useful to fabricate electronic devices, graphene-on-insulators are very attractive because the current can be confined along the graphene sheet. It should be highlighted that this situation is not given in *h*-BN research, as it is an insulator. In many devices like capacitors and resistive random access memories, the *h*-BN doesn't need to be transferred, which is a very important advantage compared to graphene. For example, P.S. Lee fabricated a transfer-free metal/BN/metal resistive memory by directly evaporating electrodes on CVD-grown BN, in which the metal substrate played the role of bottom electrode [97]. Despite the BN films shows to be amorphous rather than layered (and therefore it cannot be considered a 2D device), this work points out the ability of BN to create electronic devices without the assistance of a transfer process.

The second critical step towards device fabrication is the pattern of the *h*-BN sheet. In 2015, J.I. Wang reported that *h*-BN layer can be effectively removed by reactive ion etching in the rate of ~1.4 nm/s with the main etching gases mixed of CF₄, CHF₃, and H₂, which greatly boosted the development of nano-electronic devices based on the h-BN films [98]. In our investigation, the exfoliated *h*-BN film on the SiO₂ wafer was firstly patterned by e-beam lithography in order to define the etching region; then, the exposed area was etched as above mentioned; and finally the polymer mask remaining on the sample was removed by immersing the whole device in the acetone. The last step was to further clean the sample by thermal annealing in the Ar/H₂ atmosphere at 350 °C for 3 h. It should be highlighted that photolithography instead electron beam lithography could be also used, as the only characteristic process is how to etch the *h*-BN (for example to open electrical contacts to underlying materials, i.e. conductive channel). Therefore, using this standard methodology, electronic devices with complex structures can be effectively fabricated.

Finally, it should be highlighted that *h*-BN provides a superb chemical and thermal stability that avoids interaction with adjacent layers, which was a strong reliability problem in devices using high-k dielectrics. Therefore, from this point of view, the combination of *h*-BN with other materials should not be a concern. If the adjacent layers (i.e. substrates, electrodes) are made of other 2D materials, the interaction will take place by van der Waals forces, which will provide a very good interaction. On the contrary, if the adjacent material is a 3D (non-layered) material, interface defects may appear (as commented in the introduction section). In this case, buffer layers may be used when necessary.

## 4. Introduction of *h*-BN in electronic devices

Unlike graphene, *h*-BN is an insulator with a band gap of ~6 eV [81, 99] and a dielectric constant that ranges from 2 to 4 [70], two properties that enable its use as dielectric in logic devices. Compared to traditional dielectrics (like SiO₂, HfO₂, TiO₂ and Al₂O₃). *h*-BN shows advantages in many different applications. First, *h*-BN with uniform thickness and atomically flat surface free of dangling bonds can be easily obtained by different methods (see Section 2), which could effectively reduce the scattering effects in, for example, FETs [100–101]. Second, its extraordinary chemical stability (which can even overcome that of graphene [102–103]) avoids unwanted reactions with adjacent layers. Third, the high thermal conductivity of *h*-BN, which is about 20 times larger than that of SiO₂, can improve the heat dissipation within the device, which will dramatically enhance its lifetime [104]. Fourth, as other two dimensional materials, *h*-BN is flexible, mechanically stable and transparent, which makes it a strong candidate for the fabrication of flexible optoelectronic devices. And fifth, *h*-BN has shown excellent combination with other 2D materials. It should be also highlighted that, despite *h*-BN holds the same dielectric constant than SiO₂, its larger chemical stability may remarkably slow down the speed for defect formation, leading to larger device lifetimes. For all these reasons, the use of *h*-BN as dielectric can be attractive in a wide range of electronic devices. The most inspiring recent advance in this direction was the development of fully 2D metal-insulator-semiconductor (MIS) devices by combining *h*-BN with graphene and 2D semiconductors (like MoS₂ [47], WS₂ [105]), as shown in Fig. 4.

Interestingly, the first use of *h*-BN in FETs was not as a dielectric, but it was as substrate for graphene channels. In 2010, C. R. Dean et al. [106] realized that in FETs, the carrier mobility and inhomogeneity of graphene channels supported on *h*-BN substrates are almost one order of magnitude better than those on SiO₂, and comparable to those reported for suspended graphene. A similar structure is displayed in Fig. 5a. Moreover, the topographic roughness, intrinsic doping and chemical reactivity of the graphene channel were also greatly improved. Mayorov et al. [107] observed ballistic transport up to ~3 μm at low temperature, F. Guinea and colleagues [108] reported that the conductivity of graphene on *h*-BN enhances the energy gap in multiterminal measurements of the fractional quantum Hall effect [109–110]. The main reason behind this performance increase is, as mentioned above, the atomically flat surface of the *h*-BN substrate. K. M. Burson et al. [111] characterized the local electrostatic potential above *h*-BN and SiO₂ by Kelvin probe microscopy, and the results showed that the *h*-BN displays potential fluctuations up to 2 orders of magnitude lower than SiO₂, as well as lower metastable trapped charge densities. This trend of using *h*-BN as substrate for high mobility 2D devices rapidly expanded to other materials. Later reports observed that transferring another piece of *h*-BN on the 2D channel (made of graphene or 2D/TMDs) provided additional protection against the environment, neutralizing the effect of adsorbates and increasing the mobility at the channel. Very recently M. W. Iqbal et al. [105] used single layer WS₂ sandwiched between CVD grown *h*-BN films, as displayed in Fig.5b, showing unprecedented high mobility at room temperature. In the case of *h*-BN encapsulated graphene devices, N. Petrone et al. [46] achieved room temperature carrier mobilities up to 10,000 cm² V⁻¹ s⁻¹ on flexible substrates, and recently T. Chari et al. [112] beat this number using the same methodology on rigid SiO₂ substrates, which lead to an improved mobility of 13,700 cm² V⁻¹ s⁻¹.

Unfortunately, only a few of these prototypes used the *h*-BN as gate dielectric, and the reports that did it, just concentrate on the properties of the channel and/or transistor, not on the performance of the *h*-BN dielectric itself. For example, G. H. Lee et al. [46] fabricated the FETs with MoS₂ as channels, *h*-BN as dielectric, and graphene as gate electrodes. Field-effect mobilities of up to 45 cm²/V·s and operating gate voltage below 10 V were achieved by using this heterostructure devices. Recent works studied the use of h-BN for the fabrication of capacitors with enhanced performance [113]. Actually, capacitance effect is one of the most required properties of dielectrics in electronic devices, as it can provide functionalities like charge storage and electronic resistance modulation. In this direction, it has been experimentally observed that h-BN shows an unusual and significant increase of the relative permittivity with decreasing the stack thickness below 5 nm [114]. Ab initio calculations indicate that this phenomenon, which can alter the performance of the FETs [112], is related to the negative quantum capacitance of the graphene channel with a top h-BN gate dielectric. Furthermore, atomically thin h-BN stacks show low electric filed screening, and this effect has relative weak dependence to the number of layers. Nevertheless, the good performance of h-BN in these devices requires understanding the essential dielectric properties, like homogeneity, variability, reliability and dielectric breakdown. In the following section these phenomena are analyzed from the knowledge available in the literature.

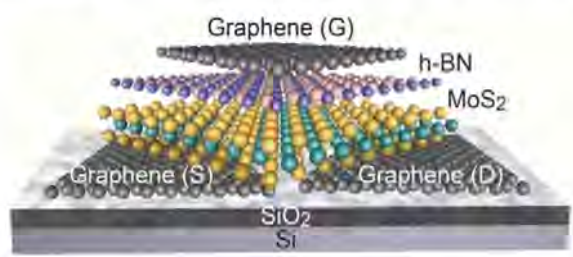

**Fig. 4.** Fully two dimensional field effect transistor synthesized by using graphene electrodes, an *h*-BN stack as gate dielectric and a MoS₂ semiconducting channel. (Reproduced with permission from [47], Copyright American Chemical Society, 2014.)

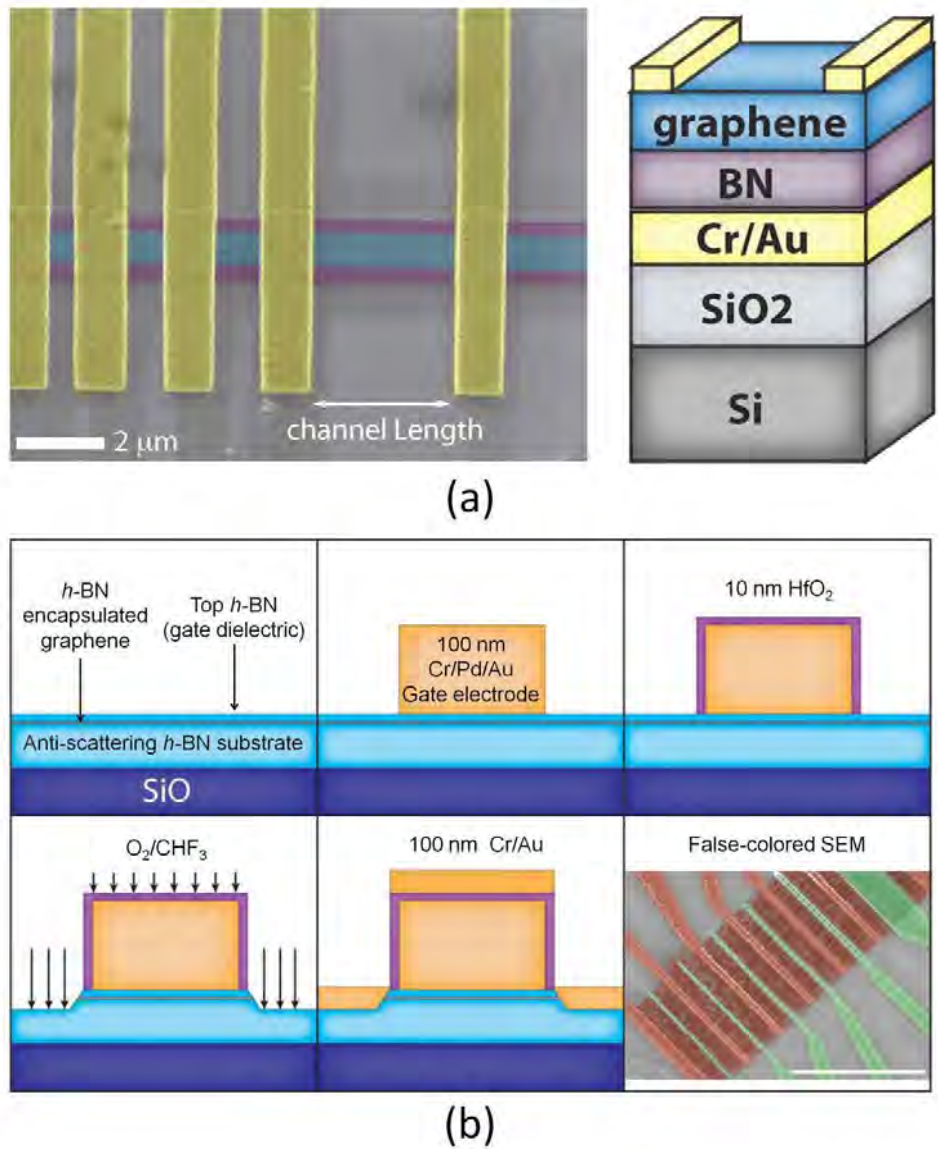

**Fig. 5.** (a) False colored scanning electron microscope image (left) and schematic (right) of single back gate graphene field effect transistors using h-BN as substrate. (b) Fabrication process of a field effect transistor with graphene channel encapsulated with h-BN. Modified and reprinted from [107,100]. (Copyright from IEEE 2015 and Nature Publishing Group, 2015.)

## 5. Use of h-BN as dielectric

Pristine h-BN free of defects presents an atomically flat structure and an extraordinary homogeneous dielectric performance, which can be characterized by monitoring the tunneling current. The most powerful techniques to conduct such nanoscale studies are the conductive atomic force microscope (CAFM) [115–119] and the scanning tunneling microscope (STM) [120–121]. As an example, CAFM studies scanning mechanically exfoliated h-BN sheets with different thickness revealed one of the most homogeneous tunneling currents ever reported in a dielectric (Fig. 6) [122]. Unfortunately, mechanically exfoliation is not a scalable technique and it doesn't allow controlling the 2D material thickness. For these reasons, the use of alternative growth techniques like CVD [74–80] and sputtering [68–69] are preferred in industrial applications, even if the quality of the *h*-BN stacks produced is lower. Some of the most common defects in large area *h*-BN stacks are: i) lattice distortions (including dangling bonds [123], non-hexagonal bonding [124] and impurities or undesired doping [125]), ii) thickness fluctuations [122], iii) wrinkles [126] and iv) cracks [113]. All these local defects in *h*-BN can notably alter its electronic properties, impoverishing the variability and performance of the whole device.

### 5.1. Nanoscale homogeneity and variability

Lattice distortions is by far the most difficult defect to detect, as it takes place at the atomic scale. However, both STM and transmission electron microscopy (TEM) [126–127] hold enough resolution to successfully characterize the morphology of the stacks with atomic resolution (see Fig. 7b and 7c). Thanks to these techniques, it has been demonstrated that lattice distortions in CVD-grown *h*-BN tend to accumulate at the grain boundaries (GBs) of the polycrystalline 2D stack [128]. This behavior has been also previously reported in other 2D materials, including graphene [129], $MoS_2$ [130] and WS2 [131]. It should be noted that, in the field of 2D materials, the concept of grain boundary has been also often referred as domain boundary in the literature. In this work we will consistently always refer to them as grain boundary. Both theoretical calculations and experimental studies demonstrated that two different types of GBs are prone to form in *h*-BN grown on Cu(111) by CVD [127]: square-octagon pairs (4/8 GBs) or pentagon-heptagon pair (5/7 GBs). Scanning tunneling microscope (STM) was used to monitor the GBs structure in CVD-grown *h*-BN, and the observed results confirm the coincidence site lattice (CSL) theory [132]. In contrast to graphene, the binary composition of *h*-BN shows a very complicated configuration of GBs. Fig. 7a displays different possible atomic configurations of the boron (pink spheres) and nitrogen (blue spheres) atoms in 4/8 and 5/7 GBs, respectively. Heteroelemental (type I) and homoelemental (type II) bonding structures are possible atomic models existing in the 4/8 GB, while for 5/7 GBs, there are three different arrangements: B—B bonds shared by 5 and 7 rings (type III), N-rich (type IV) and B-rich (type V). However, in terms of the simulation of local density of states (LDOS, Fig. 7d), the types with higher probability for 4/8 GBs and 5/7 GBs are type I and III, respectively. One of the main consequences of lattice distortions in *h*-BN is the apparition of in-gap states, which increases the local conductivity through the 2D layered dielectric stack. Further scanning tunneling spectroscopy (STS) measurements and density functional theory (DFT) calculation discovered that the band gap of *h*-BN with dislocation GBs has decreased compared to defect-free locations [127]. At the same time, deep-in-gap states with greatly reduced band gap is confined in the grain boundaries of the *h*-BN sheet. Interestingly, this observation shows a strong parallelism with those made previously in ultra-thin 3D dielectrics, mainly transition metal oxides (TMO), like $HfO_2$ [133], $Al_2O_3$ and ZrO [134]. Moreover, this finding is in line with the behavior previously observed in $MoS_2$ sheets, which showed electrical conductivity increase at the GBs area due to a decreased band gap [135]. Recent reports in the field of graphene also demonstrated that the presence of GBs can strongly influence the electronic [136,137], magnetic [138] and mechanical [139] performance of the devices. Consequently, the local leakage current increase at the GBs of *h*-BN should be considered during the design of the devices. Fortunately, CVD growth process develops very fast, and 2D materials with very large grain sizes have been reported. S. Bae et al. [140] reported 30 in. graphene films on copper foil for transparent roll-to-roll production, and G. Y. Lu et al. [141] synthesized monolayer *h*-BN with grain sizes above 80 μm. Nevertheless, the presence of GBs in dielectrics could be even useful in different configurations, as they may allow some GB-driven additional properties, such as resistive switching [142].

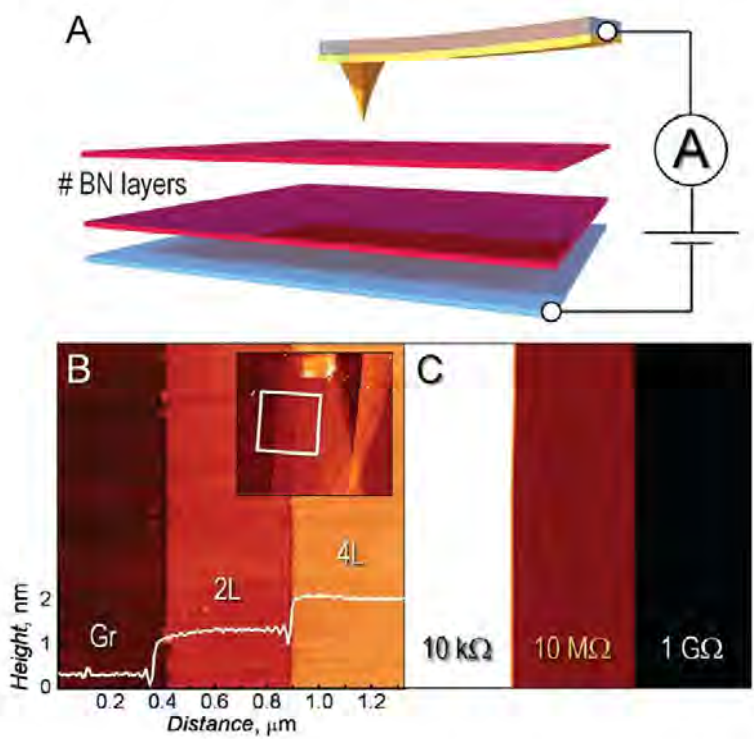

**Fig. 6.** (a) Schematic of a multilayer *h*-BN stack characterized by CAFM. (b) and (c) are the topographic and current maps collected with the CAFM on multilayer *h*-BN. Reprinted with permission from [117].
(Copyright American Chemical Society 2012.)

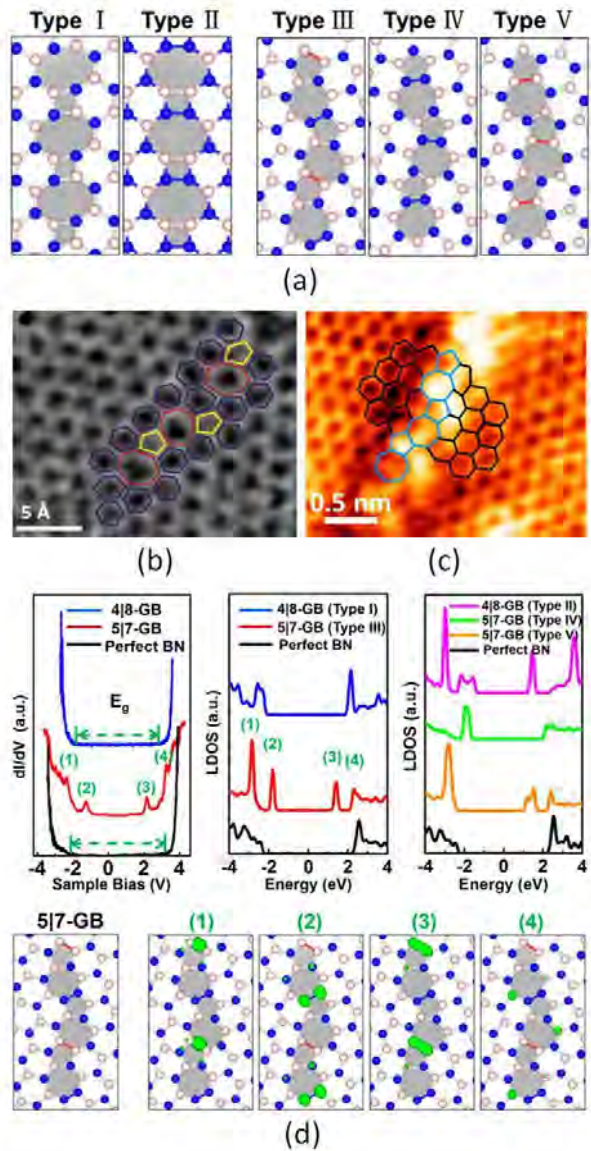

**Fig. 7.** Atomic scale defects in hexagonal boron nitride. (a) Schematic of different possible atomic configurations at the grain boundaries of *h*-BN. The boron and nitrogen atoms are represented with pink and blue spheres, respectively. (b) and (c) show scanning tunneling microscope and transmission electron microscope images of a grain boundary in *h*-BN (respectively). (d) Local density of states and schematic of a 5|7 grain boundary (made of pentagons and heptagons). Modified and reprinted from [122–123].
(Copyright from American Chemical Society 2015 and American Chemical Society 2013.)

Thickness fluctuations is another principal source of variability in ultra-thin dielectrics. The tunneling current through *h*-BN dielectric stacks with different thicknesses has been studied by several groups [63,143]. Y. F. Ji et al. [142] reported that the presence of multilayer islands in CVD-grown monolayer *h*-BN sheets (Fig. 8a) produces the apparition of insulating areas current maps collected by CAFM (Fig. 8a and 8b, respectively). G. H. Lee et al. [63] quantified the local conductivity of mechanically exfoliated *h*-BN nanosheets with different thicknesses, using also a CAFM. His I-V curves indicate that the tunneling current across the *h*-BN decreases as the thickness of the stack increases (Fig. 8c). Interestingly, at low electrical fields the tunneling current through *h*-BN stacks with thicknesses of one, two and three layers fit the direct tunneling (DT) model. This conduction is masked at high voltages, which show clear Fowler-Nordheim tunneling (FNT). The transition between the two conduction regimes can be observed from the I-V curves (Fig. 8c), which shape transforms from the linear (at low bias) into exponential (at high bias). These observations were later confirmed by L. Britnell et al. [122] who studied the tunneling current in mechanically exfoliated *h*-BN nanosheets, but in this case using strategically patterned electrodes by electron beam lithography. It is worth noting that the current densities observed by Britnell are about two orders of magnitude larger than those reported by G. H. Lee et al. [63] and Y. F. Ji et al. [142], probably due to the lower electrical contact resistance of the patterned electrodes compared to that of the CAFM setup. After that, N. Guo et al. [113] reported that the DT conduction at low fields could be maintained in thick (>18 nm) *h*-BN stacks, but the I-V curves and atomically flat surface of the stack in this work are not conclusive, making necessary a corroboration of this finding.

It should be highlighted that synthesizing *h*-BN stacks with an exact controlled amount of layers at all locations of its surface is extremely challenging. For this reason, when a 2D material is synthesized and its thickness is reported, it is convenient to indicate the amount (percentage) of area that fits the theoretical thickness. For example, X. S. Li et al. [143] reported the fabrication of graphene sheets that are monolayer in at least the 95% of their surface. Even 2D materials suppliers rarely state an exact number of layers within a multilayer sample, and they usually indicate a range in the specifications. Graphene Supermarket supplies multilayer *h*-BN films with thicknesses between 10 and 13 nm [144]. The quantification of the number of layers at a single location of the sample, as well as the percentage of area for a specific thickness can be easily performed using a standard CAFM [142].

Another defect very difficult to control during the growth process of the *h*-BN film is the formation of wrinkles, which can be also easily observed with an optical microscope, SEM and/or AFM (Fig. 8a). The origin of the wrinkles in *h*-BN is related to the large temperatures required by the CVD growth process, which are almost always above 750 °C. After the *h*-BN growth, the setup containing the *h*-BN/Cu stack needs to be cooled down, which produces the apparition of compressive strains in the *h*-BN due to the mismatch of the thermal expansion coefficients of both materials. In order to relax this compressive strain, the 2D sheet tends to delaminate from the substrate, leading to the formation of wrinkles. Oliveira et al. reported that the wrinkles in *h*-BN tend to form threefold origami-type junctions throughout the 2D sheet. The wrinkles in *h*-BN can shift the Raman peak frequency of the $E_{2g}$ mode to higher frequencies [145], and alter the secondary diffraction spots in TEM [82]. From an electrical point of view, the presence of a gap between the *h*-BN and the substrate (Fig. 8d and 8e) produces an effective increase of the dielectric thickness at that location. This phenomenon can be corroborated from CAFM current maps (Fig. 8f), which display the wrinkles as long insulating lines. The more insulating nature of the wrinkled sites could reduce the amount of charge stored in a capacitor [146] and induce threshold voltage inhomogeneities in FETs [147]. It is worth noting that the wrinkles formed during the CVD growth are maintained during the transfer process of the 2D sheet onto random substrates. N. Liu et al. [148] observed that by soaking the 2D sheet in an ultrasonic bath for some hours the amount of wrinkles could be significantly reduced. A more effective methodology consists on increasing the roughness of the substrate on which the 2D material is transferred [149]. Using this methodology, the 2D sheet can relax the compressive strain by adapting to the voids present at the surface of the substrate. Nevertheless, these methodologies could bring associated other

128                    F. Hui et al. / Microelectronic Engineering 163 (2016) 119–133

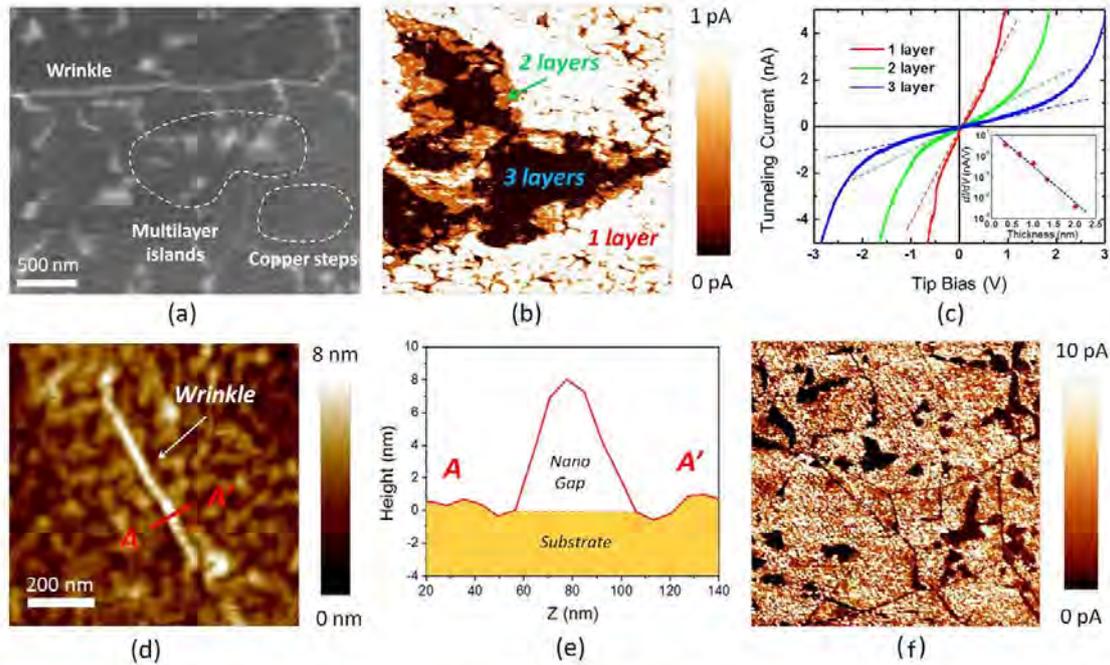

**Fig. 8.** Nanometric scale defects in hexagonal boron nitride. (a) SEM image of a CVD-grown h-BN film, showing multilayer islands and wrinkles. (b) 1 μm × 1 μm current map and (c) IV curves collected with a CAFM on h-BN. (d) topographic map and (e) schematic of a wrinkle in h-BN. (f) 10 μm × 10 μm current map of a portion of the sample shown in (a), which displays that the wrinkles and multilayer islands generate a decrease of the tunneling current though the h-BN dielectric. Modified and reprinted with permission from [63,137]. (Copyrights from American Institute of Physics 2011 and 2016.)

undesired effects, like the formation of cracks during the ultrasonic bath or the presence of voids between the substrate and the 2D sheet. Therefore, completely avoiding the formation of wrinkles or getting totally rid of their effect is, to date, an almost impossible task.

The last h-BN defect commented in this work is the presence of cracks. A crack can be understood as a physical discontinuity of the 2D sheet, and it is an extremely harmful defect in 2D insulators used as dielectric. For example, when graphene sheets are used as conductive channel, cracks represent a reduction of the effective area through which the current can flow and, therefore, an increase of the electronica resistance. Despite this defect may alter the performance of the device, unless the channel is completely broken, the device can still be operated. On the contrary, an atomic scale crack within a dielectric sandwiched by two electrodes can produce the dielectric breakdown (BD) of the whole structure, and the irreversible failure of the device. As mentioned, cracks can be generated during the CVD growth process if the substrate is not fully covered with at least one layer of h-BN. Moreover, cracks can be also generated when a continuous 2D sheet is physically broken due to an external strain. The most common source of cracks in h-BN and other 2D materials is the use of a transfer process, which can locally break the sheets due to high mechanical strains [150]. Other sources of cracks in 2D materials could be produced by electromigration due to extremely large current densities [151–152].

### 5.2. Reliability and dielectric breakdown

For a dielectric, reliability is defined as the ability of keeping unaltered its insulating properties during device operation [153]. In ultra-scaled technologies, dielectric reliability is even more important because, despite operation voltages are usually reduced with the scaling down of the devices, it is true that thinner dielectrics usually have to withstand larger electrical fields [154], which can lead to prohibitive leakage currents and power consumption. The electrical field can generate different types of defects in the microstructure of the dielectric [155], leading to the apparition of stress induced leakage currents (SILC) though it [156]. In some materials with high densities of native defects (such as high-k dielectrics) strong leakage current fluctuations during the degradation process have been observed, which is known to be a consequence of charge trapping and detrapping [157]. When the density of defects prohibitively increases, many partially formed defective paths can propagate across the dielectric, leading to a remarkable increase of the leakage current at very low voltages: this is the onset of the soft breakdown [158]. Finally, if the electrical stress still persists one of these conductive paths becomes dominant, forming a defect-related conductive filament that physically connects the top and bottom electrodes. This situation is called hard dielectric breakdown (BD), and it usually produces the failure of the device [154]. The whole degradation process of an amorphous 2 nm thick HfO₂ dielectric film is displayed in Fig. 9a using sequences of I-V curves collected with a CAFM.

Recent investigations have reported important differences between the degradation process of h-BN dielectric stacks and traditional dielectrics. Two different sources reported that the failure of the h-BN occurs layer-by-layer. This unprecedented behavior was presented for the first time in 2015 by Y. Hattori et al. [159], who used a CAFM connected to a semiconductor parameter analyzer (SPA) to monitor the dielectric breakdown of (mechanically exfoliated) multilayer h-BN nanosheets with different thicknesses. Interestingly, his results show the BD of one/two layers of h-BN in each I-V curve, until the complete stack was broken. These surprising observations were recently corroborated by Y. F. Ji et al. [142], who collected sequences of CAFM I-V curves on CVD-grown h-BN, and fitted them using charge transport modeling. More specifically, she compared the performance of HfO₂ and h-BN dielectric films with similar equivalent oxide thicknesses (EOT) [160]; therefore, comparisons between the I-V curves collected under similar ramped voltage stress (RVS) are meaningful. The results indicate that, while HfO₂ films suffer from charge trapping/detrapping, SILC and

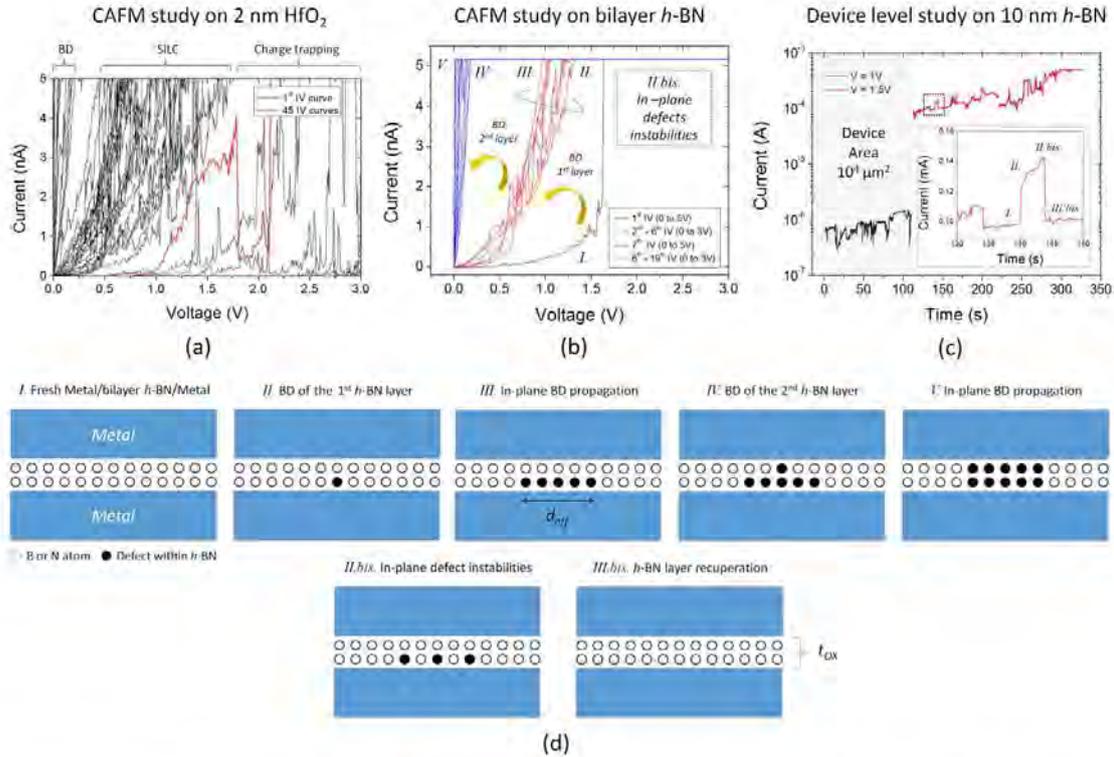

**Fig. 9.** Dielectric breakdown process in 3D and 2D layered dielectrics. (a) Current-voltage (I-V) curves collected with the CAFM on a 2 nm thick HfO$_2$ dielectric film (single location). The plot clearly displays three regions: charge trapping/detrapping, SILC and Ohmic conduction (BD). (b) Sequence of 19 IV curves measured on a bilayer spot of a CVD-grown h-BN stack (single location). The layer-by-layer BD process can be clearly observed. (c) Current-time (I-t) curves collected with a probestation in squared 100 μm × 100 μm capacitors with multilayer CVD-grown h-BN, using different low voltages. (d) Proposed schematic of the dielectric breakdown process in 2D layered dielectrics. Modified and reprinted with permission of [137].

[Copyright American Institute of Physics 2016.]

premature BD within 45 RVS (from 0 V to 3 V, Fig. 9a), the h-BN showed unaltered conduction during > 112 RVS. Only when larger voltages were applied, the BD of a layer within the h-BN stack was triggered, leading to a sudden shift of the IV curve towards lower potentials (Fig. 9b). It is worth noting that such jump was not preceded by the large current fluctuations typically observed in 3D dielectrics (Fig. 9a) [161], and only small fluctuations were recorded.

Hattori and co-workers attributed his unusual degradation mechanism to the unique anisotropic speed for defect formation in layered 2D layered dielectrics [159]. Unlike 3D insulators (i.e. HfO$_2$, Al$_2$O$_3$ and TiO$_2$), in which the bonding is similar in all directions, 2D layered h-BN stacks form covalent bonding in-plane, while the interaction between adjacent layers is dominated by Van der Waals forces. This geometry-dependent atomic interaction generates different speeds for defect formation in-plane and out-of-plane. The increase of leakage current produced by the generation of new defects distributed parallel and transversely to the applied electrical field is different (as in conventional 3D dielectrics). The total currents measured in the I-V curves can be described as $I = J (t_{OX}) \cdots A_{eff} | $ [162–163], where $J$ is the current density, $t_{OX}$ is the thickness of the dielectric and $A_{eff}$ is the effective area though which $J$ flows. The nomenclature $t_{OX}$ is used for similarity with the dielectric breakdown literature [116,164], although h-BN is not an oxide film. In-plane defects don't modify the effective thickness of the h-BN stack, and just produce a linear increase of area for current flow ($A_{eff}$), represented with $d_{eff}$ (diameter of $A_{eff}$) in the schematic of Fig. 9d. On the contrary, the formation of defects in adjacent layers implies a reduction of $t_{OX}$, which strongly alters the value of $J$ (see also Fig. 9d, drawing

IV.). Due to the exponential dependence between $J$ and $t_{OX}$ in most conduction mechanisms though a dielectric [165], it can be concluded that the current increase produced by the formation of out-of-plane defects is larger than that of in-plane ones. Therefore, the sudden current shifts displayed in Fig. 9b should be related to the BD of a layer within the h-BD stack, while the small current fluctuations within each current step should be related to the formation of in-plane defects.

It should be highlighted that, when measuring random locations with the CAFM, the probability of placing the tip on a GB within the CVD-grown h-BN is very low, because the area covered by the grains is much larger than that covered by GBs, indicating that nanoscale studies are most likely displaying the electrical behavior of the grains [166]. This is probably the reason why the results from Y. F. Ji et al. [142] in CVD-grown h-BN agree so well with those reported by Hattori [159] and/or Lee [63], who used mechanically exfoliated h-BN nanosheets. On the contrary, device level experiments are very susceptible to hot spots within the total area under tests, as the BD process is a stochastic phenomenon that always takes place at the weakest location of the sample. As explained in the previous subsection, h-BN dielectric stacks could contain many imperfections that alter their electronic properties and, despite intrinsic homogeneities in principle only present a variability concern, in traditional dielectrics it has been demonstrated that locations with different electronic properties have also shown different degradation speeds [167–168]. Therefore, the study of the degradation process of polycrystalline CVD-grown h-BN stacks also requires device level studies. Interestingly, I-t curves collected with the probestation in multilayer CVD-grown h-BN stacks show random telegraph noise

 

(RTN) signals (Fig. 9c), which are typical of defective materials [169]. Therefore, the RTN signals should be attributable to the GBs in h-BN, as they are known to hold a larger conductivity and higher density of defects and [170]. In any case, the progressive BD process observed at the device level in h-BN capacitors (Fig. 9c) contrasts with the sharp BD observed in similar (polycrystalline) HfO₂ based devices [171], indicating that even at the GBs of CVD-grown h-BN the degradation of the stack occurs layer-by-layer. Furthermore, the observation of both stepped and progressive current increases in the I-t curves (from I. to II. and from II. to II.bis, respectively, in Fig. 9c) further supports the anisotropic nature of defect formation. Another interesting behavior revealed by the device level tests is the partial recuperation of the resistance of a layer. As Fig. 9c shows, the I-t curves not only show step-up current increases related to the BD of an h-BN layer within the stack, but also steps-down of a similar magnitude, indicating its partial recuperation. This behavior was not observed at the nanoscale, indicating that it is probably a GBs-driven effect. Successive device level stresses reveal resistance recovery of more than two orders of magnitude, indicating that this material may be suitable or resistive switching applications. It is worth noting that defect-rich GBs are known to be the resistive switching driving feature in many polycrystalline TMO dielectrics [167–168,172–173].

Among all common defects in h-BN stacks, boron vacancies are the most likely to be formed, as they show the lowest activation energy. Using density-functional tight-binding (DFTB) calculations, A. Zobelli et al. [174] reported that boron vacancies are first thermally activated and can overcome the energy barriers, allowing migration and forming stable h-BN divacancies. However, nitrogen vacancies do not undergo thermally activation below the melting point of h-BN. More specifically, the energy threshold for knock-on damage of boron atoms in h-BN is 74 keV, which is much lower than the nitrogen atoms (84 keV).

Ultimately, the dielectric hard breakdown of the whole stack is reached when a high enough electrical field is applied. Several reports measured the dielectric strength of h-BN stacks synthesized following any of the processes described in Section 2, and the values reported range between 1.5 and 12 MV/cm [63,113,122,175–177] (see Table 1). Unfortunately, to our knowledge, only the work by Y. Hattori et al. [159] reported statistical information about the BD phenomenon in multilayer h-BN, which revealed an interesting phenomenon: thicker h-BN stacks show lower slopes in the Weibull distribution (Fig. 10c). This phenomenon is opposed to the observations traditionally made in amorphous oxides [178]. In dielectrics with an amorphous structure (like SiO₂ or as-grown high-k dielectrics, Fig. 10a), in which inhomogeneities can be neglected, the BD process is characterized by: i) the randomness of defect formation within the volume of the dielectric, and ii) the absence of defect-to-defect interaction [154]. Under these assumptions, the time-to-breakdown (also called time dependent dielectric breakdown, or TDDB) follows the Weibull statistical distribution, which has been used during the last two decades to perform technology reliability assessments. With the introduction of polycrystalline high-k dielectrics in microelectronics this model needed to be deeply revised [179], due to the different conductivities of both grains and GBs (Fig. 10b). Basically the grain boundaries introduce short TDDBs in the

Weibull slope, which reduces its slope. Due to the almost uncontrollable nature of the amount and position of the defects at the GBs, these new points are difficult to evaluate and predict. This is the main reason that makes reliability studies of polycrystalline dielectrics much more complex and inaccurate than in amorphous ones. The data in Fig. 10c suggest that the BD process in multilayer h-BN stacks may not follow the percolation theory, which would have important implications in electronic devices containing 2D layered dielectrics. Nevertheless, the experiments presented in Fig. 10c were performed using a CAFM, which may accelerate the BD due to the lower stability of the metallic coating on the CAFM tip, as it can melt at the large current densities measured during the BD. Moreover, that study has been developed on mechanically exfoliated h-BN nanosheets, which do not contain GBs. Therefore, more reliability studies based on nanoscale and device level characterization using different types of h-BN, as well as atomistic simulations and current modeling are necessary to accurately describe the degradation process of h-BN films.

Finally, it should be highlighted that BN may also show some post-BD properties, such as resistive switching. As mentioned above, P. S. Lee [97] fabricated a transfer-free metal/BN/metal resistive memory devices using a polycrystalline 3D film made of BN grown by CVD on copper, followed by top metal evaporation. The interesting point is that the authors attributed this phenomenon to the presence of grain boundaries in the BN film. Despite the validity of these results is still intriguing because the films presented doesn't correspond to 2D layered h-BN, this report opens the door to the use of BN for the fabrication of non-volatile memory devices.

## 6. Conclusions

The use of two dimensional insulators as dielectrics is an attractive methodology to enhance the performance of electronic devices. Hexagonal boron nitride is until now the 2D layered dielectric that has been more intensively investigated, due to its good compatibility with graphene. Initially, h-BN was used as anti-scattering substrate in graphene transistors, but the material started rapidly to attract attention as functional dielectric. On one hand, it has been already demonstrated that layered h-BN is much more stable vs. electrical stresses than other widely used dielectrics, such as HfO₂. Furthermore, h-BN exhibits a very unusual layer-by-layer dielectric breakdown due to an anisotropic speed for defect formation laterally and vertically, a behavior that may have a deep influence in the reliability of future electronic devices. And on the other hand, depending on the synthesis method, the h-BN layered dielectric can contain different amounts of local defects, being grain boundaries, thickness fluctuations, wrinkles and cracks the more common. Recent studies reported that these defects can alter the variability and reliability of the materials and devices in which they are introduced which enhance the local conductivity of the material. It is also expected that these defects may drive some interesting phenomena, i. e. the grain boundaries in polycrystalline CVD-grown h-BN may exhibit resistive switching, as it already happens in transition metal oxides. Therefore, more investigations to clarify the impact of

**Table 1**
Dielectric strength of BN films reported in the literatures.

| Fabrication method | Setup | Area | BN field (MV/cm) | Reference |
|---|---|---|---|---|
| Mechanical exfoliation | CAFM | ~100 nm² | 7.94 | [63] |
| Mechanical exfoliation | CAFM | ~100 nm² | ~12 | [159] |
| Mechanical exfoliation | CAFM | ~100 nm² | 10 | [122] |
| Mechanical exfoliation | Probestation | Ti/Au electrodes (1000 μm²) | 6.4–9.0 | [175] |
| CVD | Probestation | Au electrodes (18 mm × 200 μm) | ~9.0 | [133] |
| CVD | Probestation | Ti/Au electrodes (5 μm × 5 μm) | 1.5–2.5 | [70] |
| Magnetron sputtering | Probestation | Ru electrodes (3 μm²) | * | [68] |

The symbol "*" indicates that the paper doesn't give a value for the dielectric strength, but it shows that the tunneling conductance of sputtered BN is similar to that of exfoliated one, which is also a relevant information for comparing the quality of the films.

F. Hui et al. / Microelectronic Engineering 163 (2016) 119–133    131

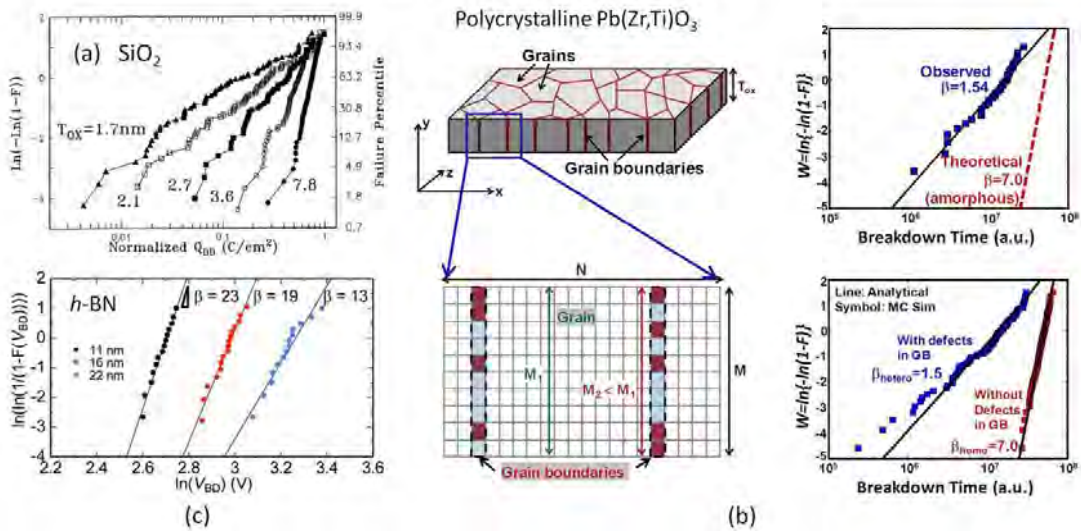

**Fig. 10.** Statistics of the dielectric breakdown in thin dielectrics. (a) Typical normalized Weibull distributions of the charge-to-breakdown in $SiO_2$ films with different thicknesses. (b) Weibull statistics of amorphous and polycrystalline dielectrics with highly conductive grain boundaries. (c) Weibull distributions for multilayer $h$-BN stacks with different thicknesses. Unlike for $SiO_2$, in $h$-BN the slope decreases with the thickness. Modified and reprinted from [155,174–175]. (Copyrights from American Chemical Society 2015, Elsevier Ltd. 2005 and IEEE 2011.)

2D layered dielectrics in microelectronics, as well as their variability and reliability are necessary.

The route for the fabrication of $h$-BN based devices should include: i) synthesis of $h$-BN films on different substrates, not only on metallic foils but also on metal-coated Si wafers; ii) reducing the growth temperatures to avoid excessive thermal budget, iii) further controlling the amounts of defects, specially grain boundaries and thickness fluctuations, iv) developing more methodologies to build the devices, especially in terms of $h$-BN transfer and etching, and v) developing and demonstrating new transfer-free device architectures, such as resistive random access memories. Nevertheless, the current ability to grow the material and the scarce knowledge their dielectric performance conforms a privileged situation for reliability engineers and scientists, as standard analyses are required, for example: i) local electronic characterization (using scanning tunneling microscopy, conductive atomic force microscopy or kelvin probe force microscopy), ii) first principles simulations of the grain boundaries (three dimensional, not only in plane), iii) tunneling current simulations and fitting to the quantum point contact model (among others), iv) validity of the percolation model for the dielectric breakdown, v) breakdown statistics based on Weibull analyses (time-to-breakdown and maximum charge accumulated in the dielectric), vi) compatibility with adjacent layers (migration of boron ions and effect of boron vacancies on the conductivity of the material), and vi) resistive switching tests are required. In summary, all the reliability analyses developed during the last decade for high-k dielectrics need to be performed in these new 2D layered dielectrics before their introduction in real products.

## Acknowledgments

This work has been supported by the Young 1000 Global Talent Recruitment Program of the Ministry of Education of China, the National Natural Science Foundation of China (grant nos. 61502326, 41550110223, 11661131002), the Government of Jiangsu Province (grant no. BK20150343), the Ministry of Finance of China (grant no. SX21400213) and the Young 973 National Program of the Chinese Ministry of Science and Technology (grant no. 2015CB932700). The Collaborative Innovation Center of Suzhou Nano Science & Technology, the Jiangsu Key Laboratory for Carbon-Based Functional Materials & Devices and the Priority Academic Program Development of Jiangsu Higher Education Institutions are also acknowledged.

## References

[1] Report contents and summaries: section 6: memory market overview, http://www.icinsights.com/services/mcclean-report/report-contents/
[2] Lilienfeld Julius Edgar U.S. Patent 1,745,175 "Method and apparatus for controlling electric currents", October 22th. (1925).
[3] G.I. Meijer, Science 319 (2008) 1625–1626.
[4] New outlook: the most exciting Nehalem EP processor, 2008, online available at http://www.cnetnews.com.cn/2008/1226/2026930.shtml on December 26th of 2008.
[5] S. Thompson, N. Anand, M. Armstrong, C. Auth, B. Arcot, M. Alavi, P. Bai, J. Bielefeld, R. Bigwood, J. Brandenburg, M. Buehler, S. Cea, V. Chikarmane, C. Choi, R. Frankovic, I. Ghani, G. Glass, W. Han, T. Hoffmann, M. Hussein, P. Jacob, A. Jain, C. Jan, S. Joshi, C. Kenyon, J. Klaus, S. Klopcic, J. Luce, Z. Ma, B. Mcintyre, K. Mistry, A. Murthy, P. Nguyen, H. Pearson, T. Sandford, R. Schweinfurth, R. Shaheed, S. Sivakumar, M. Taylor, B. Tufts, C. Wallace, P. Wang, C. Weber, M. Bohr, IEDM Tech. Dig. (2002) 61–64, http://dx.doi.org/10.1109/IEDM.2002.1175779.
[6] R. Chau, S. Datta, M. Doczy, J. Kavalieros, M. Metz, Proceedings of International Workshop on Gate Insulator, 62013 124–126.
[7] R. Choi, R. Onishi, C.S. Kang, S. Gopalan, R. Nieh, Y.H. Kim, J.H. Han, S. Krishnan, H.J. Cho, A. Shahriar, J.C. Lee, IEDM Tech. Dig. (2002) 613–617, http://dx.doi.org/10.1109/IEDM.2002.1175914.
[8] G. Lucovsky, B. Rayner, Y. Zhang, J. Whitten, IEDM Tech. Dig. (2002) 617–620, http://dx.doi.org/10.1109/IEDM.2002.1175915.
[9] S. Inumiya, K. Sekine, S. Niwa, A. Kaneko, M. Sato, T. Watanabe, H. Fukui, Y. Kamata, M. Koyama, A. Nishiyama, M. Takayanagi, K. Eguchi, Y. Tsunashima, Symp. on VLSI Technol. 17 (2003) http://dx.doi.org/10.1109/VLSIT.2003.1221064.
[10] V. Kim, C. Lim, C.D. Young, K. Matthews, J. Barrett, B. Foran, A. Agarwal, G.A. Brown, G. Bersuker, P. Zeitzoff, M. Gardner, R.W. Murto, L. Larson, C. Metzner, S. Kher, H.R. Huff, Symp. on VLSI Technol. (2003) 167–168, http://dx.doi.org/10.1109/VLSIT.2003.1221138.
[11] M. Lanza, M. Porti, M. Nafria, X. Aymerich, G. Benstetter, E. Lodermeier, H. Ranzinger, G. Jaschke, S. Teichert, L. Wilde, P. Michalowski, Microelectron. Eng. 86 (2009) 1921–1924.
[12] J. A. Kittl, K. Opsomer, M. Popovici, N. Menou, B. Kaczer, X.P. Wang, C. Adelmann, M.A. Pawlak, K. Tomida, A. Rothschild, B. Govoreanu, R. Degraeve, M. Schaekers, M. Zahid, A. Delabie, J. Meersschaut, W. Polspoel, S. Clima, G. Pourtois, W. Knaepen, C. Detavernier, V. Afanas'ev, T. Blomberg, D. Pierreux, J. Swerts, P. Fischer, J. W. Maes, D. Manger, W. Vandervorst, T. Conrad, A. Franquet, P. Favia, H. Bender, B. Brijs, S. Van Elshocht, M. Jurczak, J. Van Houdt, D.J. Wouters, ECS Trans. 19 (2009) 29–40.

[13] S. Swaminathan, M. Shandalov, Y. Oshima, P.C. McIntyre, Appl. Phys. Lett. 96 (2010) 082904.

[14] R.M. Wallace, G. Wilk, MRS Bulletin. 27 (2002) 186–191.

[15] V. Mistra, G. Lucovsky, G. Parsons, MRS Bulletin. 27 (2002) 212–216.

[16] C. Hobbs, L. Fonseca, V. Dhandapani, S. Samavedam, B. Taylor, J. Grant, L. Dip, D. Triyoso, R. Hegde, D. Gilmer, R. Garcia, D. Roan, L. Lovejoy, R. Rai, L. Hebert, H. Tseng, B. White, P. Tobin, Symp. on VLSI Technol. (2003) 9–10, http://dx.doi.org/10.1109/VLSIT.2003.1221060.

[17] S.J. Pearton, F. Ren, A.P. Zhang, K.P. Lee, Mater. Sci. Eng. 30 (2000) 55–212.

[18] C.T. Lee, H.W. Chen, H.Y. Lee, Appl. Phys. Lett. 82 (2003) 4304–4306.

[19] A.K. Geim, K.S. Novoselov, Nat. Mater. 6 (2007) 183–191.

[20] V.E. Dorgan, M.H. Bae, E. Pop, Appl. Phys. Lett. 97 (2010) 082112.

[21] B. Radisavljevic, A. Radenovic, J. Brivio, V. Giacometti, A. Kis, Nat. Nanotechnol. 6 (2011) 147–150.

[22] M.C. Lemme, T.J. Echtermeyer, M. Baus, H. Kurz, IEEE Electron Device Lett. 28 (2007) 282–284.

[23] Y. Lin, K.A. Jenkins, A. Valdes-Garcia, J.P. Small, D.B. Farmer, P. Avouris, Nano Lett. 9 (2009) 422–426.

[24] E. Guerriero, L. Polloni, M. Bianchi, A. Behnam, E. Carrion, L.G. Rizzi, E. Pop, R. Sordan, ACS Nano 7 (2013) 5588–5594.

[25] Y.M. Lin, K.A. Jenkins, A. Valdes-Garcia, J.P. Small, D.B. Farmer, P. Avouris, Nano Lett. 9 (2009) 422–426.

[26] Y. Wu, Y. Lin, A.A. Bol, K.A. Jenkins, F. Xia, D.B. Farmer, Y. Zhu, P. Avouris, Nature 472 (2011) 74–78.

[27] Y.M. Lin, C. Dimitrakopoulos, K.A. Jenkins, D.B. Farmer, H.Y. Chiu, A. Grill, P. Avouris, Science 327 (2010) 662.

[28] International Technology Roadmap for Semiconductors of 2008, online available at www.itrs.net/Links/2008ITRS/Home2008.htm on February 15th of 2016.

[29] H. Wang, L.L. Yu, Y.H. Lee, Y.M. Shi, A. Hsu, M.L. Chin, L.J. Li, M. Dubey, J. Kong, T. Palacios, Nano Lett. 12 (2012) 4674–4680.

[30] T. Georgiou, R. Jalil, B.D. Belle, L. Britnell, R.V. Gorbachev, S.V. Morozov, Y.J. Kim, A. Gholinia, S.J. Haigh, O. Makarovsky, L. Eaves, L.A. Ponomarenko, A.K. Geim, K.S. Novoselov, A. Mishchenko, Nat. Nanotechnol. 8 (2013) 100–103.

[31] H.S.S.R. Matte, A. Gomathi, A.K. Manna, D.J. Late, R. Datta, S.K. Pati, C.N.R. Rao, Angew. Chem. 122 (2010) 4153–4156.

[32] M.S. Fuhrer, J. Hone, Nat. Nanotechnol. 8 (2013) 146–147.

[33] B. Radisavljevic, A. Kis, Nat. Nanotechnol. 8 (2013) 147–148.

[34] Y. Yoon, K. Ganapathi, S. Salahuddin, Nano Lett. 11 (2011) 3768–3773.

[35] A.D. Franklin, Science 349 (2015) 2750.

[36] A.C. Ferrari, F. Bonaccorso, V. Fal'ko, K.S. Novoselov, S. Roche, P. Bøggild, S. Borini, F.H. Koppens, V. Palermo, N. Pugno, J.A. Garrido, R. Sordan, A. Bianco, L. Ballerini, M. Prato, E. Lidorikis, J. Kivioja, C. Marinelli, T. Ryhänen, A. Morpurgo, J.N. Coleman, V. Nicolosi, L. Colombo, A. Fert, M. Garcia-Hernandez, A. Bachtold, G.F. Schneider, F. Guinea, C. Dekker, M. Barbone, Z. Sun, C. Galiotis, A.N. Grigorenko, G. Konstantatos, A. Kis, M. Katsnelson, L. Vandersypen, A. Loiseau, V. Morandi, D. Neumaier, E. Treossi, V. Pellegrini, M. Polini, A. Tredicucci, G.M. Williams, B.H. Hong, J.H. Ahn, J.M. Kim, H. Zirath, B.J. van Wees, H. van der Zant, L. Occhipinti, A. Di Matteo, I.A. Kinloch, T. Seyller, E. Quesnel, X. Feng, K. Teo, N. Rupesinghe, P. Hakonen, S.R. Neil, Q. Tannock, T. Löfwander, J. Kinaret, Nanoscale 7 (2015) 4598–4810.

[37] H. Yang, T. Shin, M.M. Ling, K. Cho, C.Y. Ryu, Z.N. Bao, J. Am. Chem. Soc. 127 (2005) 11,542–11,543.

[38] D.W. Wang, Q. Wang, A. Javey, R. Tu, H.J. Dai, H. Kim, P.C. McIntyre, T. Krishnamohan, K.C. Saraswat, Appl. Phys. Lett. 83 (2003) 2432–2434.

[39] A.A. Demkov, O.F. Sankey, Phys. Rev. Lett. 83 (1999) 2038.

[40] J.L. Alay, M. Hirose, J. Appl. Phys. 81 (1997) 1606.

[41] B. Brar, G.D. Wilk, A.C. Seabaugh, Appl. Phys. Lett. 69 (1996) 2728.

[42] H. Liu, K. Xu, X. Zhang, P.D. Ye, Appl. Phys. Lett. 100 (2012) 152115.

[43] S. McDonnell, B. Brennan, A. Azcatl, N. Lu, H. Dong, C. Buie, J. Kim, C.L. Hinkle, M.J. Kim, R.M. Wallace, ACS Nano 7 (11) (2013) 10354–10361.

[44] S. Kim, J. Nah, I. Jo, D. Shanbhogerdi, L.G. Colombo, Z. Yao, E. Tutuc, S.K. Banerjee, Appl. Phys. Lett. 94 (2009) 062107.

[45] D.J. Late, B. Liu, H.S.S. Ramakrishna Matte, V.P. Dravid, C.N.R. Rao, ACS Nano 6 (2012) 5635–5641.

[46] G.H. Lee, Y.J. Yu, X. Cui, N. Petrone, C. Lee, M.S. Choi, D.Y. Lee, C. Lee, W.J. Yoo, K. Watanabe, T. Taniguchi, C. Nuckolls, P. Kim, J. Hone, ACS Nano 7 (2013) 7931–7936.

[47] T. Roy, M. Tosun, J.K. Kang, A.B. Sachid, S.B. Desai, M. Hettick, C.C. Hu, A. Javery, ACS Nano 8 (2004) 6259–6264.

[48] R. Zan, Q.M. Ramasse, R. Jalil, U. Bangert, Nanotechnol. Nanomaterials (2013) http://dx.doi.org/10.5772/56640.

[49] G. Giovannetti, P.A. Khomyakov, G. Brocks, P.J. Kelly, J. van den Brink, Phys. Rev. B 76 (2007) 73103.

[50] A. Gupta, T. Sakthivel, S. Seal, Prog. Mater. Sci. 73 (2015) 44–126.

[51] M. Xu, T. Liang, M. Shi, H.Z. Chen, Chem. Rev. 113 (2013) 3766–3798.

[52] K.S. Novoselov, A.K. Geim, S.V. Morozov, D. Jiang, Y. Zhang, S.V. Dubonos, I.V. Grigorieva, A.A. Firsov, Science 306 (2004) 666–669.

[53] K.S. Novoselov, A.H.C. Neto, Phys. Scr. 146 (2012) 014006.

[54] L.H. Li, Y. Chen, B.M. Cheng, M.Y. Lin, S.L. Chou, Y.C. Peng, Appl. Phys. Lett. 100 (2012) 261108.

[55] A.G.F. Garcia, M. Neumann, F. Amet, J.R. Williams, K. Watanabe, T. Taniguchi, D.G. Gordon, Nano Lett. 12 (2012) 4449–4454.

[56] D. Pacilé, J.C. Meyer, Ç.Ö. Girit, A. Zettl, Appl. Phys. Lett. 92 (2008) 133107.

[57] C. Li, Y. Bando, C. Zhi, Y. Huang, D. Golberg, Nanotechnology 20 (2009) 385707.

[58] K. Watanabe, T. Taniguchi, Int. J. Appl. Ceram. Technol. 8 (2011) 977–989.

[59] W. Gannett, W. Regan, K. Watanabe, T. Taniguchi, M.F. Crommie, A. Zettl, Appl. Phys. Lett. 98 (2011) 242105.

[60] Y. Kubota, K. Watanabe, O. Tsuda, T. Taniguchi, Science 317 (2007) 932–934.

[61] K.J. Watanabe, T. Taniguch, T. Niiyam, K. Miya, M. Taniguchi, Nat. Photonics. 3 (2009) 591–594.

[62] Z. Liu, Y.J. Gong, W. Zhou, L.L. Ma, J.J. Yu, J.C. Idrobo, J. Jung, A.H. MacDonald, R. Vajtaii, J. Lou, P.M. Ajayan, Nat. Commun. 4 (2013) 254.

[63] G.H. Lee, Y.J. Yu, C.G. Lee, C. Dean, K.L. Shepard, P. Kim, J. Hone, Appl. Phys. Lett. 99 (2011) 243114.

[64] L.H. Li, Y. Chen, G. Behan, H.Z. Zhang, M. Petravicc, A.M. Glushenkov, J. Mater. Chem. 21 (2011) 11862–11866.

[65] K.S. Novoselov, D. Jiang, F. Schedin, T.J. Booth, V.V. Khotkevich, S.V. Morozov, A.K. Geim, PNAS 102 (2005) 10451–10453.

[66] X.L. Li, X.P. Hao, M.W. Zhao, Y.Z. Wu, J.X. Yang, Y.P. Tian, G.D. Qian, Adv. Mater. 25 (2013) 2200–2204.

[67] F. Torrisi, T. Hasan, W. Wu, Z. Sun, A. Lombardo, T.S. Kulmala, G.W. Hsieh, S. Jung, F. Bonaccorso, P.J. Paul, D. Chu, A.C. Ferrari, ACS Nano 6 (2012) 2992–3006.

[68] P. Sutter, J. Lahiri, P. Zahl, B. Wang, E. Sutter, Nano Lett. 13 (2013) 276–281.

[69] J.M. Caicedo, G. Bejarano, G. Zambrano, E. Bacal, O. Morán, P. Prieto, Phys. Stat. Solidi B 242 (2005) 1920–1923.

[70] K.K. Kim, A. Hsu, X.T. Jia, S.M. Kim, Y.M. Shi, M. Dresselhaus, T. Palacios, J. Kong, ACS Nano 6 (2012) 8583–8590.

[71] M. Goto, A. Kasahara, M. Tosa, T. Kimura, K. Yoshihar, Appl. Surf. Sci. 185 (2002) 172–176.

[72] S. Nakhaie, J.M. Wofford, T. Schumann, U. Jahn, M. Ramsteiner, M. Hanke, J.M.J. Lopes, H. Riechert, Appl. Phys. Lett. 106 (2015) 213108.

[73] A.T. Barton, R. Yue, S. Anwar, H. Zhu, X. Peng, S. McDonnell, N. Lu, R. Addou, L. Colombo, M.J. Kim, R.M. Wallace, C.L. Hinkle, Microelectron. Eng. 147 (2015) 306–309.

[74] A.S. Rozenberg, Y.A. Sinenko, N.V. Chukanov, J. Mater. Science. 28 (1993) 5528–5533.

[75] V. Cholet, L. Vandenbulcke, J.P. Rouan, J. Mater. Sci. 29 (1994) 1417–1435.

[76] F. Rebillat, A. Guette, C.R. Brosse, Acta Mater. 47 (1999) 1685–1696.

[77] C.C. Tang, Y. Bando, Y. Huang, S.L. Yue, C.Z. Gu, F.F. Xu, D. Golberg, J. Am. Chem. Soc. 127 (2005) 6552–6553.

[78] T. Matsuda, J. Mater. Sci. 24 (1989) 2353–2358.

[79] J.S. Beck, C.R. Albani, A.R. McGhie, J.B. Rothman, L.G. Sneddon, Chem. Mater. 1 (1989) 433–438.

[80] S. Chatterjee, Z.T. Luo, M. Acerce, D.M. Yates, A.T.C. Johnson, L.G. Sneddon, Chem. Mater. 23 (2011) 4414–4416.

[81] L. Song, L.J. Ci, H. Lu, P.B. Sorokin, C.H. Jin, J. Ni, A.G. Kvashnin, D.G. Kvashnin, J. Lou, B.I. Yakobson, P.M. Ajayan, Nano Lett. 10 (2010) 3209–3215.

[82] Y.M. Shi, C. Hamsen, X.T. Jia, K.K. Kim, A. Reina, M. Hofmann, A.L. Hsu, K. Zhang, H. Li, Z.Y. Juang, M.S. Dresselhaus, L.J. Li, J. Kong, Nano Lett. 10 (2010) 4134–4139.

[83] Y.H. Lee, K.K. Liu, A.Y. Lu, C.Y. Wu, C.T. Lin, W.J. Zhang, C.Y. Su, C.L. Hsu, T.W. Lin, K.H. Wei, Y. Shid, L.J. Li, RSC Adv. 2 (2012) 111–115.

[84] W. Auwarter, M. Muntwiler, J. Osterwalder, T. Greber, Surf. Sci. 545 (2003) L735–L740.

[85] W. Auwarter, H.U. Suter, H. Sachdev, T. Greber, Chem. Mater. 16 (2004) 343–345.

[86] K.K. Kim, A. Hsu, X.T. Jia, S.M. Kim, Y.M. Shi, M. Hofmann, D. Nezich, J.F.R. Nieva, M. Dresselhaus, T. Palacios, J. Kong, Nano Lett. 12 (2012) 161–166.

[87] K.H. Lee, H.J. Shin, J. Lee, I. Lee, G.H. Kim, J.Y. Choi, S.W. Kim, Nano Lett. 12 (2012) 714–718.

[88] L.F. Wang, B. Wu, J.S. Chen, H.T. Liu, P.A. Hu, Y.Q. Liu, Adv. Mater. 26 (2014) 1559–1564.

[89] J. Han, J.Y. Lee, H. Kwon, J.S. Yeo, Nanotechnology 25 (2014) 145604.

[90] P.R. Kidambi, R. Blume, J. Kling, J.B. Wagner, C. Baehtz, R.S. Weatherup, R. Schloegl, B.C. Bayer, S. Hofmann, Chem. Mater. 26 (2014) 6380–6392.

[91] G. Kim, A.R. Jang, H.Y. Jeong, Z. Lee, D.J. Kang, H.S. Shin, Nano Lett. 13 (2013) 1834–1839.

[92] R.Y. Tay, S.H. Tsang, M. Loeblein, W.L. Chow, G.C. Loh, J.W. Toh, S.L. Ang, E.H.T. Teo, Appl. Phys. Lett. 106 (2015) 101901.

[93] A. Ismach, H. Chou, D.A. Ferrer, Y.P. Wu, S. McDonnell, H.C. Floresca, A. Covacevich, C. Pope, R. Piner, M.J. Kim, R.M. Wallace, L.G. Colombo, R.S. Ruoff, ACS Nano 6 (2012) 6378–6385.

[94] A.B. Preobrajenski, A.S. Vinogradov, N. Martensson, Surf. Sci. 582 (2005) 21–30.

[95] G.Y. Lu, T.R. Wu, Q.H. Yuan, H.S. Wang, H.M. Wang, F. Ding, X.M. Xie, M.H. Jiang, Nat. Commun. 6 (2015) 1–8.

[96] X.S. Li, Y.W. Zhu, W.W. Cai, M. Borysiak, B.Y. Han, D. Chen, R.D. Piner, L. Colombo, R.S. Ruoff, Nano Lett. 9 (2009) 4359–4363.

[97] K. Qian, R.Y. Tay, V.C. Nguyen, J.K. Wang, G.F. Cai, T.P. Chee, E. Hang, T. Teo, P.S. Lee, Adv. Funct. Mater. 26 (2016) 2176–2184.

[98] J.J.J. Wang, Y.F. Yang, Y.A. Chee, K. Watanabe, T. Taniguchi, H.O.H. Churchill, P. Jarillo-Herrero, Nano Lett. 15 (2015) 1898–1903.

[99] K. Watanabe, T. Taniguchi, H. Kanda, Nat. Mater. 3 (2004) 404–409.

[100] M.S. Bresnehan, M.J. Hollander, M. Wetherington, M. LaBella, K.A. Trumbull, R. Cavallero, D.W. Snyder, J.A. Robinson, ACS Nano 6 (2012) 5234–5241.

[101] E. Kim, T.H. Yu, E.S. Song, B. Yu, Appl. Phys. Lett. 98 (2011) 262103.

[102] K.K. Kim, A. Hsu, X.T. Jia, S.M. Kim, Y.M. Shi, M. Hofmann, D. Nezich, J.F. Rodriguez-Nieva, M. Dresselhaus, T. Palacios, J. Kong, Nano Lett. 12 (2012) 161–166.

[103] L.H. Li, J. Cervenka, K. Watanabe, T. Taniguchi, Y. Chen, ACS Nano 8 (2014) 1457–1462.

[104] A. Lipp, K.A. Schwetz, K. Hunold, J. Eur. Ceram. Soc. 5 (1989) 3–9.

[105] M.W. Iqbal, M.Z. Iqbal, M.F. Khan, M.A. Shehzad, Y. Seo, J.H. Park, C. Hwang, J. Eom, Sci. Rep. 5 (2015) 10699.

[106] C.R. Dean, A.F. Young, I. Meric, C. Lee, L. Wang, S. Sorgenfrei, K. Watanabe, T. Taniguchi, P. Kim, K.L. Shepard, J. Hone, Nat. Nanotechnol. 5 (2010) 722–726.

[107] A.S. Mayorov, R.V. Gorbachev, S.V. Morozov, L. Britnell, R. Jalil, L.A. Ponomarenko, P. Blake, K.S. Novoselov, K. Watanabe, T. Taniguchi, A.K. Geim, Nano Lett. 11 (2011) 2396–2399.

[108] J. Schiefele, F. Sols, F. Guinea, Phys. Rev. B 85 (2012) 195420.

[109] C.R. Dean, A.F. Young, P. Cadden-Zimansky, L. Wang, H. Ren, K. Watanabe, T. Taniguchi, P. Kim, J. Hone, K.L. Shepard, Nat. Phys. 7 (2011) 693–696.

[110] T. Taychatanapat, K. Watanabe, T. Taniguchi, P. Jarillo-Herrero, Nat. Phys. 7 (2011) 621–625.

[111] K.M. Burson, M. Kristen, W.G. Cullen, S. Adam, C.R. Dean, K. Watanabe, T. Taniguchi, P. Kim, M.S. Fuhrer, Nano Lett. 13 (2013) 3576–3580.

[112] T. Chari, I. Meric, C. Dean, K. Shepard, IEEE Trans. Electron Devices 62 (2015) 4322–4326.

[113] N. Gao, J.Q. Wei, Y. Jia, H.H. Sun, Y.H. Wang, K.H. Zhao, X.L. Shi, L.W. Zhang, X.M. Li, A.Y. Cao, H.W. Zhu, K.L. Wang, D.H. Wu, Nano Res. 6 (2013) 602–610.

[114] G. Shi, Y. Hanlumyuang, Z. Liu, Y.J. Gong, W.L. Gao, B. Li, J. Kono, J. Lou, R. Vajtai, P. Sharma, P.M. Ajayan, Nano Lett. 14 (2014) 1739–1744.

[115] M. Nafria, R. Rodriguez, M. Porti, J. Martin-Martinez, M. Lanza, X. Aymerich, Electron Devices Meeting (IEDM), 2011 IEEE International, 6.3. 1–6.3. 4

[116] M. Lanza, M. Porti, M. Nafria, G. Benstetter, W. Frammelsberger, H. Ranzinger, E. Lodermeier, G. Jaschke, Microelectron. Reliab. 47 (9) (2007) 1424–1428.

[117] M. Lanza, V. Iglesias, M. Porti, M. Nafria, X. Aymerich, Nanoscale Res. Lett. 6 (2011) 108.

[118] M. Lanza, M. Porti, M. Nafria, X. Aymerich, G. Benstetter, E. Lodermeier, H. Ranzinger, G. Jaschke, S. Teichert, L. Wilde, P.P. Michalowski, Nanotechnol. IEEE Trans. Electron Devices 10 (2) (2011) 344–351.

[119] V. Iglesias, M. Lanza, K. Zhang, A. Bayerl, M. Porti, M. Nafria, X. Aymerich, G. Benstetter, Z.Y. Shen, G. Bersuker, Appl. Phys. Lett. 99 (10) 103510.

[120] K. Shubhakar, K.L. Pey, M. Bosman, R. Thamankar, S.S. Kushvaha, Y.C. Loke, Z.R. Wang, N. Raghavan, X. Wu, S.J. O'Shea, IEEE Int. Symp. on Physical and Failure Analysis of Integrated Circuits (IPFA), 2012 1–7.

[121] K. Shubhakar, N. Raghavan, K.L. Pey, Int. J. Sci. Technol. 2 (2014) 81–86.

[122] L. Britnell, R.V. Gorbachev, R. Jalil, B.D. Belle, F. Schedin, M.I. Katsnelson, L. Eaves, S.V. Morozov, A.S. Mayorov, N.M.R. Peres, A.H.C. Neto, J. Leist, A.K. Geim, L.A. Ponomarenko, K.S. Novoselov, Nano Lett. 12 (2012) 1707–1710.

[123] G. Shi, Y. Hanlumyuang, Z. Liu, Y.J. Gong, W.L. Gao, B. Li, J. Kono, J. Lou, R. Vajtai, P. Sharma, P.M. Ajayan, Nano Lett. 14 (2014) 1739–1744.

[124] Y.Y. Liu, X.L. Zou, B.I. Yakobson, ACS Nano 6 (2012) 7053–7058.

[125] T.W. Lin, C.Y. Su, X.Q. Zhang, W.J. Zhang, Y.H. Lee, C.W. Chu, H.Y. Lin, M.T. Chang, F.R. Chen, L.J. Li, Small 8 (2012) 1384–1391.

[126] A. Singh, U.V. Waghmare, Phys. Chem. Chem. Phys. 16 (2014) 21664–21672.

[127] Q.C. Li, X.L. Zou, M.X. Liu, J.Y. Sun, Y.B. Gao, Y. Qi, X.B. Zhou, B.I. Yakobson, Y.F. Zhang, Z.F. Liu, Nano Lett. 15 (2015) 5804–5810.

[128] A.L. Gibb, N. Alem, J.H. Chen, K.J. Erickson, J. Ciston, A. Gautam, M. Linck, A. Zettl, J. Am. Chem. Soc. 135 (2013) 6758–6761.

[129] K. Kim, Z. Lee, W. Regan, C. Kisielowski, M.F. Crommis, A. Zettl, ACS Nano 5 (2011) 2141–2146.

[130] W. Zhou, X.L. Zou, S. Najmaei, Z. Liu, Y.M. Shi, J. Kong, J. Lou, P.M. Ajayan, B.I. Yakobson, J.C. Idrobo, Nano Lett. 13 (2013) 2615–2622.

[131] Y. Zhang, Y.F. Zhang, Q.Q. Ji, J. Ju, H.T. Yuan, J.P. Shi, T. Gao, D.L. Ma, M.X. Liu, Y.B. Chen, X.J. Song, H.Y. Hwang, Y. Cui, Z.F. Liu, ACS Nano 7 (2013) 8963–8971.

[132] J.M. Carlsson, L.M. Ghiringhelli, A. Fasolino, Phys. Rev. B 84 (2011) 165423.

[133] W.J. Zhu, T. Tamagawa, M. Gibson, T. Furukawa, IEEE Electron. Device Lett. 23 (2002) 649–651.

[134] T. Yamaguchi, H. Satake, N. Fukushima, A. Toriumi, IEDM Tech. Dig. (2000) 19–22.

[135] Y.L. Huang, Y. Chen, W. Zhang, S.Y. Quek, C.H. Chen, L.J. Li, W.T. Hsu, W.H. Chang, Y.J. Zheng, W. Chen, A.T.S. Wee, Nat. Commun. 6 (2015) 6298.

[136] O.V. Yazyev, S.G. Louie, Nat. Mater. 9 (2010) 806–809.

[137] C.X. Ma, H.F. Sun, Y.L. Zhao, B. Li, Q.X. Li, A.D. Zhao, X.P. Wang, Y. Luo, J.L. Yang, B. Wang, J.G. Hou, Phys. Rev. Lett. 112 (2014) 226802.

[138] J. Cervenka, M.I. Katsnelson, C.F. Flipse, J. Nat. Phys. 5 (2009) 840–844.

[139] R. Grantab, V.B. Shenoy, R.S. Ruoff, Science 30 (2010) 946–948.

[140] S. Bae, H. Kim, Y.B. Lee, X.F. Xu, J.S. Park, Y. Zheng, J. Balakrishnan, T. Lei, H.R. Kim, Y. Song, Y.J. Kim, K.S. Kim, B. Ozyilmaz, J.H. Ahn, B.H. Hong, S. Iijima, Nat. Nanotechnol. 5 (2010) 574–578.

[141] G.Y. Lu, T.R. Wu, Q.H. Yuan, H.S. Wang, H.M. Wang, F. Ding, X.M. Xie, M.H. Jiang, Nat. Commun. 6 (2015) 6160.

[142] Y.F. Ji, C.B. Pan, M.Y. Zhang, S.B. Long, X.J. Lian, F. Miao, F. Hui, Y.Y. Shi, L. Larcher, E. Wu, M. Lanza, Appl. Phys. Lett. 108 (2016) 012905.

[143] X.S. Li, W.W. Cai, J. An, S.Y. Kim, J. Nah, D.X. Yang, R. Piner, A. Velamakanni, I. Jung, E. Tutuc, S.K. Banerjee, L. Colombo, R.S. Ruoff, Science 324 (2009) 1312–1314.

[144] Graphene Supermarket, https://graphene-supermarket.com/home.php, consulted on May 8th of 2016.

[145] L. Song, L.J. Ci, H. Lu, P.B. Sorokin, C.H. Jin, J. Ni, A.G. Kvashnin, D.G. Kvashnin, J. Lou, B.I. Yakobson, P.M. Ajayan, Nano Lett. 10 (2010) 3209–3215.

[146] J.Y. Luo, H.D. Jang, J.X. Huang, ACS Nano 7 (2013) 1464–1471.

[147] Y. Nanishi, S. Ishida, T. Honda, H. Yamazaki, S. Miyazawa, J. Appl. Phys. 21 (1982) L335.

[148] N. Liu, Z.H. Pan, L. Fu, C.H. Zhang, B.Y. Dai, Z.F. Liu, Nano Res. 4 (2011) 996–1004.

[149] M. Lanza, Y. Wang, A. Bayerl, T. Gao, M. Porti, M. Nafria, H. Liang, G. Jing, Z. Liu, Y. Zhang, Y. Tong, H. Duan, J. Appl. Phys. 113 (2013) 104301.

[150] S. Vinod, C.S. Tiwary, P.A.S. Autreto, J.T.S. Ozden, A.C. Chipara, P. Vajtai, D.S. Galva, T.N. Narayanan, P.M. Ajayan, Nat. Commun. 5 (2014) 4541.

[151] X.Y. Chen, D.H. Seo, S. Seo, H. Chung, H.S.P. Wong, IEEE Electron. Device Lett. 33 (2012) 1604–1606.

[152] M. Lanza, Y. Wang, T. Gao, A. Bayerl, M. Porti, M. Nafria, Y.B. Zhou, G.Y. Jing, Y.F. Zhang, Z.F. Liu, D.P. Yu, H.L. Duan, Nano Res. 6 (2013) 485–495.

[153] C.H. Ho, S.Y. Kim, K. Roy, Microelectron. Reliab. 55 (2015) 308–317.

[154] R. Degraeve, B. Kaczer, G. Groeseneken, Microelectron. Reliab. 39 (1999) 1445–1460.

[155] D.J. DiMaria, E. Cartier, D. Arnold, J. Appl. Phys. 73 (1993) 3367–3384.

[156] D.J. Dimaria, E. Cartier, J. Appl. Phys. 78 (1995) 3883–3894.

[157] L. Vandelli, A. Padovani, L. Larcher, G. Bersuker, IEEE Trans. Electron Devices 60 (2013) 1754–1762.

[158] E. Miranda, J. Sune, R. Rodriguez, M. Nafria, X. Aymerich, Appl. Phys. Lett. 73 (1998) 490–492.

[159] Y. Hattori, T. Taniguchi, K. Watanabe, K. Nagashio, ACS Nano 9 (2015) 916–921.

[160] T. Ando, Materials 5 (2012) 478–500.

[161] L. Aguilera, M. Porti, M. Nafria, X. Aymerich, IEEE Electron Device Lett. 27 (2006) 157–159.

[162] M. Lanza, M. Porti, M. Nafria, X. Aymerich, E. Whittaker, B. Hamilton, Microelectron. Reliab. 50 (2010) 1312–1315.

[163] M. Lanza, M. Porti, M. Nafria, X. Aymerich, E. Wittaker, B. Hamilton, Rev. Sci. Instrum. 81 (2010) 106110.

[164] W. Frammelsberger, G. Benstetter, J. Kiely, R. Stamp, Appl. Surf. Sci. 253 (2007) 3615–3626.

[165] X. Blasco, Universitat Autonoma de Barcelona(PhD Thesis) May 2005.

[166] M. Lanza, K. Zhang, M. Porti, M. Nafria, Z.Y. Shen, L.F. Liu, J.F. Kang, D. Gilmer, G. Bersuker, Appl. Phys. Lett. 100 (2012) 123508.

[167] Y. Shi, Y. Ji, F. Hui, M. Nafria, M. Porti, G. Bersuker, M. Lanza, Adv. Electron. Mater. 1 (2015) 1400058.

[168] Y.Y. Shi, Y.F. Ji, F. Hui, V. Iglesias, M. Porti, M. Nafria, E. Miranda, G. Bersuker, M. Lanza, ECS Trans. 64 (2014) 19–28.

[169] X. Li, C.H. Tung, K.L. Pey, V.L. Lo, Appl. Phys. Lett. 94 (2009) 132904.

[170] G. Bersuker, J. Yum, L. Vandelli, A. Padovani, L. Larcher, V. Iglesias, M. Porti, M. Nafria, K. McKenna, A. Shluger, P. Kirsch, R. Jammy, Solid State Electron. 65 (2011) 146–150.

[171] X. Wu, K.L. Pey, N. Raghavan, W.H. Liu, X. Li, P. Bai, G. Zhang, M. Bosman, Nanotechnology 22 (2011) 455702.

[172] M. Lanza, Materials 7 (2014) 2155–2182.

[173] M. Lanza, G. Bersuker, M. Porti, E. Miranda, M. Nafria, X. Aymerich, Appl. Phys. Lett. 101 (2012) 193502.

[174] A. Zobelli, C.P. Ewels, A. Gloter, G. Seifert, Phys. Rev. B 75 (2007) 094104.

[175] N. Jain, T. Bansal, C.A. Durcan, Y. Xu, B. Yu, Carbon 54 (2013) 396–402.

[176] K.K. Kim, A. Hsu, X.T. Jia, S.M. Kim, Y.M. Shi, M. Dresselhaus, T. Palacios, J. Kong, ACS Nano 6 (2012) 8583–8590.

[177] P. Sutter, J. Lahiri, P. Zahl, B. Wang, E. Sutter, Nano Lett. 13 (2013) (R).

[178] E.Y. Wu, J. Sune, Microelectron. Reliab. 45 (2005) 1809–1834.

[179] N. Raghavan, K.L. Pey, K. Shubhakar, M. Bosman, IEEE Electron Device Lett. 32 (2011) 78–80.

# Chapter 3:

# Electrical homogeneity and reliability of *h*-BN grown on different substrates

As mentioned, the main measurements defining the performance of an insulating material when used as dielectric in electronic devices are: *i)* tunneling current, *ii)* charge trapping and de-trapping, *iii)* trap-assisted tunneling, *iv)* SILC, *v)* dielectric strength, *vi)* soft/hard BD, and *vii)* RS. Electrical homogeneity is defined as the variability of these properties from one location of the *h*-BN to another. Therefore, the study of electrical homogeneity requires the use of electrical characterization techniques with high lateral resolution, like the CAFM. Reliability is defined as the ability of a dielectric to keep its insulating properties when exposed to an electrical stress in a device. In this case the reliability of the entire *h*-BN stack is defined by the weakest location, as the BD event is a stochastic process. In this chapter these two properties are analyzed for *h*-BN stacks grown on different metallic substrates.

## 3.1. Growth *h*-BN on Pt, Cu and Fe substrates by CVD

The *h*-BN stacks used in this PhD thesis have been grown via CVD approach in two different laboratories. The first one is the laboratory of Prof. Xiaoming Xie at the Shanghai Institute of Microsystem and Information Technology, and the substrate used was always Ni-doped Cu. The second one is the laboratory of Prof. Jing Kong at the Massachusetts Institute of Technology, and the substrates used were mainly Pt and Fe, although some tests with Cu foils have been also performed. The only growth method we used in this thesis is CVD because it has been proved that is the scalable approach that provides the best quality (i.e. the largest amount of layered area with the lowest density of defects).

Cu was the substrate initially used by the 2D materials community to grow *h*-BN stacks, due to its low cost and good catalytic activity. However, controlling the thickness of *h*-BN stacks grown on Cu is not easy. Some works even claimed (erroneously) the growth of multilayer *h*-BN on Cu while showing evident amorphous structure (i.e. bad quality) in their own TEM images [65-66]. For this reason, here we explore: *i)* the introduction of new treatments to growth *h*-BN on Cu substrates, and *ii)* the use of different metallic substrates for the growth of *h*-BN.

First of all, we doped the Cu substrates with Ni. The cleaning process of Cu substrates (25 µm thick) is different from that of Pt and Fe substrates (1 mm thick). Cu substrates are first immersed in nitric acid (purity 19%) and sonicated for 30 s. Then they are immersed in pure water and sonicated for 5 minute to clean the surface. Finally, the Cu substrates are dried using $N_2$. Pt and Fe substrates were cleaned using the electro-polishing method. The Pt/Fe substrate and a counter electrode are immersed in a solution containing 940 ml acetic acid and 60 ml perchloric acid connected to the positive and negative terminals of an electrochemical workstation. The two electrodes are polarized under a bias 30 V for 30 s, which removed the oxides on the surface of the metallic substrates. This process is repeated three times, and finally the substrates are cleaned in pure water (under sonication) for 5 minutes and dried using a $N_2$ gun.

After cleaning, the substrates are placed in the center of the tube furnace, which is connected to two gas lines: the first one (line 1) drives $H_2$ inside the tube, and the second one (line 2) drives the precursor. For the growth of graphene the precursor is typically a gas (methane, Argon), but in the case of *h*-BN the precursor we used is liquid borazine. Other works used solid ammonia borane powder heated [67]. Therefore, in the precursor line (line 2) a gas was necessary to drive the borazine molecules (seeds) inside the tube. We also used $H_2$ because it has provided good results in previous literature [68]. Both lines are controlled by a valve, and both of them are closed initially.

After introducing the substrates in the center of the tube furnace, the system was closed and a pump working at a certain pressure ($P$) was used to remove the air molecules inside the tube. Once the desired vacuum level is reached, a constant flow of $H_2$ was immersed in the tube using line 1 (namely $F_1$), and the temperature of the tube furnace was ramped up from room temperature (RT) to the annealing temperature ($T_A$). $T_A$ was maintained constant during the annealing time ($t_A$). Despite the flow of $H_2$ in the tube furnace during the annealing step may not be strictly necessary, it has been demonstrated that this helps to the formation of 2D materials [68]. After the annealing, a constant flow of $H_2$ was introduced in the tube furnace using line 2 ($F_2$), which carried borazine molecules into the tube for reaction with the metallic substrate. The time that the metallic substrate was exposed to the precursor is called growth time ($t_G$), and during $t_G$ the temperature of the furnace was set at a suitable value for $h$-BN growth ($T_G$), which depends on the metallic substrate used. After that, the furnace was cooled to RT. When using Pt substrate the cooling down time ($t_C$) was quite fast, and when using Fe substrates this time needed to be enlarged.

The value of these parameters strongly affects the quality of the $h$-BN stack, and needed to be found experimentally. We are aware that world leading scientists in the field of CVD-grown 2D materials suffered strong difficulties to tune these parameters, especially when moving to a new lab. By empirical experience and collaborations with other groups we have corroborated that the use of identical parameters in two different labs may lead to different thicknesses and amount of defects. Most probably experimental factors like the brand of the equipment used and the purity and cleanness of the tube lines and metallic foils may play a role. Table 3.1 summarizes the values of the CVD parameters that we used to produce the best *h*-BN quality on Cu, Pt and Fe substrates. In order to find them, I repeated the process > 20 times for each material.

**Table 3.1**. CVD growth parameters of *h*-BN on different substrates.

| Substrate | Annealing process (℃/min) | Growth temperature (℃) | Flow rate ratio of borazine/$H_2$ | Growth time (min) | Cooling down rate (℃/min) |
|---|---|---|---|---|---|
| Cu | 1000/30 | 1000 | 1/55 | 60 | 30 |
| Pt | 1000/30 | 1000 | 1/700 | 60 | 30 |
| Fe | 1100/30 | 1100 | 1/70 | 120 | 5 |

Mainly two different chemical mechanisms producing the growth of layered *h*-BN at the surface of the metallic foils have been detected. The first one is surface-mediated growth mechanism, which happens in the metallic substrates with high solubility of boron but with/low nitrogen solubility under high temperature, such as Pt and Cu substrates [69]. The *h*-BN layers form on the surface of the catalytic substrates by decomposition of precursor and boron and nitrogen atoms deposited on the catalytic substrate. The second one is precipitation reaction, in which the metallic substrates have low solubility of boron and nitrogen; then both atoms will segregate from the substrate and form the *h*-BN layers during the cooling down process [69]. This mechanism is related to the use of Fe substrates. However, in some cases, mixed mechanisms with both surface-mediated and precipitation may contribute to the formation of thick *h*-BN stacks [70].

Generally speaking, the use of larger $F_2$, $t_G$ and $t_C$ result in thicker $h$-BN stacks. $F_2$ tunes the number of seeds on the metallic substrate, which will also control the domain size. Ref. [71] achieved centimeter scale single crystal graphene using a single seed deposited suing an external pipette but, to the best of our knowledge, this strategy has never been tried using $h$-BN. $T_A$ and $T_G$ are normally the same, and if $T_G$ is too low the borazine seeds cannot decompose properly to achieve the conformal growth of $h$-BN layers. Therefore, $T_G$ has a deep influence on the amount of defects within the $h$-BN stack. $T_G$ is selected mainly depending on the melting temperature of the substrate used, i.e. the substrate needs to be heated at a relative high temperature (lower that its melting point) in order to facilitate atomic rearrangements. It is also worth noting that under high temperature, the $H_2$ gas could dissolve into the substrate, which facilitates the $h$-BN growth. The pressure used during the entire process was set at a constant value between 35 and 70 mtorr. This modest vacuum level is already enough to extract the air inside the tube and provide a reasonable clean atmosphere for the CVD process. Some groups used a CVD system working in ultra high vacuum (UHV) [72], but the complexity of the process increased dramatically, and the quality improvement is not remarkable. Independently of the material used, the growth of $h$-BN has been always carried out on metallic foils. Moreover, it has been reported that monolayer $h$-BN stacks can also be synthesized on wafers coated with 500 nm Fe metal films [73]. However, this has been only reported by one group, which didn't repeat the experiments.

## 3.2. Characterization of the samples and devices

The characterization of the samples has been mostly carried out at the laboratory of Prof. Mario Lanza at Soochow University, although few data have been also collected at the laboratory of Prof. Jing Kong at Massachusetts Institute of Technology.

As-grown $h$-BN stacks have been usually transferred on a 300 nm $SiO_2$/Si wafer for morphological characterization. The optical microscopy and SEM images have been used to analyze the continuity of the $h$-BN sheet, as well as the density and shape of cracks, wrinkles and multilayer islands. Several different positions of the samples were analyzed with Raman spectroscopy to evaluate the thickness and quality of the $h$-BN stack. Topographic AFM maps have been also collected to quantify the roughness of the $h$-BN surface, the thickness of the $h$-BN stacks, and the accurate width and height of the wrinkles. All these analyses can provide valuable information about the $h$-BN stacks, but it should be highlighted that the only technique that can 100% ensure the layered structure of the $h$-BN stacks (and other 2D materials) is cross sectional transmission electron microscopy (TEM). Therefore, cross sectional TEM images must be provided in all experimental works claiming the growth and/or characterization of any 2D layered material. Unfortunately, not all authors follow this good recommendation, and some others show cross sectional TEM images with no signal of layered structure [64-65]. In this work we always used cross sectional TEM images to confirm the thickness, layered structure and amount of defects in our samples. To do so, we followed two different procedures depending on lab availability: $i)$ thin lamellas have been cut using focused ion beam (FIB), and placed later on the target TEM grids using vacuum tweezers. During the FIB, a protective Ti/Au/Cr stack was deposited on the surface of the $h$-BN stack in order to protect it from high impact energies, which may produce defects. And $ii)$ the $h$-BN can be directly transferred on a TEM grid and the TEM user needs to find a fold, so that the cross section can be monitored from the top view. Despite this second approach brings associated more randomness, the number of folds on the TEM grid is usually large, and it allows rapid localization. On the other hand, direct transfer on TEM grids is cheaper than FIB.

The most innovative characterization method presented in this PhD thesis is the use of CAFM. CAFM consists on placing a very sharp and conductive probe tip on the surface of the *h*-BN, which needs to be placed on a conductive substrate (e.g. the *h*-BN can be studied directly on the substrate where it was grown, or transferred onto a different conductive substrate). When a voltage is applied between the tip and the sample holder an electrical field will be generated, leading to a vertical current flow across the *h*-BN stack. CAFM is a versatile tool that can collect the topographic and current maps simultaneously; based on this, we can make a spatial connection between the conductive or insulating properties and the morphological features. Moreover, the CAFM can apply electrical stresses to study the entire BD process of the *h*-BN sheets. Furthermore, for the first time we analyze the BD spot using adhesion and deformation maps, which can give information about the sign of the traps inside the dielectric, as well as their mobility.

Finally, we also carry out the device level electrical characterization using probestation. To do so, we fabricated real devices by depositing top electrodes on the *h*-BN stacks (as grown on the metallic substrates) using a laser-patterned shadow mask and an electron beam evaporator. The use of a mask patterned by laser is essential to avoid large variability on the electrode size. Using this process, we fabricated Au/Ti/*h*-BN/Pt and Au/Ag/*h*-BN/Fe devices.

### 3.3. Results and discussion

By using the parameters described in Table 3.1, in **Articles 2**, **3** and **4** we successfully grew premium quality monolayer and/or multilayer *h*-BN on Pt, Cu, and Fe (respectively). Optical images indicate that the *h*-BN is continuous and that it has a low density of cracks. Raman spectra show the characteristic $h$-BN peak between 1366 cm$^{-1}$ and 1370 cm$^{-1}$, and we are able to distinguish between monolayer, bilayer and multilayer $h$-BN depending on the position of the Raman peak. The number of $h$-BN layers can be identified TEM images, which also prove that our $h$-BN contains low amount of defects.

**Article 2** shows that the thickness of multilayer $h$-BN grown on polycrystalline Pt substrates depends on the crystallographic orientation of the surface of the Pt substrate, which is different for each Pt grain. CAFM studies reveal remarkable different tunneling currents across the $h$-BN stack from one Pt grain to another, due to the different thicknesses. However, when measuring within the same Pt grain (which diameters are typically 60-200 μm) the tunneling current fluctuations are very low, much lower than that across amorphous HfO$_2$ and TiO$_2$. These observations are corroborated at the device level, i.e. Au/Ti/$h$-BN/Pt devices within the same grain show very similar pre-BD currents and dielectric strength. Therefore, $h$-BN grown on single crystalline metallic substrates may enable the fabrication of low variability electronic devices. In **Article 3** we show that the tunneling currents across $h$-BN stacks grown on Ni-doped Cu substrates do not remarkably change from one Cu grain to another, and compare this observation with the data obtained on Pt.

**Article 4** compares the BD process in monolayer and multilayer $h$-BN stacks grown on polycrystalline Cu substrates. We observe that multilayer $h$-BN shows a BD event that is followed by the formation of large hillocks, which is a behavior that also happens in ultra-thin SiO$_2$, HfO$_2$ and Al$_2$O$_3$. However, monolayer $h$-BN does not show such hillocks (i.e. it can keep the structural properties unaltered) even after more severe BD. This behavior is attributed to the high thermal conductivity of monolayer $h$-BN, which should be able to spread the local heat through the adjacent metals, avoiding surface modification. In this work we introduce two very novel analyses: *i)* by collecting adhesion maps at the BD location, we find out that the BD spot contains negative charges, which are surrounded by an area rich in positive charges. And by collecting deformation maps, we find out that the negative charges are fixed at the center and mobile at the surroundings, while the positive charges are just fixed. And *ii)* we use ionic liquid electrical stress combined with CAFM and cross sectional TEM to characterize structure of the *h*-BN stack without the need of removing a solid top electrode. Our experiments reveal the presence of mainly two types of defects promoting high currents in the *h*-BN stack: lattice distortions (e.g. B vacancies) and metal particles/clusters penetration from the substrate.

In **Article 5**, *h*-BN stacks have been successfully grown on Fe foils, and Au/Ag/*h*-BN/Fe structures have been used to fabricate memristive devices. Memristors are elements that can change their electrical resistivity depending on the history of electrical impulses previously applied. The concept of memristor was firstly suggested in 1971 [74], and developed in 2008 [75]. Most of the memristors studied in the literature use a metal/insulator/metal (MIM) structure, which is in most cases vertically aligned to reduce space. These memristive MIM cells can switch their resistivity cyclically between (at least) two stable resistive states, namely high resistive state (HRS) and low resistive state (LRS). The performance of a memristor is defined by different performances: *i)* switching speed, *ii)* switching energy, *iii)* endurance (i.e. number of cycles that a memristor can switch before one of the states becomes permanent), *iv)* retention (i.e. minimum time that the memristor stays in the desired state without spontaneous state change), *v)* device size, *vi)* device integration. Despite the best memristive performances have been achieved using transition metal oxides (TMOs) as insulator in the MIM cell, the integration of 2D mateirals in this structure has started to show interesting properties, and novel performances that even the most advanced metal/TMO/metal devices cannot achieve have been detected.

In **Article 5** we fabricate Au/Ag/$h$-BN/Fe memristors, and they have been analyzed by applying electrical stresses with opposed polarities. We observe that the BD induced by applying positive bias to the top Ag electrode is weak and recovers when the electrical field vanishes, leading to volatile BD that can be used to emulate threshold RS devices. On the contrary, if the BD is induced by applying negative bias to the top Ag electrode the BD event is non-volatile, meaning that the insulating properties of the $h$-BN stack can be only recovered by applying an additional stress of opposed polarity. This behavior can be used to emulate bipolar RS devices. The different RS behaviors may be related to the different compositions at the BD spot, which is formed by Ag/Fe ions when positive/negative bias is applied to the top Ag electrode. Our study concludes that $h$-BN may be suitable for the fabrication of memristors, especially those dedicated to emulate electronic synapses that require both volatile and non-volatile switching.

# Electrical Homogeneity of Large-Area Chemical Vapor Deposited Multilayer Hexagonal Boron Nitride Sheets

Fei Hui,[‡,†,§] Wenjing Fang,[†,§] Wei Sun Leong,[§,⊥] Tewa Kpulun,[∥] Haozhe Wang,[§] Hui Ying Yang,[⊥] Marco A. Villena,[‡,#] Gary Harris,[∥] Jing Kong,[§] and Mario Lanza*[‡]

[‡]Institute of Functional Nano & Soft Materials, Collaborative Innovation Center of Suzhou Nano Science and Technology, Soochow University, Suzhou 215123, China

[§]Department of Electrical Engineering and Computer Science, Massachusetts Institute of Technology, Cambridge, Massachusetts 02139, United States

[⊥]Pillar of Engineering Product Development, Singapore University of Technology and Design, 8 Somapah Road, Singapore 487372, Singapore

[∥]Department of Electrical and Computer Engineering, Howard University, Washington, D.C. 20059, United States

[#]Department of Materials Science and Engineering, Stanford University, Stanford, California 94305, United States

**S** Supporting Information

**ABSTRACT:** Large-area hexagonal boron nitride (*h*-BN) can be grown on polycrystalline metallic substrates via chemical vapor deposition (CVD), but the impact of local inhomogeneities on the electrical properties of the *h*-BN and their effect in electronic devices is unknown. Conductive atomic force microscopy (CAFM) and probe station characterization show that the tunneling current across the *h*-BN stack fluctuates up to 3 orders of magnitude from one substrate (Pt) grain to another. Interestingly, the variability in the tunneling current across the *h*-BN within the same substrate grain is very low, which may enable the use of CVD-grown *h*-BN in ultra scaled technologies.

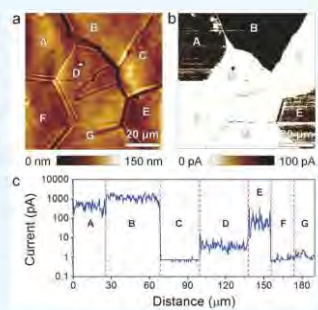

**KEYWORDS:** hexagonal boron nitride, chemical vapor deposition, electrical homogeneity, conductive AFM, polycrystalline

Hexagonal boron nitride (*h*-BN) is a layered insulator (direct band gap ~5.9 eV),[1] in which boron and nitrogen atoms arrange in a sp² hexagonal structure by covalent bonding, whereas the layers stick to each other by van der Waals attraction. Given its high in-plane mechanical strength (500 N/m),[2] large thermal conductivity (600 Wm⁻¹K⁻¹),[3] and high chemical stability (up to 1500 °C in air),[4] *h*-BN has attracted much attention for a wide range of potential applications. For example, thanks to their ultraflat surface free of dangling bonds, *h*-BN substrates can increase the mobility of graphene-based FETs up to ~140 000 cm²V⁻¹ s⁻¹·⁵ (on SiO₂ substrates it is lower, 15 000 cm² V⁻¹ s⁻¹). But its use as dielectric in different types of devices (e.g., field effect transistors memristors) is much more promising, as *h*-BN has shown enhanced reliability (compared to HfO₂),[6] characteristic layer-by-layer dielectric breakdown process,[7] and resistive switching.[8,9]

Chemical vapor deposition (CVD) in ultra high vacuum (UHV),[10,11] atmospheric pressure (APCVD) and low pressure (LPCVD)[12-15] is an attractive approach to synthesize large-area *h*-BN stacks with low density of defects on metallic foils (e.g., Cu, Fe, Pt). Because of the high temperatures required for the CVD growth of *h*-BN (>800 °C), the metallic substrates become polycrystalline. ref.[15] reported that the thickness of the *h*-BN stack grown by CVD on polycrystalline Ni depends on the size of the Ni crystal underneath, and it was also suggested that *h*-BN grows faster on the surface of Ni (100) than Ni (111), because of the different catalytic reaction activities.[16] Similar observations have been recently reported for *h*-BN stacks grown via CVD on polycrystalline Pt substrates.[17] However, the impact of these thickness fluctuations on the local electrical properties of the *h*-BN stacks and their effect on the performance of electronic devices is still unknown. This information is essential to understand and control the device-to-device variability, which has been reported as one of the major problems of ultrascaled technologies.[18]

In this work, the electrical homogeneity of continuous, large-area and high-quality *h*-BN stacks (grown by LPCVD on Pt substrates) is investigated via conductive atomic force microscopy (CAFM) and probe station. We find that thicker *h*-BN preferably grows on Pt grains with (101) crystallographic orientation. The excellent topographic-current correlation observed in CAFM maps indicates that the tunneling current across the *h*-BN is homogeneous within each Pt grain, but very different from grain to grain. Device level tests revealed that the variability of the devices fabricated within the same Pt grain is very small, and that the properties of devices grown on different Pt grains are strongly different to each other. The results here presented provide new insights on the electrical homogeneity of large-area *h*-BN stacks, and contribute to understanding the variability of *h*-BN-based electronic devices.

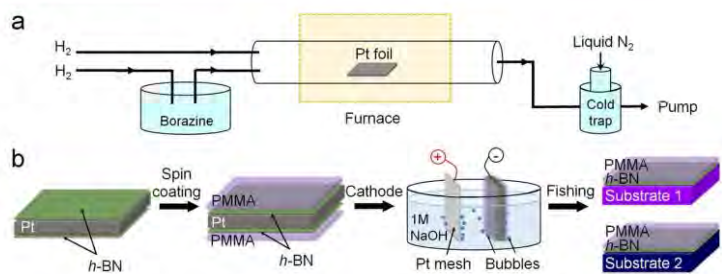

**Figure 1.** Hexagonal boron nitride growth and transfer. (a) Schematic of LPCVD process to grow *h*-BN on Pt foils. (b) Electrochemical (bubble) method to transfer *h*-BN from Pt foil to the target substrates.

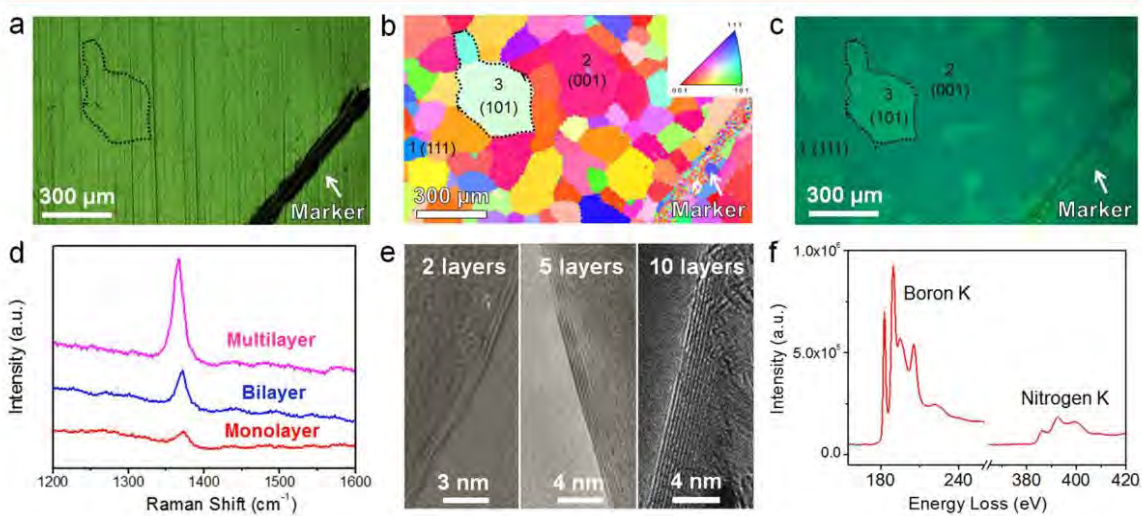

**Figure 2.** Characterization of *h*-BN films. (a) Optical microscope and (b) EBSD images of a polycrystalline Pt substrate before *h*-BN growth (annealed). The two images correspond to the same location of the sample, as highlighted by the marker in the right-bottom part. (c) Optical microscope image of a *h*-BN/300 nm-SiO$_2$/Si sample. The *h*-BN was grown on the area displayed in (a) and subsequently transferred onto a SiO$_2$/Si substrate. The granular pattern with different dark-green and light-green colors (thicknesses) observed in c perfectly matches the shape of the grains in (a). Therefore, it is possible to know which crystallographic orientation produces dark-green (thinner) and light-green (thicker) *h*-BN stacks. (d) Raman spectra collected on different locations of the *h*-BN/300 nm-SiO$_2$/Si sample. Monolayer/multilayer have been collected at dark-green/light-green locations of panel (c). (e) High-resolution TEM images demonstrating the different thicknesses and the good layered structure of the *h*-BN stacks. (f) EELS spectrum of *h*-BN showing the typical boron and nitrogen peaks.

Figure 1a shows the schematic of the LPCVD process for *h*-BN growth. A 1 cm × 2.5 cm Pt substrate was cleaned via thermal annealing (see the Supporting Information) and introduced in the center of the CVD tube (see Figure 1a). We use Pt as substrate because, despite being more expensive than Cu, Ni, or Fe, the quality of the *h*-BN grown on Pt is higher (i.e., it holds a better layer structure with less randomly oriented crystallites).[19] Liquid-phase borazine precursor was kept in a commercial cold container at 3 °C to avoid it is self-decomposition, and a cold trap filled with liquid nitrogen was used to prevent the damage of the pump. The borazine molecules (0.1 sccm) were delivered to the Pt substrate on H$_2$ carrier gas (70 sccm) at 950 °C, which produced their absorption and decomposition on the surface of the Pt substrate, and the self-mediated growth of *h*-BN. After the LPCVD growth, the *h*-BN stack was transferred onto 300 nm-SiO$_2$/Si for Raman spectroscopy and optical microscopy inspection, and on metallic grids for transmission electron microscopy (TEM) characterization. The transfer of the *h*-BN was carried out following the bubbling approach based on water electrolysis (see Figure 1b and Supporting Information).[19] This method is beneficial because it allows recycling the Pt foil for unlimited times (i.e., it is cost-effective).

Figure 2a and 2b show the optical image and electron backscatter diffraction (EBSD) map of a Pt substrate after the thermal annealing (before *h*-BN growth). Different Pt grains and crystallographic orientations can be distinguished. A long marker at the bottom right part of the image was made with a razor blade to identify this location in subsequent analyses. After that, multilayer *h*-BN was grown directly on the surface of the Pt substrate and transferred onto flat 300 nm-SiO₂/Si. The optical image of the *h*-BN/300 nm-SiO₂/Si (Figure 2c) reveals regions with different colors (dark-green to light-green) perfectly matching the shapes of the Pt grains before *h*-BN growth (Figures 2a and 2b). Figure 2c also proves that the *h*-BN stack is continuous, as well as the nondestructive nature of the bubbling transfer method. Interestingly, the *h*-BN stacks grown on Pt grains with crystallographic orientations near (101) show light-green colors in the optical microscope image (see Figure 2c), which has been attributed to a larger thickness.[15,17] This hypothesis has been verified by collecting Raman spectra at different locations on the surface of the *h*-BN/300 nm-SiO₂/Si sample (see Figure 2d). The dark-green grains in the optical microscope image (Figure 2c) always showed an $E_{2g}$ peak near 1370 cm⁻¹, which corresponds to monolayer *h*-BN; on the contrary at the light-green grains the $E_{2g}$ peak shifted toward ∼1366 cm⁻¹, which is the characteristic value of bulk *h*-BN. High-magnification TEM images (Figure 2e) present the definitive corroboration of grain-dependent thickness in the *h*-BN stack. The thicknesses observed via TEM at multiple locations of the sample (Figure 2e) always ranged between 1 and 2 and 10−13 layers, meaning that the dark and light regions in Figure 2c should correspond to thicknesses of 1−2 and 10−13 layers, respectively. It is worth noting that, for all thicknesses, the TEM images reveal defect-free layered structure. The electron energy loss spectroscopy (EELS) spectrum (Figure 2f) shows two edges at around 180 and 390 eV, which correspond to the characteristic *k*-shell ionization edges of boron and nitrogen. This verifies that the stoichiometric ratio of boron and nitrogen is 1:1, which is characteristic of layered *h*-BN with hexagonal lattice.[2]

The electrical performance of as-grown *h*-BN/Pt (without transfer) has been tested via CAFM. Figures 3a, b show the simultaneously collected topographic and current maps (respectively) measured on the surface of the *h*-BN stack when applying a bias of −2 V to the Pt substrate (CAFM tip grounded). In order to analyze the electrical properties of *h*-BN grown on different Pt grains, we used a large (80 μm × 80 μm) scan size and strategically selected an area of the sample containing several Pt grains. The different Pt grains detected have been named with letters from A to G. Despite the *h*-BN cannot be detected in the topographic map (it is very flat compared to the Pt surface, see Figure S1), its effect can be clearly seen in the current map (Figure 3b), which reveals sharp conductivity changes from grain to grain. The different currents collected on each Pt grain must be related to the presence of *h*-BN stacks with different thicknesses because the underlying Pt grains, despite having different crystal orientations, hold similar conductivities.[20] It should be highlighted that the *h*-BN adhesion to the Pt surface is always by van der Waals forces for any Pt crystalline orientation, which means that the electronic coupling between the *h*-BN and Pt from one grain to another does not change. Therefore, *h*-BN/Pt electronic coupling is not one factor producing the conductivity changes from one grain to another observed in Figure 3b. The fact that the conductivity changes have been detected in the same image discards tip wearing from one grain to the other, and the perfect topography-current correlation undoubtedly demonstrate that *h*-BN grown on different grains hold different conductivities (because of their different thicknesses, as shown in Figure 2). It should be highlighted that, sporadically, the

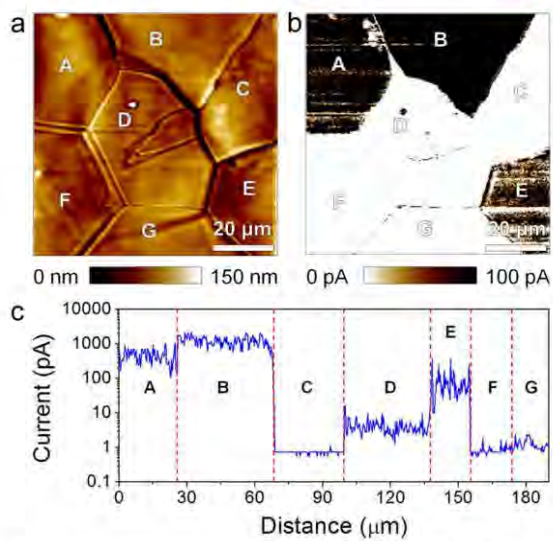

**Figure 3.** CAFM Characterization of as-grown *h*-BN/Pt. (a) Topographic and (b) current maps simultaneously collected on *h*-BN grown on a polycrystalline Pt substrate, under a bias of −2 V (applied to the substrate, tip grounded). (c) Assembly of cross-sections collected at the different grains of the current map in b.

regions close to the Pt grain boundaries have shown higher currents (see for example the grain boundaries between grains D-G and F-G). The explanation for this observation is as follows: when the metallic substrate is exposed to large temperatures it becomes polycrystalline; the grain boundaries of the metal substrate are rough and may contain asperities; these topographic accidents alter the *h*-BN growth, and at that location the *h*-BN may be thinner or even cracked, displaying large currents in the current maps. Cross sections have been collected (offline) at all the grains of the current image (Figure 3b) using the AFM software (NanoScope Analysis) and assembled one after the other (using OriginPro 8 software). The result is displayed in Figure 3c. Within each grain the current is homogeneous, and sharp changes are detected from grain to grain. As can be observed, the highest currents are detected on grain B, indicating that it holds the thinnest *h*-BN stack on its surface. Grain C shows negligible currents similar to the electrical noise of the CAFM, indicating that the −2 V applied were not enough to generate tunneling current across the *h*-BN stack. It should be emphasized that the current deviations within each grain are below 1 order of magnitude for all grains (compare maximum and minimum peaks within each grain in Figure 3c). This value is smaller than that observed in other thin insulating films (of similar thickness) being currently used in the industry, such as HfO₂ and TiO₂ (see Figure S3). Therefore, the electrical properties of *h*-BN within the same grain seem to be very homogeneous, which shows great potential to mitigate device-to-device variability problems of ultra scaled devices. Further electrical information about the grains has been obtained by measuring the onset voltage ($V_{ON}$) of the *h*-BN stacks on each Pt grain. The onset voltage is defined as the minimum voltage that needs to be applied between the CAFM tip and the substrate of the sample (Pt) in order to observe tunneling currents above the noise level.[21] Despite the noise level of our CAFM being ∼1 pA, we selected

- 51 -

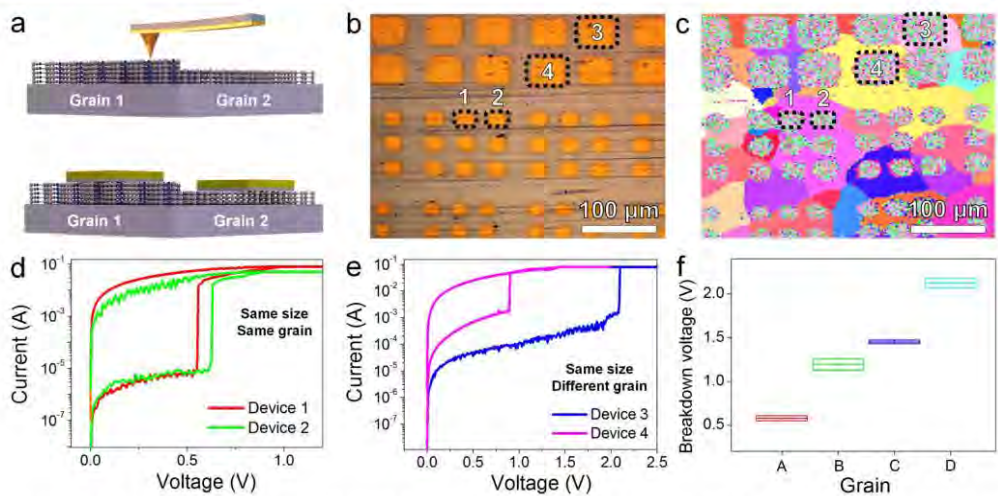

**Figure 4.** Device level characterization of $h$-BN stacks. (a) Schematic of as-grown $h$-BN/Pt (top) and $h$-BN-based devices (bottom). The device cell is an Au/Ti/$h$-BN/Pt structure. (b) Optical image and (c) EBSD map of matrixes of Au/Ti/$h$-BN/Pt cells with different lateral sizes. As this map has been collected with the presence of $h$-BN, it is only meaningful to distinguish different grains, not to assess the real crystallographic orientation of each grain. For this reason, the color scale has been intentionally removed. (d) $I$–$V$ curves collected on two devices with the same size (25 $\mu$m × 25 $\mu$m) within the same Pt grain. (e) $I$–$V$ curves collected on two devices with the same size (50 $\mu$m × 50 $\mu$m) located at different Pt grains. (f) Analysis of the BD voltage for Au/Ti/$h$-BN/Pt devices patterned on 4 different Pt grains (located outside the region of c).

$V_{ON} = V$ ($I = 5$ pA) in order to be completely sure that non-negligible current is flowing across the $h$-BN stack. For this experiment, $V_{ON}$ has been extracted by measuring individual current maps on each grain (1 $\mu$m × 1 $\mu$m). The results obtained (Table S1), strongly support the observations in Figure 3: the smallest $V_{ON}$ (0.1 V) was detected on grain B, and the highest (6 V) on grain C. By fitting current vs voltage ($I$–$V$) curves collected on the $h$-BN at different Pt grains to the tunneling model equations it can be concluded that the real thickness of the $h$-BN stack fluctuates between 1 and 13 layers (see modeling section in the Supporting Information), in agreement with Figure 2e.

Finally, the electrical properties of $h$-BN stacks grown on polycrystalline Pt substrates have been analyzed at the device level using matrixes of metal/insulator/metal (MIM) cells (i.e., Au/Ti/$h$-BN/Pt). Figure 4a shows the schematic of as-grown $h$-BN on polycrystalline Pt foil before and after electrodes deposition, in which $h$-BN film serves as insulating layer between the top (Au/Ti) and bottom (Pt) metal electrodes. The optical microscope image of the sample after Au/Ti electrode deposition is shown in Figure 4b. The corresponding EBSD map of the same area of the sample has been also collected (Figure 4c). It should be highlighted that, unlike in Figure 2b, in Figure 4c the EBSD map has been collected in the presence of $h$-BN on the Pt substrate. As the $h$-BN is an insulator, this may distort the signal related to the local crystallographic orientation. For this reason, the EBSD map in Figure 4d will be only used to distinguish different grains (in that case the contrast is large) but not to identify which is the real crystallographic orientation of the Pt within each grain. For this reason, the color scale in Figure 4c has been intentionally removed. Several devices on the same and different Pt grains have been tested in the probe station by applying ramped voltage stresses (RVS), and the resulting current vs voltage ($I$–$V$) curves have been compared. As an example, Figure 4d shows the $I$–$V$ curves collected on two devices with the same size (25

$\mu$m × 25 $\mu$m) that belong to the same Pt grain (devices 1 and 2 in Figure 4c). As it can be observed, the $I$–$V$ characteristics for devices 1 and 2 are strikingly similar: i) the BD voltages ($V_{BD}$) show very small deviation (0.55 V and 0.61 V); (ii) the pre- and post-BD currents ($I_{PRE\text{-}BD}$ and $I_{POST\text{-}BD}$, respectively) almost overlap, and (iii) the $I_{POST\text{-}BD}/I_{PRE\text{-}BD}$ ratio is identical. These results are indeed indicating that the device-to-device variability within the same Pt grain is very small. Any potential variability of the prebreakdown $I$–$V$ curves in devices within the same Pt grain should be related to nanoscale inhomogeneities within the $h$-BN stack, such as thickness fluctuations (see Figure S2), local defects, $h$-BN domain boundaries, and/or wrinkles. Nevertheless, as displayed in Figures 4d and 3c and Figure S3 the variability of the electrical properties of the $h$-BN within the same Pt grain are very small. On the contrary, the devices with the same size but patterned on different Pt grains show very inhomogeneous $I$–$V$ characteristics. As an example, Figure 4e shows the $I$–$V$ curves collected on devices 3 and 4 (see Figure 4c). First, the pre-BD currents are very different; second, $V_{BD}$ for both devices are remarkably different: 2.09 V for device 3 and 0.89 V for device 4; and third, the $I_{POST\text{-}BD}/I_{PRE\text{-}BD}$ ratio is also slightly different. The different electrical properties of devices 3 and 4 are related to the different thicknesses of the $h$-BN stack, due to the different crystallographic orientation of the underlying Pt grain. These observations have been corroborated by measuring additional MIM devices at different Pt grains. As Figure 4f shows, the deviation of $V_{BD}$ within each grain is very small (from $\pm$0.013 V for grain C to $\pm$0.054 V for grain B), but the deviations of $V_{BD}$ from one Pt grain to another are large (from 0.58 V in grain A to 2.13 V in grain D). The breakdown event observed for all devices (displayed as a sharp current increase in Figure 4d, e) further demonstrates that the $h$-BN sheet grown on the Pt substrate is continuous, otherwise a larger current under smaller voltage (i.e., $1 \times 10^{-2}$ A @ 0.1 V) typical of shorted devices should be observed.

 Letter

Fortunately, cutting-edge electronic devices based on MIM cells cover ultra scaled areas. In the case of FETs, the current technology node is 7 nm, and the total length of current FETs never exceeds 50 nm. In the case of memristors and other nonvolatile memories, such as resistive random access memories (RRAM), phase change random access memories (PCRAM) and ferroelectric random access memories (FeRAM), device areas down to 10 nm × 10 nm are preferred.[22] Therefore, as the diameter of the Pt grains easily surpass 100 $\mu$m, the fabrication of $h$-BN based electronic devices with very low variability is feasible. More efforts toward the growth of large-area single-crystalline $h$-BN stacks should conduct to ultralow variability technologies.

In conclusion, the electrical homogeneity of $h$-BN stacks grown via CVD on Pt substrates has been analyzed by CAFM and a probe station. We observe that $h$-BN grows thicker on Pt grains with crystallographic orientations close to (101). In situ CAFM characterization reveals that the tunneling current across the $h$-BN grown on the same Pt grain is very homogeneous (i.e., more homogeneous than that observed in other insulators being currently used in the industry, such as $HfO_2$ and $TiO_2$), but sharp conductivity changes are detected from grain to grain. Device level tests in the probestation reveal that the variability of Au/Ti/$h$-BN/Pt devices within each grain is strikingly low in terms of $I_{PRE-BD}$, $I_{POST-BD}$, and $V_{BD}$. These results contribute to the understanding of the electrical properties of $h$-BN and variability of $h$-BN-based electronic devices.

## ■ ASSOCIATED CONTENT

### ⬢ Supporting Information

The Supporting Information is available free of charge on the ACS Publications website at DOI: 10.1021/acsami.7b09417.

Additional experimental explanations, AFM characterization of the $h$-BN/300 nm-$SiO_2$/Si sample, electrical measurements, and modeling details (PDF)

## ■ AUTHOR INFORMATION

### Corresponding Author
*E-mail: mlanza@suda.edu.cn.

ORCID
Wei Sun Leong: 0000-0001-8131-2468
Hui Ying Yang: 0000-0002-2244-8231
Mario Lanza: 0000-0003-4756-8632

### Author Contributions

†F.H. and W.F. contributed equally to this work. The manuscript was written through contributions of all authors. All authors have given approval to the final version of the manuscript.

### Notes

The authors declare no competing financial interest.

## ■ ACKNOWLEDGMENTS

F.H. acknowledges the support from the Young 1000 Global Talent Recruitment Program of the Ministry of Education of China, the National Natural Science Foundation of China (Grants 61502326, 41550110223, 11661131002), the Jiangsu Government (Grant BK20150343), the Ministry of Finance of China (Grant SX21400213) and the Young 973 National Program of the Chinese Ministry of Science and Technology (Grant 2015CB932700). The Collaborative Innovation Center of Suzhou Nano Science & Technology, the Jiangsu Key Laboratory for Carbon-Based Functional Materials & Devices, the Priority Academic Program Development of Jiangsu Higher Education Institutions, and the Opening Project of Key Laboratory of Microelectronic Devices & Integrated Technology (Institute of Microelectronics, Chinese Academy of Sciences) are also acknowledged. W.F., T.K., G.H., and J.K. acknowledge the support from the STC Center for Integrated Quantum Materials, NSF Grant DMR-1231319. H.W. and J.K. acknowledge the support from NSF DMR/ECCS−1509197. W.S.L. acknowledges the support from SUTD-MIT Postdoctoral Fellows Program.

## ■ REFERENCES

(1) Watanabe, K.; Taniguchi, T.; Kanda, H. Direct-bandgap Properties and Evidence for Ultraviolet Lasing of Hexagonal Boron Nitride Single Crystal. *Nat. Mater.* **2004**, *3*, 404−409.
(2) Song, L.; Ci, L. J.; Lu, H.; Sorokin, P. B.; Jin, C. H.; Ni, J.; Kvashnin, A. G.; Kvashnin, D. G.; Lou, J.; Yakobson, B. I.; Ajayan, P. M. Large Scale Growth and Characterization of Atomic Hexagonal Boron Nitride Layers. *Nano Lett.* **2010**, *10*, 3209−3215.
(3) Lindsay, L.; Broido, D. A. Enhanced Thermal Conductivity and Isotope Effect in Single-Layer Hexagonal Boron Nitride. *Phys. Rev. B: Condens. Matter Mater. Phys.* **2011**, *84*, 155421.
(4) Liu, Z.; Gong, Y. J.; Zhou, W.; Ma, L. L.; Yu, J. J.; Idrobo, J. C.; Jung, J.; MacDonald, A. H.; Vajtai, R.; Lou, J.; Ajayan, P. M. Ultrathin High-Temperature Oxidation-Resistant Coatings of Hexagonal Boron Nitride. *Nat. Commun.* **2013**, *4*, 2541.
(5) Dean, C. R.; Young, A. F.; Meric, I.; Lee, C.; Wang, L.; Sorgenfrei, S.; Watanabe, K.; Taniguchi, T.; Kim, P.; Shepard, K. L.; Hone, J. Boron Nitride Substrates for High-Quality Graphene Electronics. *Nat. Nanotechnol.* **2010**, *5*, 722−726.
(6) Ji, Y. F.; Pan, C. B.; Zhang, M. Y.; Long, S. B.; Lian, X. J.; Miao, F.; Hui, F.; Shi, Y. Y.; Larcher, L.; Wu, E.; Lanza, M. Boron Nitride as Two Dimensional Dielectric: Reliability and Dielectric Breakdown. *Appl. Phys. Lett.* **2016**, *108*, 012905.
(7) Hui, F.; Pan, C. B.; Shi, Y. Y.; Ji, Y. F.; Grustan-Gutierrez, E.; Lanza, M. On the Use of Two Dimensional Hexagonal Boron Nitride as Dielectric. *Microelectron. Eng.* **2016**, *163*, 119−133.
(8) Pan, C. B.; Ji, Y. F.; Xiao, N.; Hui, F.; Tang, K. C.; Guo, Y. Z.; Xie, X. M.; Puglisi, F. M.; Larcher, L.; Miranda, E.; Jiang, L. L.; Shi, Y. Y.; Valov, I.; McIntyre, P. C.; Waser, R.; Lanza, M. Coexistence of Grain-Boundaries-Assisted Bipolar and Threshold Resistive Switching in Multilayer Hexagonal Boron Nitride. *Adv. Funct. Mater.* **2017**, *27*, 1604811.
(9) Hui, F.; Grustan-Gutierrez, E.; Long, S.; Liu, Q.; Ott, A. K.; Ferrari, A. C.; Lanza, M. Graphene and Related Materials for Resistive Random Access Memories. *Adv. Electron. Mater.* **2017**, *3*, 1600195.
(10) Nagashima, A.; Tejima, N.; Gamou, Y.; Kawai, T.; Oshima, C. Electronic States of Monolayer Hexagonal Boron Nitride Formed on the Metal Surfaces. *Surf. Sci.* **1996**, *357*, 307−311.
(11) Nagashima, A.; Tejima, N.; Gamou, Y.; Kawai, T.; Oshima, C. Electronic Dispersion Relations of Monolayer Hexagonal Boron Nitride Formed on the Ni(111) Surface. *Phys. Rev. B: Condens. Matter Mater. Phys.* **1995**, *51*, 4606−4613.
(12) Kim, S. M.; Hsu, A.; Park, M. H.; Chae, S. H.; Yun, S. J.; Lee, J. S.; Cho, D. H.; Fang, W. J.; Lee, C. G.; Palacios, T.; Dresselhaus, M.; Kim, K. K.; Lee, Y. H.; Kong, J. Synthesis of Large-Area Multilayer Hexagonal Boron Nitride for High Material Performance. *Nat. Commun.* **2015**, *6*, 8662.
(13) Kim, K. K.; Hsu, A.; Jia, X. T.; Kim, S. M.; Shi, Y. M.; Hofmann, M.; Nezich, D.; Rodriguez-Nieva, J. F.; Dresselhaus, M.; Palacios, T.; Kong, J. Synthesis of monolayer hexagonal boron nitride on Cu foil using chemical vapor deposition. *Nano Lett.* **2012**, *12*, 161−166.
(14) Gao, Y.; Ren, W. C.; Ma, T.; Liu, Z. B.; Zhang, Y.; Liu, W. B.; Ma, L. P.; Ma, X. L.; Cheng, H. M. Repeated and controlled growth of monolayer, bilayer and few-layer hexagonal boron nitride on Pt foil. *ACS Nano* **2013**, *7*, 5199.

- 53 -

(15) Shi, Y. M.; Hamsen, C.; Jia, X. T.; Kim, K. K.; Reina, A.; Hofmann, M.; Hsu, A. L.; Zhang, K.; Li, H.; Juang, Z. Y.; Dresselhaus, M. S.; Li, L. J.; Kong, J. Synthesis of Few-Layer Hexagonal Boron Nitride Thin Film by Chemical Vapor Deposition. *Nano Lett.* **2010**, *10*, 4134–4139.

(16) Lee, Y. H.; Liu, K. K.; Lu, A. Y.; Wu, C. Y.; Lin, C. T.; Zhang, W. J.; Su, C. Y.; Hsu, C. L.; Lin, T. W.; Wei, K. H.; Shi, Y. M.; Li, L. J. Growth Selectivity of Hexagonal-Boron Nitride Layers on Ni with Various Crystal Orientations. *RSC Adv.* **2012**, *2*, 111–115.

(17) Park, J. H.; Park, J. C.; Yun, S. J.; Kim, H.; Luong, D. H.; Kim, S. M.; Choi, S. H.; Yang, W.; Kong, J.; Kim, K. K.; Lee, Y. H. Large-Area Monolayer Hexagonal Boron Nitride on Pt Foil. *ACS Nano* **2014**, *8*, 8520–8528.

(18) Allan, A. *International Technology Roadmap for Semiconductors of 2008*. https://cseweb.ucsd.edu/classes/wi09/cse242a/itrs/ORTC.pdf, accessed February, 2016.

(19) Kim, G.; Jang, A. R.; Jeong, H. Y.; Lee, Z.; Kang, D. J.; Shin, H. S. Growth of High-Crystalline, Single-Layer Hexagonal Boron Nitride on Recyclable Platinum Foil. *Nano Lett.* **2013**, *13*, 1834–1839.

(20) Powell, R. W.; Tye, R. P.; Woodman, M. J. The Thermal Conductivity and Electrical Resistivity of Polycrystalline Metals of the Platinum Group and of Single Crystals of Ruthenium. *J. Less-Common Met.* **1967**, *12*, 1–10.

(21) Lanza, M.; Porti, M.; Nafría, M.; Aymerich, X.; Ghidini, G.; Sebastiani, A. Trapped Charge and Stress Induced Leakage Current (SILC) in Tunnel SiO2 Layers of De-processed MOS Non-Volatile Memory Devices Observed at the Nanoscale. *Microelectron. Reliab.* **2009**, *49*, 1188–1191.

(22) Govoreanu, B.; Kar, G. S.; Chen, Y. Y.; Paraschiv, V.; Fantini, A.; Radu, I. P.; Goux, L.; Clima, S.; Degraeve, R.; Jossart, N. 10 × 10nm2 Hf/HfOx Crossbar Resistive RAM with Excellent Performance, Reliability and Low-Energy Operation. *IEEE Int. Electron Dev. Meeting* **2011**, *11*, 729–732.

- 54 -

# Electrical homogeneity of large-area chemical vapor deposited multilayer hexagonal boron nitride sheets

*Fei Hui,* *§,‡ *Wenjing Fang,*§,‡ *Wei Sun Leong,* §,& *Tewa Kpulun,*ξ *Haozhe Wang,*§ *Hui Ying*

*Yang,*& *Marco A. Villena\*, Gary Harris,*ξ *Jing Kong,*§ *Mario Lanza\**

\*Institute of Functional Nano & Soft Materials, Collaborative Innovation Center of Suzhou

Nano Science and Technology, Soochow University, Suzhou, 215123, China.

§Department of Electrical Engineering and Computer Science, Massachusetts Institute of

Technology, Cambridge, MA 02139, USA.

&Pillar of Engineering Product Development, Singapore University of Technology and

Design, 8 Somapah Road, Singapore 487372, Singapore.

ξDepartment of Electrical and Computer Engineering, Howard University, Washington DC,

20059, USA.

δDepartment of Materials Science and Engineering, Stanford University, CA 94305, USA

Corresponding author Email: mlanza@suda.edu.cn (Mario Lanza)

### Experimental Section

*Growth of h-BN Pt substrates*: High purity (99.997%) 100 μm thick Pt foil purchased from Alfa Aesar (item no. 12059) was employed as substrate for the *h*-BN growth. First, the as-received Pt foil was cleaned in acetone and isopropanol (IPA) for 10 minutes to remove the surface contamination, and dried with a nitrogen gun. Then, the Pt foil was introduced in the center of the quartz tube, and the temperature ramped up to 950 °C in 70 sccm $H_2$ atmosphere. The time required to increase the temperature from ~20 °C to 950 °C was ~ 40 minutes, and following by the annealing process under the temperature of 950 °C in 70 sccm $H_2$ for 30 minutes, in order to remove the contamination contains carbon or oxygen. This pre-growth heating step is called annealing, and it is useful to clean impurities on the Pt surface. Then, the valve of the tube coming from the Borazine container (which used a $H_2$ flow rate of 0.1 sccm) was opened, allowing the introduction of borazine molecules in the quarz tube containing the Pt substrate. This process was kept for 1 hour at 950 °C. After that time, the temperature controller was set at room temperature and the CVD system was cooled down.

*Transfer process for the h-BN*: After sample fabrication, the *h*-BN stacks were transferred on $SiO_2$/Si wafers and TEM grids for analysis. To do so, liquid poly(methyl methacrylate) (PMMA) from MicroChem was spin-coated on both sides of the *h*-BN/Pt/*h*-BN sample at 2500 rpm for 1 min. The sample was backed in the oven at 70 °C for 10 min to solidify the PMMA, and the resulting sample (PMMA/*h*-BN/Pt/*h*-BN/PMMA) was immersed in a container filled with 1 M NaOH. In the same container, a Pt mesh was also introduced, and a potential difference of 3 V between it and the PMMA/*h*-BN/Pt/*h*-BN/PMMA sample was applied using a source meter. The Pt mesh served as anode, and the PMMA/*h*-BN/Pt/*h*-BN/PMMA sample as cathode. The application of voltage lead to the formation of hydrogen bubbles at the *h*-BN/Pt interface, leading to the effective PMMA/*h*-BN detachment from the Pt foil [1] in less than 10 minutes. Afterwards, the PMMA/*h*-BN stack was cleaned in deionized water three times to remove the residual NaOH solution. Finally, PMMA/*h*-BN was transferred on the target substrates ($SiO_2$/Si or TEM grids). When transferred on the $SiO_2$/Si substrate the PMMA was removed by soaking the entire sample in acetone for 2 hours, followed by an annealing at 400 °C for 2 h in a mixed $H_2$ (200 sccm) and Ar (200 sccm) atmosphere. When transferred on the TEM grids, only the annealing step was used (no sample soaking, that could damage the *h*-BN suspended on the TEM grid).

*h-BN and Pt characterization*: The different Pt grains and their crystallographic orientations were analyzed using a standard optical microscope, and an EBSD system (from Oxford Technology) integrated in a scanning electron microscope (Zeiss Merlin HRSEM). The topographic maps in Figures S1-S2 were collected using a Dimension 3000 AFM working in air atmosphere. These experiments were performed in tapping mode using Si probe tips from Budgetsensors (model: Tap300-G). The CAFM experiments were carried out in a Multimode VI equipment from Veeco working inside a nitrogen chamber (relative humidity ~ 0.5%) [2]. The use of a nitrogen atmosphere is beneficial to stabilize the current signal and achieve larger lateral resolution [3]. For this experiment we used PtIr varnished Si probes from Nanosensors (model: CONTPT). The cross section in the current CAFM maps have been calculated using the NanoScope Analysis software of the AFM (Bruker) and assembpled using OriginPro 8 software. Atomic scale information about the thickness and quality of the *h*-BN stacks was collected using a JEOL 2010 HRTEM with EELS capability integrated, and a LabRAM Raman spectrometer from Horiba.

*Device fabrication and characterization*: Squared metallic top electrodes with lateral sizes ranging from 10 μm × 10 μm to 100 μm × 100 μm have been deposited on as-grown *h*-BN/Pt samples. First 20 nm Ti and second 60 nm Au have been deposited via E-beam evaporator (Ajaint AJA-ATC) using a laser-patterned shadow mask. The electrical measurements were carried out in a Summit 11000 AP probe station connected to an Agilent 4155C semiconductor parameter analyzer. The RVS was applied to the top electrodes and the Pt substrates were grounded. MIM cells are used in several devices, including FETs [4] capacitors [5], and memristors [6], as well as in test structures to evaluate essential parameters of the insulator, such as charge trapping, stress induced leakage current, dielectric breakdown (BD) and resistive switching (RS) [7]. Therefore, despite holding an easy structure, the MIM cells fabricated are representative of $h$-BN based microelectronic devices.

**Additional Characterization**

The surface roughness of the $h$-BN grown on Pt grains with different orientations has been analyzed via atomic force microscopy (AFM). Figure S1 shows the topographic AFM maps measured on the surface of the 300-nm-SiO$_2$/Si sample shown in Figure 2c; the root mean square (RMS) roughness of each image (calculated via AFM software) is also displayed. As it can be observed, the flattest surface is detected for the $h$-BN that was grown on Pt (111), and the roughest corresponds to the $h$-BN that was grown on Pt (101). As the surface roughness of 2D materials increase with their thickness [8], Figure S1 further supports that the thinnest $h$-BN stack grows on Pt (111), and that the thickest grows on Pt (101). Moreover, multilayer islands have been detected on the surface of the $h$-BN grown on Pt (001) and Pt (101), as shown in **Figure S2**, further suggesting the growth of thicker $h$-BN stacks. Alternative methods to evaluate the thickness of the $h$-BN stack are low energy electron microscopy (LEEM) and low energy electron diffraction (LEED) [9].

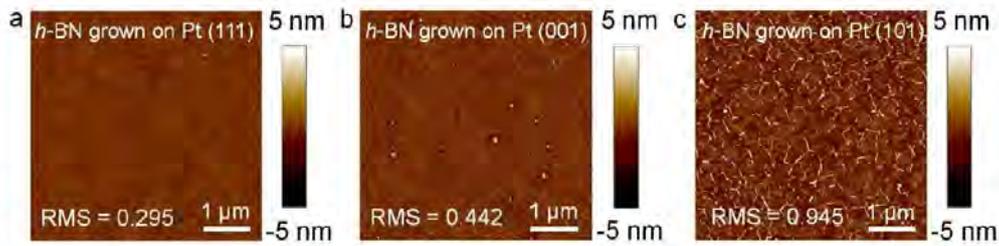

**Figure S1**. AFM characterization of the $h$-BN/300-nm-SiO$_2$/Si sample displayed in Figure 2c. Panels (a)-(c) in Figure S1 show that the roughness and density of wrinkles in the $h$-BN depend on the crystallographic orientation of the Pt substrate on which it was grown.

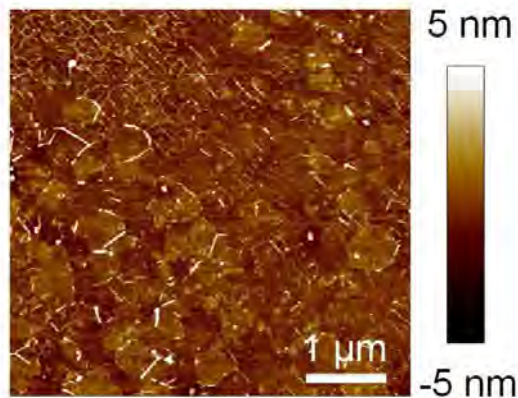

**Figure S2**. AFM characterization of a $h$-BN/300-nm-SiO$_2$/Si sample, on a region on which the $h$-BN was previously grown on Pt(101). This image shows multilayer $h$-BN islands.

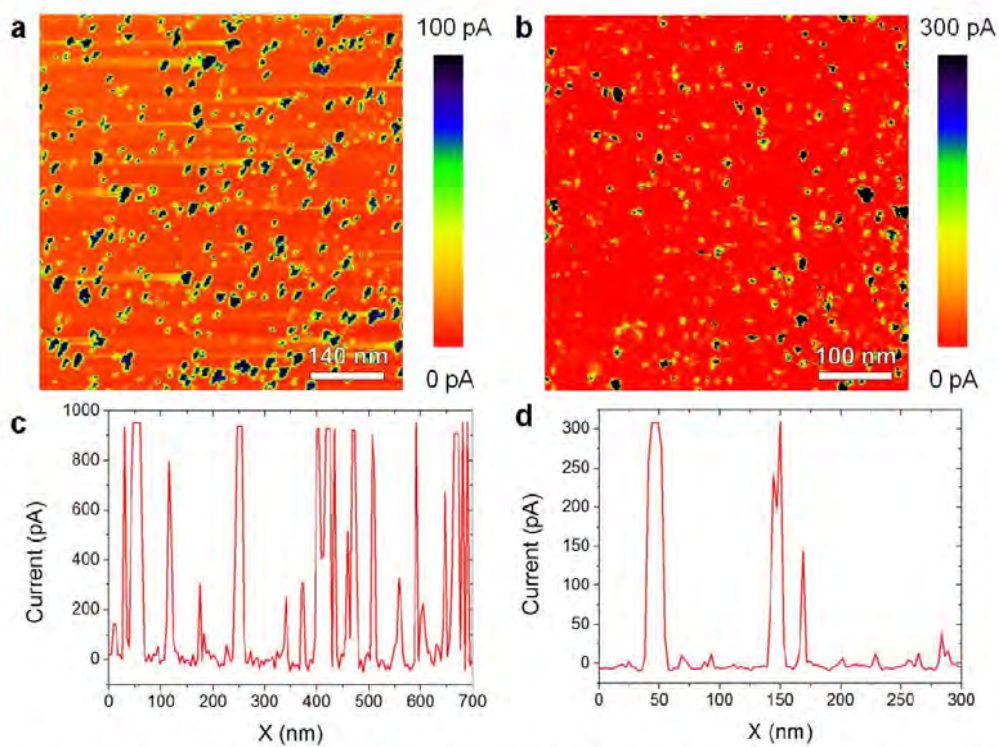

**Figure S3.** Current maps collected with the tip of the CAFM on the surface of (a) 4 nm thick $HfO_2$ and (b) 2 nm thick $TiO_2$ films, both of them grown via atomic layer deposition on a conductive substrate (Zr below the $HfO_2$ and $n^{++}Si$ below the $TiO_2$). The current fluctuations are 2-3 orders of magnitude. (c) and (d) show the cross sections of the maps in (a) and (b) respectively.

**Table S1** Electrical measurements are conducted on each grain with the constant voltage and constant 5 pA, respectively.

| Grain | Current (for a voltage of -1 V) | Voltage (for a current of 5 pA) |
|-------|-------------------------------|--------------------------------|
| A | 5 nA | 0.5 V |
| B | 5 nA | 0.1 V |
| C | < 1pA | 6 V |
| D | < 1pA | 2 V |
| E | < 1pA | 1 V |
| F | < 1pA | 5 V |
| G | < 1pA | 5 V |

**Modeling section**

In Figure 3c it is difficult to quantify the real thickness of the *h*-BN on each Pt grain, and correlate it with the current levels observed. The reason is that best techniques used for physical thickness characterization of 2D materials (i.e. TEM) are destructive. In order to provide more insights to this point here we perform an additional analysis, consisting on measuring I-V curves at different locations of each grain via CAFM. Based on the shape of the I-V curves showing the tunneling current across the insulating stack, its physical thickness can be calculated very accurately (this was done before for ultra thin SiO₂ films with sub-nanometer resolution [10]. In previous works [11] it was determined that the dominant conduction across multilayer *h*-BN stacks was by Forler-Nordheim Tunneling (FNT). Therefore, here we use the FNT equation to calculate the tunneling current for different *h*-BN thicknesses:

$$I = \frac{A_{eff}\sqrt{m\phi_B}q^2V}{h^2 d} \exp\left[\frac{-4\pi\sqrt{m\phi_B}d}{h}\right] \qquad (1)$$

where $V$ is the applied voltage, $d$ is the thickness of the *h*-BN, $A_{eff}$ is the effective contact area, $\phi_B$ is the barrier height. The parameters $m$, $q$ and $h$ are the free electron mass, the electron charge and the Planck constant, respectively. This calculation has been repeated for different values of d depending on the number of layers (N), that is $d = 0.33$ nm for N=1, $d = 0.66$ nm for N=2, etc... having increments of 0.33 nm for each layer until N=24. By comparing the calculations with the I-V curves experimentally collected (see Figure S4), it can be concluded that the tunneling currents across grain B fits well the conduction of monolayer *h*-BN, while the tunneling currents across grain D fit well the conduction across 11-13 layers (see Figure S4). This result is interesting because provides an indirect quantification of the thickness at each grain, something that the CAFM maps (Figure 3b) cannot provide.

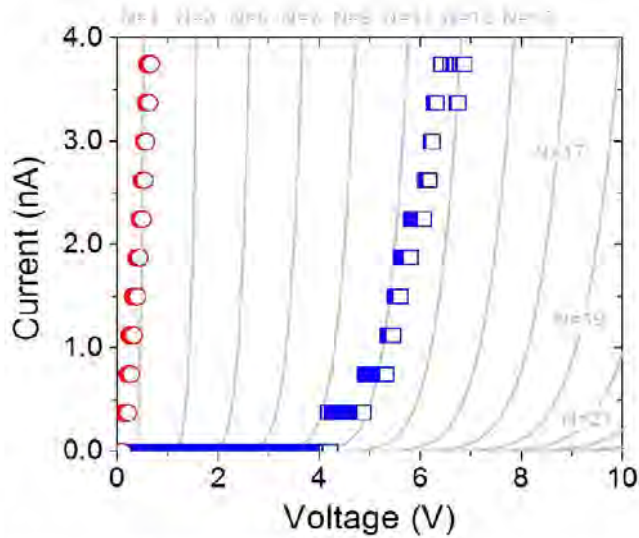

**Figure S4.** Calculation of FNT conduction for different thicknesses (N is the number of layers). Typical experimental IV curves measured on grain B (red symbols) and grain D (blue symbols) of Figure 3b of the manuscript.

**Supporting Information References**

[1]  Gao, L. B.; Ren, W. C.; Xu, H. L.; Jin, L.; Wang, Z.; Ma, T.; Ma, L. P.; Zhang, Z. Y.; Fu, Q.; Peng, L. M.; Bao, X. H.; Cheng, H. M. Repeated Growth and Bubbling Transfer of Graphene with Millimeter-Size Single-Crystal Grains Using Platinum. *Nat. Commun.* **2012**, *2*, 699-705.

[2]  Ji, Y. F.; Hui, F.; Shi, Y. Y.; Han, T. T.; Song, X. X.; Pan, C. B.; Lanza, M. Fabrication of a Fast-Response and User-Friendly Environmental Chamber for Atomic Force Microscopes. *Rev. Sci. Inst.* **2015**, *86*, 106105.

[3]  Lanza, M.; Porti, M.; Nafría, M.; Aymerich, X.; Whittaker, E.; Hamilton, B. Electrical Resolution During Conductive Atomic Force Microscopy Measurements under Different Environmental Conditions and Contact Forces. *Rev. Sci. Inst.* **2010**, *81*, 106110.

[4]  Kim, B. J.; Lee, S. K.; Kang, M. S.; Ahn, J. Y.; Cho, J. H. Coplanar-Gate Transparent Graphene Transistors and Inverters on Plastic. *ACS Nano*, **2012**, *6*, 8646-8651.

[5]  Guo, N.; Wei, J.; Jia, Y.; Sun, H.; Wang, Y.; Zhao, K.; Shi, X.; Zhang, L.; Li, X.; Cao, A. Fabrication of Large Area Hexagonal Boron Nitride Thin Films for Bendable Capacitors. *Nano Res.* **2013**, *6*, 602-610.

[6]  Puglisi, F. M.; Larcher, L.; Pan, C.; Xiao, N.; Shi, Y.; Hui, F.; Lanza, M. 2D h-BN Based RRAM Devices. *IEEE Int. Electron Dev. Meeting* **2016**, *16*, 874-877.

[7]  Miranda, E.; Suñé, J.; Rodríguez, R.; Nafría M.; Aymerich, X. Soft Breakdown Fluctuation Events in Ultrathin $SiO_2$ Layers. *Appl. Phys. Lett.* **1998**, *73*, 490-492.

[8]  Fang, W. J. Synthesis of Bilayer Graphene and Hexagonal Boron Nitride by Chemical Vapor Deposition Method. Ph.D. Thesis, Massachusetts of Institute Technology, Cambridge, US, 2015.

[9]  Ismach, A.; Chou, H.; Mende, P.; Dolocan, A.; Addou, R.; Aloni, S.; Wallace, R.; Feenstra, R.; Ruoff, R.S.; Colombo, L. Carbon-Assisted Chemical Vapor Deposition of Hexagonal Boron Nitride. *2D Mater.* **2017**, *4*, 025117.

[10]  Frammelsberger, W.; Benstetter, G.; Kiely, J.; Stamp, R. CAFM-Based Thickness Determination of Thin and Ultra-Thin $SiO_2$ Films by Use of Different Conductive-Coated Probe Tips. *Appl. Surf. Sci.* **2007**, *253*, 3615-3626.

[11]  Lee, G.H.; Yu, Y.J.; Lee, C.; Dean, C.; Shepard, K.L.; Kim, P.; Hone, J. Electron Tunneling Through Atomically Flat and Ultrathin Hexagonal Boron Nitride. *Appl. Phys. Lett.* **2011**, *99*, 243114.

# Uniformity of multilayer hexagonal boron nitride dielectric stacks grown by chemical vapor deposition on platinum and copper substrates

Fei Hui[1,2*], Xianhu Liang[1], Wenjing Fang[2], Wei Sun Leong[2,3], Haozhe Wang[2], Hui Ying Yang[3], Yuanyuan Shi[1], Marco A. Villena[1,4], Jing Kong[2] and Mario Lanza[1*]

[1]Institute of Functional Nano & Soft Materials, Collaborative Innovation Center of Suzhou Nano Science and Technology, Soochow University, Suzhou 215123, China; [2]Department of Electrical Engineering and Computer Science, Massachusetts Institute of Technology, Cambridge, Massachusetts 02139, United States; [3]Pillar of Engineering Product Development, Singapore University of Technology and Design, 8 Somapah Road, Singapore 487372, Singapore; [4]Department of Materials Science and Engineering, Stanford University, Stanford, California 94305, United States.
Phone: (+86) 18801544070 Fax: (+86) 18801544070 *Email: huifei324@126.com; mlanza@suda.edu.cn

## Abstract

Large-area multilayer hexagonal boron nitride (*h*-BN) dielectric stacks can be grown on different metallic substrates via chemical vapor deposition (CVD). The high temperatures used during the *h*-BN growth produce the polycrystallization of the metallic substrate (leading to different crystallographic orientations at the surface of each grain), which may influence the catalytic activity of the CVD process on different grains, and the properties of the *h*-BN stacks grown on them. In this work we compare the uniformity of multilayer *h*-BN dielectric stacks grown via CVD on two different metallic substrates: Pt and Cu. Our study indicates that the use of Pt substrates leads to severe *h*-BN thickness fluctuations from one Pt grain to another, while this effect remarkably reduced when the h-BN is grown on Cu substrates. Therefore, the use of Cu substrates seems to be more convenient for *h*-BN production and integration at the wafer level.

## Introduction

Hexagonal boron nitride (*h*-BN) is a layered insulator (direct band gap ~5.9 eV), in which boron and nitrogen atoms arrange in a sp² hexagonal structure by covalent bonding, whereas the layers stick to each other by Van der Waals attraction [1]. In the field of electronics *h*-BN dielectric stacks are very attractive given their very high thermal conductivity (600 Wm⁻¹K⁻¹), which is expected to slow down the dielectric breakdown (BD) process. However, the integration of *h*-BN dielectric stacks in electronic devices is still problematic. The first reports studied mechanically exfoliated thick (>20 layers) *h*-BN nano flakes (diameter <10 μm) [2]. Recent studies have successfully synthesized *h*-BN stacks via CVD on metallic substrates [3-5]. However the effect of substrate inhomogeneity on the quality of the *h*-BN dielectric stacks is still unclear. Here we show that the use of Pt substrates to grow *h*-BN via CVD leads to severe thickness inhomogenities, a behavior that is minimized when the *h*-BN are grown on Cu.

## Experimental

Multilayer *h*-BN stacks have been grown on polycrystalline Pt foils in a low pressure CVD furnace with two gas lines: one driving 70 sccm H₂ and the other inputting the liquid borazine precursor (assisted by 1 sccm H₂). The *h*-BN growth on Pt was carried out at 900 °C for 1 hour (after the typical annealing step for cleaning the substrate). The *h*-BN/Cu samples were prepared in a similar way by tuning the growth parameters. The *h*-BN stacks were transferred on SiO₂/Si wafers and metallic grids via bubbling method [3]; the Pt substrate was recycled. The *h*-BN/SiO₂/Si wafers and *h*-BN/grids were analyzed via Raman spectroscopy and transmission electron microscopy (TEM), respectively. The *h*-BN/Pt stacks were analyzed by conductive atomic force microscopy (CAFM), and the obtained data have been treated using Nanoscope software. Au/Ti top electrodes were deposited on the *h*-BN/Pt and *h*-BN/Cu samples for probe station analysis.

## Characterization of CVD grown *h*-BN on Pt

After growth, the *h*-BN stacks were transferred on SiO₂/Si wafers for Raman spectroscopy analysis (laser λ = 532 nm). We find out that the color of the *h*-BN sheet on the SiO₂/Si substrate (observed with an optical microscope) changed on different areas following patterns identical to the shape of the grains in the polycrystalline Pt foil on which the *h*-BN was grown (not shown). Raman spectroscopy revealed that the different colors correspond to different *h*-BN thicknesses (see Fig.1a). The typical Raman peak (~1370 cm⁻¹) appeared at all the spots tested, and it has a slightly shift towards ~1366 cm⁻¹ with the *h*-BN layers increase (bulk signal). Furthermore, high magnification cross-section TEM images collected on several *h*-BN/grids also proved thickness fluctuations within the *h*-BN stacks (from 1-2 to 10-13 layers). The amount of defects in the *h*-BN grown in our lab (observed via TEM) is much lower than in commercial samples (i.e. compare to Ref. [5]).

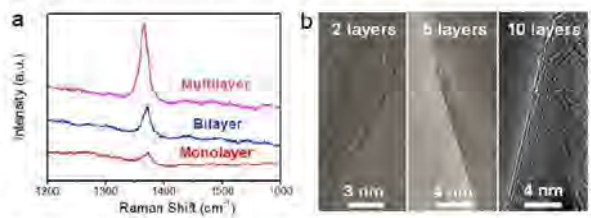

Fig.1: (a) Raman spectra of *h*-BN/SiO₂/Si on spots that correspond to different Pt grains (Pt is the substrate on which the *h*-BN was grown). (b) High resolution TEM images demonstrating different thicknesses and low amount of defects in the layered structure of the *h*-BN stacks.

Afterwards, the topographic and electrical performances of as-grown *h*-BN were characterized via CAFM under ambient atmosphere. Fig.2 displays the large-area (80 μm × 80 μm) topographic and current maps collected on as-grown *h*-BN/Pt stack simultaneously under a Pt substrate potential of -2V (CAFM tip grounded). For this experiment CAFM tips made of Silicon coated with 23 nm of a Pt-Ir alloy from Nanoworld (ref. ContPt) were used. The images show that the current collected on grains (A-G) can differ a lot from one to another (Fig.2b). Therefore, it can be concluded that the different thicknesses of the $h$-BN grown on different Pt grains (see Fig.1) produce severe current fluctuations from one grain to another. In particular, Fig.2 shows that grain B drives the largest currents (i.e. thinnest $h$-BN), while grain C drives the lowest currents (i.e. thickest $h$-BN). It is also worth noting that the conductivity within each single grain shows to be very homogeneous.

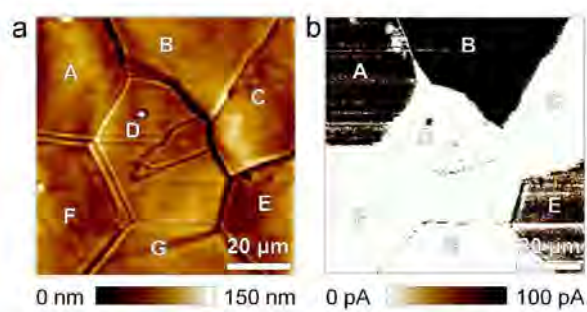

Fig. 2 (a) Topographic and (b) current maps simultaneously collected on as-grown $h$-BN/Pt, under a bias of -2 V applied to the Pt substrate (CAFM tip grounded).

In order to further display the differences on the currents registered in each grain, we statistically analyze Fig.2b using the software of the CAFM (called Nanoscope). The current at each single pixel of the current map in each grain is analyzed statistically in Fig.3. This plot displays much better the current deviations within each grain, as well as the current fluctuations from one grain to another. The mean current value among these grains varies from 0.74 pA (grain B) to 1.2 nA (grain C). Interestingly, the standard deviation of the currents all the grain (e.g. 0.067 pA for grain B and 398.14 pA for grain C) is comparable to that of other dielectrics currently used in microelectronics industry (e.g. $HfO_2$), indicating that the growth of large grains might enable the fabrication of electronic devices with a low device-to-device variability problems of ultra-scaled devices.

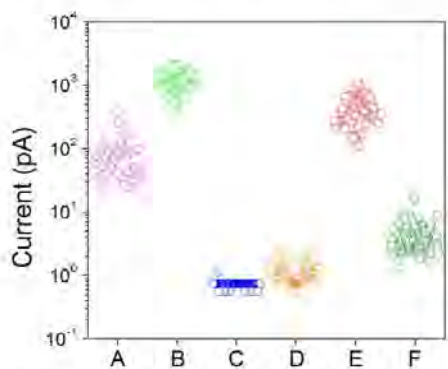

Fig.3. Statistical analysis of current value on each grain.

Device level characterization of Au/Ti/$h$-BN/Pt samples was carried out in the probe station. Fig.4 displays the current-voltage (I-V) sweeps collected on different devices located in same (a) and different (b) Pt grains. The results further support the CAFM observations: the variability in terms of pre-BD current and BD voltage is small/high in same/different grains.

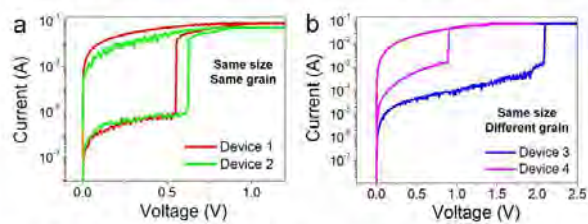

Fig.4. I-V sweeps collected on 25 μm × 25 μm and 50 μm × 50 μm Ti/$h$-BN/Pt devices located in the same (a) and different (b) Pt grains.

#### Characterization of CVD grown $h$-BN on Cu

The device level characterization has been repeated in $h$-BN stacks grown on Cu foils. Fig.5 shows I-V sweeps collected on different fresh devices. The lower currents (~$10^{-13}$A) and larger BD voltage (~8V) compared to Fig.4 (~$10^{-6}$A, <3V) are related to the thicker nature of the $h$-BN stack grown on the Cu foil. As it can be observed, in this case the variability of the pre-BD currents and BD voltage is minimal.

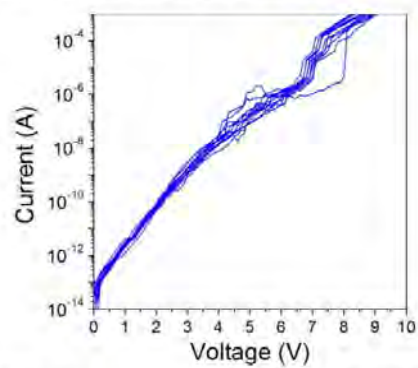

Fig.5. I-V curves collected on different fresh 50 um × 50 um devices in which the $h$-BN has been grown on Cu foils

#### Conclusion

Electrical characterization demonstrates that the multilayer $h$-BN stacks grown on Cu foils show a more homogeneous conductivity than those grown on Pt foils. The reason is that the thickness of the $h$-BN fluctuates on different Pt grains, most probably due to the different crystallographic orientation of the Pt surface on each grain, which may lead to different catalytic activity and $h$-BN thicknesses.

#### References

[1] Hui *et al.*, *Microelectron. Eng.*,163, 119-133, 2016.
[2] Hattori *et al. ACS Nano*, 9 (1), 916–921, 2015.
[3] Hui *et al. ACS Appl. Mater. Interfaces*, 9, 39895, 2017.
[4] Jiang *et al, ACS Appl. Mater. Interfaces*, 9, 39758, 2017.
[5] Pan *et al., Adv. Funct. Mater*, 27, 1604811, 2017.

# Dielectric Breakdown in Chemical Vapor Deposited Hexagonal Boron Nitride

Lanlan Jiang,[†] Yuanyuan Shi,[†,‡] Fei Hui,[†,§] Kechao Tang,[‖] Qian Wu,[†] Chengbin Pan,[†] Xu Jing,[†,⊥] Hasan Uppal,[#] Felix Palumbo,[¶] Guangyuan Lu,[∇] Tianru Wu,[∇] Haomin Wang,[∇] Marco A. Villena,[†] Xiaoming Xie,[∇,°] Paul C. McIntyre,[‖] and Mario Lanza[*,†]

[†]Institute of Functional Nano and Soft Materials, Collaborative Innovation Center of Suzhou Nanoscience & Technology, Soochow University, 199 Ren-Ai Road, Suzhou 215123, China

[‡]Department of Electrical Engineering and [‖]Department of Materials Science and Engineering, Stanford University, Stanford, California 94305, United States

[§]Department of Electrical Engineering and Computer Sciences, Massachusetts Institute of Technology, Cambridge, Massachusetts 02139, United States

[⊥]Microelectronics Research Center and Department of Electrical and Computer Engineering, The University of Texas at Austin, Austin, Texas 78758, United States

[#]Microelectronics and Nanostructures, The University of Manchester, Sackville Street, Manchester M13 9PL, U.K.

[¶]National Scientific and Technical Research Council (CONICET), UTN-CNEA, Godoy Cruz 2290, Buenos Aires, Argentina

[∇]State Key Laboratory of Functional Materials for Informatics, Shanghai Institute of Microsystem and Information Technology, Chinese Academy of Sciences, 865 Changning Road, Shanghai 200050, China

[°]School of Physical Science and Technology, ShanghaiTech University, 319 Yueyang Road, Shanghai 201210, China

**S** *Supporting Information*

**ABSTRACT:** Insulating films are essential in multiple electronic devices because they can provide essential functionalities, such as capacitance effects and electrical fields. Two-dimensional (2D) layered materials have superb electronic, physical, chemical, thermal, and optical properties, and they can be effectively used to provide additional performances, such as flexibility and transparency. 2D layered insulators are called to be essential in future electronic devices, but their reliability, degradation kinetics, and dielectric breakdown (BD) process are still not understood. In this work, the dielectric breakdown process of multilayer hexagonal boron nitride (h-BN) is analyzed on the nanoscale and on the device level, and the experimental results are studied via theoretical models. It is found that under electrical stress, local charge accumulation and charge trapping/detrapping are the onset mechanisms for dielectric BD formation. By means of conductive atomic force microscopy, the BD event was triggered at several locations on the surface of different dielectrics (SiO$_2$, HfO$_2$ Al$_2$O$_3$, multilayer h-BN, and monolayer h-BN); BD-induced hillocks rapidly appeared on the surface of all of them when the BD was reached, except in monolayer h-BN. The high thermal conductivity of monolayer h-BN combined with the one-atom-thick nature are genuine factors contributing to heat dissipation at the BD spot, which avoids self-accelerated and thermally driven catastrophic BD. These results point to monolayer h-BN as a sublime dielectric in terms of reliability, which may have important implications in future digital electronic devices.

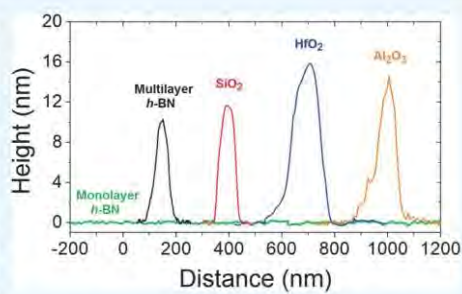

**KEYWORDS:** *dielectric breakdown, 2D materials, insulator, hexagonal boron nitride, CAFM*

## 1. INTRODUCTION

Insulators are key elements in most digital electronic devices because they can provide essential functionalities, such as capacitance effects in field effect transistors (FETs).[1] During device operation, insulating films are usually exposed to electrical fields in metal−insulator−semiconductor (MIS) and/or metal−insulator−metal (MIM) structures, which causes degradation of their microstructure and partial/complete loss of their insulating properties.[2] This phenomenon is known as dielectric breakdown (BD) and has been widely studied in several insulators for electronic devices (e.g., SiO$_2$, HfO$_2$ and Al$_2$O$_3$).[2−6] In these oxides, the percolation model is the most accepted theory for BD formation, and it states that the insulating capability is lost because of the formation of a defective conductive nanofilament (CNF) connecting the two sides of the dielectric.[7] When the last defect that forms the filament is trapped, the local currents increase sharply by several orders of magnitude, leading to the accumulation of thermal heat at the BD site.[2] This supplies nearby atoms (in both the oxide and adjacent metallic electrodes) with a high energy that produces avalanche currents,[3] lateral BD spot propagation,[4] and electromigration,[5] which ultimately results in irreversible surface extrusion (hillock formation)[6,8−10] and dramatic failure of the entire device. Avoiding BD-induced irreversible damage in dielectrics is highly desirable to enhance the reliability and lifetime of digital electronic devices,[11] but until now, all dielectrics known (e.g., $SiO_2$, $HfO_2$, and $Al_2O_3$) show severe hillock formation when they reach BD.[8−10] A number of authors[12−15] have shown that in poly-Si/$SiO_xN_y$/Si, the BD spot is characterized by the formation of a Si-rich region in the $SiO_xN_y$ dielectric.[12,14,15] In silicon-based technologies, BD-induced surface extrusion is also known as BD-induced epitaxy because the hillock forms by the nucleation of silicon atoms at the BD site (either at the silicon substrate or at the polysilicon gate interfaces), similar to epitaxial growth.[16,17] Similarly, the BD event in metal/$HfO_2$/$SiO_xN_y$/Si leads to a metal-rich region in the high-$k$ dielectric at the BD spot.[18] This is a clear evidence of atomic diffusion and electromigration, which results in the formation of CNFs in the dielectric. In traditional dielectrics (e.g. $SiO_2$, $HfO_2$, and $Al_2O_3$), the size of the BD-induced hillock depends on the polarity of the voltage applied[16] and on the magnitude of the currents generated during the BD event, which are directly related to the local temperature at the BD spot.[4] Within this framework, the thermal conductivity of the insulating material plays a fundamental role in the BD growth.

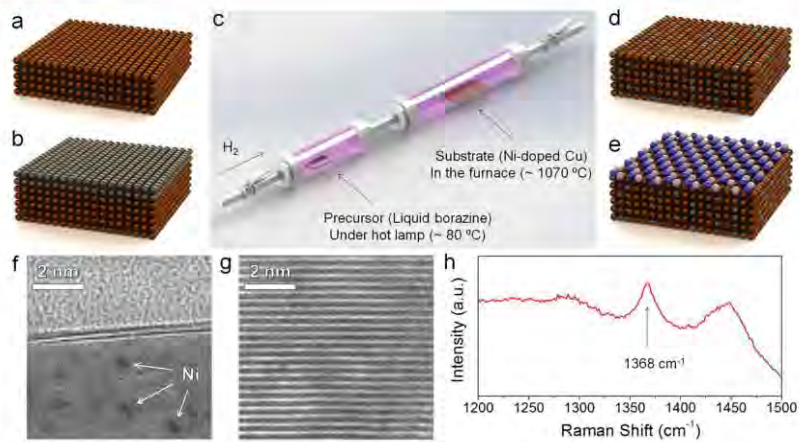

**Figure 1.** Schematic of the (a) as-received Cu substrate and (b) Cu substrate coated with a thin Ni film. (c) Schematic of the CVD furnace used for the annealing and h-BN growth. (d) Schematic of the resulting CuNi substrate after thermal annealing. (e) Schematic of the h-BN/CuNi sample. In (a), (b), (d), and (e), the brown, dark gray, blue, and light gray balls represent Cu, Ni, N, and B atoms, respectively. Cross-sectional TEM images of (f) the monolayer h-BN sheets on the CuNi substrate and (g) the multilayer h-BN stack. In panel (f), the top part corresponds to the chromium protective layer (only for TEM) and the bottom part is the CuNi substrate; the dark areas in the bottom part of (f) are the Ni dopants in the Cu substrate. (h) Raman spectrum of h-BN; for this measurement, the h-BN was transferred on a 300 nm $SiO_2/Si$ substrate.

With the introduction of two-dimensional (2D) materials in the structure of micro- and nanoelectronic devices, the concept of BD needs to be revised as 2D materials hold special physical, chemical, and mechanical properties. For example, recent reports successfully fabricated MISFET devices using exclusively 2D layered materials [i.e., graphene as the conductive electrode, hexagonal boron nitride (h-BN) as the insulator, and $MoS_2$ as the semiconductor],[19,20] but because the point of the BD in such type of material systems were not reported. One clear advantage of 2D materials from a device reliability point of view is their superb thermal conductivity (3080−5150 W/m K in graphene,[21] 83 W/m K in $MoS_2$,[22] and 360 W/m K in h-BN),[23] which may dissipate local thermal heat, reduce avalanche currents, and slow down electromigration, enhancing the overall reliability of the entire device. As the insulator is the most determinant layer defining the kinetics of the BD event in a device, understanding the BD process in 2D layered insulators (e.g., h-BN) is crucial to assess the reliability of 2D material-based electronic devices. Unfortunately, these kind of studies are very scarce. It is known that the dielectric strength of h-BN (12 MV/cm)[24] is larger than those of traditional oxides (7−9 MV/cm in $SiO_2$[25] and 2−4.5 MV/cm in $HfO_2$),[26] and that the BD process in 2D layered insulators (e.g., h-BN) takes place layer-by-layer because of the anisotropic speed of defect formation, which is related to the different atomic interactions in the layered stack: covalent bonding in-plane and van der Waals attraction plane-to-plane.[24,27,28] Here, the BD process in multilayer h-BN stacks and monolayer h-BN sheets is analyzed on both the nanoscale and on the device level, and the experimental results are further studied via theoretical BD modeling. Our experiments indicate that the degradation of the h-BN stacks takes place because of the local accumulation of defects and charge trapping/detrapping at weak sites; this means that although the BD in 2D layered insulators may be reached layer-by-layer, the degradation via local defects formation is a universal behavior that also applies to 2D materials. By means of conductive atomic force microscopy, the BD event is triggered at several locations on the surface of different dielectrics ($SiO_2$, $HfO_2$, $Al_2O_3$, multilayer h-BN, and monolayer h-BN); BD-induced surface extrusion rapidly appeared on the surface of all of them when the BD was reached, except in monolayer h-BN. The high thermal

 

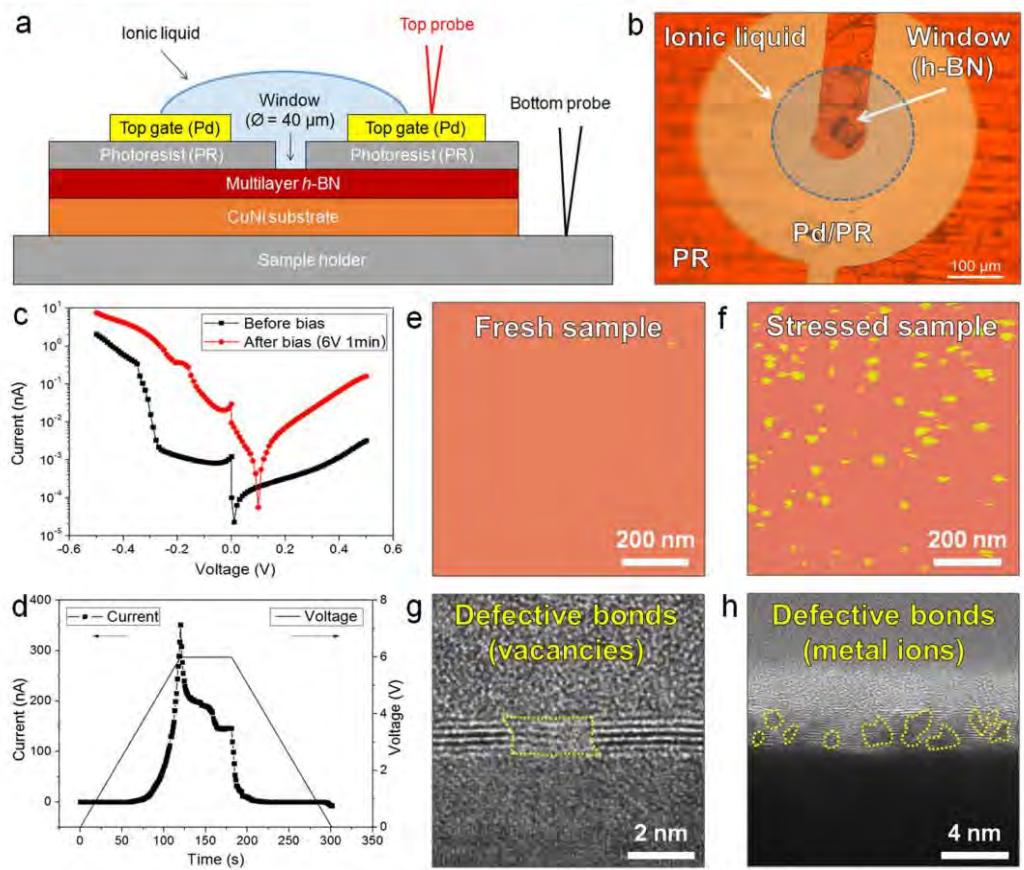

**Figure 2.** (a) Schematic and (b) photograph of the setup used for IL electrical stress in the h-BN/Cu sample. (c) I−V curves collected in the IL/h-BN/CuNi test structures before and after the electrical tests. The voltage sweeps from −0.5 V to 0.5 V. The sweep rate is sufficiently slow (∼0.01 V/s) to ensure that the current is measured at a steady state. (d) Electrical tests applied to the IL/h-BN/CuNi samples. Current maps collected on the surface of the h-BN/CuNi samples before (e) and after (f) the IL stress (rose = 0 pA; yellow = 1 nA). Cross-sectional TEM images of (g) Ti/h-BN/CuNi and (h) Ti/h-BN/Au devices exposed to electrical stress. In both cases, the h-BN shows stress-induced defective regions, which are attributed to missing bonding (B-vacancies generation in panel g) and penetration of impurities from adjacent layers (metallic ions in panel h). These defective regions are highlighted with dashed yellow lines.

conductivity of h-BN in the basal plane[23] combined with the one-atom-thick nature are genuine factors contributing to heat dissipation (probably through the adjacent electrodes),[29] which avoids self-accelerated and thermally driven catastrophic BD. Our results point to monolayer h-BN as a sublime dielectric in terms of reliability, which may have important implications in future digital electronic devices.

## 2. RESULTS AND DISCUSSION

Monolayer h-BN sheets and multilayer h-BN stacks with different thicknesses ranging between 5 and 25 layers were grown by chemical vapor deposition (CVD) on Ni-doped Cu (CuNi) substrates following the process developed in our recent work (see Methods and Figure 1a−e).[30] The advantage of using Ni doping in Cu substrates is that the grain size in the CVD-grown polycrystalline h-BN sheet/stack is larger; this results in a better layered structure and less number of defects in the h-BN because of the minimization of the number of grain boundaries, which are highly defective.[30] The presence of h-BN on the CuNi substrates after the CVD growth process was corroborated via cross-sectional transmission electron microscopy (TEM, see Figure 1f,g) and Raman spectroscopy (Figure 1h). The TEM images reveal an excellent layered structure, which is essential to ensure large thermal conductivity in the h-BN. The surface roughness of the h-BN/CuNi samples was analyzed on the nanoscale by means of atomic force microscopy (AFM); the images show the typical steps in the CuNi substrate beneath the h-BN stack, and the surface roughness of the h-BN on the CuNi plateaus is very low [root mean square < 0.2 nm, Figure S1]. This further confirms the excellent morphology of the samples fabricated in this investigation.

The degradation of 5−7 layer thick h-BN/CuNi stacks was induced by applying an homogeneous electrical field in a circular area of 40 μm diameter using the probe station; after that, the same area was scanned using a conductive atomic force microscope (CAFM, working in the contact mode) to map the degradation (increase of conductivity) induced in the insulating h-BN stack. Normally, this kind of test is performed by depositing a top metallic electrode on the insulator, and this top electrode needs to be removed before the CAFM characterization.[31−33] Different methods to remove the top electrode have been suggested, including wet etching,[31] dry etching,[32] and even CAFM-tip-induced etching.[33] However, all of them provide poor controllability on the etching and can easily damage the surface of the insulator. In this investigation, we use the approach recently reported in ref 34, in which the top electrode is replaced by an ionic liquid (IL, see Figure 2a,b, Methods section, and Figure S2). Using this method, the IL can be easily rinsed after the probe station electrical stress, and then the surface of the sample is exposed and scanned using conductive atomic force microscopy.

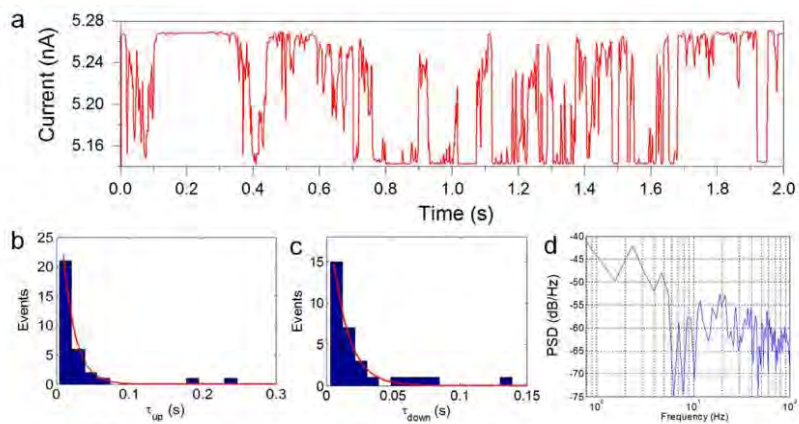

**Figure 3.** RTN signal analysis. (a) $I$–$t$ curve collected with the tip of the CAFM on the multilayer h-BN showing the RTN signal. (b,c) Calculation of the time for trap capture ($\tau_{up}$) and emission ($\tau_{down}$) for the RTN signal displayed in (a). The values obtained are 20 and 12 ms. (d) Calculation of the power spectral density for the same RTN signal (21.22 Hz).

The IL electrical test consisted of the following: (i) A fresh IL droplet was first placed over the window region, and a spectroscopic ramped voltage stress (RVS) using a very low voltage from −0.5 to +0.5 V was applied to a fresh h-BN/CuNi sample to characterize its initial conductivity. The corresponding current versus voltage ($I$–$V$) curve is shown in Figure 2c (black squares); (ii) The device was transferred to a vacuum probe station and pumped down to below 1 mTorr. Then a constant voltage stress (CVS) at +6 V for 1 min was applied with the aim of degrading the microstructure of the h-BN stack. The evolution of the current versus time ($I$–$t$ curve) is displayed in Figure 2d; and (iii) Following the transfer of the device out of the vacuum, the IL was rinsed off and replaced by a fresh droplet of IL. Then, another RVS from −0.5 to +0.5 V was applied after the CVS to characterize the conductivity of the stressed h-BN/CuNi sample. The corresponding $I$–$V$ curve is shown in Figure 2c (red circles). Figure 2c clearly shows that the overall conductivity of the h-BN stack increased after the CVS. Figure 2e,f shows the typical current maps collected with the CAFM on the surface of the fresh and stressed h-BN/CuNi samples, respectively. In this experiment, the CAFM was operated in the contact mode under a tip bias of +1 V (CuNi substrate grounded). As Figure 2f shows, after the electrical stress, several highly conductive (yellow) spots appeared randomly distributed along the surface of the sample. Using the software of the CAFM, the density of the conductive spots, their size, and the currents driven are statistically analyzed, and they are 105.34 spots/$\mu m^2$, 459.43 ± 432.42 $nm^2$, and 4.15 ±

1.74 nA, respectively (see Figures S3 and S4). The atomic rearrangements produced by the electrical field in h-BN-based MIM devices were analyzed via cross-sectional TEM (see Figure 2g,h). The experiments reveal that the highly conductive spots are related to the formation of defective bonds within the microstructure of the h-BN stack, probably because of the migration of boron vacancies[35] and/or penetration of species from the adjacent electrodes.[36,37] The content of impurities (carbon oxygen) was low and did not change with the application of a bias, meaning that these species are not related to the resistive switching mechanism. This degradation mechanism is very similar to that observed in three-dimensional (3D) insulators[38] and indicates that although the degradation kinetics of multilayer 2D insulators may be different (layer-by-layer), the physical mechanism producing the degradation of the material is the same.

The local formation of defects within the h-BN stacks was further investigated via $I$–$t$ curves collected with the tip of the CAFM on the surface of fresh 5−7 layer thick h-BN/CuNi stacks (see Figure 3a). The $I$–$t$ curves collected show abrupt random fluctuations between different well-defined conduction levels. This behavior is typical of the random telegraph noise (RTN) signal[39,40] and indicates the trapping and detrapping of charges in the multilayer h-BN stack. For the $I$–$t$ curve in Figure 3a, the time for trap capture ($\tau_{up}$) and emission ($\tau_{down}$) were statistically calculated, and they are 20 and 12 ms, respectively (see Figure 3b,c). Figure 3d shows the power spectral density plot, which is 21.22 Hz. Because the number of conduction levels in Figure 3a is only two, most probably the RTN signal in that plot corresponds to the trapping and detrapping of a single trap. Other locations of the sample showed up to four discrete conduction levels, indicating that multitrap RTN is also possible. Therefore, the local trapping and detrapping of charges in the h-BN during its degradation process is a universal behavior that can be extrapolated to 2D layered insulators.

In the next step, the effect of the BD event in the h-BN stack was analyzed using the CAFM. Nine spectroscopic RVSs from 0 V to $V_{MAX}$ were applied at different locations on the bare surface of a 5−7 layers thick h-BN/CuNi sample (locations A−I in Figure 4a); the value of $V_{MAX}$ was 8 V at positions A−C, 4

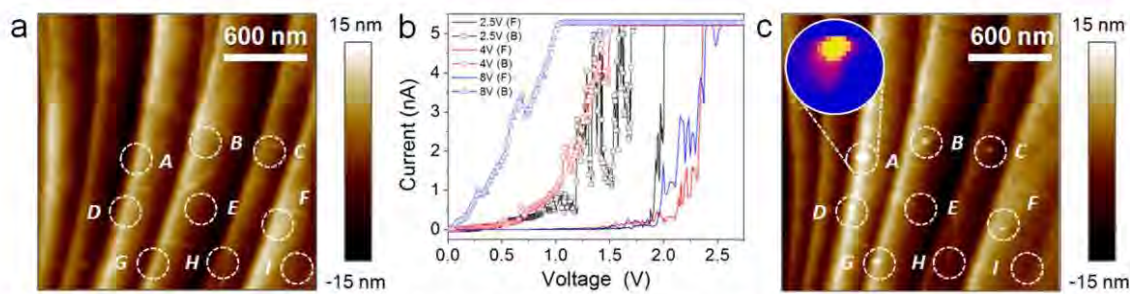

**Figure 4.** (a) AFM topographic map obtained on a 5−7 layer thick h-BN grown on CuNi substrates. The CuNi steps can be observed, and the surface of the h-BN within each CuNi terrace is atomically flat. Nine RVSs ranging from 0 V to $V_{MAX}$ were applied at the area shown in (a). The value of $V_{MAX}$ was 8 V for locations A−C, 4 V for locations D−F, and 2.5 V for locations G−I. Panel (b) shows the typical forward (F) and backward (B) $I−V$ curves measured in each type of RVS. (c) Topographic map measured at the same location after the RVS, under 1 V biasing. Hillock formation can be observed at most locations. The number of locations and the width/height of the hillocks increase with $V_{MAX}$. The inset in (c) is the current map of one hillock (blue = 0 pA, yellow = 10 pA).

V at positions D−F, and 2.5 V at positions G−I. Figure 4b shows the typical forward (F) and backward (B) current versus voltage ($I−V$) curves measured when using different $V_{MAX}$ values. In all cases, the currents driven during the forward curves are very small, and they increase remarkably during the backward curves, confirming the presence of an insulating material (h-BN) on the CuNi substrate. The voltage at which the forward (F) $I−V$ curves start to show currents (from now, onset potential, $V_{ON}$) is ∼1.5 V, which agrees well with the values previously reported in similar experiments and calculations (for h-BN sheets of similar thicknesses).[27,39] Interestingly, $V_{ON}$ is very similar for all forward $I−V$ curves, indicating that this h-BN sample is intrinsically very homogeneous; this is always desirable to reduce the device-to-device variability in patterned nanodevices. At these low voltages, the localized currents measured correspond to direct and/or Fowler−Nordheim tunneling across the h-BN stack.[27,39] At around 1.9 V, all forward $I−V$ curves show a sudden increase in the current, probably related to the generation of defects within the h-BN stack. When the current reaches 5.5 nA, the $I−V$ curves become horizontal, indicating that the saturation level of the CAFM has been reached. All backward (B) $I−V$ curves shift to lower potentials, corroborating the generation of defects that favor the leakage current. The magnitude of this shift is proportional to the value of $V_{MAX}$. The backward ramp when using $V_{MAX} = 2.5$ V shows abundant current fluctuations, indicating severe charge trapping and detrapping. Nevertheless, the stress voltage applied was not large enough to induce a consistent percolation path across the h-BN as $V_{ON}$ is well above zero (it is >0.5 V). The backward ramp using $V_{MAX} = 4$ V shows slightly larger currents that are less noisy, indicating a more severe degree of degradation. Again, the percolation path is not completely formed because the stressed location still needs non-negligible voltages (>0.5 V) to display currents above the noise level. Finally, the backward ramp using $V_{MAX} = 8$ V shows a near-zero $V_{ON}$, indicating that an effective CNF has been completely formed.[9] This is also supported by the change in the shape of the $I−V$ curve: exponential for the RVSs with $V_{MAX}$ of 2.5 and 4 V and linear for those with $V_{MAX} = 8$ V.

After the spectroscopic RVS (Figure 4b), the same area of the sample was scanned again, and the resulting topographic map is shown in Figure 4c. As it can be observed, all RVSs with $V_{MAX} = 8$ V show a BD-induced hillock formation (spots A, B, and C). For the other six RVSs ($V_{MAX}$ of 4 and 2.5 V), only three hillocks appeared in the topographic map (Figure 4c, spots D, F, and G), and they are shorter (in average), indicating a smaller degree of degradation at lower $V_{MAX}$. This observation is in agreement with the larger shifts observed for RVS using higher $V_{MAX}$ in Figure 4b. The current maps show that these hillocks drive much larger currents compared to the unstressed locations (see the inset in Figure 4c), corroborating the degradation of the multilayer h-BN stack. These experiments were repeated at 23 locations of 2 multilayer h-BN samples with thicknesses ranging between 5 and 25 layers, and similar results were observed.

To compare the formation of BD-induced hillocks in different stoichiometric dielectrics, these experiments were repeated on the surface of 4 nm $HfO_2$, 10 nm $Al_2O_3$, and 1 nm $SiO_2$ films, all of them grown by atomic layer deposition on silicon (see Methods). Figure 5a,b shows the topographic AFM maps collected on the surface of 4 nm $HfO_2$/Si and 10 nm $Al_2O_3$/Si samples, on which the BD event was previously triggered via RVS at one and four different locations, respectively. These two experiments were carried out with the CAFM working in the contact mode and in normal air atmosphere. Figure 5c shows the topographic AFM map collected on the surface of a 1 nm $SiO_2$/Si sample, on which the BD event was triggered at six different locations. This experiment was carried out in ultrahigh vacuum (UHV) atmosphere and by applying different current limitations. Figure 5d shows the horizontal cross-section at the central-upper part of Figure 5c. In all cases, profound electrical-field-driven surface extrusion (hillock formation) was observed, which was much more dramatic than in h-BN. In the multilayer h-BN (Figure 4c) as well as in $SiO_2$ and transition metal oxides,[8−10] larger thickness, $V_{MAX}$, and/or current limitation always resulted in a larger surface extrusion (see Figure 5d). It should be highlighted that the BD-induced hillocks observed in Figure 5a,b cannot be related to the presence of water molecules on the sample when measuring in normal air atmosphere (i.e., local anodic oxidation)[41] because in all cases, the RVSs were applied by injecting electrons from the substrate (see Methods).[42,43] This is further corroborated by the formation of BD-induced hillocks on the surface of $SiO_2$ when measuring under UHV conditions (Figure 5c,d).

The hillocks generated on the surface of the multilayer h-BN stack during the BD event were analyzed in-depth from zoom-in topographic, adhesion, and deformation maps collected in the PeakForce TUNA[44] mode under a tip bias of 1 V (see Figure 6). This mode collects one force versus distance (F–Z) curve at each pixel of the image. Whereas the topographic map (Figure 6a) only displays a central protrusion ~11.2 nm in height and ~92 nm in diameter, the adhesion and deformation maps (Figure 6b,c) show concentric ringlike structures that overlap very well with the hillock observed in the topographic map. The adhesion map refers to the interaction force between the tip and the sample just before the tip detaches from the sample in each F–Z curve. This force ($F_{ad}$) depends on several parameters,[43] including capillary forces ($F_{cap}$), van der Waals forces ($F_{vdW}$), forces related to chemical bonds or acid−base interactions ($F_{chem}$), and electrostatic forces ($F_{el}$):

$$F_{ad} = F_{cap} + F_{vdW} + F_{chem} + F_{el} \qquad (1)$$

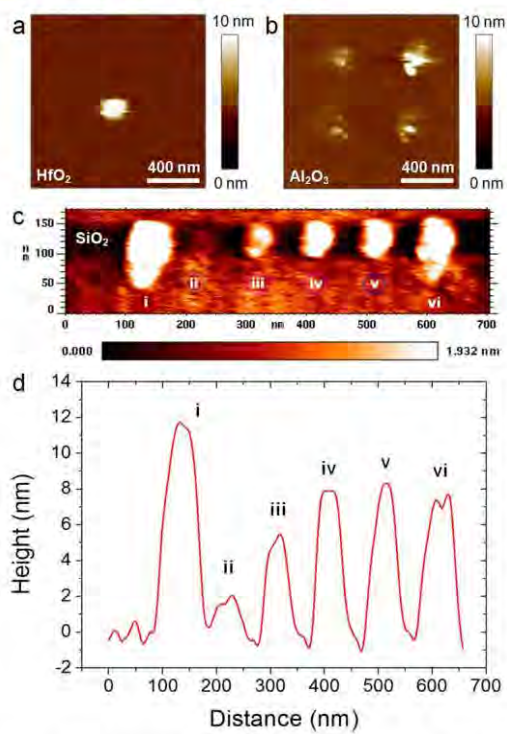

**Figure 5.** Comparison of the BD-induced hillock formation in different materials. Topographic AFM maps collected at the BD location/s for (a) 4 nm HfO₂, (b) 10 nm Al₂O₃, and (c) 1 nm SiO₂. The BD events were triggered at different locations of the sample via RVS before the scans. (a) and (b) have been collected in a CAFM working in air, whereas (c) has been collected in UHV. In (c), the RVS (from 0 V to $V_{MAX}$ of ~7.5/8 V) had been previously applied at six different locations. The RVS at location (i) did not use any current limitation (CL), and the RVS at locations (ii), (iii), (iv), (v), and (vi) used a CL of 50, 100, 500 pA, 1, and 10, respectively. (d) Cross-section at the upper-central part of (c), which displays the size of the hillocks and demonstrates quantitative control of the size of the BD-induced hillocks via CL.

As the only difference between a stressed and unstressed location in Figure 6b is the amount of charges trapped in the h-BN stack during the BD (which could alter $F_{ad}$), in this experiment, changes in the adhesion force can be attributed to the different distributions of charge trapped in the dielectric. The typical adhesion force between the unstressed h-BN and the CAFM tip (under a tip bias of 1 V) can be deduced from the outer area in Figure 6b (blue color), and it is near zero. At the BD spot (dark central area), the adhesion map shows negative (attractive) forces up to ~−50 nN. Most probably, the atomic rearrangements in the h-BN dielectric stack at the BD location altered the $F_{ad}$ contribution in eq 1: it has been reported that the amount of charge in nanoparticles can strongly modify the interaction force in F–Z curves.[45] As the tip bias during the scan was 1 V, the large attractive force indicates that the sign of the charges trapped at the BD location during the RVS is negative, being consistent with the kinetics of the BD event. When a positive RVS is applied to the Pt-coated CAFM tip in contact with the h-BN/CuNi structure, Cu⁺ ions cannot penetrate in the h-BN stack because they are dragged by the electrical field in the opposite direction, and the Pt coating from the CAFM tip is a noble, stable, and inert material that requires higher energies for electromigration.[46] On the contrary, abundant migration of boron toward the anode during the BD (reservoir formation) has been readily observed via electron energy loss spectroscopy (EELS);[36,37] it is known that the activation energy of boron vacancies is much lower than that of nitrogen ones.[35] This observation of boron movement toward the positive electrode implies that the boron ions need to be negatively charged. Although boron is often considered to be an electron donor (it can lose three electrons to become stable), boron atoms can also accept electrons to become stable. Moreover, the electronic affinity of boron is 27 kJ/mol, which indicates facility for becoming ionized. In addition, the local energy generated during the BD event is very high (the current density can easily reach $J \approx 10^6$ A/cm²), facilitating boron ionization. Therefore, the high attractive forces observed at the central part of Figure 6b should be related to the accumulation of negatively charged B⁻ ions at the BD spot. Interestingly, the adhesion map shows a ringlike structure (yellow/green colors) surrounding the BD spot. This area, which is masked in the topographic map, shows repulsive forces (~50 nN), indicating that the charges within the h-BN stack at these locations may have an opposite polarity (positive) compared to the center of the BD spot (negative). Probably the negative charges within the h-BN stack at the BD location repeal/attract the negative/positive mobile charges nearby, generating a ringlike area with inverse polarity surrounding the BD spot.

Additional information can be gained from the deformation map (Figure 6c), which can be understood as the modification of the contact forces between the tip and the sample.[39] As displayed in Figure 6b, the contact forces at the BD location are governed by the charges trapped in the dielectric; therefore, a high deformation signal can be understood as a change in the amount of charges trapped in the dielectric during the measurement. It is known that the BD event in a dielectric can generate both deep and superficial traps;[47] the first type is normally immobile (also called fixed), whereas the second can get self detrapped with time and/or when another body (such as the CAFM tip) gets in contact with them. In Figure 6c, the very center of the BD spot shows a low deformation (smallest circle, yellow color); this is an indication that the central part of the filament is stable and made of fixed charges. On the contrary, the surrounding areas within the BD spot region (middle circle, black/pink/purple colors) reveal mobile charges that get detrapped during the scan as the deformation signal is larger. Finally, the red area surrounding the BD spot in Figure 6c, which as mentioned above corresponds to the presence of positive charges within the h-BN stack (yellow-green ring in Figure 6b), shows almost negligible deformation. This observation is consistent with the presence of fixed negative charges at the BD location. Figure 6d shows the cross-sectional schematic of the conductive filament structure. Multiple investigations have reported the in situ observation of CNFs through different kinds of dielectrics via scanning probe microscopy (SPM);[48,49] however, to the best of our knowledge, the charge separation at the BD location shown in Figure 6 has never been reported before. Similarly, we are not aware of other works analyzing the amount of charge trapped in a dielectric using adhesion and deformation images collected via SPM. This new methodology can complement very well the information about the BD spot traditionally collected via conductive atomic force microscopy and Kelvin probe force microscopy.

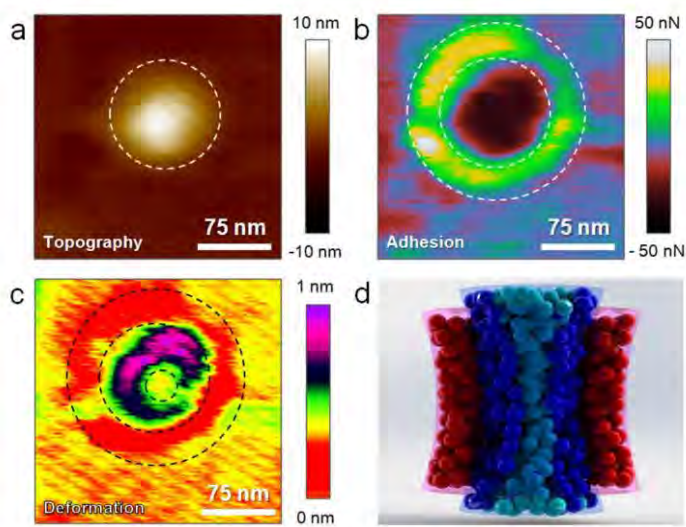

**Figure 6.** Nanoscale characterization of the BD-induced hillock formation in the multilayer h-BN. (a) Topography, (b) adhesion, and (c) deformation maps collected with the CAFM in the PeakForce TUNA mode. (d) Cross-sectional schematic of a CNF in the multilayer h-BN upon analysis of panels (a), (b), and (c). Light-blue balls represent fixed negative charges, dark-blue balls represent mobile negative charges, and red balls represent fixed positive charges.

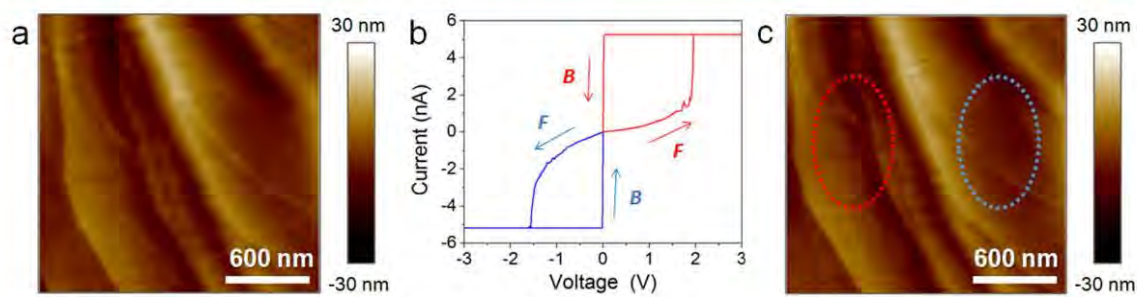

**Figure 7.** Absence of BD-induced surface extrusion in the monolayer h-BN. (a) AFM topographic map collected on the surface of the monolayer h-BN. The dielectric BD has been induced at eight locations of this area, by applying four RVSs ranging from 0 to +8 V and four RVSs ranging from 0 to −8 V. (b) Typical forward (F) and backward (B) $I-V$ curves collected during these RVSs. The observation of a transition from nonlinear to linear conduction corroborates the presence of the insulating h-BN on the CuNi substrate. The presence of h-BN on the CuNi substrate has also been proved via TEM (see Figure 1f). (c) AFM topographic map collected after the application of RVS. The red and blue ellipses indicate the areas where the RVSs from 0 to +8 V and 0 to −8 V were applied (respectively). No signal of BD-induced hillock formation has been detected.

The surprising observation came when these experiments were repeated in monolayer h-BN sheets. As in the case of multilayer h-BN (Figure 4a), the surface of fresh monolayer h-BN samples is very flat and displays the typical steps of the CuNi foil (Figure 7a). Several RVSs from 0 to ±8 V are applied at different locations of the sample. The typical $I-V$ curves collected are displayed in Figure 7b. Interestingly, the currents driven in both polarities during the forward (F) ramps fit well with the previous experimental and modeled observations in the monolayer h-BN;[27,39] this, together with the cross-sectional TEM image displayed in Figure 1f, confirms the presence of the monolayer h-BN on the CuNi substrate. From an electrical point of view, the BD in the monolayer h-BN sheet was even stronger than the BD in the multilayer h-BN, as corroborated by the higher slope of the backward (B) curve rising from 0 V. Contrarily to what was expected, subsequent topographic maps collected at the BD locations never showed any signal of surface modification (see Figure 7c). These experiments were repeated at 32 different locations of the monolayer h-BN/CuNi samples, and BD-induced hillock formation was never observed. To discard any influence of the different substrates (in Figure 5, the SiO₂, HfO₂ and Al₂O₃ materials were grown on nSi, not on CuNi), these experiments were further repeated after transferring the h-BN on a nSi substrate (without its native oxide). The results are displayed in Figures S6 and S7. The data collected prove that (i) the surface of both nSi and h-BN/Si samples is atomically flat; (ii) the BD event is reached (the backward plot is shifted toward lower potentials); and (iii) there is no electrical-field-driven surface extrusion (hillock formation) after the BD event.

Figure 8 compares the height of the BD-induced hillocks triggered on the surface of all materials studied in this work (for

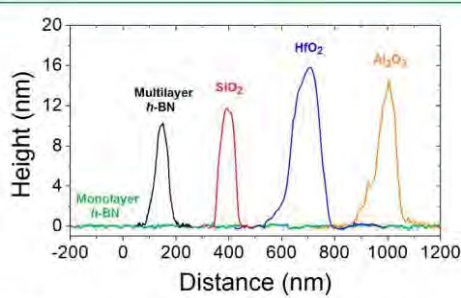

**Figure 8.** Cross-sectional analyses of BD-induced hillock formation in multilayer h-BN, SiO₂, HfO₂, and Al₂O₃ (extracted from Figures 4c, 5c,a,b, respectively). The profile of monolayer h-BN at the BD location (extracted from Figure 7c) has also been plotted for comparison. Multilayer h-BN shows BD-induced surface extrusion (hillock formation) comparable to that of traditional 3D dielectrics. Monolayer h-BN shows no signal of BD-induced surface extrusion.

all materials, the median hillock height of all experiments has been displayed; for all samples, only hillocks induced without the use of current limitation during the RVS curves have been considered). As it can be observed, the monolayer h-BN is the only dielectric capable of maintaining its flat surface after the BD, even if the magnitude of the currents measured during the BD was much larger (compare the backward $I-V$ curve for $V_{MAX}$ = 8 V in Figures 4b and 7b). It should be highlighted that the atoms that form the hillock not only come from the insulator but also come from the substrate (often in an even larger proportion) because of thermal electromigration.[16,17] Therefore, despite being the thinnest dielectric, the monolayer h-BN protects more effectively the MIM structure from thermal electromigration and surface extrusion. The high thermal conductivity of h-BN[11] combined with the one-atom-thick structure of monolayer sheets should be the genuine factors promoting thermal heat dissipation at the BD spot (most probably through the electrodes), which results in an unaltered surface and a superior electronic reliability. Table 1 shows the thermal conductivity of different 2D materials and thin dielectrics[21−23,50−56] as well as the dielectric strength for those materials that are insulators.[24−26,57−59] As it can be observed, the h-BN shows the highest thermal conductivity among all insulators, which correlates with the largest dielectric strength, pointing to the thermal conductivity as the main factor behind superior dielectric reliability. Interestingly in Table 1, the thermal conductivity of the monolayer h-BN (>600 W/m K) is much higher than that of the multilayer h-BN (∼230−300 W/m K), further supporting the different behaviors observed in Figures 4c and 7c. Some works[60] analyzed the tunneling current in exfoliated atomically thin h-BN samples (1−30 layers) but not in the BD event. Other works studied the BD process in thick (>33 layers) exfoliated h-BN samples via conductive atomic force microscopy.[61] In that case, the BD formed one hole on the surface of the h-BN

**Table 1. Thermal Conductivity (at Room Temperature, ∼300 K) and Dielectric Strength of Different 2D Materials and Traditional 3D Insulators**[a]

| materials | material classification | thermal conductivity (W/m K) | sample description | ref | dielectric strength (MV/cm) | ref |
|---|---|---|---|---|---|---|
| graphene | 2D conductor | 4840 ± 440−5300 ± 480 | suspended single layer | 50 | NA | NA |
| | | 3080−5150 | suspended single layer | 21 | NA | NA |
| | | 2500−5300 | suspended single layer | 51 | NA | NA |
| | | 600−5000 | suspended single layer | 52 | NA | NA |
| | | 5000 | suspended single layer | 53 | NA | NA |
| MoS₂ | 2D semiconductor | 83 | monolayer | 22 | NA | NA |
| | | 34.5 ± 4 | monolayer | 55 | NA | NA |
| | | 52 | suspended few layers | 53 | NA | NA |
| phosphorene | 2D semiconductor | 10−35 | suspended few layers | 52 | NA | NA |
| h-BN | 2D insulator | 250 | 5 layers thick | 23 | 12 | 24 |
| | | 360 | 11 layers thick | | 12 | 24 |
| | | 250−360 | 11 layers thick | 52 | 12 | 24 |
| | | >600 | single layer | 54 | 12 | 24 |
| | | 230 | few layers | 53 | 12 | 24 |
| SiO₂ | 3D insulator | 0.69−1.4 | thickness 20−1560 nm | 56 | 10 | 57 |
| | | | thickness 20−1560 nm | 56 | 5−10 | 59 |
| | | | thickness 20−1560 nm | 56 | 7−9 | 25 |
| Al₂O₃ | 3D insulator | 0.49−2.3 | thickness 5−55 000 nm | 56 | 10 | 57 |
| HfO₂ | 3D insulator | 0.3−2.55 | thickness 3−500 nm | 56 | 2−4.5 | 26 |
| TiO₂ | 3D insulator | 0.35−3 | thickness 110−2000 nm | 56 | 0.4 | 58 |

[a]NA indicates not applicable because conductive and semiconducting materials do not have the property of dielectric strength. 2D indicates that this material has a layered structure, with only atomic bonds in plane and layer-to-layer attraction via van der Waals forces.

 

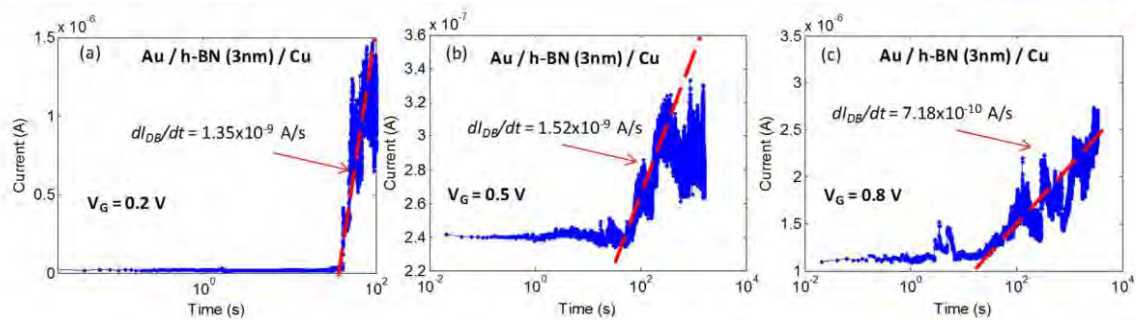

**Figure 9.** $I–t$ curves collected in three different Ti/3 nm h-BN/Cu capacitors when stressed at different constant voltages of (a) 0.2 V, (b) 0.5 V, and (c) 0.8 V during different periods of time. In all cases, progressive BD can be distinguished.

(material removal), as detected by subsequent topographic AFM maps, in contrast to our study. However, those samples are much thicker than the ones studied here, and thicker samples may block electromigration effects. Therefore, comparisons between our work and ref 61 are not meaningful. In any case, in this study, we want to concentrate only in CVD-grown h-BN because these samples are scalable and competitive for mass device fabrication, whereas mechanical exfoliation is not a synthesis method suitable for the industry.

To understand the influence of thermal conductivity of the h-BN layer on the BD event, additional electrical characterization was conducted on the device level. Figure 9 shows the $I–t$ curves collected at different voltages in three different Au/Ti/h-BN/CuNi devices. Interestingly, the slope of the $I–t$ curves is similar independent of the current level. This observation is different from what was expected, and it is in contrast to what has been previously reported in traditional dielectrics (e.g., SiO$_2$, HfO$_2$, and Al$_2$O$_3$); in these materials, larger currents produce larger local thermal heat,[62] which promotes additional defect generation.[63] This self-accelerated process results in a faster increase in the current (higher slope in the $I–t$ curve).[64,65] Moreover, the degradation process is very progressive, which further suggests that the breakdown process is influenced by the high thermal conductivity of the h-BN layers in the BD event. It should be highlighted that by means of EELS profiles collected at the BD spot locations, a recent paper clearly shows that the BD in Ti/h-BN/CuNi capacitors is related to the migration of B toward the Ti electrode and at the same time penetration of Ti into the h-BN layer.[57]

In the next step, the BD process in Ti/h-BN/CuNi devices was analyzed using the model recently developed in refs 64 and 66, in which the BD growth rate d$I_{BD}$/dt is described. The boron vacancy migration is considered because it is the first thermally activated defect. The diffusion coefficients obtained for the migration processes in the h-BN are obtained from ref 35. The d$I_{BD}$/dt is described by the following equation reported in ref 42.

$$\frac{dI_{BD}}{dt} = \frac{q \cdot V}{k_B \cdot T} \cdot \frac{f_1}{t_{ox}^2} \cdot D \cdot I_{BD} \qquad \text{with: } f_1 = n_e \cdot \lambda_e \cdot \sigma_e$$

where $q$ is the elementary charge, $V$ is the stress voltage, $k_B$ is the Boltzman constant, $T$ is the progressive BD (PBD) spot temperature, $t_{ox}$ is the oxide thickness, $D$ is the bottle neck diffusivity of the atomic species among those participating in the PBD spot growth, and $f_1$ represents the probability of collision between the electron and the atom, producing electromigration with $n_e$, $\lambda_e$, and $\sigma_e$ being the electron density, electron mean free path, and cross-section for atom-electron collision, respectively. The values of $D$ and $T$ are given by the following equations:

$$D = D_0 \exp(-E_a/k_B \cdot T) \qquad \text{with: } T = \frac{f_2 V I_{BD}}{2\pi t_{ox}\kappa} + T_{amb}$$

where $E_a$ is the activation energy for atom diffusion, $f_2$ is the fraction of the energy $qV$ per electron lost at the BD spot, $\kappa$ is the thermal conductivity, and $T_{amb}$ is the ambient temperature. Figure 10 shows the rate of the BD current increase, d$I_{BD}$/dt, as a function of the stress voltage measured in metal oxide semiconductor (MOS) stacks with different dielectrics and h-BN-based MIM stacks. The experimental data correspond to poly-Si/SiO$_2$ (2 nm)/Si from refs 67 and 68; Au/Ti/HK/n-InGaAs with Al$_2$O$_3$ (9 nm), Si$_3$N$_4$ (9 nm), HfO$_2$ (10 nm) from ref 64, and h-BN (3 nm)-based MIM stacks. The calculations according to the model are also included. The model provides a good quantitative account for the observed BD growth rate in many systems, including h-BN layers. The order of magnitude of the predicted BD current growth rate, d$I_{BD}$/dt, is close to the experimental data. This result confirms that the thermal conductivity of ultrathin h-BN layers plays a relevant role in the BD event.

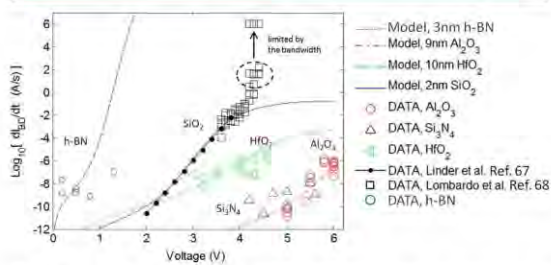

**Figure 10.** Rate of the BD current increase (d$I_{BD}$/dt) as a function of the stress voltage measured in MOS and MIM stacks with different dielectric layers. For modeling the d$I_{BD}$/dt in h-BN layers, the following parameters have been used; $t_{ox}$ = 3 nm, $k$ = 300 W/m K, and $E_a$ = 1.1 eV.

## 3. CONCLUSIONS

In conclusion, the dielectric BD event in monolayer h-BN sheets and multilayer h-BN stacks was analyzed on the nanoscale and on the device level. Although the multilayer h-BN reaches BD in a characteristic layer-by-layer manner, the BD process is characterized by abundant local charge trapping (as confirmed by the observation of RTN-like current signals and local charge accumulation), indicating that this is a universal behavior that takes place in both 2D (layered) and 3D dielectrics. When the BD is triggered, multilayer h-BN stacks show severe BD-induced surface extrusion (hillock formation) very similar to that of traditional 3D dielectrics ($SiO_2$, $HfO_2$, and $Al_2O_3$). On the contrary, the monolayer h-BN never shows BD-induced surface extrusion, even when the BD event is stronger. The enhanced reliability of h-BN is related to its superior thermal conductivity, which may dissipate local thermal heat, reduce avalanche currents, and slow down electromigration, enhancing the overall reliability of the entire device. These hypotheses have been demonstrated via device level (probe station) analysis and fittings to the BD theoretical models. Our work provides new insights into the reliability and BD of 2D layered insulators, which will be in high demand in future digital electronic nanodevices.

## 4. METHODS

**4.1. h-BN Synthesis.** The h-BN was grown via the CVD approach on a CuNi foil, following the methodologies reported in our previous work.[32] First, a 25 $\mu$m thick Cu foil (with 99.9 purity, purchased at Alfa Aesar) was electrochemically polished using a current of 10 A for 1.5 min to decrease its surface roughness. The electrolyte used here was a mixture of 500 mL of water, 250 mL of ethanol, 250 mL of orthophosphoric acid, 50 mL of isopropyl alcohol, and 5 g of urea. After that, to further enhance the flatness and its grain size, the Cu foil was annealed at 1050 °C for 2 h in a mixed flow (400 sccm and 100 sccm for Ar and $H_2$, respectively) under atmosphere pressure. The next step was to electroplate the Ni layer on the Cu foil; the electrolytic solution used here contained 1 L of water, 280 g of $NiSO_4$·$6H_2O$, 8 g of $NiCl_2$·$6H_2O$, 4 g of NaF, and 30 g of $H_3BO_3$. During this process, the current density was set to 0.01 A·cm$^{-2}$ to maintain a constant Ni deposition rate of 200 nm/min. Then, the Ni-coated Cu stack was annealed at 1050 °C for 2 h under a $H_2$ flow with a pressure of 5 kPa to drive these two atomic species (Cu and Ni) completely mixed, leading to a homogeneous CuNi alloy. The atomic proportion of Ni here is determined by the thickness of the deposited Ni layer.

After that, the h-BN growth on the CuNi substrate was carried out using borazane as the precursor. Borazane was located 60 cm away from the catalytic substrate (outside the main heated area of the tube furnace) and surrounded by a heating belt at 70–90 °C. The temperature and pressure in the substrate region for the h-BN growth process was 1070 °C and 50 Pa, respectively. The gas carried the precursor molecules and deposited them on the surface of the CuNi substrate. These seeds led to the growth of the mono/multi layer h-BN. By tuning the growth time, we controlled the thickness of the multilayer h-BN, and longer growth times result in a thicker h-BN layer.[32] The lateral size and growth speed of the h-BN can also be controlled by tuning the amount of Ni in the CuNi foil. In our previous work, we observed that when the Ni atomic proportion ranges from 10 to 20%, the h-BN grain shows the largest lateral grain size and growth speed.

**4.2. h-BN Characterization.** Monolayer and multilayer h-BN stacks were characterized on the nanoscale using a MultiMode VIII AFM from Bruker working in the PeakForce TUNA mode.[33] When the surface of the sample is scanned using this mode, a force–distance ($F$–$Z$) curve is collected at each pixel of the image, which allows plotting not only the topography (as in the tapping mode) but also other magnitudes, such as adhesion and deformation forces. This mode can also collect electrical information of the sample at each pixel by reading (and averaging) the current flowing through the tip/sample junction in different periods of time during the $F$–$Z$ curve. The $I$–$V$ curves were collected by stopping the tip at specific locations of the sample using the tool named *Point & Shoot*. When displaying the $I$–$V$ curves (Figures 4b and 7b), no average of all curves per row was plotted because that would distort the fluctuations of the current signal; this information is very valuable to understand charge trapping and detrapping phenomena in the h-BN dielectric stack. In Figures 4b and 7b, the horizontal $X$-axis refers to the tip voltage, whereas the sample substrate was kept grounded (electron injection from the substrate).

The samples were scanned using Pt-coated silicon tips from Olympus (model AC240TM, item number 4B4035), which have a spring constant of 2 N/m, a resonance frequency of 70 Hz, and a tip radius of 15 nm (all nominal values). The force and deformation values given in the Z-scale of Figure 8b,c of the main text (respectively) should be considered as typical as they were calculated using the nominal spring constant given by the manufacturer (which allows variations up to ±30%). All electrical measurements (both $I$–$V$ curves and current maps) were collected by injecting electrons from the substrate, which avoids local anodic oxidation[41,43] and electrode position.[42] The thickness and morphology of the h-BN stacks were studied via cross-sectional TEM. The samples were first processed in a focus ion beam (model HELIOS NANOLAB 450S) to extract ~40 nm thick lamellas and then placed on a transmission electron microscope copper grid for inspection. The transmission electron microscope tool used was JEOL JEM-2100.

**4.3. IL Gating Biasing.** For the test structures under IL, the h-BN/CuNi stack was spin-coated with S1813 photo resist at 5000 rpm for 1 min, and small circular windows of 40 $\mu$m diameter were opened by standard photolithography to expose the surface of the h-BN (see Figure 2a). The device was then heated at 180 °C for 2 h to cross-link the photoresist into a stable film. On the photoresist and close to the edge of the window, a Pd top electrode was made by thermal evaporation. The h-BN surface was electrically connected to the top Pd electrode using a drop of IL (DEME-BF4 with formula $C_8H_{20}NOBF_4$). The electrical stresses were applied to the top IL electrode, keeping the CuNi substrate grounded, so that comparisons with the probe station measurements are allowed.

**4.4. h-BN Device Fabrication and Characterization.** The h-BN-based devices were fabricated by evaporating 100 $\mu$m × 100 $\mu$m top 40 nm Au/20 nm Ti electrodes on the surface of the as-grown h-BN/CuNi sample. The CuNi substrate served as the bottom electrode, and no annoying manual transfer process using polymers was needed. The metal deposition was made using the PVD75 evaporator from Kurt J. Lesker using a shadow mask with squared holes patterned via laser (Tecan, UK). The deposition rate in the evaporator was 0.5 Å s$^{-1}$. The resulting Au/Ti/h-BN/CuNi devices were measured in a Cascade TRIAX probe station connected to a Keithley 2636B semiconductor parameter analyzer. One set of Au/Ti/h-BN/Au devices (Figure 2h) was fabricated by transferring one h-BN stack on an Au-coated 300 nm $SiO_2$/Si wafer and depositing Au/Ti electrodes on it following the same approach.

**4.5. Experiments with $HfO_2$ and $Al_2O_3$.** The 4 nm $HfO_2$ films were grown on a 1 nm $SiO_2$/n-Si substrate, and the 10 nm $Al_2O_3$ films were grown on a 1 nm $SiO_2$/p-Si substrate. The thickness of each film was corroborated via cross-sectional TEM. Both samples were characterized on the nanoscale using a Veeco Dimension 3100 CAFM working in the contact mode. The RVS on the $HfO_2$ films was collected using Co–Cr-coated Si tips, and the RVS on $Al_2O_3$ was collected using PtIr-coated Si tips. For the RVS, negative biases were applied to the substrate while keeping the tip grounded (electron injection from the substrate). This ensures that the hillocks observed are not related to local anodic oxidation at the tip/sample junction.[41,43]

**4.6. Experiments with $SiO_2$.** The 1 nm $SiO_2$ films were chemically grown on p-Si substrates. The CAFM characterization was carried out using an Omicron CAFM (model SPM 1000) working in UHV (10$^{-9}$ Torr), using PtIr-coated Si tips. The samples were heated at 120 °C for 20 min in a vacuum prechamber to remove the rest of the moisture. Although in the previous conductive atomic force microscopy experiments on h-BN and high-$k$ dielectrics (carried out in air atmosphere), local anodic oxidation and electro-deposition could be discarded because of the use of electron injection from the substrate, the acquisition of conductive atomic force microscopy data in UHV further corroborates that the hillocks observed in the topographic maps are indeed generated during the BD event (discards the involvement of water molecules on the sample).[43] The RVS were collected using an Agilent 4156C connected directly to the CAFM tip, which allows the observation of extended current range and variable current limitations. The stress voltages were applied under negative substrate voltages and a grounded atomic force microscope tip (electron injection from the substrate).

## ■ ASSOCIATED CONTENT

### ⑤ Supporting Information

## ■ AUTHOR INFORMATION

### Corresponding Author
*E-mail: mlanza@suda.edu.cn.

### ORCID ⓘ
Kechao Tang: 0000-0003-4570-0142
Mario Lanza: 0000-0003-4756-8632

### Author Contributions
L.J., Y.S., and F.H. have contributed equally.

### Notes
The authors declare no competing financial interest.

## ■ ACKNOWLEDGMENTS

This work has been supported by the Young 1000 Global Talent Recruitment Program of the Ministry of Education of China, the National Natural Science Foundation of China (grants nos. 61502326, 41550110223, and 11661131002), the Jiangsu Government (grant no. BK20150343), the Ministry of Finance of China (grant no. SX21400213), the Young 973 National Program of the Chinese Ministry of Science and Technology (grant no. 2015CB932700), the National Council for Scientific and Technical Research (CONICET) under Project PIP-11220130100077CO, and the National Technological University (UTN.BA) under Project PIDUTN2014/UTI2423. The Collaborative Innovation Centre of Suzhou Nano Science & Technology, the Jiangsu Key Laboratory for Carbon-Based Functional Materials & Devices, and the Priority Academic Program Development of Jiangsu Higher Education Institutions are also acknowledged. Professors H.-S. Philip Wong and Eric Pop (Stanford University) are acknowledged for useful discussions.

## ■ REFERENCES

(1) Stathis, J. H.; DiMaria, D. J. Reliability Projection for Ultra-thin Oxides at Low Voltage. *IEDM Tech. Dig.* **1998**, *98*, 167–170.

(2) Seo, S.-H.; Hwang, J.-S.; Yang, J.-M.; Hwang, W.-J.; Song, J.-Y.; Lee, W.-J. Failure Mechanism of Copper Through-Silicon Vias under Biased Thermal Stress. *Thin Solid Films* **2013**, *546*, 14–17.

(3) Obreja, V. V. N.; Codreanu, C.; Poenar, D.; Buiu, O. Edge Current Induced Failure of Semiconductor PN Junction during operation in the Breakdown Region of Electrical Characteristic. *Microelectron. Reliab.* **2011**, *51*, 536–542.

(4) Uppal, H. J.; Mitrovic, I. Z.; Hall, S.; Hamilton, B.; Markevich, V.; Peaker, A. R. Breakdown and Degradation of Ultrathin Hf-based

(HfO₂)ₓ(SiO₂)₁₋ₓ Gate Oxide Films. *J. Vac. Sci. Technol., B: Microelectron. Nanometer Struct.—Process., Meas., Phenom.* **2009**, *27*, 443.

(5) Hwang, S.-S.; Jung, S.-Y.; Joo, Y.-C. Characteristics of Leakage Current in the Dielectric Layer due to Cu Migration during Bias Temperature Stress. *J. Appl. Phys.* **2008**, *104*, 044511.

(6) Raghavan, N.; Pey, K. L.; Shubhakar, K.; Bosman, M. Modified Percolation Model for Polycrystalline High-κ Gate Stack with Grain Boundary Defects. *IEEE Electron Device Lett.* **2011**, *32*, 78–80.

(7) Suñé, J.; Placencia, I.; Barniol, N.; Farrés, E.; Martín, F.; Aymerich, X. On the Breakdown Statistics of very Thin SiO₂ Films. *Thin Solid Films* **1990**, *185*, 347–362.

(8) Porti, M.; Nafria, M.; Aymerich, X.; Olbrich, A.; Ebersberger, B. Post-breakdown Electrical Characterization of Ultrathin SiO₂ Films with Conductive Atomic Force Microscopy. *Nanotechnology* **2002**, *13*, 388–391.

(9) Lanza, M.; Bersuker, G.; Porti, M.; Miranda, E.; Nafria, M.; Aymerich, X. Resistive Switching in Hafnium Dioxide Layers: Local Phenomenon at Grain Boundaries. *Appl. Phys. Lett.* **2012**, *101*, 193502.

(10) Magtoto, N. P.; Niu, C.; Ekstrom, B. M.; Addepalli, S.; Kelber, J. A. Dielectric Breakdown of Ultrathin Aluminum Oxide Films Induced by Scanning Tunneling Microscopy. *Appl. Phys. Lett.* **2000**, *77*, 2228–2230.

(11) *International Technology Roadmap for Semiconductors*, 2013th ed., Process Integration, Devices, and Structures section, http:www.itrs.net, last accessed online February 11th 2015.

(12) Lombardo, S.; Stathis, J. H.; Linder, B. P.; Pey, K. L.; Palumbo, F.; Tung, C. H. Dielectric Breakdown Mechanisms in Gate Oxides. *J. Appl. Phys.* **2005**, *98*, 121301.

(13) Condorelli, G.; Lombardo, S. A.; Palumbo, F.; Pey, K.-L.; Tung, C. H.; Tang, L.-J. Structure and Conductance of the Breakdown Spot During the Early Stages of Progressive Breakdown. *IEEE Trans. Device Mater. Reliab.* **2006**, *6*, 534–541.

(14) Tung, C. H.; Pey, K. L.; Tang, L. J.; Radhakrishnan, M. K.; Lin, W. H.; Palumbo, F.; Lombardo, S. Percolation Path and Dielectric-Breakdown-Induced-Epitaxy Evolution during Ultrathin Gate Dielectric Breakdown Transient. *Appl. Phys. Lett.* **2003**, *83*, 2223–2225.

(15) Palumbo, F.; Condorelli, G.; Lombardo, S.; Pey, K. L.; Tung, C. H.; Tang, L. J. Structure of the Oxide Damage under Progressive Breakdown. *Microelectron. Reliab.* **2005**, *45*, 845–848.

(16) Tung, C. H.; Pey, K. L.; Lin, W. H.; Radhakrishnan, M. K. Polarity-dependent Dielectric Breakdown-Induced Epitaxy (DBIE) in Si MOSFETs. *IEEE Electron Device Lett.* **2002**, *23*, 526–528.

(17) Ranjan, R.; Pey, K. L.; Selvarajoo, T. A. L.; Tang, L. J.; Tung, C. H.; Lin, W. H. Dielectric-Breakdown-Induced Epitaxy: a universal Breakdown Defect in Ultrathin Gate Dielectrics. *Proceedings of 11th IPFA, Taiwan*, 2004; pp 53–56.

(18) Privitera, S.; Bersuker, G.; Butcher, B.; Kalantarian, A.; Lombardo, S.; Bongiorno, C.; Geer, R.; Gilmer, D. C.; Kirsch, P. D. Microscopy Study of the Conductive Filament in HfO₂ Resistive Switching Memory Devices. *Microelectron. Eng.* **2013**, *109*, 75–78.

(19) Roy, T.; Tosun, M.; Kang, J. S.; Sachid, A. B.; Desai, S. B.; Hettick, M.; Hu, C. C.; Javey, A. Field-Effect Transistors Built from All Two-Dimensional Material Components. *ACS Nano* **2014**, *8*, 6259–6264.

(20) Lee, G.-H.; Yu, Y.-J.; Cui, X.; Petrone, N.; Lee, C.-H.; Choi, M. S.; Lee, D.-Y.; Lee, C.; Yoo, W. J.; Watanabe, K.; Taniguchi, T.; Nuckolls, C.; Kim, P.; Hone, J. Flexible and Transparent MoS₂ Field-Effect Transistors on Hexagonal Boron Nitride-Graphene Heterostructures. *ACS Nano* **2013**, *7*, 7931–7936.

(21) Ghosh, S.; Calizo, I.; Teweldebrhan, D.; Pokatilov, E. P.; Nika, D. L.; Balandin, A. A.; Bao, W.; Miao, F.; Lau, C. N. Extremely High Thermal Conductivity of Graphene: Prospects for Thermal Management Applications in Nanoelectronic Circuits. *Appl. Phys. Lett.* **2008**, *92*, 151911.

(22) Li, W.; Carrete, J.; Mingo, N. Thermal Conductivity and Phonon Linewidths of Monolayer MoS₂ from First Principles. *Appl. Phys. Lett.* **2013**, *103*, 253103.

(23) Jo, I.; Pettes, M. T.; Kim, J.; Watanabe, K.; Taniguchi, T.; Yao, Z.; Shi, L. Thermal Conductivity and Phonon Transport in Suspended Few-Layer Hexagonal Boron Nitride. *Nano Lett.* **2013**, *13*, 550−554.

(24) Hattori, Y.; Taniguchi, T.; Watanabe, K.; Nagashio, K. Layer-by-Layer Dielectric Breakdown of Hexagonal Boron Nitride. *ACS Nano* **2015**, *9*, 916−921.

(25) Ang, S.; Wilson, S. Rapid Thermal Annealed Low Pressure Chemical-Vapor-Deposited SiO$_2$ as Gate Dielectric in Silicon MOSFET's. *J. Electrochem. Soc.* **1987**, *134*, 1254−1258.

(26) Balog, M.; Schieber, M.; Michman, M.; Patai, S. Chemical Vapor Deposition and Characterization of HfO$_2$ Films from Organo-Hafnium Compounds. *Thin Solid Films* **1977**, *41*, 247−259.

(27) Ji, Y.; Pan, C.; Zhang, M.; Long, S.; Lian, X.; Miao, F.; Hui, F.; Shi, Y.; Larcher, L.; Wu, E.; Lanza, M. Boron Nitride as Two Dimensional Dielectric: Reliability and Dielectric Breakdown. *Appl. Phys. Lett.* **2016**, *108*, 012905.

(28) Hui, F.; Pan, C.; Shi, Y.; Ji, Y.; Grustan-Gutierrez, E.; Lanza, M. On the use of Two Dimensional Hexagonal Boron Nitride as Dielectric. *Microelectron. Eng.* **2016**, *163*, 119−133.

(29) Wang, C.; Guo, J.; Dong, L.; Aiyiti, A.; Xu, X.; Li, B. Superior Thermal Conductivity in Suspended Bilayer Hexagonal Boron Nitride. *Sci. Rep.* **2016**, *6*, 25334.

(30) Lu, G.; Wu, T.; Yuan, Q.; Wang, H.; Wang, H.; Ding, F.; Xie, X.; Jiang, M. Synthesis of Large Single-Crystal Hexagonal Boron Nitride Grains on Cu−Ni Alloy. *Nat. Commun.* **2015**, *6*, 6160.

(31) Wu, Q.; Bayerl, A.; Porti, M.; Martin-Martinez, J.; Lanza, M.; Rodriguez, R.; Velayudhan, V.; Nafria, M.; Aymerich, X.; Gonzalez, M. B.; Simoen, E. A Conductive AFM Nanoscale Analysis of NBTI and Channel Hot-Carrier Degradation in MOSFETs. *IEEE Trans. Electron Devices* **2014**, *61*, 3118−3124.

(32) Singh, B.; Mehta, B. R.; Varandani, D.; Savu, A. V.; Brugger, J. CAFM Investigations of Filamentary Conduction in Cu$_2$O ReRAM Devices Fabricated using Stencil Lithography Technique. *Nanotechnology* **2012**, *23*, 495707.

(33) Celano, U.; Chen, Y. Y.; Wouters, D. J.; Groeseneken, G.; Jurczak, M.; Vandervorst, W. Filament Observation in Metal-Oxide Resistive Switching Devices. *Appl. Phys. Lett.* **2013**, *102*, 121602.

(34) Tang, K.; Meng, A. C.; Hui, F.; Shi, Y.; Petach, T.; Hitzman, C.; Koh, A. L.; Goldhaber-Gordon, D.; Lanza, M.; McIntyre, P. C. Distinguishing Oxygen Vacancy Electromigration and Conductive Filament Formation in TiO$_2$ Resistive Switching Using Liquid Electrolyte Contacts. *Nano Lett.* **2017**, *17*, 4390−4399.

(35) Zobelli, A.; Ewels, C. P.; Gloter, A.; Seifert, G. Vacancy Migration in Hexagonal Boron Nitride. *Phys. Rev. B: Condens. Matter Mater. Phys.* **2007**, *75*, 094104.

(36) Puglisi, F. M.; Larcher, L.; Pan, C.; Xiao, N.; Shi, Y.; Hui, F.; Lanza, M. 2D h-BN based RRAM Devices. *IEDM Technical Digest*, 2016.

(37) Pan, C.; Ji, Y.; Xiao, N.; Hui, F.; Tang, K.; Guo, Y.; Xie, X.; Puglisi, F. M.; Larcher, L.; Miranda, E.; Jiang, L.; Shi, Y.; Valov, I.; McIntyre, P. C.; Waser, R.; Lanza, M. Coexistence of Grain-Boundaries-Assisted Bipolar and Threshold Resistive Switching in Multilayer Hexagonal Boron Nitride. *Adv. Funct. Mater.* **2017**, *27*, 1604811.

(38) Satake, H.; Toriumi, A. Dielectric Breakdown Mechanism of Thin-SiO$_2$ Studied by the Post-Breakdown Resistance Statistics. *IEEE Trans. Electron Devices* **2000**, *47*, 741−745.

(39) Puslisi, F. M.; Pavan, P.; Larcher, L.; Padovani, A. Statistical Analysis of Random Telegraph Noise in HfO$_2$-Based RRAM Devices in LRS. *Solid-State Electron.* **2015**, *113*, 132−137.

(40) Thamankar, R.; Raghavan, N.; Molina, J.; Puglisi, F. M.; O'Shea, S. J.; Shubhakar, K.; Larcher, L.; Pavan, P.; Padovani, A.; Pey, K. L. Single Vacancy Defect Spectroscopy on HfO$_2$ using Random Telegraph Noise Signals from Scanning Tunneling Microscopy. *J. Appl. Phys.* **2016**, *119*, 084304.

(41) Garcia, R.; Martinez, R. V.; Martinez, J. Nano-Chemistry and Scanning Probe Nanolithographies. *Chem. Soc. Rev.* **2006**, *35*, 29−38.

(42) Polspoel, W.; Vandervorst, W. Evaluation of Trap Creation and Charging in Thin SiO$_2$ using both SCM and C-AFM. *Microelectron. Eng.* **2007**, *84*, 495−500.

(43) Lanza, M.; Porti, M.; Nafria, M.; Aymerich, X.; Whittaker, E.; Hamilton, B. UHV CAFM Characterization of High-k Dielectrics: Effect of the Technique Resolution on the Pre- and Post-Breakdown Electrical Measurements. *Microelectron. Reliab.* **2010**, *50*, 1312−1313.

(44) Li, C.; Minne, S.; Pittenger, B.; Mednick, A.; Guide, M.; Nguyen, T. *Simultaneous Electrical and Mechanical Property Mapping at the Nanoscale with PeakForce TUNA.* http:www.brukerafmprobes.com in Application Note #132 (accessed August, 2015).

(45) Price, R.; Tobyn, M. J.; Staniforth, J. N.; Thomas, M.; Davies, M. B. Variation in Particle Adhesion due to Capillary and Electrostatic Forces. *Respiratory Drug Delivery* VII, 2000.

(46) Yang, Y.; Gao, P.; Li, L.; Pan, X.; Tappertzhofen, S.; Choi, S.; Waser, R.; Valov, I.; Lu, W. D. Electrochemical Dynamics of Nanoscale Metallic Inclusions in Dielectrics. *Nat. Commun.* **2014**, *5*, 4232.

(47) Lanza, M.; Porti, M.; Nafria, M.; Aymerich, X.; Sebastiani, A.; Ghidini, G.; Vedda, A.; Fasoli, M. Combined Nanoscale and Device-Level Degradation Analysis of SiO$_2$ Layers of MOS Nonvolatile Memory Devices. *IEEE Trans. Device Mater. Reliab.* **2009**, *9*, 529−536.

(48) Lanza, M. A Review on Resistive Switching in High-k Dielectrics: A Nanoscale Point of View Using Conductive Atomic Force Microscope. *Materials* **2014**, *7*, 2155−2182.

(49) Sadewasser, S.; Glatzel, T. *Kelvin Probe Force Microscopy: Measuring and Compensating Electrostatic Forces.* Springer Series in Surface Sciences; Springer Verlag: Heidelberg, Germany, 2012; Vol. 48, ISBN: 978-3-642-22565-9.

(50) Balandin, A. A.; Ghosh, S.; Bao, W.; Calizo, I.; Teweldebrhan, D.; Miao, F.; Lau, C. N. Superior Thermal Conductivity of Single-Layer Graphene. *Nano Lett.* **2008**, *8*, 902−907.

(51) Ng, T. Y.; Yeo, J. J.; Liu, Z. S. A Molecular Dynamics Study of the Thermal Conductivity of Graphene Nanoribbons Containing Dispersed Stone−Thrower−Wales defects. *Carbon* **2012**, *50*, 4887−4893.

(52) Gupta, S. K.; Sonvane, Y.; Wang, G.; Pandey, R. Size and Edge Roughness Effects on Thermal Conductivity of Pristine Antimonene Allotropes. *Chem. Phys. Lett.* **2015**, *641*, 169−172.

(53) Sahoo, S.; Gaur, A. P. S.; Ahmadi, M.; Guinel, M. J.-F.; Katiyar, R. S. Temperature-Dependent Raman Studies and Thermal Conductivity of Few-Layer MoS$_2$. *J. Phys. Chem. C* **2013**, *117*, 9042−9047.

(54) Lindsay, L.; Broido, D. A. Enhanced Thermal Conductivity and Isotope Effect in Single-layer Hexagonal Boron Nitride. *Phys. Rev. B: Condens. Matter Mater. Phys.* **2011**, *84*, 155421.

(55) Yan, R.; Simpson, J. R.; Bertolazzi, S.; Brivio, J.; Watson, M.; Wu, X.; Kis, A.; Luo, T.; Walker, A. R. H.; Xing, H. G. Thermal Conductivity of Monolayer Molybdenum Disulfide Obtained from Temperature-Dependent Raman Spectroscopy. *ACS Nano* **2014**, *8*, 986−993.

(56) Wingert, M. C.; Zheng, J.; Kwon, S.; Chen, R. Thermal Transport in Amorphous Materials: A Review. *Semicond. Sci. Technol.* **2016**, *31*, 113003.

(57) Amazawa, T.; Ono, T.; Shimada, M.; Matsuo, S.; Oikawa, H. Ultrathin Oxide Films Deposited using Electron Cyclotron Resonance Sputter. *J. Vac. Sci. Technol., B: Microelectron. Nanometer Struct.−Process., Meas., Phenom.* **1999**, *17*, 2222.

(58) Szabo, J. P.; Hiltz, J. A.; Cameron, C. G.; Underhill, R. S.; Massey, J.; White, B.; Leidner, J. Elastomeric Composites with High Dielectric Constant for use in Maxwell Stress Actuators. *Proc. SPIE* **2003**, *5051*, 180−190.

(59) Chou, N. J.; Eldridge, J. M. Effects of Material and Processing Parameters on the Dielectric Strength of Thermally Grown SiO$_2$ Films. *J. Electrochem. Soc.* **1970**, *117*, 1287−1293.

(60) Lee, G.-H.; Yu, Y.-J.; Lee, C.; Dean, C.; Shepard, K. L.; Kim, P.; Hone, J. Electron Tunneling through Atomically Flat and Ultrathin Hexagonal Boron Nitride. *Appl. Phys. Lett.* **2011**, *99*, 243114.

(61) Hattori, Y.; Taniguchi, T.; Watanabe, K.; Nagashio, K. Layer-by-Layer Dielectric Breakdown of Hexagonal Boron Nitride. *ACS Nano* **2015**, *9*, 916−921.

(62) Villena, M. A.; González, M. B.; Jiménez-Molinos, F.; Campabadal, F.; Roldán, J. B.; Suñé, J.; Romera, E.; Miranda, E. Simulation of Thermal Reset Transitions in Resistive Switching Memories including Quantum Effects. *J. Appl. Phys.* **2014**, *115*, 214504.

(63) Menzel, S.; Kaupmann, P.; Waser, R. Understanding Filamentary Growth in Electrochemical Metallization Memory Cells using Kinetic Monte Carlo Simulations. *Nanoscale* **2015**, *7*, 12673−12681.

(64) Palumbo, F.; Lombardo, S.; Eizenberg, M. Physical Mechanism of Progressive Breakdown in Gate Oxides. *J. Appl. Phys.* **2014**, *115*, 224101.

(65) Pazos, S.; Aguirre, F.; Miranda, E.; Lombardo, S.; Palumbo, F. Comparative Study of the Breakdown Transients of Thin $Al_2O_3$ and $HfO_2$ films in MIM Structures and their Connection with the Thermal Properties of Materials. *J. Appl. Phys.* **2017**, *121*, 094102.

(66) Palumbo, F.; Eizenberg, M.; Lombardo, S. General Features of Progressive Breakdown in Gate Oxides: a Compact Model. *IEEE International Reliability Physics Symposium*, 2015; pp 5A.1.1−5A.1.6.

(67) Linder, B. P.; Lombardo, S.; Stathis, J. H.; Vayshenker, A.; Frank, D. J. Voltage Dependence of Hard Breakdown Growth and the Reliability Implication in Thin Dielectrics. *IEEE Electron Device Lett.* **2002**, *23*, 661−663.

(68) Lombardo, S.; Stathis, J. H.; Linder, B. P. Breakdown Transients in Ultrathin Gate Oxides: Transition in the Degradation Rate. *Phys. Rev. Lett.* **2003**, *90*, 167601.

# Supporting information

# Dielectric Breakdown in Chemical Vapor Deposited

# Hexagonal Boron Nitride

Lanlan Jiang[1]‡, Yuanyuan Shi[1,2]‡, Fei Hui[1,3]‡, Kechao Tang[4], Qian Wu[1], Chengbin Pan[1], Xu Jing[1,5],

Hasan Uppal[6], Felix Palumbo[7], Guangyuan Lu[8], Tianru Wu[8], Haomin Wang[8], Marco A. Villena[1],

Xiaoming Xie[8,9], Paul C. McIntyre[4], Mario Lanza[1]*

[1] Institute of Functional Nano and Soft Materials, Collaborative Innovation Center of Suzhou Nanoscience &

Technology, Soochow University, 199 Ren-Ai Road, Suzhou, 215123, China

[2] Department of Electrical Engineering, Stanford University, Stanford, CA 94305, USA

[3] Department of Electrical Engineering and Computer Sciences, Massachusetts Institute of Technology,

Cambridge, MA 02139, USA

[4] Department of Materials Science and Engineering, Stanford University, California, USA

[5] Microelectronics Research Center and Department of Electrical and Computer Engineering, The University

of Texas at Austin, Austin, Texas 78758, USA

[6] Microelectronics and nanostructures, The University of Manchester,

Sackville Street, Manchester M13 9PL, UK

[7] National Scientific and Technical Research Council (CONICET), UTN-CNEA, Godoy Cruz 2290, Buenos

Aires, Argentina

[8] State Key Laboratory of Functional Materials for Informatics, Shanghai Institute of Microsystem and

Information Technology, Chinese Academy of Sciences, 865 Changning Road, Shanghai 200050, China

[9] School of Physical Science and Technology, Shanghai Tech University, 319 Yueyang Road, Shanghai

201210, China

* Corresponding author Email: mlanza@suda.edu.cn - ‡ Equal contribution

S-1

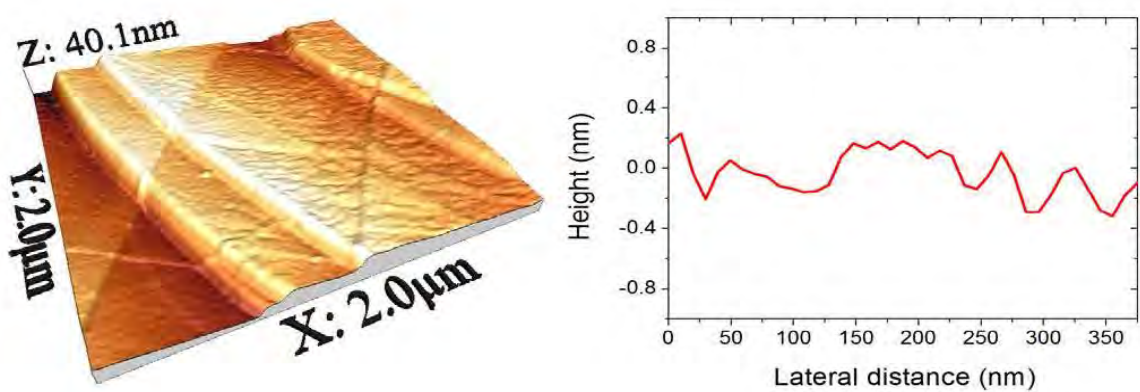

Figure S1: (left) AFM topographic map collected on the surface of a multilayer *h*-BN/CuNi sample. The steps in the CuNi substrate can be clearly distinguished. (right) Surface profile of multilayer *h*-BN performed on one of the CuNi terraces. Sub-nanometer distances between maximums and minimums along a cross section of more than 350 nm corroborate the atomically flat nature of the *h*-BN stack.

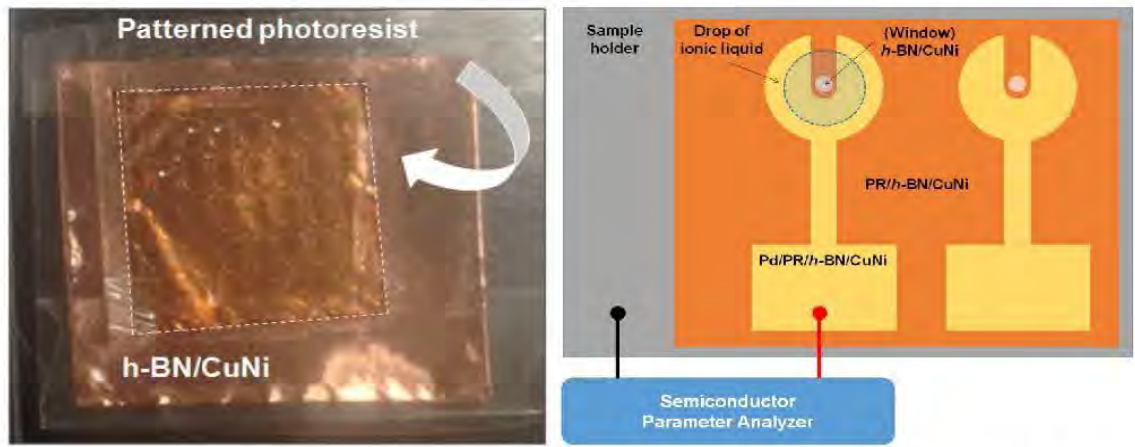

Figure S2: (left) Photograph of the *h*-BN/CuNi sample patterned with photoresist for ionic liquid gating. (right) Schematic of the ionic liquid stress for the *h*-BN/CuNi sample.

S-2

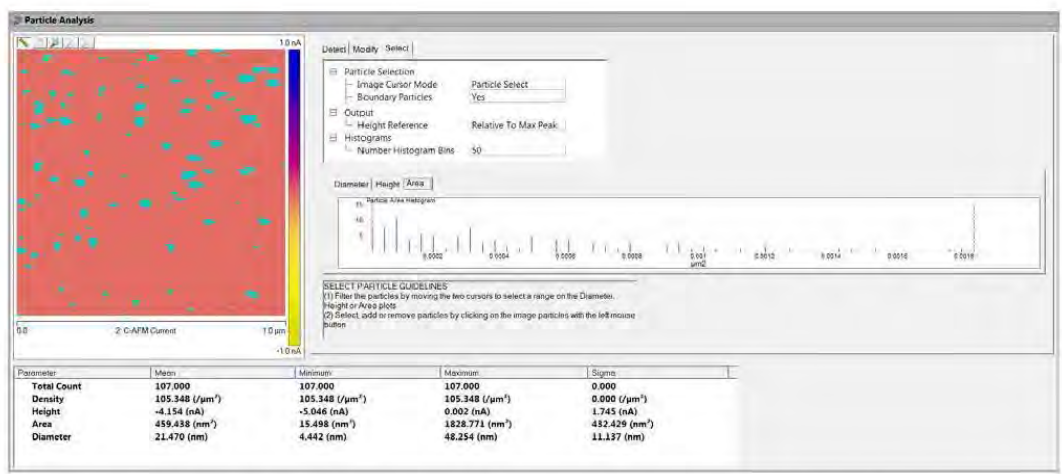

Figure S3: Screen capture of the Nanoscope software showing the calculation of the density, size and current of the conductive spots generated via ionic liquid and detected via CAFM.

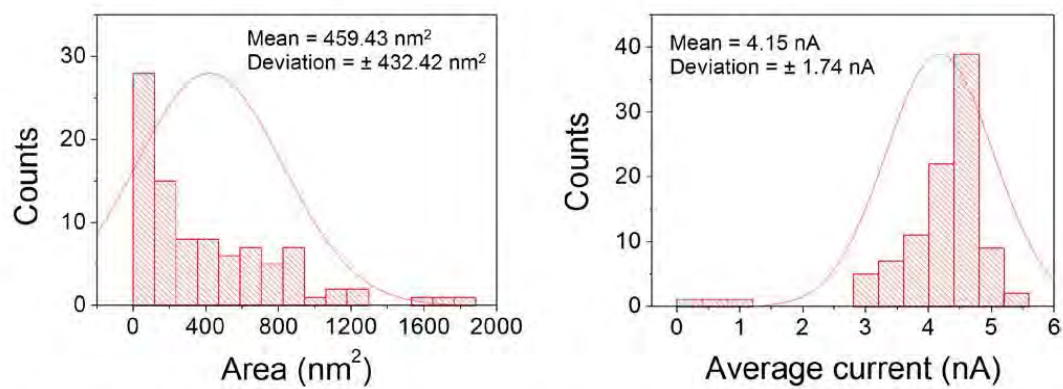

Figure S4: Statistical analyses of the size and current of the conductive spots generated via ionic liquid and detected via CAFM (in Figure 2e of the main text).

S-3

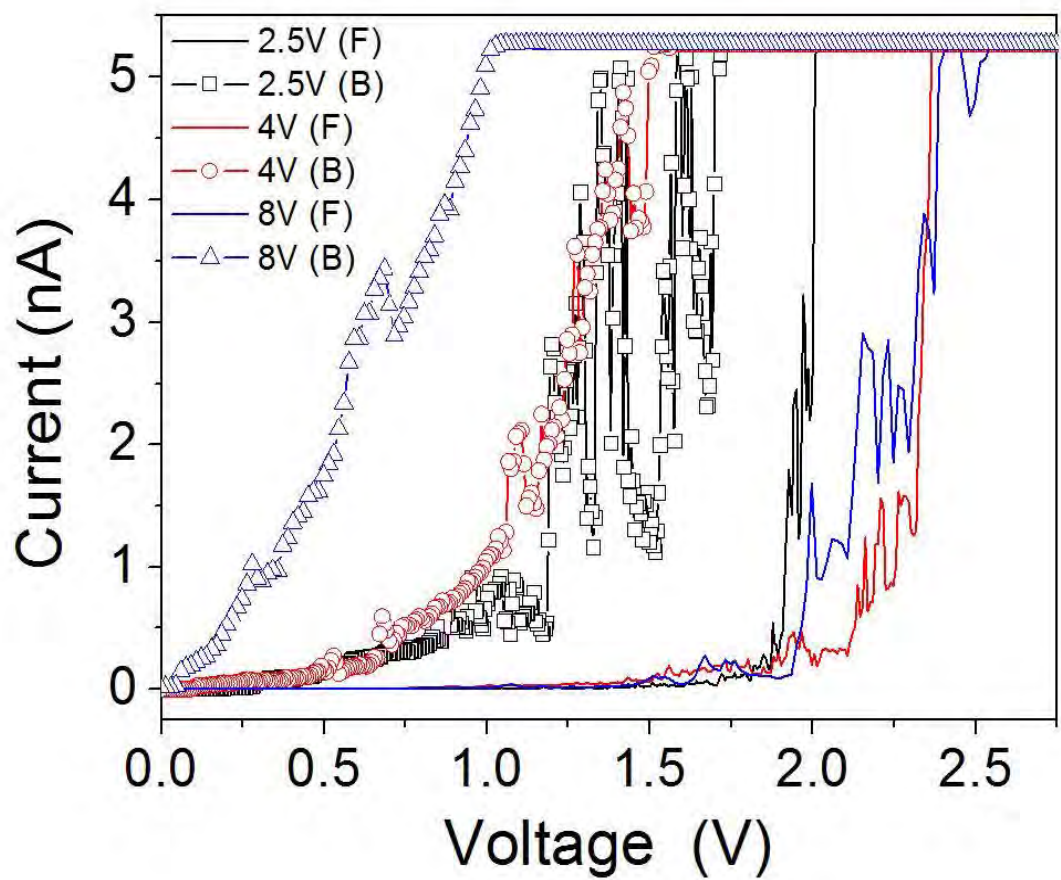

Figure S5: Larger version of Figure 4(b) shown in the main text. We plot this image just for clarity.

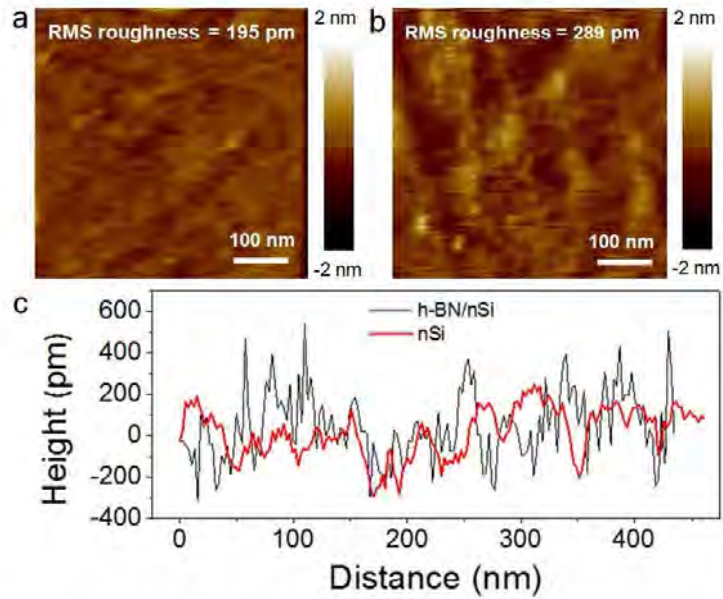

Figure S6: Topographic AFM maps on the surface of nSi (without native oxide) and *h*-BN/nSi, in panels (a) and (b), respectively. (c) Cross section taken from the AFM maps in (a) and (b). The surface of the *h*-BN/nSi sample is a bit rougher than the one of nSi, but still atomically flat. The transfer is successful and the samples of good quality.

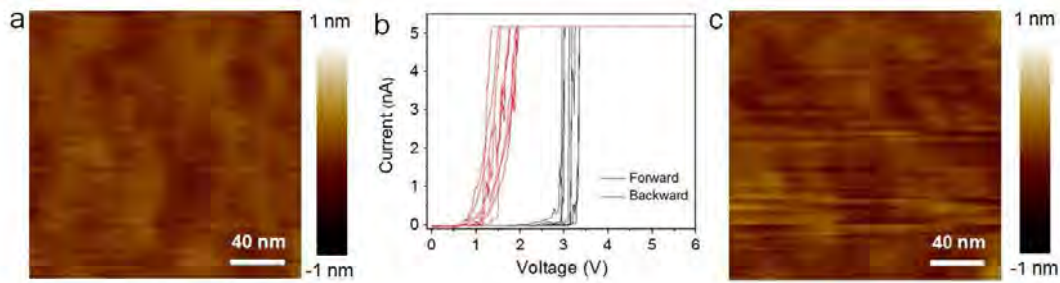

Figure S7: (a) Topographic map of the surface of the *h*-BN/nSi sample before the I-V curves. (b) I-V curves collected at several different locations of the surface of the *h*-BN/nSi sample. The I-V curves show clear BD event due to the shift of the backward I-V curve towards lower potentials. It should be highlighted that the backward I-V curve is not linear (differently from Figure 7), and the current does not starts to increase at 0V (it starts to increase at 0.5V). The reason is the different work functions of the metal and semiconductor and the voltage drop in the semiconductor (as described in Ref.[S1]). (c) Topographic map of the surface of the *h*-BN/nSi sample at a location on which an I-V curve until the BD was previously triggered; that location is the same than in panel (a). No electrical-field-driven surface extrusion (hillock formation) has been observed at any location analyzed.

**Supplementary references**

[S1] Frammelsberger, W.; Benstetter, G.; Kiely, J.; Stamp, R. C-AFM-based Thickness Determination of Thin and Ultra-thin $SiO_2$ Films by use of Different Conductive-coated Probe Tips. *Appl. Surf. Sci.* **2007**, *253*, 3615–3626.

# Synthesis of large-area multilayer hexagonal boron nitride sheets on iron substrates and its use in resistive switching devices

Fei Hui,[1,2] Marco A. Villena,[1,3] Wenjing Fang,[2] Ang-Yu Lu,[2] Jing Kong,[2]
Yuanyuan Shi[1], Xu Jing[1], Kaichen Zhu[1], Mario Lanza[1*]

[1] Institute of Functional Nano & Soft Materials, Collaborative Innovation Center of
Suzhou Nano Science and Technology, Soochow University, Suzhou, 215123, China.

[2] Department of Electrical Engineering and Computer Science, Massachusetts
Institute of Technology, Cambridge, MA 02139, USA.

[3] Department of Materials Science and Engineering, Stanford University, CA 94305,
USA.

## Abstract

Hexagonal boron nitride (h-BN) is an attractive insulating material for
nanoelectronic devices due to its high reliability as dielectric and excellent
compatibility with other two dimensional (2D) materials (e.g. graphene, $MoS_2$).
Multilayer h-BN stacks have been readily grown on Cu and Pt substrates via chemical
vapor deposition (CVD) approach, confirming its potential for wafer scale integration.
However, the growth of h-BN on other substrates needs to be also achieved in order to
expand the use of this material. Recently, the CVD growth of monolayer h-BN on Fe
substrates was reported, but it just focused on material structure characterization. Here
we present the first fabrication of electronic devices using multilayer h-BN dielectric
stacks grown on Fe foils. We fabricate and characterize resistive switching (RS)

devices based on Ag/h-BN/Fe nanojunctions, and observe the coexistence of both volatile and non-volatile RS depending on the electrode to which the bias is applied. The characteristics measured agree well with those simulated via $SIM^2RRAM$ software, and the cycle-to-cycle variability is slightly lower than that of transition metal oxide based RS devices.

## MAIN TEXT

Two dimensional (2D) hexagonal boron nitride ($h$-BN) is an insulating material made of boron and nitrogen atoms arranged in an $sp^2$ hexagonal lattice via covalent bonding [1]. Initially, h-BN attracted interest due to its great compatibility with graphene, and it was used as anti-scattering substrate in graphene field effect transistors (GFETs) to enhance the charge carriers' mobility at the channel region [2]. Recent studies demonstrated that multilayer h-BN stacks can also be used as dielectric in a wide range of optoelectronic devices, e.g. transistors [3], capacitors [4], sensors [5], resistive switching (RS) devices [6-8]. Initially, most works studied thick (>20 layers) h-BN nanoflakes (diameter <2μm) using experimental and prototypic techniques, such as conductive atomic force microscopy [9] and electron beam lithography [10]. Using these setups it was observed that one layer h-BN can block the current in a factor 50 [10], that it holds a very high and anisotropic dielectric strength [11], and that the BD process takes place layer-by-layer [12]. Now the interest for h-BN stacks grown via scalable approaches, such as chemical vapor deposition (CVD) [13] and molecular beam epitaxy (MBE) [14], increased a lot. It is known that h-BN stacks grown by this method contain larger amounts of defects [6] (specially at the grain boundaries [15]), which produces that the devices reach the BD in a softer manner. This behavior may be exploited for the fabrication of RS devices [6-8], something that is not possible using exfoliated h-BN.

During the past two years some reports claimed the observation of RS in metal/h-BN/metal devices. Refs. [16] and [17] reported the observation of bipolar RS in Ag/h-BN/Cu and ITO/h-BN/graphene/Cu devices (respectively), and attributed the switching mechanism to the migration of metal ions across the h-BN stack to form and disrupt one/few conductive filaments. Unfortunately, the cross sectional transmission electron microscopy (TEM) images in those reports do not show a layered structure of the h-BN (just amorphous BN), meaning that those devices may not hold the genuine properties of the 2D materials. Truly layered h-BN produced via CVD approach was used in Refs. [6-7] to fabricate RS devices, which demonstrated the coexistence of volatile and non-volatile RS —this performance was not achieved in amorphous BN—. Ref. [18] even observed RS in monolayer h-BN combined with Au electrodes. Despite to date h-BN based RS devices still didn't reach the endurance of transition metal oxide (TMO) based ones [19], they may be suitable for other applications. For example, the fabrication of electronic synapses using h-BN is attractive because these devices require the use of both volatile and non-volatile RS to emulate short-term and long-term plasticity operations. This is something that cannot be done with most TMO-based RS devices, as they don't show competitive threshold RS.

So far, RS in multilayer h-BN has been only demonstrated when using Ti and Cu electrodes [6-7]; therefore, studying this phenomenon using different materials combinations is necessary to expand their applications. Here we show the growth of h-BN on Fe substrates and the fabrication of Ag/h-BN/Fe RS devices, which exhibit both volatile and non-volatile RS depending on the electrode to which the bias is applied. Interestingly, the cycle-to-cycle variability observed is comparable to that of TMO-based RS devices, and SIM$^2$RRAM indicates that the switching mechanism is based on the formation and rupture of conductive nanofilaments (CNF) across the h-BN stack.

Multilayer h-BN stacks have been grown on 2 cm × 2 cm × 100 μm Fe foils using a LPCVD furnace. The size of the samples was limited by the diameter of the tube, but larger furnaces may allow wafer-scale growth [20]. The as-received Fe foils were cleaned via electrochemical method. To do so, the Fe foils and a counter electrode were immersed in a solution consisting on 940 mL acetic acid and 60 mL perchloric acid, and a potential difference of 30 V was applied between them for 30 s. After that, the Fe foil was cleaned in deionized water and introduced into the CVD tube for h-BN growth, a process that consisted in five steps: *i)* First, the cleaned Fe foil was exposed to 70 sccm H$_2$ (valve 1 was opened), and the temperature was raised from room temperature to 1100 °C in 40 min. *ii)* Second, the temperature was kept at 1100 °C (under 70 sccm H$_2$) during 30 minutes for annealing/cleaning the Fe foil. *iii)* Third, the valve controlling the amount of precursor (valve 2) was opened for 60 minutes to grow the h-BN at 1100°C. The precursor consisted on liquid borazine, which was transported into the CVD tube using 1 sccm H$_2$ as carrier gas. Therefore, the Fe foil was exposed to 70 sccm H$_2$ plus the 1 sccm H$_2$ carrying the liquid borazine. *iv)* Fourth, under the same atmosphere the temperature was decreased slowly 700 °C at a rate of ~5°C/min, a process that took 80 minutes. This step was critical for the correct growth of h-BN when using Fe substrates, and helped to improve the molecular stability of the h-BN. Other attempts to grow h-BN on Fe substrates without this step failed, and this step was not necessary to grow h-BN on Pt or Cu substrates. *v)* And fifth, once it arrived to 700 °C the furnace was cooled down to room temperature, a process that took 15 minutes.

The h-BN stacks were transferred on a 300 nm $SiO_2$ / Si wafer for quality inspection via optical microscopy and Raman spectroscopy (model LabRAM). Scanning electron microscopy images (SEM, model Zeiss Merlin) of the h-BN stacks were taken right after the growth (on the Fe substrate) and after transfer to the $SiO_2$/Si wafers. Some pieces of h-BN were also transferred on metallic grids for high resolution transmission electron microscopy (TEM, model JEOL 2010) analysis. The transfer was carried out using the electrochemical delamination method described in Ref. [21] (often called *bubble transfer*). It should be noted that the transfer step was only required to characterize the quality and thickness of the h-BN stacks, but not for the fabrication of the devices. The Au/Ag/h-BN/Fe devices were fabricated by patterning squared top electrodes (with sizes ranging from 100 μm × 100 μm down to 10 μm × 10 μm) directly on the surface of the as-grown h-BN/Fe stacks (the Fe substrate served as bottom electrode). The top electrodes consisted on 20 nm Ag and 60 nm Au, and they were deposited using a thermal evaporator (model Inficon SQM-160) and a laser-patterned shadow mask. The Au/Ag/h-BN/Fe devices were tested using a Summit 11000 AP probe station connected to a Keithley 707B semiconductor parameter analyzer. The electrical stresses were applied to the top Ag electrode, keeping the Fe substrate grounded.

Figure 1a shows the top view SEM image of an as-grown *h*-BN/Fe stack. The successful growth of a continuous h-BN sheet on the Fe foil is confirmed by the observation of wrinkles (network of white lines) [22]. The cross sectional TEM images (see Figure 1b) reveal excellent layered structure, and the number of h-BN layers is ~15. By collecting more than 9 TEM images with sub-nanometer resolution we detected that the h-BN stacks contain some local defects within the layered structure (see highlighted area in Figure 1b and the schematic in Figure 1c). Local defects in the h-BN are typically related to lattice distortions (pentagonal/heptagonal bonding) [15], boron vacancies [23], and interstitial atoms and bonding between the layers [6]. In our samples, the defective locations are typically <1nm wide, and they are separated from each other by distances of ~100 nm, representing <1% of the total area of the h-BN stack. Previous works using CVD-grown h-BN just claimed perfect layered structure everywhere, which is not realistic; they ignored the size, density and effect of these defects in the h-BN stacks. Despite the presence of some local defects, the Raman spectra show a peak at ~1367 cm$^{-1}$ (see Figure 1d), further indicating that the multilayer h-BN stacks are of high quality [24]. Figure 1e shows the top view SEM image where the different device sizes can be distinguished. In total, more than 20 devices have been characterized.

Figure 2a shows the current vs. voltage (I-V) sweeps collected on a 25 μm × 25 μm Au/Ag/h-BN/Fe device, which exhibits stable bipolar RS during more than 150 cycles. The first cycle is highlighted in red, and the next 149 in grey. Plotting the data in this way gives us an idea about cycle-to-cycle (temporal) variability. At around -1.3 V the plot shows a sudden current increase until reaching the current limitation, which was applied in order to limit the size of the dielectric breakdown (BD) spot. When the polarity of the I-V sweep is inverted, the current signal drops progressively, reaching the initial values. The sudden set indicates that the RS is governed by the formation of one/few/several CNFs across the h-BN stack [25] (i.e. it is not a distributed effect [26]). Interestingly, the Au/Ag/h-BN/Fe devices didn't require the use of large voltages to reach the initial BD (also called forming step); this is a clear advantage compared to other transition metal oxides (TMO) based RS devices, such as $HfO_2$ [27], $TiO_2$ [28], $Al_2O_3$ [29]). In these materials forming-free RS can be also achieved, but that requires the implantation of atomic species [30], which represents an additional (complex and expensive) processing step. In the case of h-BN this process is not necessary, and as-grown devices exhibit forming-free RS (Figure 2a). The reason should be the presence of some local defects in the structure of the h-BN stack (see Figures 1b and 1c), which act as weak spots promoting current flow across the devices at low potentials. In fact, the presence of local defects in the h-BN stack may be necessary for the switching of the devices, as RS has never been demonstrated in exfoliated h-BN samples (which contain few/no lattice defects). Furthermore, it is known that the presence of defects in dielectrics is an important factor enabling RS, as they allow softer BD and the formation of narrower CNFs at lower potentials, which can be easily disrupted. If the dielectric material contains no defects, normally RS evolves towards an irreversible BD spot, and in the case of defect-free exfoliated h-BN even dramatic material removal (formation of a hole) has been observed [12].

The very progressive reset suggests the presence of several filaments within the device (across the h-BN stack), and also that the switching mechanism takes place by the diffusion of atomic species due to the electric field —reset process related to self-accelerated thermal diffusion of the atoms by Joule effect is discarded because that would produce a very sharp reset process [31]—. This hypothesis has been corroborated in Figure 2b, which shows that the reset process (size of the CNF) can be tuned by selecting the end voltage of the reset (positive) I-V sweep. Another observation supporting that the RS is driven by few/several CNFs is the reduction of the set and reset voltages ($V_{SET}$ and $V_{RESET}$, respectively) for larger devices (see

Figure 2c).

The I-V sweeps in LRS have been simulated using the macroscopic SIM$^2$RRAM software described in Ref. [32]. In brief, this tool works following a multi-filamentary model, i.e. the insulator is crossed by one or several CNFs electrically coupled. These CNFs are described as nanowires that connect the top and bottom electrodes [33]. For the physical description, this model takes into account its macroscopic electrical and thermal properties such as thermal and electrical conductivity, and heat dissipation rate between the CNF and the insulator (see Ref. [34] for an accurate model description). The conduction across these CNFs is modeled in this case as Ohmic conduction but the resistance of the CNFs depends on their shape and temperature [35]. Finally, the dissolution of the CNF, i.e. its shape variation of the CNF is modeled following the thermal diffusion rules of the ions [34]. The calculations can fit very well the experimental I-V sweeps (see Figure 3a). We carried out a linear regression for all reset I-V curves in the voltage range of 0 V ~ 0.6 V, and obtained an average quadratic linear correlation coefficient $<R^2>$ = 0.98 (see Figures 3b and 3c). This behavior could be modeled as Ohmic conduction, suggesting that the CNFs across the h-BN stack are metallic.

As negative voltage is applied to the Ag top electrode during the set process (see Figure 2a), most probably positive $Fe^+$ ions from the substrate may be dragged into the h-BN stack by the electrical field, forming the CNFs. During the positive I-V sweep the $Fe^+$ ions diffuse back, disrupting the CNFs and resetting the device. This hypothesis has been analyzed by inducing the set process using a positive ramp (see Figure 4). As it can be observed, the positive $V_{SET}$ (0.25 V to 0.42 V) in Figure 4 is much smaller than the negative $V_{SET}$ (-1.3 V) in Figure 2a. Moreover, the CNF seems to be unstable, and it gets self-disrupted when the bias is switched off (at around

-0.05V in Figure 4), leading to volatile (threshold type) RS. The different set behaviors when using positive and negative set processes are related to the different top electrodes used [36]. In this case, when the set is induced by applying positive biases in the Ag electrode, $Ag+$ ions penetrate into the h-BN stack. Given the larger diffusivity of $Ag+$ ions [36] (compared to the $Fe^+$ ones), the CNF is formed faster (*i.e.* at a lower potentials), reducing $V_{SET}$. Moreover, as demonstrated in Ref. [37], CNFs made of silver inside dielectrics can self-disrupt, leading to the characteristic volatile/threshold RS mechanism displayed in Figure 4. Therefore, the switching mechanism in these devices should be by metal penetration, although severe B migration should not be discarded [6, 23]. The observation of both threshold and bipolar RS in the same RS device may be interesting for the implementation of both short-term and long-term plasticity learning rules in RS based electronic synapses [38].

Finally, we analyze one of the most recognized problems of RS devices: variability. It is very striking that the cycle-to-cycle variability of our Au/Ag/h-BN/Fe devices (fabricated in a university lab) is comparable to that of encapsulated TMO-based RS devices fabricated using industrial procedures [39, 40]. The statistical analyses of the set and reset voltages for the device shown in Figure 2a are displayed in the form of a histogram (Figure 5a) and Weibull distribution (Figure 5c). Similar statistical analyses for the set and reset currents ($I_{SET}$ and $I_{RESET}$ respectively) are shown in Figures 5b and 5d. We defined $V_{RESET}$ as the first voltage after the maximum current ($I_{MAX}$) at which the current is $0.7 \times I_{MAX}$ or lower [41]. $V_{SET}$ and $V_{RESET}$ show an acceptable dispersion which follow the Weibull distribution, and the gap between them is well determined and stable, allowing to control the RS process with enough accuracy. On the other hand, $I_{SET}$ and $I_{RESET}$ show a clear overlap; for this reason, this parameter is useless to define and/or control the RS process.

In conclusion, large-area multilayer $h$-BN (~18 layers) has been synthesized on Fe substrates via CVD method, and matrixes of Au/Ag/$h$-BN/Fe RS devices have been fabricated. The RS mechanism is related to the formation/disruption of metallic CNFs across the h-BN stack, although the formation of B vacancies should not be discarded. The devices exhibit non-volatile bipolar RS when the CNFs is formed by applying negative bias to the Ag electrode (penetration of $Fe^+$ ions into the h-BN), and volatile threshold RS when the CNFs are formed by applying the positive bias to the Ag electrode (penetration of $Ag^+$ ions into the h-BN). These conclusions are also supported by modeling using $SIM^2RRAM$ software. The set/reset voltages depend on the device size, and the cycle-to-cycle variability is comparable (if not smaller) than that of TMO-based RS devices.

**Acknowledgements**

M. Lanza acknowledges the support from the Young 1000 Global Talent Recruitment Program of the Ministry of Education of China, the National Natural Science Foundation of China (grants no. 61502326, 41550110223, 11661131002), the Jiangsu Government (grant no. BK20150343), and the Ministry of Finance of China (grant no. SX21400213). The Collaborative Innovation Center of Suzhou Nano Science & Technology, the Jiangsu Key Laboratory for Carbon-Based Functional Materials & Devices, and the Priority Academic Program Development of Jiangsu Higher Education Institutions are also acknowledged. J. Kong acknowledge support from the STC Center for Integrated Quantum Materials, NSF Grant No. DMR-1231319.

**References**

[1]   Hui F., Pan C., Shi Y., Ji Y., Grustan-Gutierrez E. and Lanza M., *Microelectron.*
      *Eng.* **2016**, *163*, 119.

[2]   Dean C. R. et al., *Nat. Nanotechnol.* **2010**, *5*, 722.

[3]   Lee G. H. et al., *ACS Nano* **2013**, *7*, 7931.

[4]   Guo N. et al., *Nano Res.* **2013**, *6*, 602.

[5]   Xu Y., Guo Z., Chen H., Yuan Y., Lou J., Lin X., Gao H., Chen H. and Yu B.,
      *Appl. Phys. Lett.* **2011**, *99*, 133109.

[6]   Pan C. B. et al., *Adv. Funct. Mater.* **2017**, *27*, 1604811.

[7]   Pan C., Miranda E., Villena M. A., Xiao N., Jing X., Xie X., Wu T., Hui F., Shi
      Y. and Lanza M., *2D Mater.* **2017**, *4*, 025099.

[8]   Puglisi F. M., Larcher L., Pan C., Xiao N., Shi Y., Hui F. and Lanza M., **2016**
      *IEEE Int. Electron Dev. Meeting* 38.8.1-34.8.4.

[9]   Lee G. H., Yu Y. J., Lee C. G., Dean C., Shepard K. L., Kim P. and Hone J.,
      *Appl. Phys. Lett.* **2011**, *99*, 243114.

[10]  Britnell L. et al., *Nano Lett.* **2012**, *12*, 1707.

[11]  Hattori Y., Taniguchi T., Watanabe K. and Nagashio K., *ACS Appl. Mater.*
      *Interfaces* **2016**, *8*, 27877.

[12]  Hattori Y., Taniguchi T., Watanabe K. and Nagashio K., *ACS Nano* **2015**, *9*, 916.

[13]  Kim K. K., Hsu A., Jia X. T., Kim S. M., Shi Y. M., Dresselhaus M., Palacios T.
      and Kong J., *ACS Nano* **2012**, *6*, 8583.

[14]  Nakhaie S., Wofford J. M., Schumann T., Jahn U., Ramsteiner M., Hanke M.,
      Lopes J. M. J. and Riechert H., *Appl. Phys. Lett.* **2015**, *106*, 213108.

[15]  Li Q., Zou X., Liu M., Sun J., Gao Y., Qi Y., Zhou X., Yakobson B. I., Zhang Y.

and Liu Z., *Nano Lett.* **2015**, *15*, 5804.

[16] Qian K., Tay R. Y., Nguyen V. C., Wang J., Cai G., Chen T., Teo E. H. T. and Lee P. S., *Adv. Funct. Mater.* **2016**, *26*, 2176.

[17] Qian K. et al., *ACS Nano* **2017**, *11*, 1712.

[18] Ge R., Wu X., Kim M., Shi J., Sonde S., Tao L., Zhang Y., Lee J. C. and Akinwande D., *Nano Lett.* **2018**, *18*, 434.

[19] Hui F., Grustan-Gutierrez E., Long S., Liu Q., Ott A. K., Ferrari A. C. and Lanza M., *Adv. Electron. Mater.* **2017**, 1600195.

[20] Lupina G. et al., *ACS Nano* **2015**, *9*, 4776.

[21] Hui F., Fang W., Leong W. S., Kpulun T., Wang H., Yang H. Y., Villena M. A., Harris G., Kong J. and Lanza M., *ACS Appl. Mater. Interfaces* **2017**, *9*, 46.

[22] Lanza M. et al., *Nano Res.* **2013**, *6*, 485.

[23] Zobelli A., Ewels C. P., Porti M., Miranda E., Nafria M. and Aymerich X., *Appl. Phys. Lett.* **2012**, *101*, 193502.

[24] Gorbachev R. V. et al., *Small* **2011**, *7*, 465.

[25] Ielmini D., Waser R., Resistive switching: from fundamentals of nanoionics redox processes to memristive device applications, **2015**, book, WILEY-VCH.

[26] Xiao N. et al., *Adv. Funct. Mater.* **2017**, *27*, 1700384.

[27] Long S. B., Perniola L., Cagli C., Buckley J., Lian X. J., Miranda E., Pan F., Liu M. and Sune J., *Sci. Rep.* **2013**, *3*, 2929.

[28] Jeong H. Y., Kim S. K., Lee J. Y. and Choi S. Y., *J. Electrochem. Soc.* **2011**, *158*, 979.

[29] Sarkar B., Lee B. and Misra V., *Semicond. Sci. Tech.* **2015**, *30*, 105014.

[30] Zhang H., Liu L., Gao B., Qiu Y., Liu X., Lu J., Han R., Kang J. and Yu B., *Appl. Phys. Lett.* **2011**, *98*, 042105.

[31] Yang Y. C. and Lu W., *Nanoscale* **2013**, *5*, 10076.

[32] Villena M. A., Jiménez-Molinos F., Roldán J. B., Suñé J., Long S., Lian X., Gamiz F. and Liu M., *J. Appl. Phys.* **2013**, *114*, 144505.

[33] Russo U., Ielmini D., Cagli C. and Lacaita A. L., *IEEE Trans. Electron Devices* **2009**, *56*, 186.

[34] Bocquet M., Deleruyelle D., Muller C. and Portal J. M., *Appl. Phys. Lett.* **2011**, *98*, 263507.

[35] González-Cordero G., Roldan J. B., Jiménez-Molinos F., Suñé J., Long S. and Liu M., *Semicond. Sci. Tech.* **2016**, *31*, 115013.

[36] Yang Y., Gao P., Li L., Pan X., Tappertzhofen S., Choi S., Waser R., Valov I. and Lu W. D., *Nat. Commun.* **2014**, *5*, 4232.

[37] Wang Z. et al., *Nat. Mater.* **2017**, *16*, 101.

[38] Shi Y., Pan C., Chen V., Raghavan N., Pey K. L., Puglisi F. M., Pop E., Wong H.-S. P. and Lanza M., IEEE Tech. Dig. 2017, 5.4.1-5.4.4.

[39] Yu M. et al., *Sci. Rep.* **2016**, *6*, 21020.

[40] Kar G. S. et al., *In Tech. Dig. VLSI Symp. Technol.* **2012**, 157.

[41] Villena M. A., Roldan J. B., Jimenez-Molinos F., Suñe J., Long S., Miranda E. and Liu M., *J. Phys. D: Appl. Phys.* **2014**, *47*, 205102.

**List of Figures**

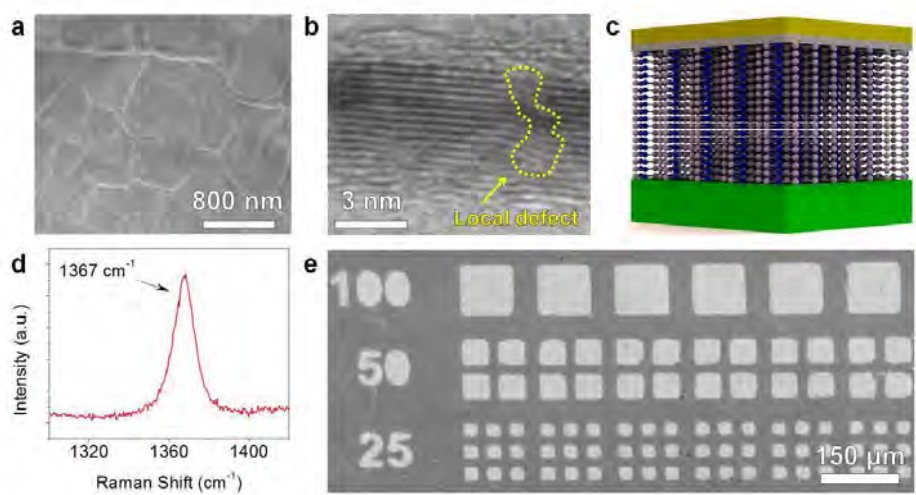

Figure 1. Characterization of CVD grown multilayer *h*-BN on Fe substrate and devices with the structure of Ag/*h*-BN/Fe. (a) Top view SEM and (b) TEM images of multilayer as-grown and transferred *h*-BN films, respectively. (c) Three-dimensional (3D) simulation of Ag/h-BN/Au. Graph (d) is the Raman spectrum of *h*-BN. (d) SEM image of Ag/*h*-BN/Fe with different sizes.

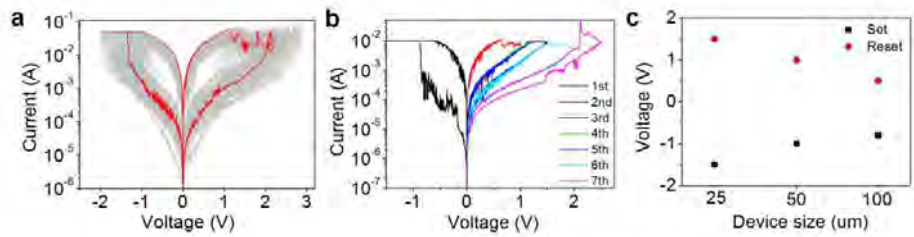

Figure 2. (a) Current vs. voltage (I-V) sweeps (>150 cycles) collected on a 25 μm × 25 μm Au/Ag/h-BN/Fe device, the red curve is the first cycle. (b) Reset (positive) I-V sweep is obtained by different end voltages. (c) Set and reset voltages are collected by different sizes of devices.

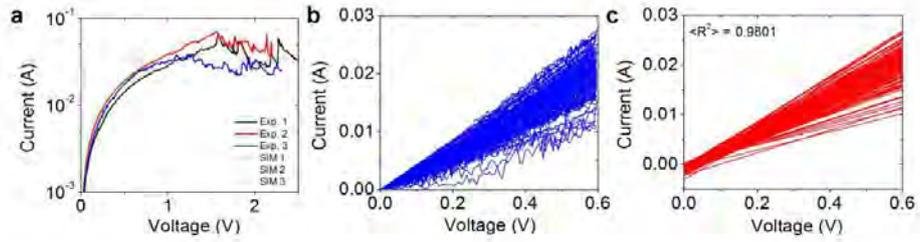

Figure 3. (a) I-V sweeps under low resistance state (LRS) are simulated by the macroscopic SIM$^2$RRAM software, the calculations (dashed line) can fit very well the experimental (solid lines) I-V sweeps. (b,c) Linear regression for all reset I-V curves in the voltage range of 0V-0.6V, the average quadratic linear correlation coefficient shows $<R^2>=0.98$.

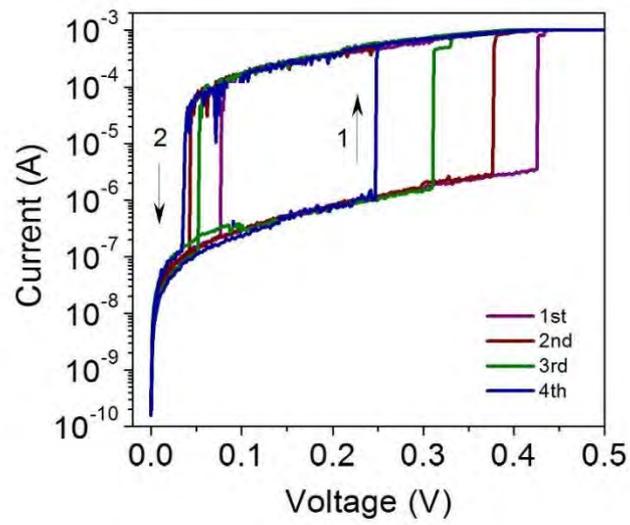

Figure 4. I-V sweeps (set process) indicate that the volatile (threshold type) resistance switching behavior is induced using a positive ramp.

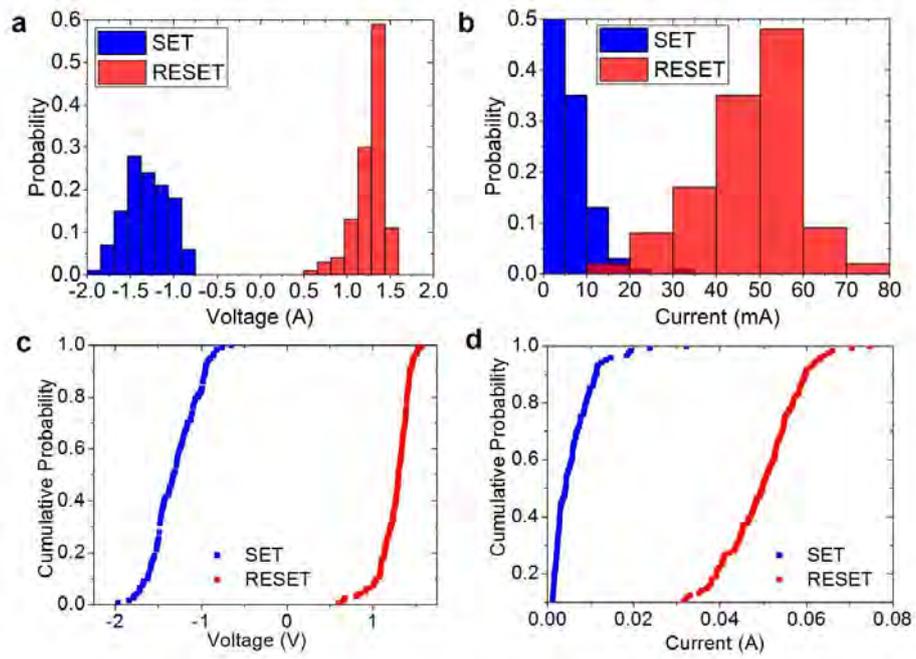

Figure 5. Statistical analysis of both set and reset process based on the Weibull distribution, including set/reset voltage (a, c) and current (b, d).

# Chapter 4:

# On the use of *h*-BN and other 2D materials in memristors

In Chapter 3 the fabrication of Au/Ag/*h*-BN/Fe memristors with very attractive performances has been described. The development of memristors using 2D materials is a strategy that has recently gained a lot of interest [76-78], as they may enhance some RS performances and provide flexibility and transparency capabilities. However, the integration of 2D materials in the structure of a memristor is not an easy task, and it brings associated several challenges [79]. Moreover, introducing a 2D materials in the structure of a memristor does not necessarily improves its performance, and we have detected some reports in which the real usefulness of the 2D material in the device is highly questionable. In this chapter we review the fabrication of memristors using 2D materials and summarize their most remarkable performances.

## 4.1. Fabrication of memristors using 2D materials

The integration of 2D materials in memristive devices is complex because their synthesis and deposition methods are remarkably different than those used in the microelectronics industry. As mentioned in Chapter 3, the scalable synthesis method that produces the best quality so far is CVD. However, the temperatures used during the growth are too high (>1000 ℃ ). If graphene or *h*-BN would be grown on wafers patterned with devices, the high temperatures would unavoidably damage all the devices due to severe atomic migration. Therefore, a transfer process is necessary in order to integrate the 2D material on the patterned wafers.

Different processes have been developed to transfer 2D materials on arbitrary substrates, such as dry transfer [50], wet transfer assisted by different solid scaffolds (like poly(methyl methacrylate) [PMMA], perfluoropolymer-hyflon or rubber film) [49, 80], roll-to-roll transfer [81], and face-to-face transfer [82], among others [83]. Among them, wet transfer and electro-chemical transfer are the two most widely used. Wet transfer method uses a polymer (PMMA) coated on the *h*-BN (or any other 2D material) as solid support, and then etches away the bottom metallic substrate. Then the polymer/*h*-BN is fished using the arbitrary substrate and the polymer support is etched away. Electrochemical method (also called bubble method) allows recycling the metallic substrate, and it is often used to transfer the *h*-BN grown on noble metals. Another possibility to integrate the 2D materials in memristors is to use LPE 2D materials spin coated on the wafers, but that may bring associated other problems, such as incontrollable thickness fluctuations that may lead to large device-to-device variability, and even uncovered areas or pinholes that reduce the yield (e.g. many devices would be initially shorted).

Another recognized problem is that graphene forms a bad interface with TMO materials, which in many cases are required to build the memristor. For example, Ref. [6] shows that, due to the absence of dangling bonds in graphene, TMOs cannot be directly deposited on it using standard methods, such as atomic layer deposition (ALD). After that, several reports claimed that this problem does not takes place on the surface of $MoS_2$ [85-87], and therefore it was believed that the mechanism for TMO growth by ALD on $MoS_2$ was different than that on graphene. However, Ref. [7] proved all them wrong by using an *in situ* characterization. Using that setup it was observed that $HfO_2$ aggregates at the local defects of the MoS$_2$, leading to a multi-island pattern instead a conformal coating with uniform thickness (i.e. similar to what happens on graphene). Ref. [7] demonstrated that the samples in Ref. [86] may have been exposed to contamination and for that reason the HfO$_2$ film could have been grown due to the presence of dangling bonds in the moisture on the MoS$_2$.

## 4.2. Status and best performances of 2D materials based memristors

The main application of memristors is as non-volatile memories (NVM) for information storage, and the companies in that field (e.g. Intel, Samsung, Micron) have been the main players boosting its research. According to the International Technology Roadmap of Semiconductor (ITRS) [54], the preformance requirements for any NVM technology are: small operating voltages (< 1 V), low power consumption (~ 10 pJ per transition), high operation speed or switching time (< 10 ns per transition), high endurance (more than $10^9$ cycles), long data retention time (> 10 years), small MIM cell sizes (< 600 nm$^2$) and high ON/OFF current ratio ($I_{ON}/I_{OFF} > 10^6$).

Different 2D materials can be used to carry out different functions in memristive devices. Generally, graphene has been used as top or bottom electrode to provide flexibility and transparency [88], and as interface layer between the metallic electrodes and the RS medium to avoid atomic diffusions, resulting in an effective decrease of the cycle-to-cycle variability [89]. Moreover, the high out-of-plane contact resistance of graphene also contributes to reduce the power consumption [90]. Other insulating 2D materials, like GO, *h*-BN and black phosphorous (BP), can serve as active RS medium to induce the RS either by migration of intrinsic species or by penetration of metallic ions from adjacent electrodes [1]. The best performances reported so far for 2D materials based memristors are summarized in Table 4.1.

**Table 4.1**. Best performances reported for 2D materials based RS devices. Reproduced with permission from Ref [1], copyright Wiley-VCH 2017. The column Ref. applies to the research article where the table was extracted (i.e. Ref. [1])

| Parameter | Technology requirements | 2D materials based RS devices | | |
|---|---|---|---|---|
| | | Best performances | Device structure | Ref. |
| Operating voltages | < 1 V | ~ -0.6 V | ITO/GO/Ag | 91 |
| | | ~ 0.4 V | Ti/$h$-BN/Cu | 18 |
| | | ~ 0.7 V | Al/GO/Al | 92 |
| Power consumption | ~ 10 pJ/transition | ~ 100 pW | Gr/TiOx/Al$_2$O$_3$/TiO$_2$/Gr | 93 |
| Switching times | < 10 ns/transition | 10 ns (set) / 1 ns (reset) | W/ta-C/W | 94 |
| | | 5 ns (set) / 5 ns (reset) | Pt/RGO–th/Pt | 95 |
| | | < 10 s | PEN/Ti/Pt/GO/Ti/Pt | 96 |
| Endurance | >10$^9$ cycles | 2 × 10$^{13}$ cycles @ 75°C | W/ta-C/W | 94 |
| | | 10$^8$ cycles | Al/PFCF/RGO/ITO | 97 |
| | | 10$^3$ cycles | Ag/MoS$_2$/Ag | 98 |
| | | > 650 cycles | Ti/$h$-BN/Cu | 18 |
| Data retention | >10 years | > 10$^7$ s (115 days) | Al/GO/ITO | 99 |
| MIM cell Size | 576 nm$^2$ | 8.5 nm$^2$ | Pt/ta-C/C-AFM tip | 100 |
| I$_{ON}$/I$_{OFF}$ ratio | 10$^6$ | ~ 10$^9$ | Ag/ZrO$_2$/SLG/Pt | 101 |
| | | > 10$^6$ | Ti/$h$-BN/Cu | 18 |

As it can be observed, 2D materials based memristors still do not fit the technological requirements for being used a NVM. RS devices using traditional metals and oxides have been investigated for already 50 years; however, 2D materials based memristors have been studied for less than 10 years. Therefore, any comparison at this stage is unfair. However, the introduction of 2D materials in the structure of memristors has already enabled several interesting functionalities that would be impossible without them. For example, graphne/SiO$_2$/graphene devices showed a transparency >92%, and achieved stable RS even after >10$^5$ bending stresses uner a radius as small as few nanometers [88]. Another example is the coexistence of bipolar and threshold type RS in a single device (see **Article 5**), which is something very complex that can only be achieved in very specific TMO/metal structures [90,102-103]. In the incoming years more studies should be carried out to determine if 2D materials based memristors could be used as NVM. However, if the high performances of NVMs are not achieved, may still be they may be used in other RS applications, such as playing the role of electronic synapses in aartificial neural networks, as the technological needs in terms of endurance, switching speed, retention, and current ON/OFF window, are more relaxed and the dynamic changes on the RS play a more important role. **Article 6** summarizes the state-of-the-art on 2D materials based memristors for their use as NVM, and discusses their prospects and main challenges.

# Graphene and Related Materials for Resistive Random Access Memories

*Fei Hui, Enric Grustan-Gutierrez, Shibing Long, Qi Liu, Anna K. Ott, Andrea C. Ferrari, and Mario Lanza\**

Graphene and related materials (GRMs) are promising candidates for the fabrication of resistive random access memories (RRAMs). Here, this emerging field is analyzed, classified, and evaluated, and the performance of a number of RRAM prototypes using GRMs is summarized. Graphene oxide, amorphous carbon films, transition metal dichalcogenides, hexagonal boron nitride and black phosphorous can be used as resistive switching media, in which the switching can be governed either by the migration of intrinsic species or penetration of metallic ions from adjacent layers. Graphene can be used as an electrode to provide flexibility and transparency, as well as an interface layer between the electrode and dielectric to block atomic diffusion, reduce power consumption, suppress surface effects, limit the number of conductive filaments in the dielectric, and improve device integration. GRM-based RRAMs fit some non-volatile memory technological requirements, such as low operating voltages (<1V) and switching times (<10 ns), but others, like retention >10 years, endurance >10⁵ cycles and power consumption ≈10 pJ per transition still remain a challenge. More technology-oriented studies including reliability and variability analyses may lead to the development of GRMs-based RRAMs with realistic possibilities of commercialization.

## 1. Introduction

The technology-driven development during the past half-century has been possible thanks to the creation of new electronic devices (computers, smart phones, vehicles, medical equipment), which allow for the performance of multiple complex operations (calculations, interpolations, statistics), leading to the appearance of new services (email, global positioning system (GPS), data mining) that create new jobs.[1] Non-volatile

F. Hui, Dr. E. Grustan-Gutierrez, Prof. M. Lanza
Institute of Functional Nano and Soft Materials (FUNSOM)
Collaborative Innovation Center of Suzhou Nanoscience and Technology
Soochow University
199 Ren-Ai Road, Suzhou 215123, China
E-mail: mlanza@suda.edu.cn
Prof. S. Long, Prof. Qi Liu
Key Laboratory of Microelectronic Devices & Integrated Technology
Institute of Microelectronics of Chinese Academy of Sciences
Beijing 100029, China
Dr A. K. Ott, Prof. A. C. Ferrari
Cambridge Graphene Centre
University of Cambridge
Cambridge CB3 0FA, UK

memories (NVMs) are essential elements in most modern electronic devices and integrated circuits, as they allow storing enormous amounts of data (5.62 × 10¹⁰ bits cm⁻²)[2] in a fast (<ns bit⁻¹)[3] and cheap (≈0.019 $ GB⁻¹)[4] way. For this reason, it is estimated that the memory market reached 47 billion USD in 2016.[5] To date, the most used NVM device is the NAND Flash.[6] It stores one bit of information in a capacitor (integrated in the floating gate of a field-effect transistor, FET).[6] The charge/discharge of the capacitor can be used to simulate the ones/zeros of the binary code, therefore storing information. Over the past 25 years technological advances have been linked to the scaling down of the NAND Flash memory, a process that improved its size (from 2 μm node in 1980 to 7 nm in 2015),[7] switching speed (from 1MB s⁻¹ in 1985 to 10GB s⁻¹ in 2012)[8] and cost (from 437,500 $ GB⁻¹ in 1980 to 0.019 $ GB⁻¹ in 2016).[4] As the size approaches the nanometer range, leakage currents in the capacitor become prohibitive, leading to severe information loss.[9,10] Therefore, in order to continue the growth of information storage, new devices using non-capacitive working principles need to be developed.

According to the International Technology Roadmap for Semiconductors (ITRS),[2] the performance requirements for any NVM technology are (see **Table 1**): i) low operating voltages (<1V); ii) low power consumption (≈10 pJ per state transition); iii) high operation speed (<10 ns per transition); iv) high endurance >10⁹ cycles (this is defined as the number of times a NVM can be switched on/off before one of the states becomes irreversible);[2] v) long state retention time >10 years (>3 × 10⁸ s; this is defined as the time before the state is lost, i.e. a state change without the application of any electrical stress);[2] vi) small size below 600 nm² (this refers to the cell that stores 1 bit of information, not the whole NVM); vii) good integration, with a capacity density larger than 10¹¹ bits cm⁻²; and viii) simple structure, which usually brings associated low fabrication costs. Several new memory concepts are being developed to achieve these targets,[1,2] including dynamic RAM (DRAM),[11,12] ferroelectric RAM (FRAM),[13,14] phase change RAM (PCRAM),[15,16] magnetoresistive RAM (MRAM),[17,18] resistive RAM (RRAM),[19–22] carbon nanotube (CNT) RAM,[23,24] spin transfer torque magnetic RAM (STTM-RAM),[25,26] molecular memories,[27,28] and Mott memories.[29,30] A comparative review of the different technologies being considered for future information storage can be found in Ref. [1]. Until now RRAMs have shown the most advanced performance (see Table 1).[2,31–33]

The RRAM is a simple and industry-compatible structure formed by a matrix of metal/insulator/metal (MIM) junctions,[34] in which the sandwiched dielectric enables reversible electrical resistance changes (see **Figure 1**). This phenomenon, called resistive switching (RS),[34] can be used to induce two logic states: the high resistance state (HRS) and the low resistance state (LRS).[35] Cyclic transitions between these two states can also be used to simulate the ones and zeros of binary code, without need for a capacitor, making the storage of digital information possible.[36] Before stable cycling between HRS and LRS can be achieved (e.g., 50 electrical pulses applied to a MIM cell produce 50 state changes without resistance mismatch), most RRAMs require a one-time activation process called forming.[34] This is defined as the first generation of a reversible dielectric breakdown (BD) in the insulator.[2] The RS phenomenon can be classified as: i) unipolar/bipolar if the electrical stresses that produce the state change are of the same/opposed polarity;[37] or ii) local/distributed if the atomic rearrangements that produce the state change take place at few/most locations within the area under stress.[36,38]

State of the art RRAMs use transition metal oxides (TMOs) as insulators, including $HfO_x$,[39–42] $Al_2O_3$,[43–46] $TiO_2$[47–50] and $TaO_x$.[51,52] In these cells, RS is related to the formation and dissolution of a nanosized conductive filament (CF) across the insulator,[35] leading to an effective connection/separation of the two electrodes. In this case RS is a local phenomenon. The physical mechanisms inducing the formation/dissolution of the CF depend on the materials that compose the MIM cell (not only the insulator, but also the metal),[53] and it is thought that mainly two phenomena are predominant.[54–58] The first is the movement of oxygen vacancies in the TMO as a consequence of the applied field, leaving behind an oxygen-free metallic path.[54] These devices have been called valence change memories (VCM)[55] and/or redox random access memory (ReRAM).[56] The second is the generation of a CF made of metallic ions from the adjacent electrodes, which can penetrate into the dielectric when the MIM structure is polarized.[57] These devices are called electrochemical metallization (ECM) memories,[55] programmable metallization cells (PMC)[58] and/or conductive bridge random access memories (CBRAM),[59] even though all these names refer to the same structure.

Over the past decade many RRAM prototypes with different characteristics have been reported.[39–52] Amongst them, the most remarkable performances are: i) ultrafast (<20 ns) logic state transitions;[31,32,44,45,60] the reset (LRS-to-HRS transition) process is usually much slower (60 ns) than the set (HRS-to-LRS transition) one.[61,62] Ref. [32] achieved sub-nanosecond (300 ps) set/reset transitions in $HfO_x$-based RRAMs. ii) Energy consumption per state transition down to 0.1 pJ,[11,44] lower than that of other technologies, such as PCRAM[63] and MRAM.[64] iii) Long cycling endurance up to $10^{12}$ cycles. Ref. [33] achieved $10^{12}$ cycles using $Ta_2O_5$/$TaO_2$ x bilayer structures coupled with Pt electrodes. iv) Long data retention.[46] Ref. [65] indicated that RRAMs can retain a resistive state even at high temperatures

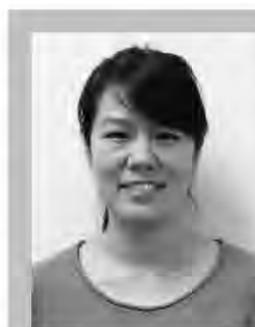

**Fei Hui** received a Bachelor's degree in Chemistry from Huanghuai University in 2013. She is currently pursuing her Ph.D. at Soochow University. Her research mainly focuses on two-dimensional materials and their integration in resistive random access memories. Fei Hui has developed a patent on graphene-coated nanoprobes that has received $1M investment. In 2016 she received the Royal Society of Chemistry Researcher Mobility Grant.

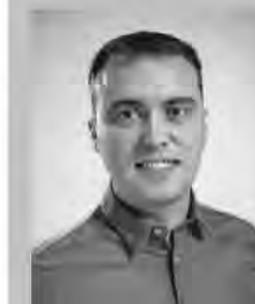

**Mario Lanza** is a Young 1000 Talent Professor and research group leader at Soochow University. He received his Ph.D. in Electronic Engineering in 2010 at the Universitat Autonoma de Barcelona. During his Ph.D., he was a visiting scholar at the Deggendorf Institute of Technology (DAAD grantee), The University of Manchester (Ministry of Education grantee), and Peking University (ICO foundation grantee). In 2010–2011 he was an NSFC postdoctoral fellow at Peking University, and in 2012–2013 he was a Marie Curie postdoctoral fellow at Stanford University. His research focuses on the development of advanced electronic devices using graphene and related materials.

(up to 200 °C). v) As the switching takes place through nanosale CFs,[35] the area of the cell is just limited by the area of the CF (typically tens of nanometers).[31] For example, Ref. [31] reported a 10 nm × 10 nm $TiN/Hf/HfO_x/TiN$ RRAM with fast ns-range on/off switching times at low voltage below 3V, switching energy <0.1 pJ bit$^{-1}$, excellent endurance >5 × $10^7$ cycles, current on/off ratios ($I_{ON}/I_{OFF}$) >50. The devices also showed 30 h retention at 200 °C. RS has been observed in even smaller areas (≈10 nm²) using the tip of a conductive atomic force microscope (CAFM).[35] vi) Simple fabrication process, as the structure basically consists of a capacitor. The materials that form the RRAM have been used in complementary metal oxide semiconductor (CMOS) technology for years. This favors their three dimensional integration.[66] RRAMs have also shown potential for multi-bit storage per unit cell,[43,67] highly desired for future NVM technologies.[2]

All these factors, summarized in Table 1, have been observed in RRAMs made of different materials (e.g. $TaO_x$ provides the highest endurance, $HfO_x$ the fastest transitions and lowest power consumption), but no single RRAM has yet shown all NVM technology requirements simultaneously.[1,2] The most critical tradeoffs are speed–retention, power–speed and endurance-retention.[1] Crossbar Inc.[68] claimed the development of RRAMs covering all these capabilities, but no details about the composition of the core cell have been revealed to date. ITRI,[65] NEC,[69] and Fujitsu[70] have also announced similar devices, with no commercial device yet available. Panasonic has commercialized the MN101L RRAM Embedded 8-bit microcontroller unit,[71] Adesto is distributing their Mavriq 45 nm CBRAM,[72] and Nantero developed a RRAM memory using MIM cells integrated on CNTs, but their use is still limited to few applications (mainly sensors).[73] More information about commercial RRAMs can be found in Ref. [1].

Table 1. Technology requirements for RRAM according to ITRS[a] compared to the best performances reported for TMO-based and GRMs-based RRAM. $I_{ON}/I_{OFF}$ is not strictly a technology requirement, but it is a reference parameter usually quoted in RRAM literature.

| Parameter | Technology requirements | TMOs based RRAMs | | | GRMs based RRAMs | | |
|---|---|---|---|---|---|---|---|
| | | Best performance | Device structure | Ref. | Best performance | Device structure | Ref. |
| Operating voltages | <1 V | 0.3 V | Ti/HfO$_2$/TiN | [41] | ≈ −0.6 V | ITO/GO/Ag | [251] |
| | | 0.1 V | Pt/Ni/Al$_2$O$_3$/SiO$_2$/Si | [43] | ≈0.4 V | Ti/h-BN/Cu | [163] |
| | | −0.2 V (set)/0.5 V (reset) | Pt/TiO$_2$/Pt | [111] | ≈0.7 V | Al/GO/Al | [253] |
| Power consumption | ≈10 pJ/transition | 0.1 pJ per transition | TiN/Hf/HfO$_2$/TiN | [31] | ≈100 pW | Gr/TiO$_x$/Al$_2$O$_3$/TiO$_2$/Gr | [185] |
| | | 0.1–7 pJ per transition | Al/Ti/Al$_2$O$_3$/p-SNT | [44] | — | | |
| Switching times | <10 ns/transition | 300 ps | TiN/TiO$_x$/HfO$_x$/TiN | [32] | 10 ns (set)/1 ns (reset) | W/ta-C/W | [286] |
| | | <10 ns | Al/Ti/Al$_2$O$_3$/p-Si-CNT | [44] | 5 ns (set)/5 ns (reset) | Pt/RGO-4h/Pt | [273] |
| | | ≈ ns level | Cu/Al$_2$O$_3$/a-Si/Ta | [45] | <10s | PEN/Ti/Pt/GO/Ti/Pt | [280] |
| Endurance | >10$^6$ cycles | 10$^{12}$ cycles | Pt/Ta$_2$O$_{5-x}$/TaO$_{2-x}$/Pt | [33] | 2 × 10$^{11}$ cycles @ 75 °C | W/ta-C/W | [286] |
| | | 5 × 10$^6$ cycles | Pt/TaO$_x$/Pt | [364] | 10$^5$ cycles | Al/PFCF/RGO/ITO | [351] |
| | | ≥10$^{10}$ cycles | Ta/TaO$_x$/TiO$_2$/Ti | [111] | 10$^5$ cycles | Ag/MoS$_2$/Ag | [316] |
| | | 10$^{10}$ cycles | Pt/Ta$_2$O$_5$/Ta | [346] | >650 cycles | Ti/h-BN/Cu | [163] |
| | | 10$^{11}$ cycles | W/AlO/Ta$_2$O$_5$/ZrO$_2$/Ru | [347] | — | | |
| Data retention | >10 years | >10 years @ 85 °C | Pt/Al$_2$O$_3$/HfO$_2$/Al$_2$O$_3$/TiN/Si | [46] | >10$^7$ s (115 days) | Al/GO/ITO | [249] |
| | | >10 years @ 85 °C | Pt/TaO$_x$/Pt | [364] | — | | |
| MIM cell Size | 576 nm$^2$ | 5 nm$^2$ | TaN/TiN/Zr/HfO$_2$/CAFM tip | [296] | 8.5 nm$^2$ | Pt/ta-C/C-AFM tip | [296] |
| | | 10 nm × 10 nm | TiN/Hf/HfO$_2$/TiN | [31] | — | | |
| $I_{ON}/I_{OFF}$ ratio | 10$^6$ | 3 × 10$^6$ | Ni/GeO/STO/TaN | [348] | ≈10$^6$ | Ag/ZrO$_2$/SLG/Pt | [349] |
| | | 2 × 10$^6$ | Pt/Gd$_2$O$_3$/Pt | [350] | >10$^6$ | Ti/h-BN/Cu | [163] |

Despite these developments, reliability issues (endurance, retention) and variability (cycle-to-cycle and device-to-device) of essential parameters like set/reset voltages (among others), as well as the understanding of failure mechanisms are still hindering large-scale RRAM manufacturing.[1,2] Therefore, the reproducibility and uniformity of RS in RRAMs still remains an area of active research, with the need to optimize the materials that form MIM cells.

One promising approach consists of replacing the metallic and/or insulating films of the MIM structures with novel materials with enhanced capabilities which, at the same time, could provide new features to the devices, such as transparency and flexibility.[74,75] Along these lines, graphene and related materials (GRMs) are at the centre of an ever-increasing research area due to their unique electronic,[76] physical,[77] chemical,[78] mechanical,[79] optical,[80] magnetic[81] and thermal[82] properties.[83] The term GRMs encompasses graphene, graphene oxide (GO), transition metal dichalcogenides (TMDs), hexagonal boron nitride (h-BN), black phosphorous (BP) and any other layered material (LM). Furthermore, a variety of thin carbon films have been considered for the implementation of RRAMs, ranging from sp$^2$ rich amorphous carbons (a-C),[84–87] to sp$^3$ rich tetrahedral amorphous carbons (ta-C).[88,89]

Here we review the use of GRMs to build RRAMs. First, we describe the fabrication process of RRAM devices using GRMs (Section 2), the advantages of using graphene as top/bottom electrode (Section 3), the performance achieved using graphene oxides (Section 4) and amorphous carbons (Section 5), as well as recent observations of RS in other LMs, including TMDs, h-BN and BP, among others (Section 6). The status and prospects of GRM-based RRAM technology are discussed in Section 7.

## 2. Fabrication of RRAMs using GRMs

A detailed description of the different approaches for the preparation of GRMs can be found elsewhere.[83,90] The aim of this section is to emphasize the methods used to implement GRM-based RRAMs, with special emphasis on those that are scalable. We also include practical information for device integration.

### 2.1. Device Architecture

Different device architectures to achieve NVM using GRMs have been suggested. The first used a NVM configuration based on graphene FETs (GFETs), such as floating gate and charge-trap memories.[91–99] RRAMs based on redox-switchable functionalized graphene nanoribbons[100] stripes of thin (<10 nm) graphitic material grown by chemical vapor deposition (CVD)[101] and graphene/metal contacts[102] have also been proposed.

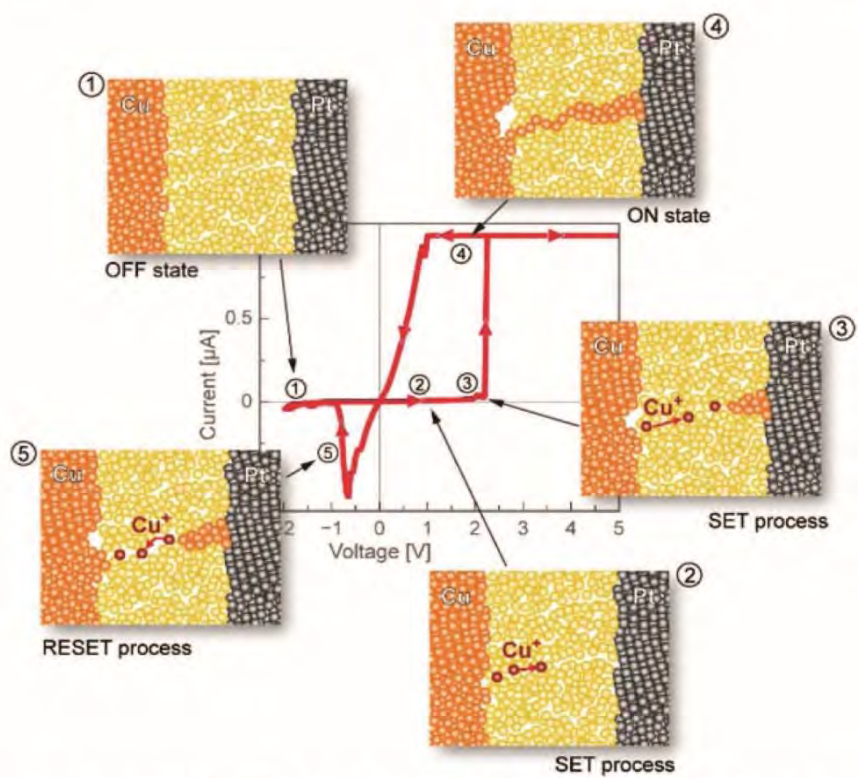

**Figure 1.** Current–voltage characteristics of ECM/CBRAM cell with schematic of the physical processes. Reproduced with permission.[34] Copyright 2014, John Wiley and Sons.

The first reports using GRMs in MIM-like RRAMs did not use the vertical MIM structure, but planar configurations containing a transversal insulating nanogap[101,105] (see **Figure** 2a and b).

Ref. [103] fabricated a planar device with two electrodes connected by a single layer graphene (SLG) placed on a 300 nm SiO$_2$/Si substrate by micromechanical exfoliation (MC), very similar to a single back-gated GFET.[106] By applying between 2.5 and 4 V, the breakdown of the SLG channel (physical fracture) was induced.[103] By applying a reverse bias from 0.1 up to 5 V, reproducible transitions between HRS and LRS could be observed. Ref. [104] improved this performance using 5–10-nm-thick films of graphitic material (consisting of discontinuous graphene sheets grown by CVD) and reported bistability in current vs. voltage ($I$–$V$) curves with $I_{ON}/I_{OFF}$ up to $10^7$ and switching times as fast as 1 µs. Ref. [105] further enhanced the capabilities of planar bilayer graphene (BLG) switching devices by coating a 10-nm layer of conducting 3-aminopropyltriethoxysilane (APTES) molecules over the surface of the insulating region (SiO$_X$). Nevertheless, the difficulty in controlling the size of the nanogap[104–106] and the likely large device-to-device variability (statistical information was not reported), made most works concentrate on the vertical MIM-like RRAMs (such as those shown in Figure 2b), which is by far the most widespread and competitive device architecture developed thus far for RS-based NVMs. [107–112]

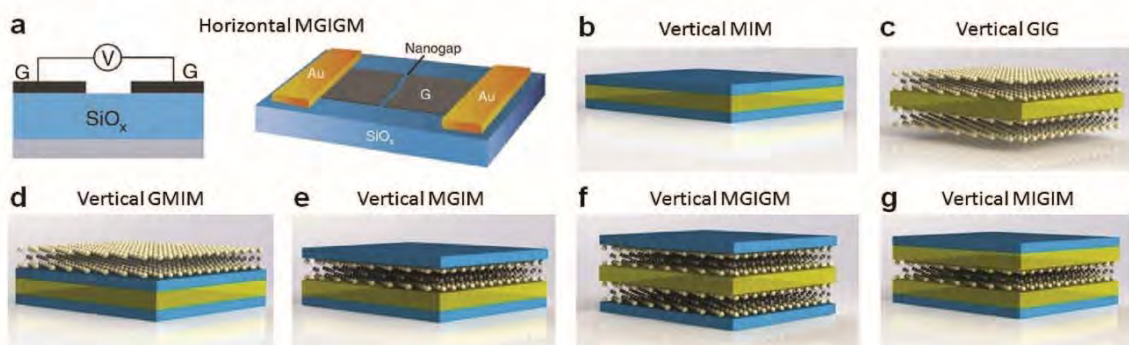

**Figure 2.** Device structures proposed in GRMs-based RRAM technology. M indicates metal, I indicates insulator, and G indicates GRM. The electric field in (b–g) is always applied between top and bottom layers.

The core cell of state-of-the-art RRAMs consists of a matrix of vertically aligned MIM structures[34,35] (see Figure 2b). These can be fabricated by sequentially depositing each material on a desired substrate, using standard industrial processes such as atomic layer deposition (ALD),[113] sputtering[107] and/or electron-beam evaporation.[114] GRMs can be used in multiple parts of RRAMs (see Figure 2c–f): i) replacing one/all layers in the MIM structure, leading to alternative configurations such as, for example, graphene/insulator/graphene (GIG) or metal/$h$-BN/metal structures; and ii) introducing one/few additional GRM layer/s within the standard MIM cell, leading to MGIM, MIGM, MGIGM, GMIM, MIMG and GMIMG (where G denotes a generic GRM). Another possible configuration is the MIGIM structure, in which the GRM is used for charge trapping purposes[115,116] (see Figure 2g). In all cases, the goal is to improve the NVM performance (i.e. switching speed, retention time, endurance, power consumption) as well as to exploit some of the distinctive properties of the GRM (i.e. transparency[74] and flexibility[75] enhanced thermal heat dissipation,[117] and chemical stability have been achieved using GRMs in other devices like FETs,[118] meaning that these properties may be also achieved in RRAMs).

## 2.2. Insertion of GRMs in the RRAM Structure

The main challenge associated with the fabrication of vertical RRAMs using GRMs is that the GRM cannot be introduced in the MIM structure using conventional fabrication tools. A large portion of the reports on GRMs used non-scalable techniques, such as micromechanical exfoliation (MC).[119,120] MC can produce flakes with a very low number of defects,[121,122] but this is not yet industrially scalable. This strongly limits its application in RRAMs, and only allows RS studies using local techniques, such as CAFM.[123] Industrially scalable GRM production methodologies,[63] such as liquid phase exfoliation (LPE)[83,90,124] and CVD[90,125] are now available, and are the most used for the fabrication of RRAMs.[126–129]

LPE gives GRMs flakes of different sizes and thicknesses (typically below 1-µm diameter and 1–20 layers thick).[83,90,124,130] They have been introduced in RRAMs by drop casting,[131] spin coating,[132] or ink-jet printing[133] which leads to 15–500-nm-thick films.[176–179] LPE is cheap and scalable,[83] but the rough surface of the samples obtained by this method (typically RMS > 20 nm)[134] may be an important source of variability,[135–137] which is one of the main concerns of RRAM technology.[1] The lack of variability analyses in all LPE-based-RRAM reports published to date[135–141] indicates the need for further studies.

CVD is the technology most widely used to produce GRMs for electronic devices, as it allows wafer-scale production.[83] GRMs can be grown by CVD on different substrates. In the case of SLG, metals with low carbon solubility (such as Cu, Ir, Co, Ni) are necessary.[107–144] Some reported direct CVD growth on SiO$_2$,[145–147] For MoS$_2$,[148–152] TiS$_2$,[153] TaS$_2$,[154] WS$_2$,[155] MoSe$_2$[156,157] and WSe$_2$[158] direct CVD growth on insulating substrates like SiO$_2$ and Al$_2$O$_3$ is preferred because their lattice constants offer a good match to that of the GRM.[148–158] CVD growth of $h$-BN was also reported on Cu,[159] Fe,[160] and Pt.[161] Ref. [162] reported the CVD growth of BP on Si using a red phosphorous powder source.

When working with insulating GRMs (such as $h$-BN) grown by CVD, the metallic substrate can be used as bottom electrode.[163] This facilitates the fabrication of RRAMs, and the top electrode can be then deposited by photolithography or shadow mask, plus metal evaporation. However, the underlying metal can have large roughness (RMS ≈ 30 nm)[164] due to the polycrystallization suffered during the CVD growth at high temperatures, usually not below 800 °C.[142–144,148–152] Therefore, the growth of insulating GRMs on ultra-flat metal-coated wafers is of utmost importance to avoid roughness-induced variability, as well as to offer better integration with industry. In general, the thermal budget may be an issue for the fabrication of GRM-based RRAMs. On the contrary, when working with conductive GRMs (such as graphene), the metallic substrate used during CVD growth is a burden for RRAM fabrication, because sometimes the presence of GRMs is required on substrates not favorable for their CVD growth, e.g. HfO$_2$ and other TMOs.[106] One approach is to transfer GRMs onto the desired substrate using polymer scaffolds, polymethyl methacrylate (PMMA) being the most commonly used.[83,90,165–167] The problems associated with this technique are: i) physical breakdown of GRMs,[168,169] producing cracks, which may locally alter the properties of the devices.[170,171] For example, a MGIM device in which the GRM contains holes may lead to local MIM structures. ii) polymer residuals on the GRM surface. Although this may not result in device failure, since the polymer is insulating, it can be understood as a decrease of the effective area of the MIM capacitor. The introduction of annealing processes (at ≈300 °C)[172] may contribute to the removal of these impurities, but may produce polycrystallization of TMOs in the RRAM (if any), leading to unwanted inhomogeneities and thickness fluctuations.[145–147] Polymer-free transfer techniques, such as electrostatic graphene/substrate attraction, can be used,[173] but this may increase the complexity of the process.[174,175]

Other methodologies to grow GRMs are physical vapor deposition,[176] growth on SiC,[177] and the hydrothermal method[178] but, to the best of our knowledge, their use in RRAM technology has not yet been reported.

The deposition of insulators on GRMs is also problematic. According to Ref. [179], TMOs cannot be deposited directly by atomic layer deposition (ALD) on defect-free SLG, due to the lack of dangling bonds or functional groups. Ref. [180] observed that, when trying to grow HfO$_2$ by ALD on MoS$_2$, HfO$_2$ did not form a homogeneous film, but instead islands on the MoS$_2$ surface, probably located at MoS$_2$ defects (where there are dangling bonds that allow HfO$_2$ agglutination).[181] One possible approach is the functionalization of the GRM surface,[182–185] which may enhance the homogeneity of the TMO film at the interface. The most common strategies to achieve a uniform SLG/high-k interface are functionalization with NO$_2$,[182] metal seed layers,[183] organic seed layers[184] and ozone (O$_3$).[182 184] An interesting method to generate a SLG-TiO$_x$/Al$_2$O$_3$/TiO$_7$-SLG cell was proposed by Ref. [185]. A seed Ti layer was first deposited on the bottom SLG electrode by e-beam evaporation and then oxidized to TiO$_x$ in air. The Al$_2$O$_3$/TiO$_2$ stack was then deposited by ALD, and the top SLG electrode was transferred. Another similar GIG device was fabricated by Ref. [186] by depositing a bilayer insulating film made of Ta$_2$O$_5$ $_x$/TaO$_x$ on a CVD-SLG using radio-frequency and reactive sputtering, followed by another CVD-SLG transfer.

For devices designed to be tested in a probe station, the use of top metallic electrodes is unavoidable, as placing the large tip on a SLG top electrode may damage it. Therefore, the GI interface is in fact a MGI. One method to measure the SLG electrodes without the need of metal deposition is the use of CAFM, which controls very accurately the tip/sample contact force and does not damage the GRM surface.[187] CAFM can also allow the investigation of ultra-scaled RRAMs.[35]

## 3. Use of Graphene as Top/Bottom Electrode

### 3.1. Transparency

One motivation for using graphene in electronic devices is to provide them with flexibility[75] and transparency.[74] For transparent devices, indium tin oxide (ITO) has been traditionally the preferred electrode material,[188–190] but its brittle nature makes it less suitable for flexible/foldable devices. One alternative is using organic materials, such as conductive polymers,[191] but in this case the NVM performance (with retention times of just 10$^4$ s and endurance below 100 cycles) is usually much lower than state-of-the-art TMO-based RRAMs. [1.31–33]

Ref. [192] fabricated transparent MLG/Dy$_2$O$_3$/ITO structures by transfer of CVD-grown multilayer graphene (MLG) patterned in a subsequent photolithography step. The devices showed forming-free unipolar RS with $I_{ON}/I_{OFF} \approx 10^5$, low set and reset voltages (0.4 and 0.2 V respectively), endurance >100 cycles, retention time >10$^4$ s and typical switching power and speed of 4.4 µW and 60 ns. Furthermore, the devices showed a transparency ≈80%. The performance as RRAMs of these devices is better than other graphene-free cells, such as ITO/ZnO/ITO,[188] ITO/AlN/ITO,[189] ITO/Gd$_2$O$_3$/ITO,[190] and other transparent prototypes like Ga-doped ZnO.[193] Ref. [105] further improved the optical performance by exploiting bi-layer graphene (BLG) in BLG/SiO$_x$/BLG structures, with a transmittance >90% (see **Figure 3a–c**). Ref. [74] also achieved good RRAM functionality with an overall light absorptance <25% in devices made of ITO/SLG/ZnO/ITO, which also showed better RS uniformity than its graphene-free counterparts. Ref. [75] used MLG with a transmittance up to 92% to fabricate a flexible organic memory device.

The characteristics of transparent and flexible graphene-based RRAM devices in literature are summarized in **Table 2**. Coupling graphene electrodes with organic RS media seems to provide the highest transparency ≈92%,[194] maintaining high $I_{ON}/I_{OFF}$ =10$^6$ and long retention ≈10$^4$ s. All graphene-based transparent devices were fabricated by CVD plus transfer (see Table 2).

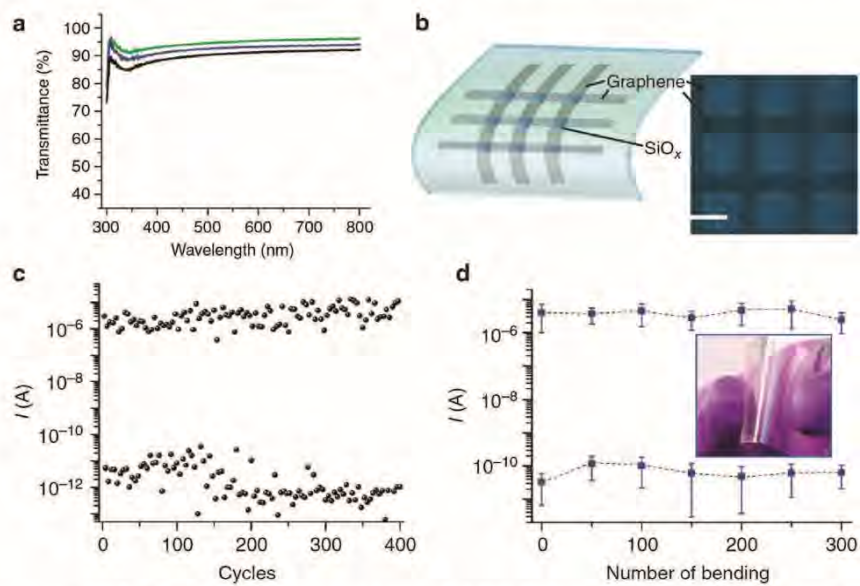

**Figure 3.** a) Transparency of SLG/SiO$_2$/SLG RRAM on glass. b) Schematic SLG/SiOx/SLG crossbar structures on a plastic (fluoropolymer) substrate (left panel) and optical image (right panel). Scale bar = 20 µm. c) Endurance measured from one of the crossbar devices using +5 and +14 V as set and reset voltages. The memory states (current) were recorded at +1 V. d) Current levels of both ON and OFF memory states (read at +1 V) from a crossbar device during repeated bending of the plastic substrate for $r_b \approx 0.6$ cm. The inset shows transparent memories using the pillar structures on plastic substrate. Reproduced with permission.[105] Copyright 2012, Nature Publishing Group.

**Table 2.** Graphene-based RRAMs with transparency and/or flexibility capability. In the column headed "Flexible", $r_b$ and $C$ are the bending radius and number of RS cycles collected during the test. In the "Transparent" column, the percentages correspond to light transmittance, and "Yes" means that the authors claim that the structure is transparent but did not quantify it.

| Device structure | Fabrication method | Device size | $I_{ON}/I_{OFF}$ | Set V [V] | Retention [s] | Endurance [cycles] | Power consumption [μW] | Switching time [ns] | Transparent | Flexible | Ref. |
|---|---|---|---|---|---|---|---|---|---|---|---|
| MLG/Dy$_2$O$_3$/ITO | CVD (Transfer) | $80 - 3 \times 10^4$ [μm$^2$] | >10$^5$ | 0.4 | >10$^4$ | >100 | 4.4 | 60 (set) 60 (reset) | 80% | No | [192] |
| ITO/SLG/ZnO/ITO | CVD (Transfer) | 200 μm in diameter | 20 | – | 10$^4$ | >100 | – | – | Yes | No | [74] |
| PS/SLG/PMMA/ SLG/PMMA$^{a)}$ | CVD (Transfer) | 500 μm in diameter | L1: 10$^4$  L2: 10$^0$ | L1: –2  L2: –4 | 10$^4$ | 1 | – | – | Yes | No | [116] |
| Al/PMMA/MLG/ PMMA/ITO/PET | CVD (Transfer) | 18–27 μm in diameter | $4.4 \times 10^2$ | 3.4 | $1 \times 10^5$ | $1.5 \times 10^5$ | – | – | Yes | $r_b = 10$ mm  $C = 1.5 \times 10^5$ | [195] |
| MLG/PEPCBM/Al | CVD (Transfer) | – | $\approx 10^6$ | 4 | $1 \times 10^4$ | >30 | – | – | 92% | $r_b = 4.2$ mm  $C = 1 \times 10^4$ | [75] |
| Ti/Pt/TiO$_2$/G/PEN (G thickness not mentioned) | CVD (Transfer) | $36 \times 36$ [μm$^2$] | 10$^3$ | 2 | 10$^4$ | 1 | 3 (set) 94 (reset) | – | Yes | $r_b = 10$ mm  $C = 100$ | [196] |
| BLG/SiO$_x$/BLG/ ITO | CVD (Transfer) | 100 μm in diameter | 10$^{5(b)}$ | 5 | – | 100 | – | – | 90% | No | [105] |
| BLG/SiO$_x$/BLG/ Polymer | CVD (Transfer) | 100 μm in diameter | 10$^6$ | 5 | – | 400 | – | – | Yes | $r_b = 12$ mm  $C = 300$ | [105] |

$^{a)}$This device shows multilevel RS. Depending on $V_{SET}$, $I_{ON}/I_{OFF}$ changes. Table 2 reports the parameters for both levels; $^{b)}$This value is not well supported in Ref. [116]; the I–V curve only shows 1–2 orders of magnitude, while the R vs. Cycle plot shows ≈10$^5$.

### 3.2. Flexibility

Graphene can be used to increase the flexibility of RRAM cells. Ref. [105] reported BLG/SiO$_x$/BLG cells with no RS degradation after bending >300 times at a bending radius ($r_b$) ≈1.2 cm (see Figure 3d). Ref. [195] presented a flexible organic device based on SLG sandwiched by two insulating poly(ethylene terephthalate) polymer (PET) layers.[195] A Ni/PMMA/SLG/PMMA/ ITO/PET cell was fabricated by transferring a CVD-SLG and spin-coating the PMMA layers. In this case SLG was used as a charge storage medium. Ref. [195] reported a good memory performance including endurance >$1.5 \times 10^5$ cycles, $I_{ON}/I_{OFF}$ >10$^6$, and retention time >$1 \times 10^5$ s. Especially significant was the lack of interference observed for scaled-down devices with SLG, as well as the ability of the devices to maintain similar switching characteristics (set/reset voltages and $I_{ON}/I_{OFF}$) even after being bent ($r_b \approx 1$ cm) over $1.5 \times 10^5$ times. Ref. [75] designed $8 \times 8$ cross-bar array-type flexible organic RRAMs on PET using MLG electrodes coupled with two different active layers: one polyimide and the other 6-phenyl-C$_{61}$ butyric acid methyl ester

(PCBM). Typical write-once-read-many characteristics and high $I_{ON}/I_{OFF}$ up to 10$^6$ were achieved; for >1000 mechanical cycles ($r_b$ between 4.2 and 27 mm) the devices maintained a retention time >10$^4$ s with <12.5% resistance fluctuations in both HRS and LRS.[75] Comparing the RS performance of all flexible RRAMs exposed to mechanical stresses is complex because these may have been applied using different $r_b$ and times. The influence of the bending time in flexible RRAMs was not reported to date, while most works report $r_b$.[75,105,195,196] Smaller $r_b$ may produce more damage to the devices, as this introduces higher stresses. Therefore, from Table 2 it can be concluded that the RRAMs with the best performance under bending are those in Ref. [195].

### 3.3. Blocking Layer for Atomic Diffusion

The most common electrode materials in RRAMs are Al, Pt, Au, Cu, Ti and Ni.[197–202] These not only serve as contacts, but play an essential role on the physics,[53] kinetics[200] and statistical distribution[203,204] of the RS. For example, different metallic electrodes can alter the number of CFs in RS media, which has an impact on the shape (sharp or progressive) of the reset process, among others.[53] One strategy to tune the switching characteristics of RRAMs is the use of active metal electrodes (like Ti or Zr) that can interact with the species from the insulator. For example, in Pt/Ti/HfO$_2$/Pt cells[205–207] oxygen atoms from the HfO$_2$ layer can interact with the Ti electrode. This allows the observation of bipolar RS thanks to the movement of oxygen in and out of the HfO$_2$ film, forming an O-vacancies-based CF with the narrower end at the cathode side.[206] On the contrary, in Pt/HfO$_2$/Pt devices[205–207] the O-vacancies movement towards the electrode is difficult, generating a CF that can only be disrupted by applying large currents,[205,206] which melt the filament by Joule effect.[208] In this case, the forming event is sharper, which leads to a higher $I_{ON}/I_{OFF} \approx 10^4$ for Pt/HfO$_2$/Pt instead of ≈12 for Pt/Ti/HfO$_2$/Pt, but the endurance may be worse due to the generation of an avalanche current.[209] BD spot propagation,[210] thermal heat,[211] insulator contamination by metal migration[212] and dielectric-breakdown-induced epitaxy.[213] Comparing the performance of Pt/Ti/HfO$_2$/Pt and Pt/HfO$_2$/Pt cells it has been observed that,[205–207] while the LRS currents in Ti-free devices were linear and the filament was symmetric, those including inserted Ti layers drove exponential currents representative of partially formed conical filaments, with the narrower end at the HfO$_2$/Ti interface. This was confirmed by fitting the experimental *I–V* curves to the quantum point contact model.[214] Moreover, at larger electric fields, the movement of metallic ions may also be activated, allowing their penetration in the TMO and producing even larger changes in the device conductivity than the motion of oxygen vacancies.[40] Therefore, as SLG is impermeable,[77,215] introducing SLG between metal and insulator alters these interactions.[108,216]

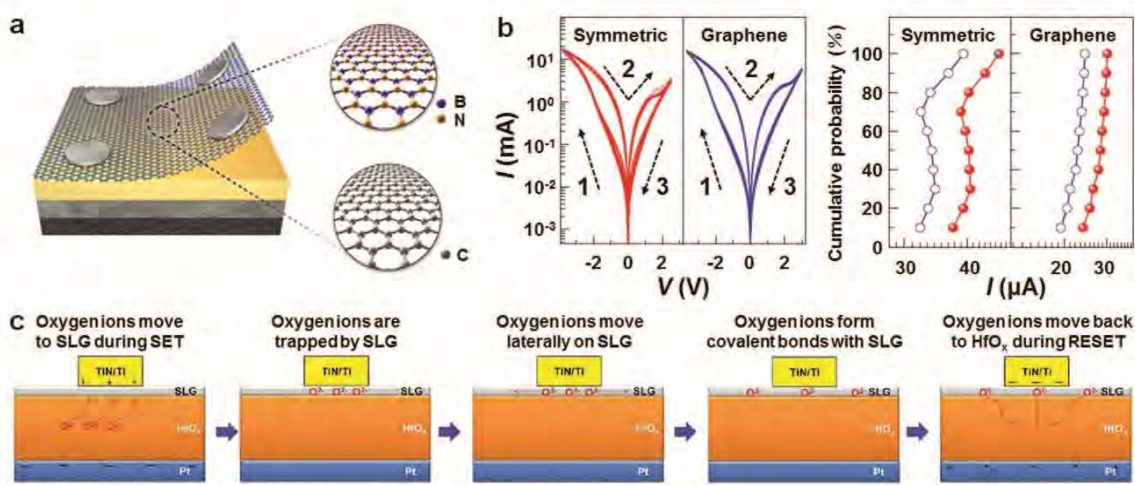

**Figure 4.** a) Schematic of SLG inserted between the top electrode and insulating film of a RRAM cell. b) Resistance switching *I–V* characteristics of a symmetric Al–WO$_3$–Al device and a Al(SLG)–WO$_3$–Al device. Cumulative probability of the HRS current at ±0.5 V for both configurations. Reproduced with permission.[107] Copyright 2015, Elsevier. c) Schematic diagrams of oxygen ions movement in MGIM structures. The diagrams represent (from left to right) steps of the process including movement of oxygen ions to SLG during SET, capture of oxygen ions by SLG, movement of oxygen ions laterally on SLG, formation of covalent bonds with SLG, followed by movement of oxygen ions back to HfO$_X$ during reset. Reproduced with permission.[217] Copyright 2013, American Chemical Society.

Ref. [107] observed that inserting CVD-SLG in Al/WO$_3$/Al structures stabilized the characteristics of the RRAM devices (see **Figure 4**a and b), reducing the variability of the set/reset voltages and currents, as well as enhancing the endurance. In SLG-free cells, when positive bias is applied to the top electrode, oxygen ions from the Al/WO$_3$ interface are pushed into the oxide bulk, leading to the formation of CFs rich in oxygen vacancies (which can be charged by electrons). During the reset process, the oxygen-deficient region is reoxidized. Ref. [107] suggested that SLG blocks the diffusion of oxygen ions into the reactive Al layer, which reduces the cycle-to-cycle variability in *I–V* curves. The dissolution of oxygen in SLG is very scarce and it presents a barrier for potential oxygen diffusion.[56] Both factors impede the diffusion of oxygen through SLG, avoiding the interaction with the metallic top electrode. Ref. [217] suggested that the electric field applied during the set operation can move the oxygen ions towards the metal/oxide interface, but they cannot penetrate into the Ti electrode due to the presence of the interfacial SLG (see Figure 4c). At most, the oxygen ions could form covalent bonds with the SLG defects (missing atoms and/or dangling bonds),[185,217] leading to a p-type doping that can be released during the reset transition. However, Ref. [55] reported the migration of metallic ions from the electrode into the dielectric in ECMs, even with the presence of interfacial SLG. Ref. [55] reported that, in ECM cells based on Ta/SLG/TaO$_X$/Pt stacks, the switching is influenced by the formation of Ta ions and their interaction with the TaO$_X$ active layer. Nevertheless, Ref. [55] used large device areas ranging between 25 × 25 and 1000 × 1000 μm$^2$. The presence of cracks and leaky grain boundaries can happen in CVD-grown and transferred SLG,[218] thus MLG may provide a better protection than SLG.

### 3.4. Lowering Power Consumption

The out-of-plane SLG contact resistance is larger than that of metallic electrodes,[219] which can be used to reduce the currents in both resistive states of the RRAMs, lowering power consumption. Ref. [217] analyzed bipolar RS in TiN/Ti/SLG/HfO$_X$/Pt RRAMs. Cyclic voltammetry indicated a reduction of the reset current by a factor ≈11 compared to SLG-free devices (**Figure 5a**), further corroborated using cumulative probability plots. Despite this improvement, the plots indicate that the HRS currents under positive polarity for the SLG-based devices increase, which is an unwanted effect. Ref. [217] pointed out that comparisons between SLG-based and SLG-free cells using similar current limitations (CL, defined as the threshold current used during the forming/set processes to limit BD) were not reliable due to the low endurance of SLG-free cells at such low (100 µA) current levels. To solve this problem, Ref. [217] compared the typical RS cycles using the optimal CL for each cell (10 µA for SLG-based cells and 100 µA for SLG-free ones), suitable to produce a lower cycle-to-cycle variability (Figure 5b), and concluded that: i) The CL needed to stabilize RS in the SLG-based device is lower, which from the power consumption point of view is an advantage. Despite the current in the HRS being the same, the LRS current was reduced more than one order of magnitude. This implies that, when the filament is completely formed in the LRS, its size (diameter) is much smaller using SLG-based electrodes. ii) The decrease of LRS current reduces $I_{ON}/I_{OFF}$. iii) SLG avoids the current overshoot during the set process, which also reduces the maximum current during the reset transition ($I_{RESET}$): in SLG-based RRAMs, $I_{RESET}$ was half CL, while in SLG-free, $I_{RESET}$ was 2–3 times larger than CL (see Figure 5b).

Ref. [98] fabricated a Pt/Ti/TiO$_x$/SLG RRAM and reported similar data as Ref. [217] (Figure 5c). $I_{ON}/I_{OFF}$ as well as both HRS and LRS currents were reduced. Therefore, from these two results,[98,217] SLG helps to stabilize RS at lower CLs, which reduces the reset current (probably due to the formation of narrower CFs) and the overall power consumption.

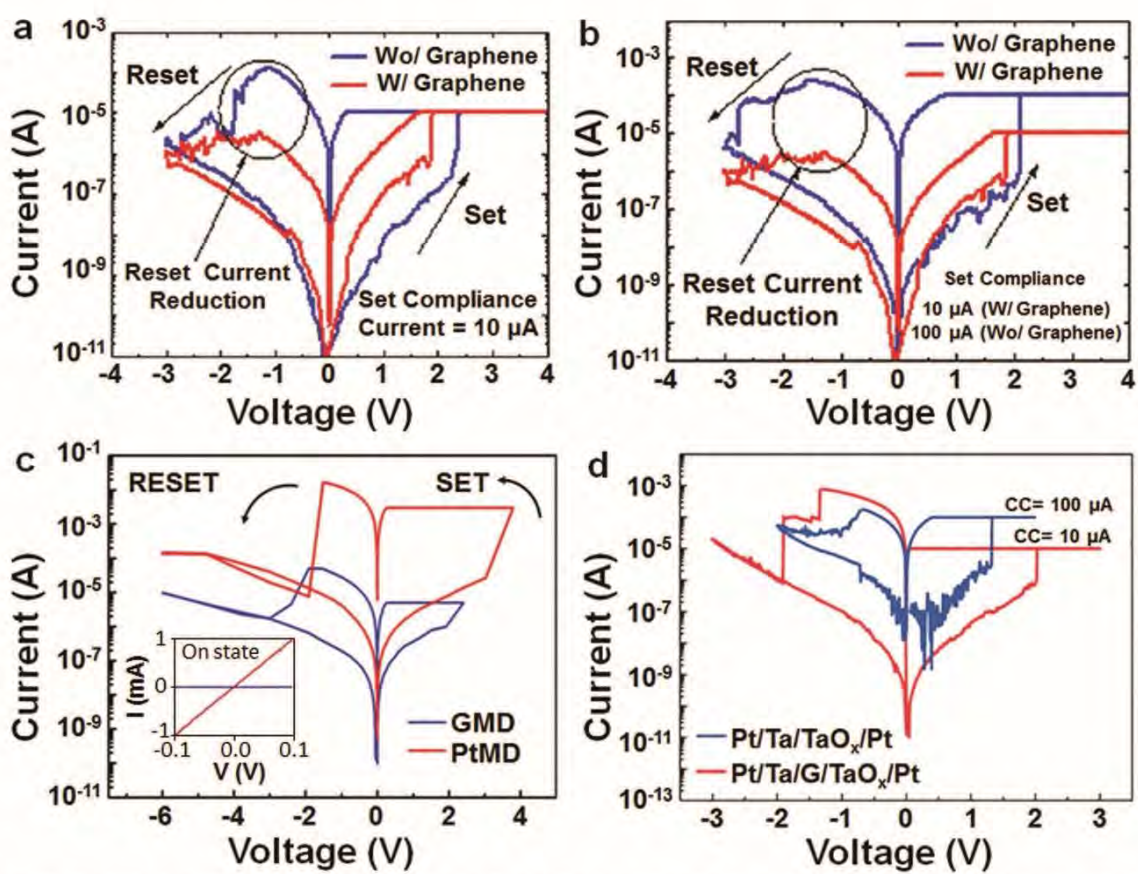

**Figure 5.** a) Typical RS behaviour of Ti/SLG/HfO$_X$/Pt (red) and Ti/HfO$_X$/Pt RRAMs under the same CL. b) Typical RS behaviour for the same devices but using optimal testing conditions. 10 and 100 µA are applied to achieve steady RS. Reproduced with permission.[217] Copyright 2013, American Chemical Society. c) RS curves of typical TiO$_2$-based memristive devices using SLG and Pt electrodes, with a SET current compliance of 5 µA and 3 mA. The arrows point to the RS directions. Inset: small-bias I–V curves for both devices in the ON state, showing different resistance. Reproduced with permission.[196] Copyright 2014, Wiley-VCH. d) I–V comparison between a Pt/Ta/TaO$_x$/Pt cell with SLG inserted between the Ta and the TaO$_x$ layers (red) and one cell without (blue). The cell without SLG needs higher HRS currents for stable RS. That with SLG offers higher $I_{ON}/I_{OFF}$ and HRS current reduction. Reproduced with permission.[55] Copyright 2015, Wiley-VCH 2015.

Ref. [55] also observed that lower CLs (10 μA) stabilize Pt/Ta/SLG/TaO$_X$/Pt RRAMs (Figure 5d), producing an increase of the reset current and $I_{ON}/I_{OFF}$ in the SLG-based cell (compared to SLG-free). These results are surprising because the CL used for the SLG-based cells was smaller, and it is usual for the reset to take place at currents similar to CL in all kinds of RRAMs (including ECMs, VCMs).[55] Indeed, Figure 5d shows a current overshoot. We cannot tell how reproducible these observations are because, unlike Ref. [217], Ref. [55] did not include the evolution with the number of cycles. On the other hand, Ref. [217] observed reset currents smaller than CL in SLG-based devices. More work is thus necessary to confirm these observations.

### 3.5. Suppression of Surface Effects

Most devices based on TMOs are influenced by surface effects,[220] including surface band bending,[221] chemisorption/photodesorption,[222] and surface roughness.[223] The barrier for species diffusion provided by SLG was used by several groups. For example, Ref. [74] inserted SLG into an ITO/ZnO/ITO stack to explore the device performance variation under different atmospheres (see **Figure 6**). O$^{2-}$ chemisorption happened at the top surface of the MIM structures (in contact with the environment), resulting in defects associated to the oxygen partial pressure. Due to oxygen ion chemisorption, the partial pressure of oxygen can influence the TMOs electrical properties, as more O$_2$ molecules are chemisorbed with increased partial pressure.[224–227] As oxygen ions are absorbed at the TMO surface defects,[224] such as oxygen vacancies,[228] acting as electron acceptors to form chemisorbed oxygen ions, which will contribute to decrease the conductivity of metal oxide. However, the introduction of SLG (forming an ITO/SLG/ZnO/ITO structure) protects the ZnO film from chemisorption of O$_2$ molecules, avoiding surface effects. The effect of oxygen ions chemisorption on the switching properties of RRAMs was analyzed by Ref. [74] by comparing the resistance of HRS and LRS with and without SLG electrodes under various ambient conditions. Without SLG, the HRS shifts to a higher resistance as it can interact with the atmospheric O$_2$[224] because the chemisorbed oxygen ions induce lower conductivity near the ZnO surface.[224,226] As the oxygen ions concentration increases, the surface band bending effect is more pronounced. However, with the SLG introduction at the ITO/ZnO interface the variation of HRS resistance is suppressed,[74] and it almost completely decouples the average variation of the HRS resistance from atmospheric conditions.[74] This improves device reliability, giving endurance >10$^2$ cycles and retention time >10$^4$ s.

### 3.6. Functionalization of Graphene Electrodes

Different functionalization strategies can be followed to achieve specific performances. For example, SLG can be used as blocking interfacial layer to avoid metal/insulator interactions.[229] If SLG is intentionally patterned with selected numbers of holes or defects, the properties of the cell at those locations can be modified, leading to specific local phenomena, such as local (instead of distributed) O-vacancy scavenging. Ref. [170] functionalized SLG in a MGIM structure by using controlled Ar$^+$-ion-assisted bombardment, which generated different amounts of defects, depending on the bombardment energy.[230] By means of CAFM Ref. [170] showed that the leakage current in functionalized samples was more confined than in pristine ones (see **Figure 7**), probably due to the lower conductivity of the SLG-free locations (i.e. the holes patterned in SLG). MGIM devices with Ar$^+$-ion-bombarded SLG had smaller variability in the set and reset voltages than those without, and more stable currents in each state.[170] This strategy was further studied by Ref. [171], who tuned ionic transport in Pd/Ta/SLG/Ta$_2$O$_5$/Pd RRAMs using SLG with engineered nanopores. SLG was grown by CVD on Cu and transferred with the assistance of a polymer scaffold.[165,167] The migration of oxygen ions in the device was controlled by opening some nanopores in SLG, which allowed to tune the properties of the devices.[171] However, since the nanopores were patterned with e-beam lithography, the process is less scalable than in Ref. [170], which used ion-assisted reaction treatment after transfer of MLG to etch residues as well as induce defects in SLG. In all, it was demonstrated that inserting a functionalized SLG in the structure of RRAMs is a good approach to tune their properties.

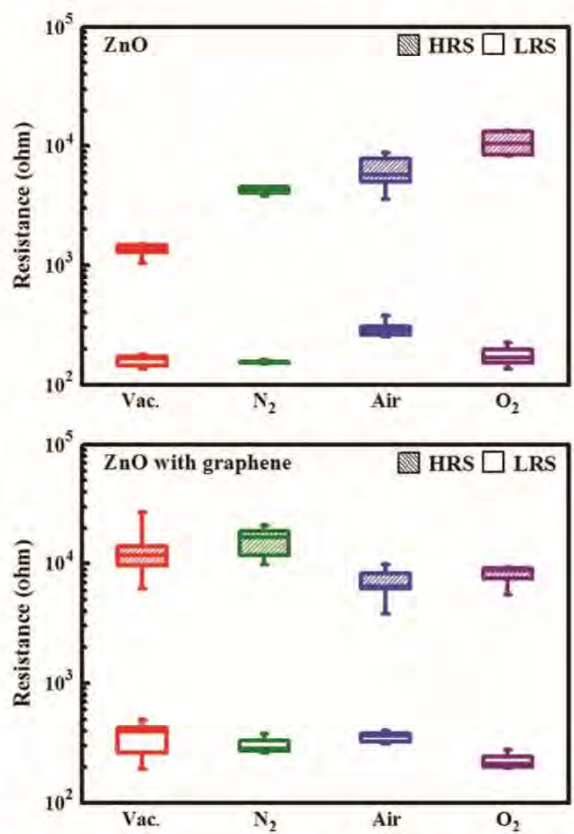

**Figure 6.** Atmosphere-dependent resistance in HRS and LRS of ZnO RRAMs with and without SLG electrodes. The bottom and the top of the box are the 25$^{th}$ percentile and the 75$^{th}$ percentile, the band near the middle of the box is the 50th percentile, and the ends of the whiskers represent the 10$^{th}$ percentile and the 90$^{th}$ percentile. Reproduced with permission.[74] Copyright 2013, IEEE.

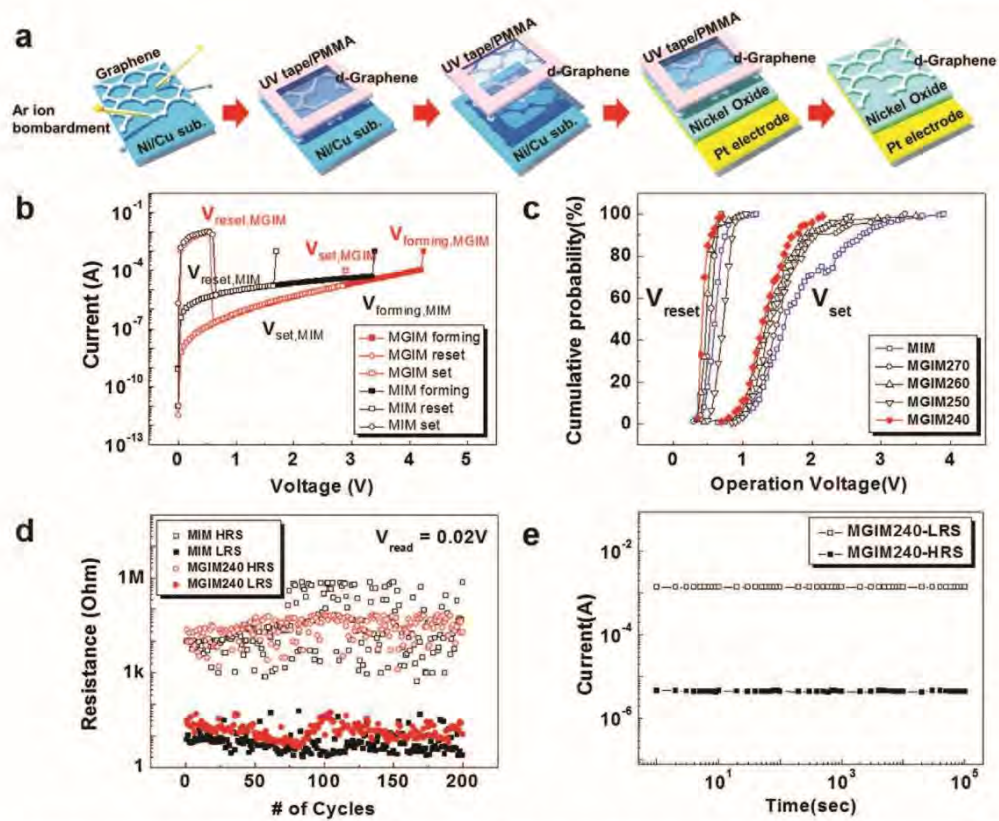

**Figure 7.** a) Fabrication process for a MGIM structure using functionalized SLG prepared before transfer. b) Initial *I–V* characteristics of MGIM and conventional MIM structures. c) Cumulative probability of RS voltages, $V_{set}$ and $V_{reset}$, for MGIMs with SLG irradiated with Ar+ ions at 240 eV (MGIM240), 250 eV (MGIM250), 260 eV (MGIM260), and 270 eV (MGIM270) as well as a MIM. d) Change in resistance states for MGIM240 and MIM, measured at room temperature and atmospheric pressure. e) Retention characteristics of MGIM240 measured at 85 °C at 1 mTorr as well as ambient atmospheric conditions, under reading voltage ≈0.1 V. Reproduced with permission.[170] Copyright 2015, Nature Publishing group.

Ref. [231] reversed the manufacturing order of the RRAM stack (from MLG/TaO$_y$/Ta$_2$O$_{5-x}$/MLG to MLG/Ta$_2$O$_{5-x}$/TaO$_y$/MLG). In this case, the conventional linear bipolar RS became highly nonlinear due to the bottom MLG electrode being oxidized at 400 °C in an Ar/O$_2$ plasma during the reactive sputtering deposition of TaO$_y$. Due to the low currents driven by these devices (0.5 mA at 8 V), they are promising as threshold switching and/or selector elements.

Another potential advantage of SLG electrode engineering is that the Fermi energy can be controlled, which is not possible in standard MIM structures. Using this approach, Ref. [98] engineered the tunneling barrier width and height at the interface of a Pt/Ti/TiO$_2$/SLG/Pt RRAM device, resulting in three orders of magnitude reduction of the switching power (from $10^{-5}$ W to $10^{-2}$ W).

### 3.7. Integration

One advantage when building NVMs using MIM structures is the potential for stackability and integration. One common approach[113,232,233] consists of fabricating a nanostructured material with alternate metallic and insulating films. Then, a vertical aperture (hole) is patterned and the RS media is deposited.[113,232,233] Finally, the rest of the hole is filled with another metal, leading to vertically aligned MIM cells in which the vertical electrode serves as common electrode, and each horizontal metallic film is the specific electrode of each (independent) MIM cell.[113,232] In this structure the thickness of each insulating film should be large enough to avoid cross-talk noise from cell to cell, therefore it cannot be reduced below a safe value (in the case of SiO$_2$ ≈ 6 nm).[113] On the contrary, the thickness of the metal should be low enough to ensure good in-plane conductivity. SLG is thus a promising building block because: i) it is only 0.34 nm thick[113] and its in-plane conductivity is excellent (≈3000 Wm$^{-1}$K$^{-1}$);[234] and ii) the lateral connection between SLG and the RS media provides a lower contact resistance (compared to metals). Ref. [235] used FETs with metallic electrodes that contacted the SLG channel laterally, and observed a mobility of 140 000 cm$^2$ Vs$^{-1}$, much higher than that of similar devices in which the SLG channel is connected vertically (40 000 cm$^2$ Vs$^{-1}$),[236] and very close to the phonon limited model.[235] The reason is that the in-plane bonding is covalent, while metallic electrodes deposited on top of SLG rely on weaker Van der Waals interactions. A similar methodology can be used in RRAMs, employing SLG as planar electrode contacted from the side (see **Figure 8**).[113] Using this principle, Ref. [233] fabricated RRAM devices with $I_{ON}/I_{OFF}$ >80, low reset currents ≈20 µA and low set/reset voltages (2 to 4 V).

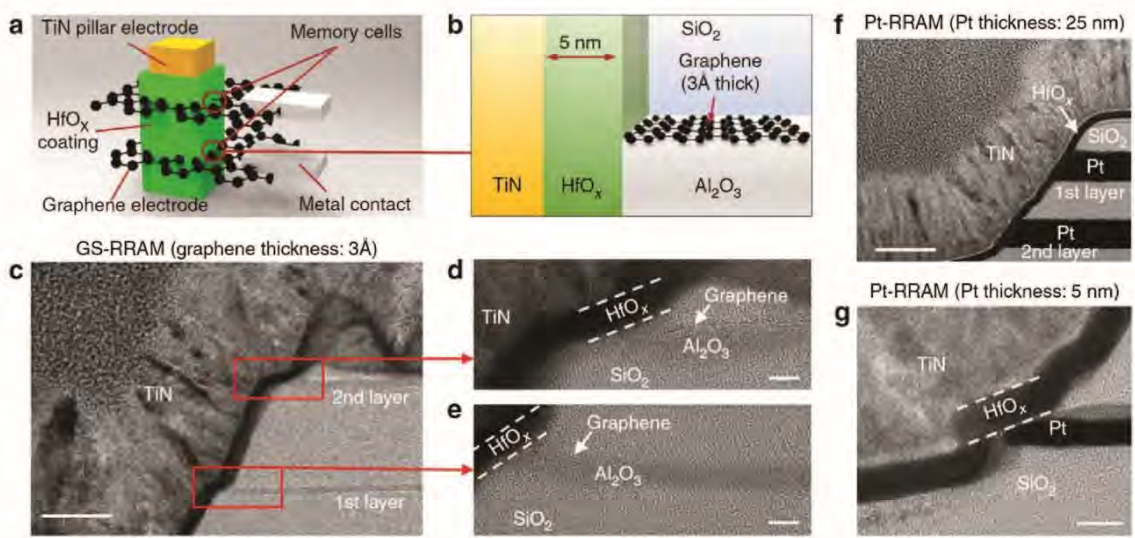

**Figure 8.** a) SLG-based RRAM in a vertical cross-point architecture. The RRAM cells are formed at the intersections of the TiN pillar electrode and the SLG plane electrode. The RS HfO$_x$ layer surrounds the TiN pillar electrode and is also in contact with SLG. b) Schematic cross-section of the SLG-based RRAM. c) High-resolution TEM image of the two-stack graphene set electrode RRAM (GS-RRAM) structure. The RRAM memory elements are highlighted in red. Scale bar = 40 nm. d,e) First and second layer of GS-RRAM with SLG on top of Al$_2$O$_3$. Scale bars = 5 nm. f, g) TEM images of the two-stack Pt-based RRAMs. Scale bars = 40 nm (f) and 5 nm (g). Reproduced with permission.[113] Copyright 2015, Nature Publishing Group.

Ref. [186] used a similar structure consisting of SLG as edge electrode to investigate the scaling limit of RRAM integration. In this case, the RS medium was a superstructure made of Ta$_2$O$_{5-x}$/TaO$_y$ and, as in Ref. [233], SLG was grown by CVD and transferred on SiO$_2$ by an electrochemical approach.[237] The Pt column and SLG serve as pillar and edge electrodes respectively. As a result, SLG edge electrodes allowed a larger density of three dimensional RRAM integration.

## 4. Graphene-Oxide-Based Switching Media for RRAM

Even though the electrodes are a crucial element defining the performance of RRAMs, the switching medium is the dominant one.[34] Apart from TMOs, a wide variety of materials has been proposed as switching media in RRAMs, including organic materials,[238] polymers,[239] perovskites,[240] GRMs[241] and amorphous carbons.[84–89] Mixtures/alloys of some of them, such as polymers with high density of graphene flakes[195] or organic polymers,[239] have also been used.

GO and reduced GO (RGO) have been widely investigated for RS applications.[132,133,241–253] GO films consisting of interconnected flakes are typically produced by LPE and spin coated on the surface of a substrate (which serves as top electrode), with subsequent deposition of top contacts on the GO surface[254] (see **Figures 9** and **10**). This contrasts with the atomically flat CVD-SLGs, and could have implications in terms of device-to-device variability.

Ref. [241] prepared a GO compound by using the Hummers method[255] and the resulting material was transferred onto Pt/Ti/SiO$_2$/Si substrates, followed by top Cu electrode evaporation. The resulting Cu/GO/Pt RRAMs contained a 30-nm-thick GO film (see Figure 10), which showed $I_{ON}/I_{OFF}$ >10, long retention times >10$^4$ s, and low switching threshold <1 V. The ability of GO to change its electrical resistance when subjected to voltage stresses was later confirmed in Al/GO/ITO cells.[249,256,257] Several authors[126,258–262] combined a GO active layer with diverse electrode metals (Pt, Au, Al), which allowed tuning the RS characteristics of the devices.[126,258–262]

Two competing hypotheses have been proposed to interpret the bipolar switching observed in GO films.[242,263] The first[242,263] resembles that of ECM cells using active metallic electrodes, in which metallic ions can diffuse from the electrodes towards the GO layer, leading to the formation/dissolution of a CF. The independence of the LRS resistance on temperature and the proportionality of the currents to the electric field support this mechanism.[242,263] An X-ray photoelectron spectroscopy (XPS) study of an Al/GO/ITO stack detected Al atoms along the GO film when the device was working in LRS, pointing to mass transfer during the cyclic switching.[249] The second[241] is similar to that of homogeneous VCM for inorganic materials, and suggests that absorption and desorption of oxygen functional groups could induce RS in the GO film.[264] In most cases, GO is associated with various oxygen groups, such as carboxyl,[265] hydroxyl, and epoxide, with their oxygen ions usually contributing to form the conduction path.[264,265] Two states—sp$^3$ and sp$^2$—exist in these oxygen groups; the latter has larger conductivity due to the introduction of $\pi$-electrons from the removed oxygen groups.[264] The change of the oxygen bonding state in the GO film usually causes a variation of the leakage currents.[241] This interpretation received partial support from the e-beam-induced current profile at the GO/metal interface and XPS depth profiles of oxygen and metals in HRS and LRS, which displayed distinct oxygen bonding near the interface.[126,266] However, the spatial distribution of the oxygen functional groups can vary in each resistive state. Furthermore, experiments on devices with different sizes indicate that the leakage current is proportional to the cell size.[105] Therefore, both results suggest that oxygen migration plays a dominant role in the switching of GO-based RRAMs.[241]

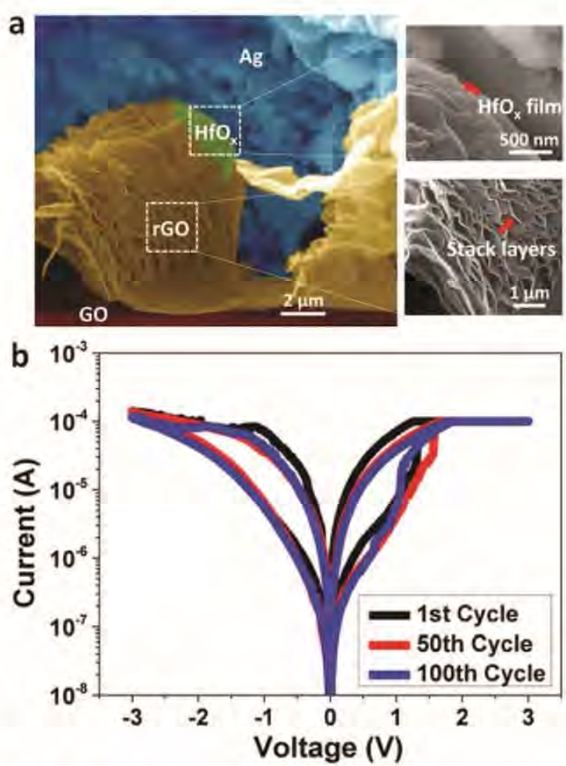

**Figure 9.** a) Cross-sectional SEM images of folded, aggregated and misaligned GRMs used in RRAM technologies. The insets highlight HfO$_x$ films and stacked RGO layers, respectively. b) RS behaviour of laser-scribed RGO (LSG)-based RRAM at the first, 50th and 100th cycle, respectively. Reproduced with permission.[254] Copyright 2014, American Chemical Society.

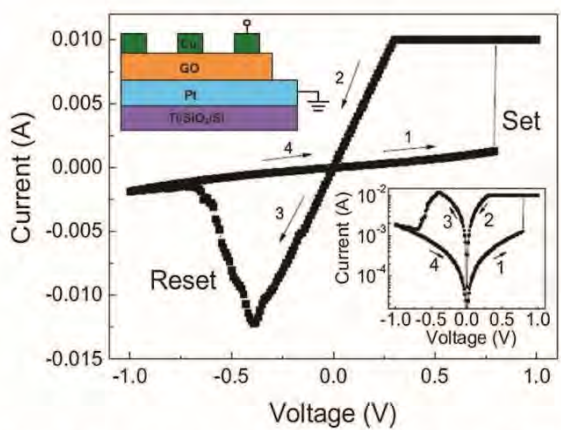

**Figure 10.** I–V characteristics of Cu/GO/Pt RRAM cell showing RS. The arrows indicate the sweep direction. The insets show the I–V characteristics in semilogarithmic scale and the schematic configuration. Adapted with permission.[241] Copyright 2009, American Institute of Physics.

Ref. [267] observed different switching polarities, switching modes or the absence of them depending on the active metallic electrode (Al, Cu, Ni, Ti). The switching directions are characterized by the different area, field and temperature dependences between them. Except for Ni electrodes (which did not show RS),[267] all the others (Al, Cu and Ti) showed bipolar switching under positive set (applied at the top electrode, bottom grounded).[267] Ti showed additional negative bipolar switching under negative set, and Al showed additional unipolar switching.[267] The bipolar RS under negative set might be related to the absorption/release of oxygen based functional groups,[241,268] while the bipolar RS under positive set may be associated with metallic ion diffusion.

The main performances shown by GO-based RRAMs are compared in **Table 3**. The highest $I_{ON}/I_{OFF}$ was achieved in ITO/GO/Ag[242] and p-Si/GO/Ag[247] structures. Ag electrodes seem to provide the lowest switching voltages,[133,153,242,245] but this contrasts with the results of Ref. [251], which show operating voltages ≈6.7 V (the thickness of the GO film in Ref. [251] was ≈15 nm, while in Refs. [133,242] it was not indicated). By comparing rows 2 to 5 in Table 3,[248] it can be concluded that Cu electrodes provide higher $I_{ON}/I_{OFF}$ than Ti, Ag and Au, probably due to the higher diffusivity of Cu atoms in the GO film, which may result in a more effective CF disruption during the reset process. It would be interesting to try ITO/GO/Cu and p-Si/GO/Ag RRAM structures. The ITO/GO/Al RRAMs from Ref. [249] show retention times >10$^7$, but they are still insufficient for RRAM technology (see Table 1).[2] By comparing the ITO/GO/Al RRAMs from Ref. [249] with the ITO/GO/Ag from Ref. [242] it looks like Ag electrodes cannot provide long retention (10$^7$ vs. 10$^3$), which is consistent with the lower operation voltages for Ag electrodes.[133,153,242,245] In any case, the long retention observed in Ref. [249] requires further corroboration (as well as the high operating voltages observed in Ref. [251]). The use of semiconductor electrodes in RRAMs, e.g. Si/GO/Al,[250] Ge/GO/Al[250] and p-Si/GO/AG[247] show (unwanted) high operation voltages of –5.5 V, –8.7 V and 3.5 V (respectively). While Si/GO/Al[250] Ge/GO/Al[250] show low $I_{ON}/I_{OFF}$ <120, p-Si/GO/AG[247] reached 10$^4$. Further confirmation of the results in Ref. [247] is necessary. Despite all papers using spin coating reporting thick >10 nm layers,[241,245,248,250] the endurance for all RRAMs in Table 3 is just ≈100 cycles. This value, which may be related to the large number of defects (missing bonds) in the GO film,[269] is very far from the technology requirements for NVMs (10⁹ cycles, see Table 1).[2] Similarly, despite all papers in Table 3 claiming that GO may be interesting for future nano RRAM devices, the RRAM size was >7500 μm², and we are not aware of any CAFM-based RS study (like those in Ref. [35]) for GO films. The data in Table 3 needs to be corroborated in smaller MIM cells.

**Table 3.** Switching in GO-based devices.

| Device Structure | Device size | $I_{ON}/I_{OFF}$ | Set V [V] | Retention [s] | Endurance [cycles] | Ref. |
|---|---|---|---|---|---|---|
| Pt/GO/Cu | 100 μm in diameter | 500 | ≈0.7 | >10⁴ | >100 | [241] |
| Pt/GO/Cu | 100 μm in diameter | ≈1250 | ≈0.8–1.2 | >10⁴ | >100 | [245] |
| Pt/GO/Ti | 100 μm in diameter | ≈650 | ≈0.8–1.2 | ≈10⁵ | >100 | [245] |
| Pt/GO/Ag | 100 μm in diameter | ≈100 | ≈0.5–1 | ≈10⁵ | ≈100 | [245] |
| Pt/GO/Au | 100 μm in diameter | ≈40 | ≈0.6–0.8 | ≈10⁵ | >100 | [245] |
| Si/GO/Al | 600 × 600 μm² | 110 | −5.5 | 10¹ | ≈100 | [250] |
| Ge/GO/Al | 600 × 600 μm² | 76 | −8.7 | 10¹ | >100 | [250] |
| Al/GO/Al | – | 10³ | 0.7 | – | – | [253] |
| ITO/GO/Al | 180 μm in diameter | 10¹ | −1.6 | 10⁷ | >100 | [249] |
| ITO/GO/Ag | – | 10⁴ | −0.6 ± 0.2 | >10¹ | – | [242] |
| ITO/GO/Ag | 80 μm in diameter | <10 | 0.6 | – | – | [133] |
| Ag/GO/Ag | | 10 | 6.7 | >10⁰ | – | [251] |
| p-Si/GO/Ag | ≈50–150 μm in diameter | 10⁴ | 3.5 | >10⁵ | >100 | [247] |
| Al/GO/Au/GO/ITO | – | 10² | – | – | 10⁴ | [243] |

The endurance can be enhanced by using RGO instead of GO, as can be observed by comparing Table 3 and **Table 4**. Ref. [262] reported unipolar RS in ITO/RGO/ITO cells (5 μm in diameter), with endurance >10⁵ cycles. The replacement of one of the ITO electrodes by Au[270] did not alter the operation voltage (2V) and retention time (10⁵), indicating that in these structures the RGO (not the electrode) plays a dominant role in the charge transport.[270] The use of one Al electrode in conjunction with the RGO/ITO stack does not significantly alter the switching time[260] (compared to ITO/RGO/ITO)[262] even in much larger cells (≈3 mm in diameter). When both electrodes are made of Al[271,272] the devices show much lower $I_{ON}/I_{OFF}$ <100. This observation correlates with a reduction of the operating voltage (≈0.6 V).[271,272] The use of Pt electrodes shows high operation voltages <1.9 V and $I_{ON}/I_{OFF}$[273] similar to those of Au electrodes. This is reasonable because both Au and Pt are noble metals with low reactivity with GO. In agreement with these observations, RGO-based RRAMs using Al electrodes showed the smallest retention times.[260,271–274] A device with Ag electrodes showed the lowest $I_{ON}/I_{OFF}$ (10) and endurance (100).[254]

GO and RGO can be combined with additional layers with the aim of further improving the performance of RRAMs.[194,264,275–278] Prototype RRAM cells combining ZnO–graphene quantum dots (GQDs),[194] metallic (Ni,[278] Au[243]) nanoparticles and nanocrystalline cellulose/GO[277] have been reported. Ref. [194] introduced ZnO–GQDs as active components and demonstrated a solution-processed organic NVM array with one-diode-one-resistor (1D1R) architecture. The switching mechanism of the ZnO–GQDs devices was governed by thermally activated transport before the turn-on process.[194] The 1D1R cell showed typical unipolar switching and low cross-talk noise. An analogous architecture of ZnO nanorods (ZnONRs) with GO displayed a significant reduction of the operating voltages (2.1 V) compared to the cell without ZnONRs (3.9V), indicating enhanced concentration of oxygen vacancies in the GO due to the incorporation of ZnONRs.[246] Ref. [278] used Ni-incorporated GO to fabricate RRAM devices with endurance >100 cycles, and Ref. [243] combined GO with Au nanoparticles, which lead to bipolar RS with retention times ≈$10^4$ s.[243]

**Table 4.** Switching in RGO-based devices.

| Device structure | Device size | $I_{ON}/I_{OFF}$ | Set V [V] | Retention [s] | Endurance [cycles] | Switching time [ns] | Ref. |
|---|---|---|---|---|---|---|---|
| PET/ITO/RGO+PVA+Au NP/Al | – | >10⁴ | 0.44 | >10⁴ | – | – | [352] |
| ITO/RGO/ITO | 50 μm in diameter | – | 2 | 10⁵ @ 85 °C | >10⁵ | 30 (set) 30 (reset) | [262] |
| Au/RGO/ITO | – | 10¹ | 2 | 10⁵ | – | – | [270] |
| Al/GO/ITO | ≈3 mm in diameter | 10¹ | – | >10⁴ | – | 25 (set) 75 (reset) | [260] |
| Al/RGO/Al | – | 10 | – | >10¹⁰ | >100 | – | [272] |
| Al/RGO/Al | 100 μm in diameter | 10¹ | 0.6 | >10⁴ | >250 | – | [271] |
| Pt/RGO-th/Pt | 100 μm in diameter | >10⁴ | −1.9–3.9 | >10⁵ | >350 | 5 (set) 5 (reset) | [273] |
| Al/PFCI/RGO/ITO | 0.4 mm in diameter | 10⁴ | −1.2 | 10⁴ | 10³ | – | [351] |
| Al/RGO-ferrocene/ITO | 0.04 mm in diameter | 10¹ | – | 10² | 10² | – | [274] |
| Ag/HfO$_x$/LSG (laser-scribed RGO) | – | 10 | – | 10⁴ | 100 | – | [254] |

The combination of GO with polymers such as poly (N-vinylcarbazole) derived GO (GO-PVK),[275] triphenylamine-based polyazomethine (TPAPAM),[274] showed typical bistable electrical conductivity and nonvolatile rewritable memory effects, with a turn-on voltage ≈−1 V and $I_{ON}/I_{OFF}$ >$10^3$. Ref. [264] presented a RRAM-based on solution-processed GO/$Pr_{0.7}Ca_{0.3}MnO_3$ forming a cell of Pt/GO/PCMO/Pt. In this structure, two active layers are necessary because GO or PCMO independently sandwiched by metal electrodes cannot reach stable RS. For example, the Pt/PCMO/Pt control sample showed no RS,[264] due to the almost Ohmic contact between each layer, and the I–V characteristics of a single Pt/GO/Pt device displayed an irreversible BD. However, the device with two active layers exhibited intrinsic and reversible bipolar RS, along with the conduction mechanisms associated with oxygen ions movement between the two active layers (see **Figure 11**). Three different phases can be detected from I–V characteristics collected in these devices: i) An initial linear behavior at low voltages. ii) A sudden current increase that switches the device to LRS, probably related to the movement of oxygen ions from GO towards the PCMO surface, which contains large amounts of oxygen vacancies compared to the bulk region. And iii) the resistance of the PCMO layer is decreased by reducing the oxygen vacancy concentration, inducing the reset and transition back to the HRS. Therefore, electrical pulses can cyclically induce a HRS to LRS transition, and vice versa.

Refs. [243,258] reported multiple stable resistive states in GO when incorporating either Au nanoparticles[243] or polyimide.[258] The presence of more than one resistive state allows for a higher information storage density as, instead of bits, multiple digits can be stored. Up to four differentiated levels and retention times of at least $10^4$ s were reported.[243] The performance of RRAMs using GO, RGO-polymer and mixed structures as RS media are summarized in **Table 5**. Outstanding performance in terms of endurance ($10^8$ cycles) is achieved[275,276] using GO-polymer composites sandwiched by ITO–Al electrodes, approaching, but not meeting, the NVM technology requirement ($10^9$). These two cells[275,276] also show high retention times >$10^4$ s and $I_{ON}/I_{OFF}$ ≈$10^3$, being surpassed by the RGO/P3HT:PCBM/Al structures shown in Ref. [279] ($10^4$–$10^5$).

GO can also provide flexibility and transparency to the devices. For example, ITO/GO/Ag RRAMs with $I_{ON}/I_{OFF}$ ≈$10^3$

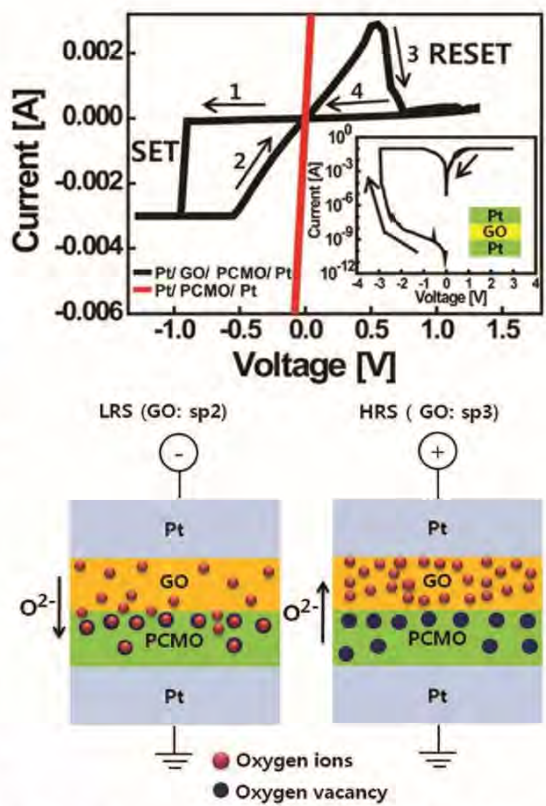

**Figure 11.** Typical I–V hysteresis curves of GO/PCMO and PCMO cells (top). The inset shows the I–V hysteresis for a Pt/GO/Pt device. Proposed switching mechanism in LRS (left) and HRS (right) for GO/PCMO devices (bottom). Reproduced with permission.[264] Copyright 2011, American Institute of Physics.

and stable retention characteristics for >$10^3$ s within 1000 cycles for $r_b$ > 4 mm have been reported.[249] Ref. [262] fabricated RGO-based RRAMs by dip-coating, and obtained ≈80% transparency from 425 to 900 nm. These devices exhibited unipolar RS characteristics with $I_{ON}/I_{OFF}$ >$10^5$, endurance ≈$10^5$ cycles for each state, retention times >$10^5$ s and multilevel capability. The performance of flexible RRAMs using GO and RGO as RS media is summarized in **Table 6**. Outstanding performance ($I_{ON}/I_{OFF}$ >$10^5$, retention >$10^5$ s and endurance >$10^5$ cycles) was achieved in PEN/Ti/Pt/GO/Ti/Pt RRAMs,[280] with MIM cells ≈100 nm × 100 nm, making these values more reliable. This is an important step towards enabling future transparent device applications based on GO and its derivatives.

## 5. Amorphous Carbon as Switching Media for RRAMs

Non-crystalline carbons are referred to as amorphous carbons. When the sp³ fraction is higher than 50%, these are called tetrahedral-amorphous carbons, ta-C.[281–284]

**Table 5.** Switching in GO and RGO polymer and mixed structures.

| Device structure | Device size | $I_{ON}/I_{OFF}$ | Set V [V] | Retention [s] | Endurance [cycles] | Ref. |
|---|---|---|---|---|---|---|
| ITO/TPAPAM-GO/Al | $0.4 \times 0.4$ [mm²] | $10^3$ | $-1$ | $>10^4$ | $10^6$ | [276] |
| ITO/GO-PVK/Al | – | $>10^3$ | $-2$ | $>10^4$ | $10^8$ | [275] |
| PET/ITO/PVK:Gr(GO)/Al | – | – | $\approx 0.2-0.4$ | – | Not reversible | [353] |
| ITO/PVA+GO/Al | 200 μm in diameter | $10^4$ | $-0.75$ | $10^4$ | $>10^4$ | [354] |
| ITO/PVDF-GO/Al | 0.0004 [cm²] | $10^4$ | $\approx 3.6-4.1$ | – | – | [355] |
| Al/CuO/GO/CuO/Al | – | – | 3.0 | – | – | [244] |
| ITO/PMMA/GO/PMMA/Al | 30 μm in diameter | $>10^1$ | $-1.7$ | $10^6$ | $>10^5$ | [356] |
| Gr/GO/ZnONR/Nb | – | $10^7$ | – | – | $>50$ | [357] |
| ITO/GOAu/Al | 200 μm in diameter | $10^6$ | $-1$ | $10^4$ | $>300$ | [358] |
| ITO/GO-FeO/Pt | – | $5 \times 10^3$ | 0.9 | $10^5$ | $>1100$ | [359] |
| ITO/TPA-rGO/Al | 0.04 [mm²] | $10^3$ | 1.6 | $>10^5$ | $>10^6$ | [274] |
| Al/GO-PFCz-ITO | $\approx 0.16-0.0225$ [mm²] | $10^3$ | 0.38 | $>10^4$ | $10^8$ | [351] |
| Au/PrGODMF/ITO | – | 100 | – | $>1000$ | 100 | [128] |
| Ag/Pt/GO:PI/Pt/ITO | – | 1000 | 5 | 1400 | 130 | [258] |
| Pt/GO/PCMO/Pt | – | $10^2$ | $-0.75$ | $10^4$ | 150 | [264] |
| Al/PS-b-P4VP-GO/ITO | $0.4 \times 0.4$ mm² | $10^4$ | $\approx 6$ | $>10^4$ | $10^8$ | [360] |
| Al/P3HT:PCBM/rGO/glass | – | $10^6$ | – | – | – | [140] |
| Pt/ZrSiOx/C:SiOx/TiN | – | 100 | – | – | – | [268] |
| rGO/P3HT:PCBM/Al | – | $10^4$-$10^1$ | – | – | – | [279] |
| PET/rGO/MoS₂-PVP/Al | – | $\approx 10^2$ | – | – | – | [320] |

Amorphous carbons can change resistance by applying unipolar electrical pulses or voltage sweeps. RS in amorphous carbons has led to their addition to the selection of emerging memory technologies in the 2014 ITRS.[7] The switching mechanism, however, is still under debate. Several mechanisms have been put forward, such as sp² clustering,[86,285] sp² filament formation,[286–289] metal filament formation[85,290] and electron trapping/detrapping.[291]

In 1972 Ref. [292] first reported RS in 10 nm thick evaporated a-C films sandwiched between Al electrodes, reporting 100 000 switching cycles. Ref. [292] found that a forming step is needed to create a CF and activate RS. Switching only occurred by applying a positive voltage to the bottom Al electrode, while opposite polarity was needed for the RESET.[292] RS was attributed to metal filament formation, since Al is a diffusive metal and amorphous carbon produced by evaporation has usually very low sp³ content and switching in sp³ rich amorphous carbon is not reversible.[285]

Non-volatile RS in doped amorphous carbon films was demonstrated by several groups.[84–89,293–300] This includes RS in nitrogen,[88,89,293,296] hydrogen,[84–87] oxygen,[294,301] silicon,[302] Co[297] and Cu[298–300] incorporated amorphous carbon films. **Table 7** summarizes the literature RS data in doped amorphous carbons.

**Table 6.** Switching in GO and RGO on flexible substrate.

| Device structure | Device area | $I_{ON}/I_{OFF}$ | Set V [V] | Retention [s] | Endurance [cycles] | Switching time [ns] | Ref. |
|---|---|---|---|---|---|---|---|
| PET/ITO/GO/Al | 300 μm in diameter | 280 | 2.2 | $10^4$ | $>100$ | – | [261] |
| PES/Al/GO/Al | 50 μm × 50 μm | $>100$ | $-2.5$ | $5 \times 10^4$ | $\approx 100$ | – | [176] |
| PES/ITO/GO/ITO | 10 | $>100$ | $\approx 0.7-1$ | $10^7$ | – | – | [361] |
| PET/ITO/GO/ZnO nanorods/Al | 200 μm in diameter | $\approx 100$ | $\approx 1-4.8$ | $10^4$ | $>200$ | – | [246] |
| PET/ITO/GO/Al | 200 μm in diameter | $\approx 100$ | 3.9 | – | – | – | [246] |
| PET/ITO/GO/Ag | 0.026 [mm²] | 5 | $-0.14$ | $\approx 10^2$ | 13 | – | [139] |
| PET/ITO/RGO+PVA+Au NP/Al | – | $>10^2$ | $-0.44$ | $>10^4$ | – | – | [352] |
| Pt/RGO-th/Pt | 100 μm in diameter | $>10^4$ | $\approx 1.9-3.9$ | $>10^5$ | $>350$ | 5 | [273] |
| Al/RGO/Al | 100 μm in diameter | $10^2$ | $-0.6$ | $>10^4$ | $>250$ | – | [271] |
| PLN/Ti/Pt/GO/Ti/Pt | 100 nm × 100 nm | – | 3.5 | $>10^3$ | $>10^6$ | $<10$ (set) | [280] |
| Al/GO/ITO | 180 μm in diameter | $10^3$ | $-1.6$ | $10^7$ | $>100$ | – | [249] |

- 122 -

Table 7. Doped amorphous carbon-based RRAMs.

| Device structure | Device area | $I_{ON}/I_{OFF}$ | Set V [V] | Retention [s] | Endurance [cycles] | Switching time [ns] | Ref. |
|---|---|---|---|---|---|---|---|
| W/a-CO$_y$/Pt,Ti,W | 100 nm in diameter | >10$^2$ | – | 10$^4$ @ 85 °C | >10$^4$ | 40 (set) | [294,301] |
| | | | | | | 4 (reset) | |
| Pt/a-CN$_{0.15}$/Cu | 100 μm in diameter | 1 | 0.6 | >10$^6$ | 10$^1$ | – | [295] |
| FTO/a-C:Co/Al | | 25 | – | >10$^5$ | – | – | [297] |
| Pt/a-C:Cu/Cu | 0.1 × 0.1 [μm$^2$] | 10$^2$ | 0.7 | 10$^4$ @ 85 °C | >10$^1$ | – | [298,299] |
| Pt/a-C:Cu/Pt | 30 × 30 [μm$^2$] | – | 0.7 | – | – | – | [300] |
| Pt/a-C:N/C-AFM tip | 12 nm in diameter | – | 3 | – | – | – | [296] |
| Pt/a-C:Si/C-AFM tip | – | – | 3.5 | – | – | – | [296] |

Refs. [88,89] reported a reversible NVM effect in nitrogen-doped tetrahedral amorphous carbon, ta-C:N, with write times down to 100 μs.[90] They attributed the switching to the promotion of electrons from acceptor states in the gap to higher donor states. However, the LRS retention was poor, only one year,[90] too short for commercial applications. Ref. [295] prepared nanoporous nitrogen-doped amorphous carbon and studied a Pt/a-C:N/Cu device structure. Set and reset occurred at opposite voltage polarities.[295] Decreasing the amount of nitrogen led to a reduction of switching voltages to +0.6 V for set and –0.5 V for reset.[295] Over 1000 switching cycles and a retention >80 days at room temperature were reported, still not good enough to meet industry requirements.[295] The switching mechanism was attributed to the formation and rupture of Cu filaments.[295] Ref. [296] reported the effect of nitrogen implantation on RS of amorphous carbon to analyze the role of sp$^3$ filamentation and clustering. Nitrogen implantation made the films more conductive with an increase in sp$^3$ bonding and clustering, facilitating the SET process.[296]

Several groups reported reversible, non-volatile switching in hydrogenated amorphous carbon, a-C:H,[85–88] with the results summarized in **Table 8**. RESET within 30 ns and SET in ≈30 ns were reported,[86] long data retention >10$^5$ s,[85] 10$^7$ switching cycles[87] and $I_{ON}/I_{OFF}$ ≈10$^3$.[86] The RS mechanism was attributed to different processes: Ref. [85] assigned RS in Pt/a-C:H/metal structures, with the metal top electrode being Cu, Ag or Au, to the formation and rupture of metal filaments, due to diffusion of the top electrode metal into the a-C:H film.[85] Ref. [86] studied RS in a-C:H with TiN as bottom and Cu, Pt or W as top electrodes, and assigned RS to thermally induced conductive sp$^2$ clusters filament formation.[86] Ref. [87] attributed the switching to a sp$^2$ carbon CF formation in a TiN/a-C:H/Pt structure. RESET was achieved by applying the opposite voltage polarity to the bottom TiN electrode and attributed to hydrogen atoms pulled from the Pt top electrode and absorbed by double bonds in sp$^2$ carbon.[87]

A limiting factor in a-C:H RS is the need of a forming step, where the material needs to be biased at the breakdown electric field.[303] The breakdown results from a capacitive discharge current, which can be 10–20 mA[86,303] and occurs within a few ns.[86,303] Therefore, an on-chip resistor or transistor is needed to limit the current during forming.[86,303] Due to the high current density during the forming step, metals from electrodes might diffuse into the carbon, if the forming is done in a dc-sweep, instead of an energy-limiting short pulse.[85,287,295,303]

The influence of other dopants, such as Co and Cu, was reported by various groups,[297–300] see Table 8. Ref. [297] studied RS in Co-doped amorphous carbon. They observed non-volatile, bipolar and reversible RS with $I_{ON}/I_{OFF}$ ≈25, but good retention >10$^5$ s at room temperature.[297] RS was attributed to a filament formed by Co ions created by an electrochemical reaction, migrating toward the top Al electrode through defects in the a-C film, forming a conductive path between top and bottom electrode.[297] Other groups[298–300] investigated RS in Cu doped carbon. They obtained $I_{ON}/I_{OFF}$ ≈ 10$^2$ and retention of 10$^4$ s at 85 °C and >10$^3$ switching cycles.[298] Ref. [300] used a slightly different device configuration with both top and bottom electrodes made of Pt. A forming step was needed. Subsequent set and reset processes could be achieved at ≈+0.7 V and –0.5 V.[300] RS was attributed to the formation and rupture of Cu filaments.[298–300]

Ref. [294] prepared oxygenated amorphous carbon, a-CO$_y$, by physical vapour deposition. Ref. [294] reported switching times ≈40 ns for SET and ≈4 ns for RESET, with opposite voltage polarity needed.[294] Ref. [294] measured cycling endurance >10$^4$ in devices with W as bottom and Pt, Ti or W as top electrodes,[294] with $I_{ON}/I_{OFF}$ ≈5 × 10$^2$ during retention measurements up to 10$^4$ s at 85 °C. The RS mechanism was attributed to an electrochemical redox reaction leading to the formation of a conductive carbon filament.[294] The choice of metal electrode material was crucial for the reset process, with strong dependence on the electron affinity of the metal electrode.[294] To make the reduction reversible, two electrode materials were needed to store and release oxygen. One with similar electron affinity to carbon, such as W, and the other with higher electron affinity, such as Pt.[296]

Table 8. Hydrogenated amorphous carbon-based RRAMs.

| Device structure | Device size | $I_{ON}/I_{OFF}$ | Set V [V] | Retention [s] | Endurance [cycles] | Critical field [V cm$^{-1}$] | Switching time [ns] | Ref. |
|---|---|---|---|---|---|---|---|---|
| C/a-C:H(B)/Au | | 10$^3$ | – | – | | 5 × 10$^5$ | – | [84] |
| Pt/a-C:H/Cu (Ag or Au)[a] | 100 μm in diameter | >100 | 1.1 | >10$^5$ | 110 | – | – | [85] |
| TiN/a-C:H/Cu (Pt, W)[b] | 49 ± 11 nm in diameter | >10$^3$ | 4.1 | 57 600 | 15 | – | 30 (set) <30 (reset) | [86] |
| TiN/a-C: H/Pt | 0.36 to 16 [μm$^2$] | 100 | 1.5 | 10 000 @ 85 °C | 10$^7$ | – | – | [87] |

[a] The Cu electrode was also replaced by Ag and Au, and the resulting devices also show resistive switching. $I_{ON}/I_{OFF}$ and switching threshold voltages (V) vary as follows: $I_{ON}/I_{OFF}$ (Cu) > $I_{ON}/I_{OFF}$ (Ag) > $I_{ON}/I_{OFF}$ (Au) and $V_{Cu}$ > $V_{Ag}$ > $V_{Au}$ [b] This report used devices with Cu, Pt or W electrodes. It is unclear which electrode corresponds to the performances indicated.

Table 9. ta-C- and a-C-based RRAMs.

| Device structure | Device size | $I_{ON}/I_{OFF}$ | Set V [V] | Retention [s] | Endurance [cycles] | Critical field [V cm$^{-1}$] | Switching time [ns] | Ref. |
|---|---|---|---|---|---|---|---|---|
| n-Si/s-C/C-AFM tip | – | 10@80 K | 5 | 3000 | 70 | – | – | [307] |
| TiN/a-C (sp$^2$-rich)/C-AFM tip (PtSi) | ≈20–30 nm in diameter | – | ≈1–2 | – | Not reversible | – | – | [285] |
| Pt/ta-C/C-AFM tip | 8.5 nm in diameter | – | 12V pulse amplitude | – | – | – | 5 (set) | [296] |
| W/ta-C/W | 150 nm in diameter | >10$^1$ | ≈1–3 | – | $2.3 \times 10^1$/@75 °C | – | 10 (set) | [286] |
| | | | | | | | 1 (reset) | |
| Ag/a-C/CNT | 0.001 [µm$^2$] | 40–200 | ≈5.4 –7.5 | >10$^6$ | 31 | – | – | [290] |
| Al/a-C/Cu | | 3 | <3 | 10$^5$ | | – | – | [291] |
| Al/ta-C/W | 2500 [µm$^2$] | 10 | <1 | >10$^2$ | 120 | – | – | [287] |
| Pt/a-C/Cu/Ag | 50 × 50 [µm$^2$] | | 0.18@0 K | 10$^4$@85 °C | – | – | – | [306] |
| Pt/a-C/Cu | 500 µm in diameter | 100 | 0.1 | 10$^4$ | – | – | – | [304] |
| Pt/a-C/Cu | 100 µm in diameter | >70 | 1 | – | 110 | – | – | [305] |
| Pt/ta-C/W | 50–500 nm in diameter | >300 | 0.8 | – | – | $5 \times 10^7$ | 50 (set) | [288,289,301] |
| | | | | | | | 4 (reset) | |

RS in amorphous carbons with different sp$^2$/sp$^3$ ratio was reported by several groups.[285,304–312] RS in sp$^3$ rich a-C was studied in a Si/TiN/a-C devices, using a CAFM as top contact.[287] The key parameters of RS devices based on a-C/ta-C are reported in **Table 9**. RS was assigned to an electrothermally (Joule heating) induced increase in the sp$^2$ cluster size and was non-reversible.[285] RS in a-C was shown to be polarity independent.[304–307] Ref. [306] studied the influence of the top metal electrode material on RS, and assigned this to metal filamentation in devices with Cu top electrodes. Pt, W and Ni top electrodes did not show switching.[306] This was attributed to the less diffusive nature of those metals.[306] Data retention> 10$^5$ s,[304] low switching voltage of 0.18 V[306] within pulses of 1 µs,[306] $I_{ON}/I_{OFF}$ ≈70,[305] endurance ≈110[305] and device structures down to 50 × 50 µm$^2$[306] were demonstrated.

$I_{ON}/I_{OFF}$ and endurance are the main challenges faced by RRAMs based on a-Cs. The issue of a low $I_{ON}/I_{OFF}$ can be overcome by using ta-C. Ref. [308] demonstrated high $I_{ON}/I_{OFF}$ in Pt/ta-C/(SLG)/Au devices. Devices with an interfacial SLG reached $I_{ON}/I_{OFF}$ ≈4 × 10$^5$ at 0.2 V, while maintaining low switching power density of 14 µW µm$^{-2}$.[308] This was attributed to the reduction of leakage currents due to the low SLG density of states near the Dirac point.[308] Refs. [288,289] explained the switching in terms of nanoscale sp$^2$ filament formation and rupture through field-induced dielectric breakdown and thermal fuse effect, i.e. an electrothermally driven set process and a thermally driven reset process. Low switching voltages of 0.4V for RESET within 10 ns and 1.2V for SET within 50 ns,[288] 10$^{13}$ read cycles at 75 °C,[286] >10$^6$ s retention[290] with device sizes of 50 nm diameter[289] and 10$^3$ switching cycles[288] were also demonstrated. The presence of multiple resistive states was reported by Ref. [308]. Multilevel storage is of particular interest as it allows to store more than one bit per cell, while the memristive behavior can be exploited to provide a range of signal processing/computing-type operations, such as implementing logic, providing synaptic and neuron-like mimics, i.e. circuits that simulate brain-like neurological functions, and performing analogue signal processing functions, paving the way for non-von-Neumann architectures, in which processing and non-volatile storage are carried out simultaneously.[313,314]

Endurance is one of the major challenges for a-C based switching devices. A comparative study by Ref. [301] of RS in ta-C and a-CO$_x$ with Pt bottom electrodes and W top electrodes suggested that, by incorporating oxygen, the endurance could be enhanced to 40 000, but at the expense of bipolar operation.[301] In ta-C devices, SET and RESET were achieved with pulses of 50 and 4 ns and switch energies of 15 and 3pJ, while a-CO$_x$ could be set and reset with 40 and 4 ns pulses with switch energy of 2 and 1pJ, respectively.[301] Both, ta-C and a-CO$_x$ showed good data retention of 10$^4$ s at 85 °C.[294,301]

Several groups[301,302,311,312] theoretically studied the switching mechanism in amorphous carbons, and assigned RS to heat driven sp$^2$ clustering and filament formation.

Ref. [303] pointed out that one of the biggest advantages of carbon-based memory devices might be the high temperature retention ≈250 °C, making them attractive for automotive and harsh conditions.[303] Another advantage of carbon-based memories is that devices do not rely on rare mineral extraction, with easier disposal/recycling, and low total energy production compared to other electronics materials.[280]

## 6. Layered Materials

Non-carbon-based LMs have also been introduced into the structure of RRAMs, mainly TMDs (like MoS$_2$[315,316] and $MoSe_2$[317]) and $h$-BN.[163] Ref. [318] reported a RRAM prototype using BP flakes.

TMDs are naturally semiconducting materials,[319] thus they are not ideally suited for RRAMs. For this reason, they need to be functionalized in order to form an insulating layer.[319] Ref. [320] suggested to combine TMDs with an insulator (such as polymers) whereby the TMD would act as dopant of the insulating layer. In Ref. [320], a stack of RGO/$MoS_2$-PVP/Al in which the PVP (polyvinylpyrrolidone), typically used to assist the exfoliation of $MoS_2$, became the dielectric RS-driving layer. The devices were fabricated using spin coating of the $MoS_2$-PVP solution on the RGO film, resulting on a thickness of 70 nm, and large $0.2 \times 0.3$ mm² electrodes were thermally evaporated. The RRAM devices show $I_{ON}/I_{OFF} \approx 10^2$. Ref. [320] claimed the switching was due to charge trap and detrap of the $MoS_2$ embedded in the PVP. However, others[321,322] reported RS also in structures with pure PVP as active layers, and Ref. [273] detected RS in RGO. We note that Ref. [320] did not provide temperature nor area analyses, which makes it difficult to discern if the RS in these devices is a local or distributed phenomenon. Therefore, the ability of $MoS_2$ to drive the RS is questionable.

Similar studies were developed by Ref. [323] in a PET/Al/PMMA/$MoS_2$/PMMA/Al stack and by Ref. [324] for Au/$MoS_2$-MoO$_x$/Ag. These have $I_{ON}/I_{OFF} \approx 10^4$ and $2.5 \times 10^3$, respectively. However the need for a TMD to be combined with PMMA is doubtful. Ref. [325] demonstrated that $MoS_2$-free Ag/PMMA/ITO devices can also achieve reproducible RS ($I_{ON}/I_{OFF} \approx 10^2$), which is driven by the penetration of metallic ions into the polymer, leading to a reversible CF through it. The combination of $MoS_2$ and GO resulted in a similar $I_{ON}/I_{OFF} \approx 10^2$ for a RRAM device.[326] Ref. [327] produced printable RRAM memories with tuneable performances using Ag/$MoS_2$-MoO$_x$/Ag stacks, with $I_{ON}/I_{OFF} > 10^6$, retention times >8000s, and non RS degradation after bending >10⁴ times. As in Refs. [324,326], the RS in Ref. [320] does not seem to be attributable to the $MoS_2$ sheets, which served to homogenize the interface between the MoO$_x$ and Ag bottom electrode. A different approach was reported by Ref. [328], who used $MoS_2$ flakes, also giving RS. In this case, the resistance changes were attributed to tunnelling across junction barriers. Very similar devices, but using MoSe$_2$ nano-islands, were studied by Ref. [317], which showed $I_{ON}/I_{OFF} \approx 12$ and low currents (>1 μA) in the LRS.

RS driven by $MoS_2$ was reported by Refs. [315,316]. Ref. [317] used three-terminal horizontal devices similar to FETs (see **Figure 12**a and b), with a grain boundary (GB) in the $MoS_2$ extending in the channel. Ref. [320] considered different GB configurations, including parallel and perpendicular to the channel, as well as intersecting. In all cases, $I_{ON}/I_{OFF} > 10^2$

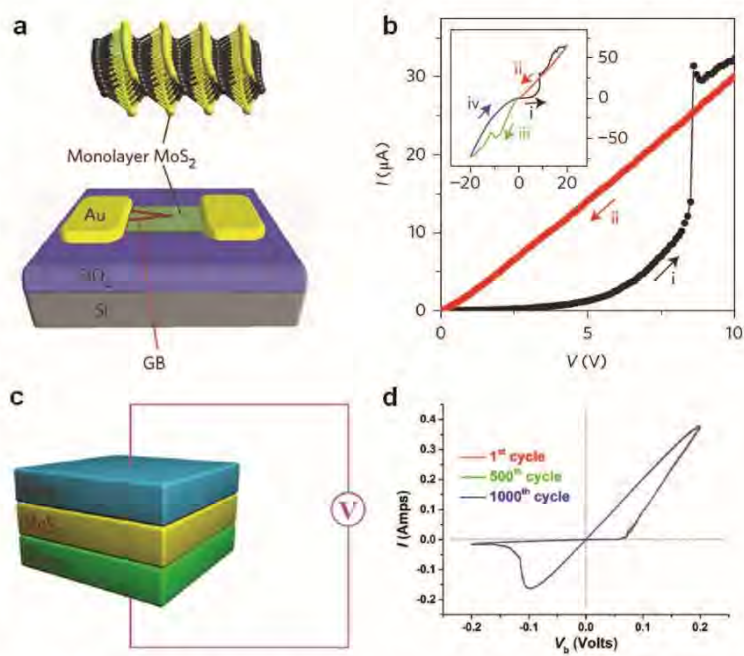

**Figure 12.** a) Schematic horizontal $MoS_2$ RRAM cell with two GBs connected to one of the electrodes and intersecting at a vertex within the channel. b) Partial $I$–$V$ characteristics of an electroformed intersecting-GB memristor (channel length, $L = 7$ μm) obtained immediately after electroforming. The set process occurs at $V_{set} = 8.3$ V with an abrupt twofold increase in current. Inset: Full $I$–$V$ characteristics of one switching cycle. Measurements were performed at a sweep rate of 1 V s⁻¹ and $V_g = 40$ V under vacuum (pressure < 2 × 10⁻⁵ torr). The voltage was swept in the order 0 V → 20 V → 0 V → −20 V → 0 V, as shown by the coloured arrows with the four sweeps labelled i, ii, iii and iv. Reproduced with permission.[315] Copyright 2015, Macmillan Publishers Limited. c) Schematic structure of vertical Ag/$MoS_2$/Ag RRAM cell. d) Typical $I$–$V$ characteristic of Ag/$MoS_2$/Ag switch at the 1st (red), 500th (green), and 1000th (blue) cycle at room temperature. Reproduced with permission.[316] Copyright 2015, American Chemical Society.

was achieved. RS in these devices was assigned to the motion of S vacancies in the $MoS_2$, which tend to accumulate at the GBs.[320] However, the reproducibility of this phenomenon was not firmly established, as Ref. [315] reported only 15 cycles. On the other hand, Ref. [316] compared two vertical Ag/$MoS_2$/Ag devices using a 550-nm $MoS_2$ film formed by $MoS_2$ flakes dispersed in propanol and spin-coated on a Ag foil, followed by a thermal treatment at 130 °C for 12h and 0.1 mm² top electrode deposition using Ag paste (Figure 12c and d). The differences between the two devices was the $MoS_2$ phase, in one case 1T flakes, and in the other 2H bulk. Despite the methodologies used not being ideal for scaling and integration (deposition of electrodes by Ag paint using a shadow mask is to be avoided because it can lead to contaminants at the interface, inhomogeneous shapes and cracks) the devices using 1T-$MoS_2$ showed good RS behaviour with $I_{ON}/I_{OFF} > 10^1$ during 100 cycles. Ref. [316] assigned RS to the migration of Mo and S ions under electrical field. Ref. [316] also included a modification of this device using Ag/$MoS_2$/Ag/$MoS_2$/Ag vertical structures, showing the possibility of reducing the current at low voltages (<0.2 V) by negative differential resistance, which may be useful to avoid sneak path

Table 10. TMD-based RRAMs.

| Device structure | Fabrication method | $I_{ON}/I_{OFF}$ | Retention [s] | Endurance [cycles] | Power consumption [μW] | Transparent | Flexible | Ref. |
|---|---|---|---|---|---|---|---|---|
| PMMA-MoS₂/Ml G/SiO₂/Si | CVD (transfer) | $2.5 \times 10^3$ | – | – | – | NO | NO | [317] |
| Ag/MoSe₂/FTO | Hydrothermal | 12 | >50 | – | – | NO | NO | [315] |
| Au/MoS₂/SiO₂/Si | CVD | $10^3$ | – | – | – | NO | NO | [326] |
| Ag/MoS₂/Ag | LPE (spin coating) | $10^3$ | – | $10^{3a)}$ | – | NO | NO | [316] |
| Ag/MoS₂-MoO₃/Ag | Modified Langmuir–Blodgett | $>10^6$ | – | >8000 | 10 nW | NO | YES | [327] |
| Al/MoS2-GO/ITO | LPE (spin coating) | $10^2$ | – | – | – | NO | NO | [326] |
| RGO/ZIF-8 coated MoS₂/RGO | LPE (spin coating) | $7 \times 10^4$ | $1.5 \times 10^3$ | – | – | NO | YES | [362] |
| RGO/MoS₂-PI23/RGO | LPE (spin coating) | $5.5 \times 10^2$ | $4 \times 10^3$ | >50 | – | NO | NO | [363] |
| PET/RGO/MoS₂-PVP/Al | Polymer-assisted exfoliation | $\approx 10^2$ | – | – | – | NO | YES | [320] |
| RGO/MoS₂/ITO/Si | Hydrothermal | $10^4$ | $5.5 \times 10^1$ | – | – | NO | NO | [328] |

a)This value is not well supported in Ref. [316]. The authors only show the 1ˢᵗ, 500ᵗʰ and 1000ᵗʰ I–V curves, and no R vs Cycle or Weibull plot is shown. The top electrodes of the devices in Ref. [316] are made by drying Ag paint on spin-coated MoS₂ using a shadow mask. More work is needed to confirm these performances.

currents in crossbar arrays. Nevertheless, none of the MoS₂-based works to date presents a conclusive memristive analysis. For example, Ref. [316] claimed 1000 cycles, but no variability analyses (like those, for example, in Refs. [315–317,324]) were presented (just 3 I–V curves are displayed). More information on the different TMDs-based RRAMs is in Table 10, including dependence on critical parameters, such as device area, working temperature, top electrode material and current limitations (among others). The RS parameters are still far from those reported for state-of-the-art TMO-based RRAM memories (see Table 1).[2,31-33]

The use of h-BN in RRAMs is even more incipient. In principle, as h-BN is an insulator,[163] if a reversible CF/BD can be induced through it, the RS behaviour should be more accentuated (larger $I_{ON}/I_{OFF}$) due to the larger resistivity in HRS (the constriction would be more insulating than in semiconducting materials). Nevertheless, this is in principle not an easy task, as the BD may become irreversible depending on the atomic structure of the h-BN stack. One should clearly distinguish between research articles using layered h-BN[163] (see Figure 13c and d), and those in which amorphous BN was used (Figure 13a and b).[329] Ref. [329] claimed the fabrication of RRAMs using multilayer h-BN stacks, but the layered nature of the film is not supported by the cross-sectional TEM images, and the layer looks more like an amorphous BN film (see Figure 13a and c). This is very important because amorphous BN may not hold the properties of the h-BN stack, such as transparency,[330] flexibility,[331] high thermal conductivity[332] and high chemical stability.[333] Ref. [316] fabricated a family of RRAMs using h-BN as RS medium. By tuning the h-BN stack thickness and the h-BN domain size, Ref. [316] achieved forming-free operation, low switching voltages down to 0.5 V, high $I_{ON}/I_{OFF}$ up to $10^6$, retention times >10 hours and low device-to-device variability (i.e. deviations of $V_{SET}/V_{RESET}$ <10%). In Ref. [163] the RS was attributed to the migration of B atoms towards the electrodes, as well as metallic ions penetration into the h-BN stack to form and disrupt one/few CFs. These atomic diffusions are more abundant at GBs, which are defect-rich locations (missing bonds, missing atoms, pentagonal/heptagonal lattices)[334,335] that can favour atomic rearrangements at lower potentials (compared to the grains), leading to a softer BD that may be easier to recover. The formation of B-vacancies at the GBs of polycrystalline h-BN stacks (see Figure 13c) presents an interesting parallelism to O vacancies at the GBs of polycrystalline TMOs. The key role of GBs in the RS is supported by the fact that the BD process in single crystalline h-BN flakes is an irreversible phenomenon that leads to the removal of the material,[336] with the formation of holes during a characteristic layer-by-layer BD process. Therefore, it is unlikely that a perfect single crystalline h-BN would offer RS capabilities. Ref. [107] investigated a RRAM comprising a monolayer CVD-grown h-BN flake inserted between the top electrode and the dielectric of an Al/WO₃/Al cell, but the performance was worse than the h-BN- free counterpart ($I_{ON}/I_{OFF}$ < 10). This is likely because it is difficult to create CFs in h-BN/WO₃ superstructures, i.e. the CF is only created at large electrical fields that produce the irreversible BD in the h-BN/WO₃ stack. Ref. [337] reported indications of RS in layered Ti/h-BN/Cu stacks (Figure 13d). The devices exploited the Cu substrate used to grow the h-BN as bottom electrode, avoiding the need for transfer.[316] When applying constant voltage stresses (CVS) at 2.5 V to the devices, the current vs. time (I-t) curves show sudden changes of the electronic resistance (up to $10^3$) similar to unipolar RS characteristics.[337] A detailed comparison of the RS capabilities of h-BN-based devices in literature is presented in Table 11. Ref. [338] also observed unipolar RS transitions in planar nanogap-based h-BN obtained by MC. However, thus far, the use of planar structures in RRAM technology is limited due to the difficulty in controlling the rupture kinetics of the nanogaps, which may result in poor RS endurance and device-to-device variability. We note that statistical information of RS in planar devices made of any GRM has not yet been reported. Moreover, MC is not a scalable technique. Ref. [337] reported layer-by-layer BD at the grains by means of CAFM, while at the same time measuring reproducible conductivity changes at the device level, which may be related to the presence of GBs. As the use of LM-dielectrics provides a flatter interface to graphene and TMDs than high-*k* dielectrics,[179,180] RS applications of *h*-BN should be deeper investigated.

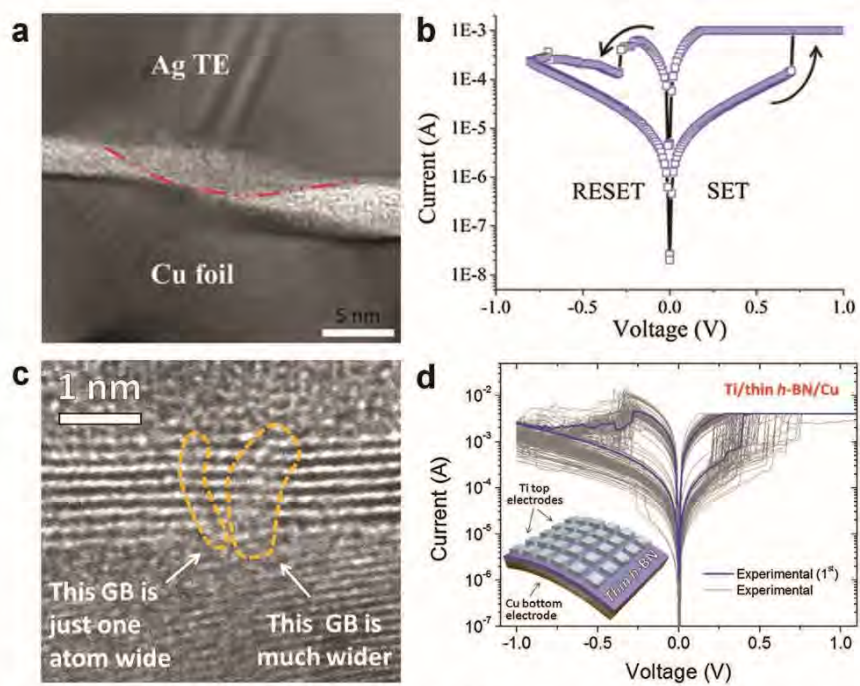

**Figure 13.** a) TEM image of a CF in amorphous BN. The thinnest region of the CF is at the amorphous BN/Cu foil interface. b) Switching characteristics of Au/amorphous BN/Cu foil/PET devices. Reproduced with permission.[329] Copyright 2016, Wiley-VCH. c) Cross-section TEM image of Ti/*h*-BN/Cu stacks. d) *I*–*V* curve collected in a Ti/*h*-BN/Cu (5–7 layers thick). Reproduced with permission.[163] Copyright 2016, Wiley-VCH.

BP is a layered semiconductor prone to degradation when exposed to atmosphere.[339] The degradation of the surface is generated by the insertion of oxygen groups, leading to a $PO_X$ structure.[339,340] This layer provides RS, as reported by Ref. [318]. They exfoliated BP using both γ-butyrolactone and isopropanol and the devices were fabricated by spin-coating on a ITO/PET flexible and transparent substrate, followed by top circular (500 μm in diameter) Ag electrode deposition by magnetron sputtering using a shadow mask. They observed that, after some days/months of exposure to atmosphere, reproducible RS with $I_{ON}/I_{OFF}$ up to $10^3$–$10^4$ could be achieved (**Figure 14**). They attributed this to the formation of Ag conductive filaments across the oxidized and insulating $PO_X$ superficial layer. However, Ref., [318] was just a proof-of-concept, lacking important RRAM parameters, specially variability. Ref. [119] reported the observation of RS in (PET)/Au/BPQD-PVP/Ag structures, with $I_{ON}/I_{OFF}$ >$10^4$ and endurance >1100 cycles. Both BP-based RRAM devices[119,318] were fabricated by LPE and spin coating and showed flexibility. The characteristics of these two prototypes are summarized in **Table 12**. None of these works show endurance analyses. They concentrate on the proof-of-concept and $I_{ON}/I_{OFF}$ ratio, which makes it difficult to know the real usefulness of this material in RRAMs.

## 7. Discussion, Challenges and Prospects

The most advanced RRAMs use MIM structures formed by metallic electrodes (Ti, Au, Ag, Cu, Ni, Pt) coupled with TMOs ($HfO_2$,[39–42] $Al_2O_3$,[43–46] $TiO_2$[47–50] and $TaO_X$).[51,52] RS in metal/TMO/metal structures was first observed in 1962.[341] After more than 50 years of research, devices with high operation speeds (≈300 ps per transition),[32,41,45,60] low power consumption (≈0.1 pJ per transition),[32,44] good endurance (above $10^{12}$ cycles),[33,111,346,347,364] long data retention times (above 10 years),[46,364] small size (down to 10 nm × 10 nm),[11,296] and high integration capacity (>$1 \times 10^{11}$ bits cm $^2$)[2] have been developed. GRMs were firstly introduced in the structure of RRAMs in 2008,[103] and in less than a decade the performance of some GRM-based RRAMs prototypes fits some of the NVM technology requirements (low operation voltages <1 V,[163,251] high switching speeds down to 1 ns,[273,280,286] endurance >$10^9$ cycles,[286] and small cell size (8.5 nm²).[296]

Table 1 compares the best performances reported for TMO-based and GRM-based RRAM devices. These are similar for both types of RRAMs, and in one case (endurance) one GRM-based RRAM achieved record values. Several GRM-based RRAMs showed low (<1 V) operating voltages,[163,248,253] and acceptable switching speeds.[273,280,286] In contrast, the number of TMO-based RRAMs that fit at least one technology requirement is much larger (Table 1 only displays a selection of them). Therefore, more work in the direction of GRMs-based RRAMs is necessary.

**Table 11.** h-BN-based RRAMs.

| Structure | Fabrication method | Bipolar RS under positive set | Forming process needed | $V_{SET}$ [V] $I_{SET}$ [A] | $V_{RESET}$ [V] $I_{RESET}$ [A] | $I_{ON}/I_{OFF}$ | Endurance cycles | Retention time | Bipolar RS Under negative set | Threshold RS | Ref. |
|---|---|---|---|---|---|---|---|---|---|---|---|
| Ti/hBN/Cu | CVD (no transfer) | YES | NO | 0.4 V $4\times10^{-4}$ A | −0.3 V $4\times10^{-3}$ A | 10 | >350 | – | NO | YES | [163] |
| Ti/hBN/Cu | CVD (no transfer) | YES | NO | 0.7 V $4\times10^{-6}$ A | −0.7 V $10^{-2}$ A | $10^4$ | >600 | – | YES | YES | [163] |
| Ti/hBN/CuNi | CVD (no transfer) | YES | YES | 0.7 V $4\times10^{-3}$ A | −0.4 V $2\times10^{-2}$ A | 15 | – | – | NO | NO | [163] |
| Ti/hBN/CuNi | CVD (no transfer) | YES | YES | 6 V $10^{-3}$ A | −2 V $10^{-1}$ A | $10^6$ | – | – | YES | NO | [163] |
| Ti/hBN/ITO | CVD (transfer) | YES | NO | 0.4 V $2\times10^{-4}$ A | −0.3 V $10^{-3}$ A | 10 | >180 | – | NO | NO | [163] |
| Ti/MLG/hBN/MLG/Au | CVD (transfer) | YES | YES | 2.3 V $10^{-3}$ A | −0.6 V $4\times10^{-2}$ A | $10^3$ | >450 | $4\times10^4$ s | NO | NO | [163] |
| Al/hBN/WO$_3$/Al | CVD (transfer) | – | – | – | – | <10 | ≈80 | $3\times10^4$ s | – | – | [107] |
| Ag/hBN/Cu/PET[a] | CVD (no transfer) | – | YES | – | – | 100 | 550 | $3\times10^3$ s | – | – | [329] |
| Au/Ti/hBN/Cu[a] | CVD (no transfer) | – | YES | – | – | – | >100 | – | – | – | [329] |
| Au/Ti/SLG/hBN/SiO$_2$/Si | MC | – | – | – | – | $10^3$ | – | $10^5$ s | – | – | [338] |

[a]The layered structure of the BN in Ref. [329] is not well supported. From their cross-sectional TEM it looks like amorphous hBN (see Figure 13a).

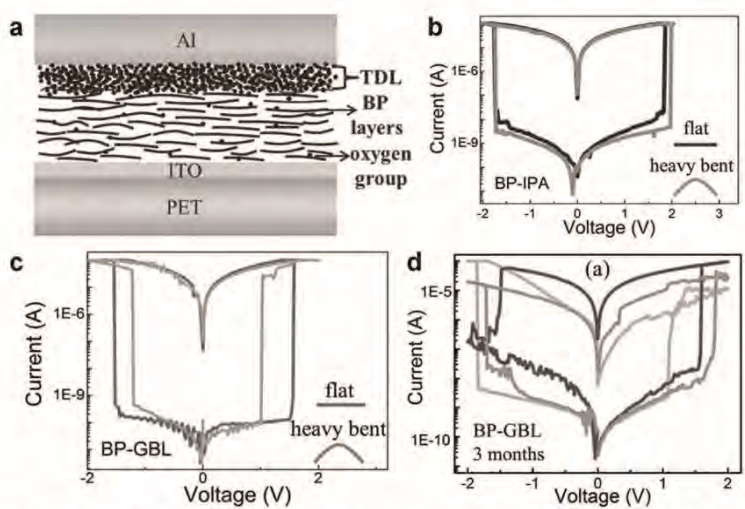

**Figure 14.** a) Schematic illustration of RRAM-BP cross-section. I–V characteristics for the RRAM-BP device for b) BP-IPA, and c) BP-GBL, in flat and bent conditions. d) Typical I–V curves obtained for repeated voltage sweeping cycles in the RRAM-BP devices fabricated with BP-GBL. Reproduced with permission.[318] Copyright 2016, Wiley-VCH.

### 7.1. Fabrication

The fabrication methods used for GRMs-based RRAMs should be improved. For example, one of the best endurances reported for RRAMs exploiting non-carbon GRMs ($10^3$ cycles in Ag/MoS$_2$/Ag)[316] was observed using Ag foils as bottom electrode, and top electrodes deposited with Ag paint and MIM cell size ≈0.1 mm$^2$. These processes/parameters are not compatible with industry, and the knowledge extracted from such works may not be applicable to ultra-scaled (state-of-the-art) RRAMs. Future work in GRMs-based RRAMs should concentrate on the use of industry-compatible methodologies (e.g., for the deposition of electrodes evaporation/and/or sputtering are preferred) and smaller device sizes (so that they can be applicable to nanosized devices).

**Table 12.** BP-based RRAMs.

| Device structure | Fabrication method | $I_{ON}/I_{OFF}$ | Retention time | Transparent | Flexible | Ref. |
|---|---|---|---|---|---|---|
| Al/TDL/BP/ITO/PET | LPE (spin coating) | $\approx 3 \times 10^5$ | $10^5$ s | NO | YES | [318] |
| (PFT)/Au/BPQD-PVP/Ag | LPE (spin coating) | $6 \times 10^4$ | 1100 s | NO | YES | [119] |

Furthermore, in most works using transferred SLG as interface electrode (see, e.g. Table 2) the device size is very large (>8000 µm²). Under such large areas, it is common that SLG layers show cracks, especially after transfer.[168,169] As the transversal electrical resistance of the MGI junction (non-cracked region) is larger than that of the MI one (at the cracked region),[170,171] and because the forming/BD is a stochastic process that takes place at the electrically weakest location of the area under stress (less insulating),[343,344] CFs in these devices are more prone to be formed at SLG cracks. Moreover, as the currents measured through the devices (especially in LRS) are mainly driven by the CFs, the *I–V* characteristics of many RRAMs using transferred SLG may refer to these nanosized MI junctions, and they may not be representative of the MGI structures under study. Ref. [217] reported that the insertion of SLG electrodes in TMO-based RRAMs reduces the HRS current by 1–3 orders of magnitude due to an increase of the out-of-plane resistance (non-cracked regions). Since the endurance and retention times are related to the CF properties, the presence of cracks should have a major influence. Future works using transferred SLG should prove that no cracks are present. One route could be to reduce the device area, which lessens the probability of finding a crack. Another option is to use MLG, which presents fewer cracks and is more resistant to mechanical fractures during transfer.[163]

Many GRMs-based RRAMs works based on polymer-scaffold-assisted transfer did not evaluate the presence of residues on the GRM surface after polymer removal. These may decrease the device size, given their higher thickness >10 nm and insulating nature.[345] This process is random, but can be reduced by using better cleaning processes,[172] which may result in device-to-device variability.[135–137] Future works should include nanoscale surface characterization techniques, such as topographic AFM maps. Including annealing treatments after transfer to remove rests of polymer may be an option.

Therefore, GRM transfer should be avoided when possible, not only due to device performance concerns, but also because it slows down the fabrication process (making it more expensive). The ideal solution would be to develop transfer-free processes, but the direct growth of GRMs on TMOs is a longer term goal. The use of insulating LMs as RS medium is preferred because, first, they do not need transfer[163] and, second, the absence of cracks can be corroborated by the observation of a forming and/or set process.[329] Recent works on CVD-grown *h*-BN report transfer-free RRAM devices.[163,329] but they still use metallic foils. The direct growth of GRMs by CVD (or any other scalable technique) on metal coated flat wafers is highly desirable. Another option is to use LPE GRM insulators that can be spin coated on arbitrary substrates, but that may present variability, given their large roughness (typically ≈20 nm),[134] much larger than flat GRMs prepared by CVD. The use of coating methods that reduce the roughness below 1 nm is necessary. Note that the roughness of TMOs for RRAM (usually grown by ALD) is ≈0.2 nm.[33,62,135]

## 7.2. Characterization

Many papers on GRM-based RRAMs only focus on RS proof-of-concept, showing acceptable >$10^2$ $I_{ON}/I_{OFF}$ in very large >1 mm² devices.[196,316] Information on the number of devices tested in each work and variability analyses is missing. Usually the reports do not concentrate on the study of the technology requirements (note that high $I_{ON}/I_{OFF}$ is not a technology requirement, i.e. just one order of magnitude is enough to reliably distinguish HRS and LRS).[2] For example, we did not find GRM-based RRAM works giving the power consumption in units of energy (Joules) per transition, which is what is demanded by industry.[2] Similarly, most GRM-based RRAMs works do not focus enough on the switching times (e.g. detailed zoomed in plots at the set/reset transition are often missing). Sometimes, endurance and retention plots are shown, but the values (<$10^3$ cycles[351] and >$10^7$ s,[749] respectively) are still insufficient to meet industry requirements.[2] The only paper showing excellent switching times and power consumption (Ref. [286], Table 1) comes from industry. Future works should study several device parameters, such as RS medium thickness, electrode material and CL, as well as provide information on endurance, retention, temperature, and variability analyses performed with a probe-station, as well as modelling and CAFM. The use of CAFM to demonstrate the switching between HRS and LRS in some GRM-based reports is very deficient, as no statistical analyses of the current/size of the CFs are provided. The methods for a correct characterization of RS using CAFM are described in Ref. [35]. Similarly, the structure of LMs-based RS is often not well supported, as explained in Figure 13. Furthermore, the electrical stresses applied to most devices (*I–V* curves) are suitable only for proof-of-concept, but real devices work under fast (<10 ns) voltage pulses.[34]

Many GRM-based RRAM reports do not present variability analyses (just typical values are shown), which raises concerns on the reliability and reproducibility of the results. Device-to-device variability was rarely reported (see, for example, Ref. [163]). In the future, more information about the dispersion of $V_{SET}$ and $V_{RESET}$ in groups of more than 20 devices is needed. The inclusion of atomistic simulations and physical modelling to further complement the experimental observations is also necessary. For example, the QPC model,[210] one of the most widespread for studying the different conductance levels in HRS and LRS in RRAMs,[205,206] has been used in very few GRM-based devices.[163,308]

### 7.3. Technology Viability

The number of TMO-based RRAMs reporting performances above the technology requirements is much larger than for GRM-based ones. For this reason, TMO-based RRAMs are more reliable and (still) superior to GRM-based RRAMs. Moreover, as for TMO-based RRAMs, there is still not a GRM-based RRAM fitting all technology requirements simultaneously, indicating that more research is required. Nevertheless, the faster optimization speed of GRM-based RRAMs as well as the superior electronic,[76] physical,[77] chemical,[78] mechanical,[79] optical,[80] magnetic[81] and thermal[82] properties[83] of GRMs (compared to TMOs) are strong arguments to further explore this technology.

## 8. Conclusions

GRMs have been introduced in the structure of RRAMs with the objectives of i) enhancing their performance as NVM (endurance, retention, switching time, power consumption, operation voltages) and ii) provide additional capabilities (flexibility, transparency, chemical stability, heat dissipation). Graphene can be used as electrode to provide flexibility and transparency, and/or as interface layer between electrodes and RS medium, to decrease the cycle-to-cycle variability, by avoiding atomic diffusion between electrode and insulator. This can reduce the power consumption due to its high out-of-plane contact resistance (compared to metallic electrodes), and suppress surface effects by avoiding chemisorption and/or photodesorption. Surface band bending may allow one to tune the properties of the devices by functionalization, reducing the thickness of the electrodes and improving the three dimensional stackability. GO, a-C, TMDs, h-BN and BP can be used as active RS media to induce the resistivity changes either by migration of intrinsic species (such as oxygen in GO and sulfur in $MoS_2$) or by penetration of metallic ions from adjacent electrodes. Graphene is usually produced by CVD and inserted in RRAMs by polymer-assisted transfer. When using h-BN as RS medium, the standard transfer can be avoided, and the catalyst substrate for CVD growth can be used as bottom electrode. GO and BP are usually produced by LPE and spin coated on a conductive wafer, which serves as bottom electrode. TMDs have been inserted in RRAMs either by CVD plus transfer or LPE plus spin coating. In all cases, top electrodes can be easily fabricated using an evaporator/sputtering coupled with standard photolithography.

GRMs-based RRAMs have shown reproducible unipolar and bipolar RS with high $I_{ON}/I_{OFF} > 10^5$, low operating voltages <1V and fast switching times (<30 ns). In most reports the switching is attributed to the formation/disruption of CFs in the RS medium, and the atomic rearrangements in each state transition are related to the movement of intrinsic species and/or penetration of metallic ions from adjacent layers, showing parallelism with TMO-based RRAMs. GRMs have also been mixed/embedded with polymers, nanoparticles, nanorods and quantum dots in order to enhance the performance (mainly retention and endurance), but in many cases it is unclear what the real need/usefulness of the GRMs are. Despite all efforts, NVMs technological requirements like endurance $>10^9$ cycles and data retention >10 years still remain a challenge. Only one report using a-C as dielectric demonstrated excellent endurance $>10^9$ cycles, and we are not aware of any GRMs-based RRAM showing retention times >10 years. From the point of view of flexibility, GRM-based RRAMs can hold RS under more than $>10^5$ bending stresses with radius down to few mm (no technological requirements in this sense have been established). Moreover, GRM-based RRAMs with transparencies >92% have been reported. The benefits of other GRM properties (such as high chemical stability and thermal heat dissipation) on the performance of RRAMs have not been discussed.

Most RS studies in GRMs concentrated on proof-of-concept demonstrations using large area (>2000 $\mu m^2$) devices, which makes it difficult to extrapolate to real ultra-scaled RRAMs. Future GRMs-based studies should use smaller sizes (<1 $\mu m^2$), focus on demonstrating performances (i.e. endurance, retention, switching time and power consumption) above the NVM technology requirements, and include reliability and variability analyses. The use of atomistic simulations and modelling to support/explain the experimental observations is also necessary.

The fact that GRMs-based devices already fit some NVM technology requirements (operating voltages, endurance and switching times) makes this field worth of further investigation.

### Abbreviations

| | |
|---|---|
| a-C | Amorphous carbon |
| a-C:H | Hydrogenated amorphous carbon |
| a-C:N | Nitrogenated amorphous carbon |
| $a-CO_x$ | Oxygenated amorphous carbon |
| ALD | Atomic layer deposition |
| APTES | 3-Aminopropyltriethoxysilane |
| BD | Dielectric breakdown |
| BLG | Bilayer graphene |
| BP | Black phosphorous |
| BP-GBL | Black phosphorous in (γ-butyrolactone) |
| BP-IPA | Black phosphorous in isopropanol |
| CAFM | Conductive atomic force microscopy |
| CBRAM | Conductive bridge random access memory |
| CF | Conductive filament |
| CL | Current limitation |
| CMOS | Complementary metal oxide semiconductor |
| CVD | Chemical vapor deposition |
| CVS | Constant voltage stresses |
| 1D1R | One diode one resistor |
| DRAM | Dynamic random access memory |

**ADVANCED
SCIENCE NEWS**
www.advancedsciencenews.com

**ADVANCED
ELECTRONIC
MATERIALS**
www.advelectronicmat.de

| | |
|---|---|
| ECM | Electrochemical metallization |
| FET | Field effect transistor |
| FRAM | Ferroelectric RAM |
| GB | Grain boundary |
| GFET | Graphene field effect transistor |
| GI | Graphene/Insulator |
| GIG | Graphene/Insulator/Graphene |
| GMIM | Graphene/Metal/Insulator/Metal |
| GMIMG | Graphene/Metal/Insulator/Metal/Graphene |
| GO | Graphene oxide |
| GQDs | Graphene quantum dots |
| GO-PVK | Poly (n-vinylcarbazole) derived graphene oxide |
| GRMs | Graphene and related materials |
| GS-RRAM | Graphene set electrode resistive random access memory |
| h-BN | Hexagonal boron nitride |
| HRS | High resistance state |
| ITO | Indium tin oxide |
| ITRS | International Technology Roadmap for Semiconductors |
| LM | Layered material |
| LPE | Liquid phase exfoliation |
| LRS | Low resistance state |
| LSG-RRAM | Laser-scribed reduced graphene oxide |
| MC | Micromechanical exfoliation |
| MGIGM | Metal/Graphene/Insulator/Graphene/Metal |
| MGIM | Metal/Graphene/Insulator/Metal |
| MIGM | Metal/Insulator/Graphene/Metal |
| MIM | Metal/Insulator/Metal |
| MIMG | Metal/Insulator/Graphene/Metal |
| MLG | Multilayer graphene |
| MRAM | Magnetoresistive random access memory |
| NVM | Non-volatile memory |
| PCBM | 6-Phenyl-C61 butyric acid methyl ester |
| PCMO | Pr0.7Ca0.3MnO3 |
| PCRAM | Phase change random access memory |
| PET | Poly(ethylene terephthalate) |
| PMC | Programmable metallization cells |
| PMMA | Polymethyl methacrylate |
| PVK | Poly (n-vinylcarbazole) derived graphene oxide |
| PVP | Polyvinylpyrrolidone |
| QPC | Quantum point contact |
| RGO | Reduced graphene oxide |
| RRAM | Resistive random access memory |
| RS | Resistive switching |
| SLG | Single-layer graphene |
| STTM-RAM | Spin transfer torque magnetic random access memory |
| ta-C | Tetrahedral amorphous carbons |
| TEM | Transmission electron microscopy |
| TMDs | Transition metal dichalcogenides |
| TMO | Transition metal oxide |
| TPAPAM | Triphenylamine-based polyazomethine |
| VCM | Valence change memory |
| ReRAM | Redox random access memory |
| XPS | X-ray photoelectron spectroscopy |
| ZnONRs | ZnO nanorods |

## Acknowledgements

We acknowledge support from the Young 1000 Global Talent Recruitment Program of the Ministry of Education of China, the National Natural Science Foundation of China (grants no. 61502326, 41550110223), the Jiangsu Government (grant no. BK20150343), the Ministry of Finance of China (grant no. SX21400213), the Young 973 National Program of the Chinese Ministry of Science and Technology (grant no. 2015CB932700), the Collaborative Innovation Center of Suzhou Nano Science & Technology, the Jiangsu Key Laboratory for Carbon Based Functional Materials & Devices, the Priority Academic Program Development of Jiangsu Higher Education Institutions, the National Natural Science Foundation of China under Grant Nos. 61521064, 61322408, 61422407, the Beijing Training Project for the Leading Talents in S&T under Grant No. ljrc201508, the Opening Project of Key Laboratory of Microelectronic Devices & Integrated Technology, Institute of Microelectronics, Chinese Academy of Sciences, the EU Graphene Flagship, CARERAMM, ERC Grants Hetero2D and Highgraink, EPSRC Grants EP/K01711X/1, EP/K017144/1, EP/N010345/1, EP/L01608//1. All members of Prof. Lanza's group are acknowledged for literature review.

[1] A. Chen, *Solid-State Electron.*, **2013**, *125*, 25.
[2] International Technology Roadmap for Semiconductors, 2013 Edition, Process Integration, Devices, and Structures section, http://www.itrs.net (accessed February 11th 2015).
[3] D. Kumar, *Preprints* **2016**, DOI: 10.20944/preprints201607.0093.v1.
[4] Average Cost of Hard Drive Storage, http://www.statisticbrain.com/average-cost-of-hard-drive-storage/(accessed September 2016).
[5] Web-Feet Research, http://www.webfeetresearch.com (accessed May 8th 2016).
[6] D. Kahng, S. M. Sze, *Bell Syst. Tech. J.* **1967**, *46*, 1288.
[7] "IBM Reports Advance in Shrinking Chip Circuitry", The Wall Street Journal, (accessed: July 9, 2015).
[8] Equipment and Tool Institute, http://eti-home.org/Newsletter-V02/The-Bus-is-Full.html (accessed: September 2016).
[9] S. Okhonin, M. Nagoga, E. Carman, R. Beffa, E. Faraoni, *IEEE Int. Electron Dev. Meeting*, Washington, DC, USA, 10–12 December, **2007**.
[10] C. Navarro, M. Bawedin, F. Andrieu, S. Cristoloveanu, *IEEE Electron Device Lett.* **2015**, *36*, 5.
[11] T. Atsumi, S. Nagatsuka, H. Inoue, T. Onuki, T. Saito, Y. Ieda, Y. Okazaki, A. Isobe, Y. Shionoiri, K. Kato, T. Okuda, J. Koyama, S. Yamazaki, *4th IEEE Int. Memory Workshop* **2012**, 99.
[12] Y. Li, Y. Z. Wang, Q. Cui, *J. Renewable Sustainable Energy* **2016**, *8*, 015901.
[13] C. T. Chu, Y. K. Wang, P. K. Liao, *5th International Symposium on Next-Generation Electronics (ISNE)*, 4–6 May, **2016**.
[14] Authenticating ferroelectric random access memory (F-RAM) device and method, US Patent 9330251 B1.
[15] Z. Xu, B. Liu, Y. F. Chen, Z. H. Zhang, D. Gao, H. Wang, Z. T. Song, C. Z. Wang, J. D. Ren, N. F. Zhu, Y. H. Xiang, Y. P. Zhan, S. L. Feng, *Solid-State Electron.* **2016**, *116*, 119.

[16] P. M. Palangappa, J. Y. Li, K. Mohanram, *IEEE Trans. Comput.* 2016, 65, 1025.

[17] P. K. Amiri, K. L. Wang, *Proc. 6th IEEE Int. Memory Workshop* 2014, DOI: 10.1109/IMW.2014.6849352.

[18] Utilization of the anomalous hall effect or polarized spin hall effect for MRAM application, US Patent 9269415 B1.

[19] A. Calderoni, S. Sills, N. Ramaswamy, *Proc. 6th IEEE Int. Memory Workshop* 2014, DOI: 10.1109/IMW.2014.6849351.

[20] W. J. Ma, *Electron. Lett.* 2016, 52, 9.

[21] A. Belmonte, A. Fantini, A. Redolfi, M. Houssa, M. Jurczak, L. Goux, *Solid-State Electron.* 2016, 125, 189.

[22] N. Gonzales, J. Dinh, D. Lewis, N. Gilbert, B. Pedersen, D. Kamalanathan, J. R. Jameson, S. Hollmer, *Proc. 8th Internation Memory Workshop* 2016, DOI: 10.1109/IMW.2016.7493566.

[23] L. Li, D. Z. Wen, *Organic Electron.* 2016, 34, 12.

[24] S. Y. Ning, T. O. Iwasaki, S. Hachiya, G. Rosendale, M. Manning, D. Viviani, T. Rueckes, K. Takeuchi, *Jpn. J. Appl. Phys.* 2016, 55, 04E01.

[25] W. Zhao, E. Belhaire, C. Chappert, P. Mazoyer, *ACM Trans. Embedded Comput. Syst.* 2009, 9, 1.

[26] J. J. Nowak, R. P. Robertazzi, J. Z. Sun, G. Hu, J. H. Park, J. Lee, A. J. Annunziata, G. P. Lauer, R. Kothandaraman, E. J. O'Sullivan, P. L. Trouilloud, Y. Kim, D. C. Worledge, *IEEE Magn. Lett.* 2016, 7, 3102604.

[27] Z. Liu, A. A. Yasseri, J. S. Lindsey, D. F. Bocian, *Science* 2003, 302, 1543.

[28] W. G. Kuhr, A. R. Gallo, R. W. Manning, C. W. Rhodine, *MRS Bull.* 2004, 29, 805.

[29] L. Carro, C. Vaju, B. Corraze, V. Guiot, E. Janod, *Adv. Mater.* 2010, 22, 5193.

[30] J. Tranchant, E. Janod, L. Carlo, B. Corraze, E. Souchier, J. L. Leclercq, P. Cremillieu, P. Moreau, M. P. Besland, *Thin Solid Films* 2013, 533, 61.

[31] B. Govoreanu, G. S. Kar, Y. Y. Chen, V. Paraschiv, A. Fantini, I. P. Radu, L. Goux, S. Clima, R. Degraeve, N. Jossart, *IEEE Int. Electron Dev. Meeting*, Washington, DC, USA, 5–7 December, 2011.

[32] H. Y. Lee, Y. S. Chen, P. S. Chen, P. Y. Gu, Y. Y. Hsu, S. M. Wang, W. H. Liu, C. H. Tsai, S. S. Shen, P. C. Chiang, W. P. Lin, C. H. Lin, W. S. Chen, F. T. Chen, C. H. Lien, M.-J. Tsai, *IEEE Int. Electron Dev. Meeting*, San Francisco, CA, USA, 6–8 December, 2010.

[33] M. J. Lee, C. B. Lee, D. S. Lee, S. R. Lee, M. Chang, J. H. Hur, Y. B. Kim, C. J. Kim, D. H. Seo, S. Seo, U. I. Chung, I. K. Yoo, K. Kim, *Nat. Mater.* 2011, 10, 625.

[34] I. Valov, *ChemElectroChem* 2014, 1, 26.

[35] M. Lanza, *Materials* 2014, 7, 2155.

[36] E. Gale, *Semicond. Sci. Technol.* 2014, 29, 10.

[37] K. Tsunoda, Y. Fukuzumi, J. R. Jameson, Z. Wang, P. B. Griffin, Y. Nishi, *Appl. Phys. Lett.* 2007, 90, 113501.

[38] A. Sawa, *Mater. Today* 2008, 11, 28.

[39] S. B. Long, L. Perniola, C. Cagli, J. Buckley, X. J. Lian, E. Miranda, E. Pan, M. Liu, J. Sune, *Sci. Rep.* 2013, 3, 2929.

[40] X. Wu, S. Mei, M. Bosman, N. Raghavan, X. X. Zhang, D. Y. Cha, K. Li, K. L. Pey, *Adv. Electron. Mater.* 2015, 1, 1500130.

[41] U. Chand, K. C. Huang, C. Y. Huang, C. H. Ho, C. H. Lin, T. Y. Tseng, *J. Appl. Phys.* 2015, 117, 184105.

[42] D. Duncan, B. Magyari Kope, Y. Nishi, *Appl. Phys. Lett.* 2016, 108, 043501.

[43] B. Sarkar, B. Lee, V. Misra, *Semiconductor Sci. Technol.* 2015, 30, 105014.

[44] C. Ahn, Z. Jiang, C. S. Lee, H. Y. Chen, J. Liang, L. S. Liyanage, H., S. P. Wong, *IEEE Trans. Electron Devices* 2015, 62, 2197.

[45] J. Zhou, F. Cai, Q. Wang, B. Chen, S. Gaba, W. D. Lu, *IEEE Trans. Electron Devices* 2016, 37, 404.

[46] L. G. Wang, X. Qian, Y. Q. Cao, Z. Y. Cao, G. Y. Fang, A. D. Li, D. Wu, *Nano Research Lett.* 2015, 10, 135.

[47] H. Y. Jeong, S. K. Kim, J. Y. Lee, S. Y. Choi, *J. Electrochem. Soc.* 2011, 158, 979.

[48] J. Shim, I. Kim, K. P. Biju, M. Jo, J. Park, J. Lee, S. Jung, W. Lee, S. Kim, S. Park, H. Hwang, *J. Appl. Phys.* 2011, 109, 033712.

[49] J. J. Huang, C. W. Kuo, W. C. Chang, T. H. Hou, *Appl. Phys. Lett.* 2010, 96, 262901.

[50] H. Y. Jeong, Y. I. Kim, J. Y. Lee, S. Y. Choi, *Nanotechnology* 2010, 21, 115203.

[51] M. X. Yu, Y. M. Cai, Z. W. Wang, Y. C. Fang, Y. F. Liu, Z. Z. Yu, Y. Pan, Z. X. Zhang, J. Tan, X. Yang, M. Li, R. Huang, *Sci. Rep.* 2016, 6, 21020.

[52] Y. D. Zhao, P. Huang, Z. Chen, C. Liu, H. T. Li, W. J. Ma, B. Gao, X. Y. Liu, J. F. Kang, *IEEE Silicon Nanoelectronics Worshop*, Kyoto, Japan, 14–15 June 2015.

[53] B. Gao, W. Y. Chang, B. Sun, H. W. Zhang, L. F. Liu, X. Y. Liu, R. Q. Han, T. B. Wu, J. F. Kang, *Int. Symp. VLSI Technol., Syst., Appl.*, Hsinchu, China, 26–28 April 2010.

[54] G. Bersuker, D. C. Gilmer, D. Veksler, P. Kirsch, L. Vandelli, A. Padovani, L. Larcher, K. McKenna, A. Shluger, V. Iglesias, M. Porti, M. Nafría, *J. Appl. Phys.* 2011, 110, 124518.

[55] M. Lubben, P. Karakolis, V. I. Sougleridis, P. Normand, P. Dimitrakis, I. Valov, *Adv. Mater.* 2015, 27, 6202.

[56] R. Waser, R. Dittmann, G. Staikov, K. Szot, *Adv. Mater.* 2009, 21, 2632.

[57] T. Hino, T. Hasegawa, K. Terabe, T. Tsuruoka, A. Nayak, T. Ohno, M. Aono, *Sci. Technol. Adv. Mater.* 2011, 12, 013003.

[58] U. Russo, D. Ielmini, A. L. Lacaita, M. N. Kozicki, *IEEE Trans. Electron Devices* 2009, 56, 5.

[59] I. Valov, R. Waser, J. R. Jameson, M. N. Kozicki, *Nanotechnology* 2011, 22, 254003.

[60] N. Xu, L. F. Liu, X. Sun, C. Chen, Y. Wang, D. D. Han, X. Y. Liu, R. Q. Han, J. F. Kang, B. Yu, *Semicond. Sci. Technol.* 2008, 23, 075019.

[61] B. Gao, W. Y. Chang, B. Sun, H. W. Zhang, L. F. Liu, X. Y. Liu, R. Q. Han, T. B. Wu, J. F. Kang, *International Symposium on VLSI Technology Systems and Applications*, Taiwan, China, 26–28 April 2010.

[62] M. Lanza, G. Bersuker, M. Porti, E. Miranda, M. Nafría, X. Aymerich, *Appl. Phys. Lett.* 2012, 101, 193502.

[63] R. Annunziata, P. Zuliani, M. Borghi, G. De Sandre, L. Scotti, C. Prelini, M. Tosi, I. Tortorelli, F. Pellizzer, *IEEE Int. Electron Dev. Meeting*, Baltimore, MD, USA, 7–9 December, 2009.

[64] K. Tsuchida, T. Inaba, K. Fujita, Y. Ueda, T. Shimizu, Y. Asao, T. Kajiyama, M. Iwayama, K. Sugiura, S. Ikegawa, T. Kishi, T. Kai, M. Amano, N. Shimomura, H. Yoda, Y. Watanabe, *IEEE Int. Solid-State Circuits Conference Digest of Technical Papers*, San, Francisco, CA, USA, 7–11 February 2010.

[65] H. Y. Lee, P. S. Chen, T. Y. Wu, Y. S. Chen, C. C. Wang, P. J. Tzeng, C. H. Lin, F. Chen, C. H. Lien, M. J. Tsai, *IEEE Int. Electron Dev. Meeting*, San Francisco, CA, USA, 25–17 December, 2010.

[66] Q. Luo, X. X. Xu, H. T. Liu, H. B. Lv, T. C. Gong, S. B. Long, Q. Liu, H. T. Sun, W. Banerjee, L. Li, J. F. Gao, N. D. Lu, S. S. Chung, J. Li, M. Liu, *IEEE Int. Electron Dev. Meeting*, Washington, DC, USA, 7–9 December, 2015.

[67] T. Breuer, A. Siemon, E. Linn, S. Menzel, R. Waser, V. Rana, *Adv. Electron. Mater.* 2015, 1, 1500138.

[68] The Crossbar RRAM Advantage: Simply Fast, Simply Scalable, Simply Reliable: http://crossbar-inc.com/technology/rram-advantages/, accessed: August 25th, 2016.

[69] Y. Sakotsubo, S. Sagarrihara, M. Terai, S. Kotsuji, Y. Saito, M. Tada, Y. Yabe, H. Hada, *VLSI Technology (VLSIT)*, Honolulu, Hawaii, USA, 15–17 June 2010.

- 132 -

[70] T. Yamamoto, T. Kubo, T. Sukegawa, E. Takii, Y. Shimamune, N. Tamura, T. Sakoda, M. Nakamura, H. Ohta, T. Miyashita, H. Kurata, S. Satoh, M. Kase, T. Sugii, *IEEE International Electron Devices Meeting*, Washington DC, USA, 10–12 December 2007.

[71] Website of Panasonic (Microcontrollers): https://na.industrial. panasonic.com/products/semiconductors/microcontrollers/8-bit-low-power-microcomputers-mn101l-series (accessed September 2016).

[72] Website of Adesto Technologies: http://www.adestotech.com/products/mavriq/ (accessed September 2016).

[73] Website of Nantero: http://nantero.com/technology/ (accessed September 2016).

[74] P. K. Yang, W. Y. Chang, P. Y. Teng, S. F. Jeng, S. J. Lin, P. W. Chiu, J. H. He, *Proc. IEEE* 2013, 101, 1732.

[75] Y. Ji, S. Lee, B. Cho, S. Song, T. Lee, *ACS Nano* 2011, 5, 5995.

[76] A. H. C. Neto, F. Guinea, N. M. R. Peres, K. S. Novoselov, A. K. Geim, *Rev. Mod. Phys.* 2009, 81, 109.

[77] E. Stolyarova, D. Stolyarov, K. Bolotin, S. Ryu, L. Liu, K. T. Rim, M. Klima, M. Hybertsen, I. Pogorelsky, I. Pavlishin, K. Kusche, J. Hone, P. Kim, H. L. Stormer, V. Yakimenko, G. Flynn, *Nano Lett.* 2009, 9, 332.

[78] S. Chen, L. Brown, M. Levendorf, W. Cai, S.-Y. Ju, J. Edgeworth, X. Li, C. W. Magnuson, A. Velamakanni, R. D. Piner, J. Kang, J. Park, R. S. Ruoff, *ACS Nano* 2011, 5, 1321.

[79] C. Lee, X. Wei, J. W. Kysar, J. Hone, *Science* 2008, 321, 385.

[80] F. Bonaccorso, Z. Sun, T. Hasan, A. C. Ferrari, *Nat. Photonics* 2010, 4, 611.

[81] N. Tombros, C. Jozsa, M. Popinciuc, H. T. Jonkman, B. J. van Wees, *Nature* 2007, 448, 571.

[82] A. Balandin, S. Ghosh, W. Bao, I. Calizo, D. Teweldebrhan, F. Miao, C. N. Lau, *Nano Lett.* 2008, 8, 902.

[83] A. C. Ferrari, F. Bonaccorso, V. Fal'ko, K. S. Novoselov, S. Roche, P. Bøggild, S. Borini, F. H. L. Koppens, V. Palermo, N. Pugno, J. A. Garrido, R. Sordan, A. Bianco, L. Ballerini, M. Prato, E. Lidorikis, J. Kivioja, C. Marinelli, T. Ryhänen, A. Morpurgo, J. N. Coleman, V. Nicolosi, L. Colombo, A. Fert, M. Garcia-Hernandez, A. Bachtold, G. F. Schneider, F. Guinea, C. Dekker, M. Barbone, Z. Sun, C. Galiotis, A. N. Grigorenko, G. Konstantatos, A. Kis, M. Katsnelson, L. Vandersypen, A. Loiseau, V. Morandi, D. Neumaier, E. Treossi, V. Pellegrini, M. Polini, A. Tredicucci, G. M. Williams, B. H. Hong, J.-H. Ahn, J. M. Kim, H. Zirath, B. J. van Wees, H. van der Zant, L. Occhipinti, A. Di Matteo, I. A. Kinloch, T. Seyller, E. Quesnel, X. Feng, K. Teo, N. Rupesinghe, P. Hakonen, S. R. T. Neil, Q. Tannock, T. Löfwander, J. Kinaret, *Nanoscale* 2015, 7, 4598.

[84] R. U. A. Khan, S. R. P. Silva, *Diamond Relat. Mater.* 2001, 10, 1036.

[85] F. Zhuge, W. Dai, C. L. He, A. Y. Wang, Y. W. Liu, M. Li, Y. H. Wu, P. Cui, R. W. Li, *Appl. Phys. Lett.* 2010, 96, 163505.

[86] L. Dellmann, A. Sebastian, P. Jonnalagadda, C. A. Santini, W. W. Koelmans, C. Rossel, E. Eleftheriou, *Proc. Eur. Solid-State Device Res. Conf.* 2013, 268.

[87] Y. J. Chen, H. L. Chen, T. F. Young, T. C. Chang, T. M. Tsai, K. C. Chang, R. Zhang, K. H. Chen, J. C. Lou, T. J. Chu, J. H. Chen, D. H. Bao, S. M. Sze, *Nanoscale Res. Lett.* 2014, 9, 52.

[88] E. G. Gerstner, D. R. McKenzie, *Diamond Relat. Mater.* 1998, 7, 1172.

[89] E. G. Gerstner, D. R. McKenzie, *J. Appl. Phys.* 1998, 84, 5647.

[90] F. Bonaccorso, A. Lombardo, T. Hasan, Z. Sun, L. Colombo, A. C. Ferrari, *Mater. Today* 2012, 15, 564.

[91] S. Jang, E. Hwang, J. H. Cho, *Nanoscale* 2014, 6, 15286.

[92] S. Jang, E. Hwang, J. H. Lee, H. S. Park, J. H. Cho, *Small* 2015, 11, 311.

[93] S. M. Kim, E. B. Song, S. Lee, J. Zhu, D. H. Seo, M. Mecklenburg, S. Seo, K. L. Wang, *ACS Nano* 2012, 6, 7879.

[94] S. Lee, E. B. Song, S. M. Kim, Y. Lee, D. H. Seo, S. Seo, K. L. Wang, *Appl. Phys. Lett.* 2012, 101, 1.

[95] Y. Park, S. Park, I. Jo, B. H. Hong, Y. Hong, *Org. Electron.* 2015, 27, 227.

[96] S. T. Han, Y. Zhou, Q. D. Yang, L. Zhou, L. B. Huang, Y. Yan, C. S. Lee, V. A. L. Roy, *ACS Nano* 2014, 8, 1923.

[97] Y. N. Kim, N. H. Lee, D. Y. Yun, T. W. Kim, *Org. Electron.* 2015, 25, 165.

[98] Y. Park, D. Gupta, C. Lee, Y. Hong, *Org. Electron.* 2012, 13, 2887.

[99] N. Zhan, M. Olmedo, G. P. Wang, J. L. Liu, *Appl. Phys. Lett.* 2011, 99, 113112.

[100] D. Selli, M. Baldoni, A. Sgamellotti, F. Mercuri, *Nanoscale* 2012, 4, 1350.

[101] A. Sinitskii, J. M. Tour, *ACS Nano* 2009, 3, 2760.

[102] X. M. Wang, W. G. Xie, J. Du, C. L. Wang, N. Zhao, J. B. Xu, *Adv. Mater.* 2012, 19, 2614.

[103] B. Standley, W. Z. Bao, H. Zhang, J. Bruck, C. N. Lau, M. Bockrath, *Nano Lett.* 2008, 8, 3345.

[104] Y. B. Li, A. Sinitskii, J. M. Tour, *Nat. Mater.* 2008, 7, 966.

[105] J. Yao, J. Lin, Y. H. Dai, G. Ruan, Z. Yan, L. Li, L. Zhong, D. Natelson, J. M. Tour, *Nat. Commun.* 2012, 3, 1101.

[106] K. S. Novoselov, A. K. Geim, S. V. Morozov, D. Jiang, Y. Zhang, S. V. Dubonos, I. V. Grigorieva, A. A. Firsov. *Science* 2004, 306, 666.

[107] J. Kim, D. Kim, Y. Jo, J. Han, H. Woo, H. Kim, K. K. Kim, J. P. Hong, H. Im, *Thin Solid Films* 2015, 589, 188.

[108] A. Asamitsu, Y. Tomioka, H. Kuwahara, Y. Tokura, *Nature* 1997, 388, 50.

[109] D. C. Kim, S. Seo, S. E. Ahn, D. S. Suh, M. J. Lee, B. H. Park, I. K. Yoo, I. G. Baek, H. J. Kim, E. K. Yim, J. E. Lee, S. O. Park, H. S. Kim, U-In Chung, J. T. Moon, B. I. Ryu, *Appl. Phys. Lett.* 2006, 88, 202102.

[110] B. J. Choi, D. S. Jeong, S. K. Kim, C. Rohde, S. Choi, J. H. Oh, H. J. Kim, C. S. Hwang, K. Szot, R. Waser, B. Reichenberg, S. Tiedke, *J. Appl. Phys.* 2005, 98, 033715.

[111] I. G. Baek, M. S. Lee, S. Seo, M. J. Lee, D. H. Seo, D. S. Suh, J. C. Park, S. O. Park, H. S. Kim, I. K. Yoo, U. I. Chung, J. T. Moon, *IEEE Int. Electron Dev. Meeting*, San Francisco, CA, USA, 13–15 December 2004.

[112] D. S. Jeong, H. Schroeder, R. Waser, *Electrochem. Solid-State Lett.* 2007, 10, G51.

[113] S. Lee, J. Sohn, Z. Jiang, H. Chen, H. S. Philip Wong, *Nat. Commun.* 2015, 6, 8407.

[114] C. A. Ross, H. I. Smith, T. Savas, M. Schattenburg, M. Farhoud, M. Hwang, M. Walsh, M. C. Abraham, R. J. Ram, *J. Vac. Sci. Technol. B* 1999, 17, 3168.

[115] Y. Ji, M. Choe, B. Cho, S. Song, J. Yoon, H. C. Ko, T. Lee, *Nanotechnology* 2012, 23, 105202.

[116] C. X. Wu, F. S. Li, T. L. Guo, *Appl. Phys. Lett.* 2014, 104, 183105.

[117] S. Ghosh, I. Calizo, D. Teweldebrhan, E. P. Pokatilov, D. L. Nika, A. A. Balandin, W. Bao, F. Miao, C. N. Lau, *Appl. Phys. Lett.* 2008, 92, 151911.

[118] Y. Chen, B. Zhang, G. Liu, X. D. Zhuang, E. T. Kang, *Chem. Soc. Rev.* 2012, 41, 4688.

[119] X. Zhang, H. Xie, Z. Liu, C. Tan, Z. Luo, H. Li, J. Lin, L. Sun, W. Chen, Z. Xu, L. Xie, W. Huang, H. Zhang, *Angew. Chem. Int. Ed.* 2015, 54, 3653.

[120] Graphene-based MIM diode and associated methods, US Patent 9202945 B2.

[121] K. S. Novoselov, D. Jiang, F. Schedin, T. J. Booth, V. V. Khotkevich, S. V. Morozov, A. K. Geim, *Proc. Natl. Acad. Sci. U. S. A.* 2005, 102, 10451.

[122] G. I. Yua, R. Jalilb, B. Belleb, A. S. Mayorova, P. Blakeb, F. Schedinb, S. V. Morozovc, L. A. Ponomarenkoa, F. Chiappinid, S. Wiedmannd, Uli Zeitlerd, M. I. Katsnelsone, A. K. Geima,

K. S. Novoselova, D. C. Eliasa, *Proc. Natl. Acad. Sci.* **2013**, *110*, 3282.

[123] L. Britnell, R. V. Gorbachev, R. Jalil, B. D. Belle, F. Schedin, M. I. Katsnelson, L. Eaves, S. V. Morozov, A. S. Mayorov, N. M. R. Peres, A. H. C. Neto, J. Leist, A. K. Geim, L. A. Ponomarenko, K. S. Novoselov, *Nano Lett.* **2012**, *12*, 1707.

[124] Y. Hernandez, V. Nicolosi, M. Lotya, F. M. Blighe, Z. Sun, S. De, I. T. Mcgovern, B. Holland, M. Byrne, Y. K. Gunko, J. J. Boland, P. Niraj, G. Duesberg, S. Krishnamurthy, R. Goodhue, J. Hutchison, V. Scardaci, A. C. Ferrari, J. N. Coleman, *Nat. Nanotechnol.* **2008**, *3*, 563.

[125] X. S. Li, W. W. Cai, J. H. An, S. Y. Kim, J. Nah, D. X. Yang, R. Piner, A. Velamakanni, I. Jung, E. Tutuc, S. K. Banerjee, L. G. Colombo, R. S. Ruoff, *Science* **2009**, *324*, 1312.

[126] H. Y. Jeong, J. Y. Kim, J. W. Kim, J. O. Hwang, J. E. Kim, J. Y. Lee, T. H. Yoon, B. J. Cho, S. O. Kim, R. S. Ruoff, S. Y. Choi, *Nano Lett.* **2010**, *10*, 4381.

[127] S. Porro, E. Accornero, C. F. Pirri, C. Ricciardi, *Carbon* **2015**, *85*, 383.

[128] S. Seo, Y. H. Yoon, J. H. Lee, Y. H. Park, H. Y. Lee, *ACS Nano* **2013**, *7*, 3607.

[129] M. Lotya, Y. Hernandez, P. J. King, R. J. Smith, V. Nicolosi, L. S. Karlsson, F. M. Blighe, S. De, Z. Wang, I. T. McGovern, G. S. Duesberg, J. N. Coleman, *J. Am. Chem. Soc.* **2009**, *131*, 3611.

[130] J. N. Coleman et al. *Science* **2011**, *311*, 568.

[131] S. C. Ray, S. K. Bhunia, A. Saha, N. R. Jana, *Microelectron. Eng.* **2015**, *146*, 48.

[132] I. Banerjee, P. Harris, A. Salimian, A. K. Ray, *IET Circuits Dev. Syst.* **2015**, *9*, 428.

[133] S. Porro, C. Ricciardi, *RSC Adv.* **2015**, *5*, 6856.

[134] A. Matkovic, I. Milosevic, M. Milicevic, T. Tomasevic-Ilic, J. Pesic, M. Music, M. Spasenovic, D. Jovanovic, B. Vasic, C. Deeks, R. Panajotovic, M. R. Belic, R. Gajic, *2D Mater.* **2016**, *3*, 015002.

[135] M. Lanza, M. Porti, M. Nafria, G. Bensetter, W. Frammelsberger, H. Ranzinger, E. Lodermeier, G. Jaschke, *Microelectron. Reliab.* **2007**, *47*, 1424.

[136] J. Petry, W. Vandervorst, O. Richard, T. Conard, P. DeWolf, V. Kaushik, A. Delabie, S. V. Elshocht, *Mater. Res. Soc. Symp. Proc.*, San Francisco, CA, USA, 13–16 April 2004.

[137] M. Lanza, M. Porti, M. Nafria, X. Aymerich, G. Bensetter, E. Lodermeier, H. Ranzinger, G. Jaschke, S. Teichert, L. Wilde, P. Michalowski, *IEEE Trans. Nanotechnol.* **2011**, *10*, 344.

[138] C. L. Tan, Z. D. Liu, W. Huang, H. Zhang, *Chem. Soc. Rev.* **2015**, *44*, 2615.

[139] L. Y. Niu, J. N. Coleman, H. Zhang, H. Shin, M. Chhowalla, Z. J. Zheng, *Small* **2016**, *12*, 272.

[140] J. Q. Liu, Z. Q. Liu, T. J. Liu, Z. Y. Yin, X. Z. Zhou, S. F. Chen, L. H. Xie, F. Boey, H. Zhang, W. Huang, *Small* **2010**, *6*, 1536.

[141] P. Blake, P. D. Brimicombe, R. R. Nair, T. J. Booth, D. Jiang, F. Schedin, L. A. Ponomarenko, S. V. Morozov, H. F. Gleeson, E. W. Hill, A. K. Geim, K. S. Novoselov, *Nano Lett.* **2008**, *8*, 1704.

[142] K. S. Kim, Y. Zhao, H. Jang, S. Y. Lee, J. M. Kime, K. S. Kim, J. H. Ahn, P. Kim, J. Y. Choi, B. H. Hong, *Nature* **2009**, *457*, 706.

[143] J. Corans, A. T. Ndiaye, M. Engler, C. Busse, D. Wall, N. Buckanie, F. J. M. Heringdorf, R. V. Gastel, B. Poelsema, I. Michely, *New J. Phys.* **2009**, *11*, 023006.

[144] Y. Z. Xue, B. Wu, Y. L. Guo, L. P. Huang, L. Jiang, J. Y. Chen, D. C. Geng, Y. Q. Liu, W. P. Hu, G. Yu, *Nano Res.* **2011**, *4*, 1208.

[145] Z. W. Peng, Z. Yan, Z. Z. Sun, J. M. Tour, *ACS Nano* **2011**, *5*, 8241.

[146] D. R. Lenski, M. S. Fuhrer, *J. Appl. Phys.* **2011**, *110*, 013720.

[147] T. Kato, R. Hatakeyama, *ACS Nano* **2012**, *6*, 8508.

[148] Y. J. Zhan, Z. Liu, S. Najmaei, P. M. Ajayan, J. Lou, *Small* **2012**, *8*, 966.

[149] S. F. Wu, C. M. Huang, G. Aivazian, J. S. Ross, D. H. Cobden, X. D. Xu, *ACS Nano* **2013**, *7*, 2768.

[150] K. K. Liu, W. J. Zhang, Y. H. Lee, Y. C. Lin, M. T. Chang, C. Su, C. S. Chang, H. Li, Y. M. Shi, H. Zhang, C. S. Lai, L. J. Li, *Nano Lett.* **2012**, *12*, 1538.

[151] Y. H. Lee, X. Q. Zhang, W. J. Zhang, M. T. Chang, C. T. Lin, K. D. Chang, Y. C. Yu, J. T. W. Wang, C. S. Chang, L. J. Li, T. W. Lin, *Adv. Mater.* **2012**, *24*, 2320.

[152] Y. F. Yu, C. Li, Y. Liu, L. Q. Su, Y. Zhang, L. Y. Cao, *Sci. Rep.* **2013**, *3*, 1866.

[153] S. Kikkawa, R. Shimanouchi-Futagami, M. Koizumi, *Appl. Phys. A* **1989**, *49*, 105.

[154] E. S. Peters, C. J. Carmalt, I. P. Parkin, D. A. Tocher, *Eur. J. Inorg. Chem.* **2005**, *20*, 4179.

[155] C. Cong, J. Shang, X. Wu, B. C. Cao, N. Peimyoo, C. Y. Qiu, L. T. Sun, T. Yu, arXiv:1312.1418.

[156] J. Xia, X. Huang, L. Z. Liu, M. Wang, L. Wang, B. Huang, D. D. Zhu, J. J. Li, C. Z. Gu, X. M. Meng, *Nanoscale* **2014**, *6*, 8949.

[157] X. Wang, Y. Gong, G. Shi, W. L. Chow, K. Keyshar, G. L. Ye, R. Vajtai, J. Lou, Z. Liu, E. Ringe, B. K. Tay, P. M. Ajayan, *ACS Nano* **2014**, *8*, 5125.

[158] J. K. Huang, J. Pu, C. L. Hsu, M. H. Chiu, Z. Y. Juang, Y. H. Chang, W. H. Chang, Y. Iwasa, T. Takenobu, L. J. Li, *ACS Nano* **2013**, *8*, 923.

[159] L. Song, L. J. Ci, H. Lu, P. B. Sorokin, C. H. Jin, J. Ni, A. G. Kvashnin, D. G. Kvashnin, J. Lou, B. I. Yakobson, P. M. Ajayan, *Nano Lett.* **2010**, *10*, 3209.

[160] S. M. Kim, A. Hsu, M. H. Park, S. H. Chae, S. J. Yun, J. S. Lee, D. H. Cho, W. J. Fang, C. G. Lee, T. Palacios, M. Dresselhaus, K. K. Kim, Y. H. Lee, J. Kong, *Nat. Commun.*, **2015**, *6*, 8662.

[161] Y. Gao, W. C. Ren, T. Ma, Z. B. Liu, Y. Zhang, W. B. Liu, L. P. Ma, X. L. Ma, H. M. Cheng, *ACS Nano* **2013**, *7*, 5199.

[162] J. B. Smith, D. Hagaman, H. F. Ji, *Nanotechnology* **2016**, *27*, 215602.

[163] C. Pan, Y. Ji, N. Xiao, F. Hui, K. Tang, Y. Guo, X. Xie, F. M. Puglisi, L. Larcher, E. Miranda, L. Jiang, Y. Shi, I. Valov, P. C. McIntyre, R. Waser, M. Lanza, *Adv. Funct. Mater.*, **2017**, *27*, 1604811.

[164] M. Lanza, Y. Wang, T. Gao, A. Bayerl, M. Porti, M. Nafria, Y. B. Zhou, G. Y. Jin, Z. F. Liu, Y. F. zhang, D. P. Yu, H. L. Duan, *Nano Res.* **2013**, *6*, 485.

[165] J. W. Suk, A. Kitt, C. Magnuson, Y. F. Hao, S. Ahmed, J. An, A. K. Swan, B. B. Goldberg, R. S. Ruoff, *ACS Nano* **2011**, *5*, 6916.

[166] M. Lanza, A. Bayerl, T. Gao, M. Porti, M. Nafria, G. Y. Jing, Y. F. Zhang, Z. F. Liu, H. L. Duan, *Adv. Mater.* **2013**, *25*, 1440.

[167] M. Lanza, T. Gao, Z. X. Yin, Y. F. Zhang, Z. F. Liu, Y. Z. Tong, Z. Y. Shen, H. L. Duan, *Nanoscale* **2013**, *5*, 10816.

[168] X. S. Li, Y. W. Zhu, W. W. Cai, M. Borysiak, B. Y. Han, D. Chen, R. D. Piner, L. Colombo, R. S. Ruoff, *Nano Lett.* **2009**, *9*, 4359.

[169] L. B. Gao, G. X. Ni, Y. P. Liu, B. Liu, A. H. Castro Neto, K. P. Loh, *Nature* **2014**, *505*, 190.

[170] K. Lee, L. Hwang, S. Lee, S. Oh, D. Lee, C. K. Kim, Y. Nam, S. Hong, C. Yoon, R. B. Morgan, H. Kim, S. Seo, D. H. Seo, S. Lee, B. H. Park, *Sci. Rep.* **2015**, *5*, 11279.

[171] J. H. Lee, C. Du, K. Sun, E. Kioupakis, W. D. Lu, *ACS Nano* **2016**, *10*, 3571.

[172] A. Pirkle, J. Chan, A. Venugopal, D. Hinojos, C. W. Magnuson, S. Mcdonnell, L. Colombo, E. M. Vogel, R. S. Ruoff, R. M. Wallace, *Appl. Phys. Lett.* **2011**, *99*, 122108.

[173] D. Y. Wang, I. S. Huang, P. H. Ho, S. S. Li, Y. C. Yeh, D. W. Wang, W. L. Chen, Y. Y. Lee, Y. M. Chang, C. C. Chen, C. T. Liang, C. W. Chen, *Adv. Mater.* **2013**, *25*, 4521.

[174] A. P. Esser-Kahn, P. R. Thakre, H. Dong, J. F. Patrick, V. K. Vlasko-Vlasov, N. R. Sottos, J. S. Moore, S. R. White, *Adv. Mater.* **2011**, *23*, 3654.

- 134 -

**ADVANCED SCIENCE NEWS**
www.advancedsciencenews.com

**ADVANCED ELECTRONIC MATERIALS**
www.advelectronicmat.de

[175] C. G. Willson, R. R. Dammel, A. Reiser, *Proc. SPIE* 1997, 3094, 28.

[176] P. Sutter, J. Lahiri, P. Zahl, B. Wang, F. Sutter, *Nano Lett.* 2013, 13, 276.

[177] W. Yang, G. R. Chen, Z. W. Shi, C. C. Liu, L. C. Zhang, G. B. Xie, M. Cheng, D. M. Wang, R. Yang, D. X. Shi, K. Watanabe, T. Taniguchi, Y. G. Yao, Y. B. Zhang, G. Y. Zhang, *Nat. Mater.* 2013, 12, 792.

[178] S. Ratha, C. S. Rout, *Appl. Mater. Interfaces* 2013, 5, 11427.

[179] X. R. Wang, S. M. Tabakman, H. J. Dai, *J. Am. Chem. Soc.* 2008, 130, 8152.

[180] S. McDonnell, B. Brennan, A. Azcatl, N. Lu, H. Dong, C. Buie, J. Kim, C. L. Hinkle, M. J. Kim, R. M. Wallace, *ACS Nano* 2013, 7, 10354.

[181] H. Y. Nan, Z. L. Wang, W. H. Wang, Z. Liang, Y. Lu, Q. Chen, D. W. He, P. H. Tan, F. Miao, X. R. Wang, J. L. Wang, Z. H. Ni, *ACS Nano* 2014, 8, 5738.

[182] A. A. Demkov, O. F. Sankey, *Phys. Rev. Lett.* 1999, 83, 2038.

[183] J. L. May, M. Hirose, *J. Appl. Phys.* 1997, 81, 1606.

[184] B. Brar, G. D. Wilk, A. C. Seabaugh, *Appl. Phys. Lett.* 1996, 69, 2728.

[185] B. Chakrabarti, T. Roy, E. M. Vogel, *IEEE Electron Device Lett.* 2014, 35, 7.

[186] Y. Bai, H. Wu, K. Wang, R. Wu, L. Song, T. Li, J. Wang, Z. Yu, H. Qian. *Sci. Rep.* 2015, 5, 13785.

[187] G. Fisichella, G. Greco, F. Roccaforte, F. Giannazzo, *Nanoscale* 2014, 6, 8671.

[188] J. W. Seo, J. W. Park, K. S. Lim, J. H. Yang, S. J. Kang, *Appl. Phys. Lett.* 2008, 93, 223505.

[189] H. D. Kim, H. M. An, Y. Seo, T. G. Kim, *IEEE Electron Device Lett.* 2011, 32, 1125.

[190] K. C. Liu, W. H. Tzeng, K. M. Chang, Y. C. Chan, C. C. Kuo, *Microelectron. Eng.* 2011, 88, 1586.

[191] J. Lee, O. Kim, *Jpn. J. Appl. Phys.* 2011, 50, 06GF01.

[192] H. Zhao, H. Tu, F. Wei, J. Du, *IEEE Electron Devices Lett.* 2014, 61, 5.

[193] K. Zheng, X. W. Sun, J. L. Zhao, Y. Wang, H. Y. Yu, H. V. Demir, K. L. Teo, *IEEE Electron Devices Lett.* 2011, 32, 6.

[194] Y. S. Ji, S. A. Lee, A. N. Cha, M. Goh, S. Bae, S. Lee, D. I. Son, T. W. Kim. *Organic Electron.* 2015, 18, 77.

[195] D. I. Son, T. W. Kim, J. H. Shim, J. H. Jung, D. U. Lee, J. M. Lee, W. I. Park, W. K. Choi, *Nano Lett.* 2010, 10, 2441.

[196] M. Qian, Y. Pan, F. Liu, M. Wang, H. Shen, D. He, B. Wang, Y. Shi, F. Miao, X. Wang, *Adv. Mater.* 2014, 26, 3275.

[197] H. Shima, F. Takano, H. Muramatsu, H. Akinaga, Y. Tamai, I. H. Inque, H. Takagi, *Appl. Phys. Lett.* 2008, 93, 113504.

[198] H. Shima, T. Nakano, H. Akinaga, *Appl. Phys. Lett.* 2010, 96, 192107.

[199] H. Lv, M. Wang, H. Wan, Y. Song, W. Luo, P. Zhou, T. Tang, Y. Lin, R. Huang, S. Song, J. G. Wu, H. M. Wu, M. H. Chi, *Appl. Phys. Lett.* 2009, 94, 213502.

[200] C. Y. Lin, D. Y. Lee, S. Y. Wang, C. C. Lin, T. Y. Tseng, *Surface Coatings Technol.* 2008, 203, 628.

[201] T. N. Fang, S. Kaza, S. Haddad, A. Chen, Y. C. Wu, Z. Lan, S. Avanzino, D. Liao, C. Gopalan, S. Choi, S. Mahdavi, M. Buynoski, Y. Lin, C. Marrian, C. Bill, M. VanBuskirk, M. Taguchi, *Tech. Dig. Int. Electron Devices Meeting*, San Francisco, CA, USA, 2006.

[202] C. B. Lee, B. S. Kang, A. Benayad, M. J. Lee, S. E. Ahn, K. H. Kim, G. Stefanovich, Y. Park, I. K. Yoo, *Appl. Phys. Lett.* 2008, 93, 042115.

[203] K. M. Kim, B. J. Choi, Y. C. Shin, S. Choi, C. S. Hwang, *Appl. Phys. Lett.* 2007, 91, 012907.

[204] S. Seo, M. J. Lee, D. C. Kim, S. E. Ahn, B.-H. Park, Y. S. Kim, I. K. Yoo, I. S. Byun, I. R. Hwang, S. H. Kim, J. S. Kim, J. S. Choi, J. H. Lee, S. H. Jeon, S. H. Hong, B. H. Par, *Appl. Phys. Lett.* 2005, 87, 263507.

[205] X. Lian, X. Cartoixa, E. Miranda, L. Perniola, R. Rurali, S. Long, M. Liu, J. Sune, *J. Appl. Phys.* 2014, 115, 244507.

[206] X. Lian, M. Lanza, A. Rodriguez, E. Miranda, J. Suñé, *IEEE Int. Conf. Solid-State Integr. Circuit Technol.*, Guilin, China, 28–31 October 2014.

[207] S. B. Long, X. J. Lian, C. Cagli, X. Cartoix_a, R. Rurali, E. Miranda, D. Jim_enez, L. Perniola, M. Liu, J. Sune, *Appl. Phys. Lett.* 2013, 102, 183505.

[208] Y. C. Yang, W. Lu, *Nanoscale* 2013, 5, 10076.

[209] V. V. N. Obreja, C. Codreanu, O. Poenar, O. Buiu, *Microelectron. Reliab.* 2011, 51, 536.

[210] M. Lanza, M. Porti, M. Nafría, X. Aymerich, E. Whittaker, B. Hamilton, *Microelectron. Reliab.* 2010, 50, 1312.

[211] S. H. Seo, J. S. Hwang, J. M. Yang, W. J. Hwang, J. Y. Song, W. J. Lee, *Thin Solid Films* 2013, 546, 14.

[212] S. S. Hwang, S. Y. Jung, Y. C. Joo, *J. Appl. Phys.* 2008, 104, 044511.

[213] N. Raghavan, K. L. Pey, K. Shubhakar, M. Bosman, *IEEE Electron Device Lett.* 2011, 32, 78.

[214] E. A. Miranda, C. Walczyk, C. Wenger, T. Schroeder, *IEEE Electron Device Lett.* 2010, 31, 609.

[215] J. S. Bunch, S. S. Verbridge, J. S. Alden, A. M. van der Zande, J. M. Parpia, H. G. Craighead, P. L. McEuen, *Nano Lett.* 2008, 8, 2458.

[216] J. B. Oostinga, H. B. Heersche, X. L. Liu, A. F. Morpurgo, L. M. K. Vandersypen, *Nat. Mater.* 2008, 7, 151.

[217] H. Tian, H. Y. Chen, B. Gao, S. M. Yu, J. L. Liang, Y. Yang, D. Xie, J. F. Kang, T. L. Ren, Y. G. Zhang, H. S. P. Wong. *Nano Lett.* 2013, 13, 651.

[218] N. T. Kirkland, T. Schiller, N. Medhekar, N. Birbilis, *Corros. Sci.* 2012, 56, 1.

[219] L. L. Zhang, R. Zhou, X. S. Zhao, *J. Mater. Chem.* 2010, 20, 5983.

[220] C. Y. Chen, J. R. D. Retamal, D. H. Lien, M. W. Chen, I. W. Wu, Y. Ding, Y. L. Chueh, C. I. Wu, J. H. He, *ACS Nano* 2012, 6, 9366.

[221] Q. H. Li, T. Gao, Y. G. Wang, T. H. Wang, *Appl. Phys. Lett.* 2005, 86, 123117.

[222] C. Y. Chen, C. A. Lin, J. H. He, *Nanotechnology* 2009, 20, 185605.

[223] W. K. Hong, G. Jo, S. S. Kwon, S. Song, T. Lee, *IEEE Trans. Electron Dev.* 2008, 55, 3020.

[224] J. H. He, C. H. Ho, C. Y. Chen, *Nanotechnology* 2009, 20, 065503.

[225] Z. Fan, D. Wang, P. C. Chang, W. Y. Tseng, J. G. Lu, *Appl. Phys. Lett.* 2004, 85, 5923.

[226] C. Y. Chen, M. W. Chen, J. J. Ke, C. A. Lin, J. R. D. Retamal, J. H. He, *Pure Appl. Chem.* 2010, 82, 2055.

[227] M. W. Chen, J. R. D. Retamal, C. Y. Chen, J. H. He, *IEEE Electron Dev. Lett.* 2012, 33, 411.

[228] V. E. Henrich, P. A. Cox, *Surface Science of Metal Oxides*, Cambridge University Press, Cambridge, UK, 1994.

[229] H. Jeon, J. Park, W. Jang, H. Kim, S. Ahn, K.-J. Jeon, H. Seo, H. Jeon, *Carbon* 2014, 75, 209.

[230] L. G. Cançado, A. Jorio, E. M. Ferreira, F. Stavale, C. A. Achete, R. B. Capaz, M. V. O. Moutinho, A. Lombardo, T. S. Kulmala, A. C. Ferrari, *Nano Lett.* 2011, 11, 3190.

[231] Y. C. Yang, J. Lee, S. Lee, C. H. Liu, Z. H. Zhong, W. Lu, *Adv. Mater.* 2014, 26, 3693.

[232] S. Yu, H. Y. Chen, B. Gao, J. Kang, H. S. P. Wong, *ACS Nano* 2013, 7, 2320.

[233] J. Sohn, S. Lee, Z. Jiang, H. Y. Chen, H. S. P. Wong, *IEEE Int. Electron Dev. Meeting (IEDM)*, San Francisco, CA, USA, 15–17 Dec, 2014.

[234] S. Stankovich, D. A. Dikin, G. H. B. Dommett, K. M. Kohlhaas, E. J. Zimney, E. A. Stach, R. D. Piner, S. T. Nguyen, R. S. Ruoff, *Nature* 2006, 442, 282.

[235] L. Wang, I. Meric, P. Y. Huang, Q. Gao, Y. Gao, H. Tran, T. Taniguchi, K. Watanabe, L. M. Campos, D. A. Muller, J. Guo, P. Kim, J. Hone, K. L. Shepard, C. R. Dean, *Science* 2013, 342, 614.

**1600195 (29 of 32)**

- 135 -

[236] J. H. Chen, C. Jang, S. D. Xiao, M. Ishigami, M. S. Fuhrer, *Nat. Nanotechnol.* **2008**, *3*, 206.

[237] L. Gao, W. Ren, H. Xu, L. Jin, Z. Wang, T. Ma, L. P. Ma, Z. Zhang, Q. Fu, L. M. Peng, X. Bao, H. M. Cheng, *Nat. Commun.* **2012**, *3*, 699.

[238] D. R. Stewart, D. A. A. Ohlberg, P. A. Beck, Y. Chen, R. S. Williams, J. O. Jeppesen, *Nano Lett.* **2004**, *4*, 133.

[239] Y. Yang, J. Y. Ouyang, L. P. Ma, R. J. H. Tseng, C. W. Chu, *Adv. Funct. Mater.* **2006**, *16*, 1001.

[240] C. C. Lin, B. C. Tu, C. C. Lin, C. H. Lin, T. Y. Tseng, *IEEE Electron Device Lett.*, **2006**, *27*, 725.

[241] C. L. He, F. Zhuge, X. F. Zhou, M. Li, G. C. Zhou, Y. W. Liu, J. Z. Wang, B. Chen, W. J. Su, Z. P. Liu, Y. H. Wu, P. Cui, R. W. Li, *Applied Physics Letters* **2009**, *95*, 232101.

[242] M. D. Yi, Y. Cao, H. F. Ling, Z. Z. Du, L. Y. Wang, T. Yang, Q. L. Fan, L. H. Xie, W. Huang, *Nanotechnology* **2014**, *25*, 185202.

[243] D. Y. Yun, T. W. Kim, *Carbon* **2015**, *88*, 26.

[244] D. H. Yoo, T. V. Cuong, S. H. Hahn, *Current Appl. Phys.* **2014**, *14*, 1301.

[245] F. Zhuge, B. Hu, C. He, X. Zhou, Z. Liu, R. W. Li, *Carbon* **2011**, *49*, 3796.

[246] G. Khurana, P. Misra, N. Kumar, R. S. Katiyar, *J. Phys. Chem. C* **2014**, *118*, 21357.

[247] P. Hazra, A. N. Resmi, K. B. Jinesh, *Appl. Phys. Lett.* **2016**, *108*, 153503.

[248] S. K. Hong, J. E. Kim, S. O. Kim, B. J. Cho, *J. Appl. Phys.* **2011**, *110*, 044506.

[249] S. L. Hong, J. E. Kim, S. O. Kim, S. Y. Choi, B. J. Cho, *IEEE Electron Device Lett.* **2010**, *31*, 1005.

[250] S. M. Jilani, T. D. Gamot, P. Banerji, S. Chakraborty, *Carbon* **2013**, *64*, 187.

[251] G. Venugopal, S. J. Kim, *J. Nanosci. Nanotechnol.* **2012**, *12*, 8522.

[252] G. Khurana, P. Misra, R. S. Katiyar, *Appl. Phys. Lett.* **2013**, *114*, 124508.

[253] G. N. Panin, O. O. Kapitanova, S. W. Lee, A. N. Baranov, T. W. Kang, *Jpn J. Appl. Phys.* **2011**, *50*, 070110.

[254] H. Tian, H. Y. Chen, T. L. Ren, C. Li, Q. T. Xue, M. A. Mohammad, C. Wu, Y. Yang, H. S. Philip Wong, *Nano Lett.* **2014**, *14*, 3214.

[255] X. Zhou, Z. Liu, *Chem. Commun.* **2010**, *46*, 2611.

[256] S. K. Hong, J. E. Kim, S. O. Kim, B. J. Cho, *2010 10th IEEE Conference on Nanotechnology* **2010**, *604*, DOI: 10.1109/NANO.2010.5697794.

[257] H. S. Ki, C. B Jin, *4th IEEE International in Nanoelectronics Conference* **2011**, DOI: 10.1109/INEC.2011.5991806.

[258] C. Wu, F. Li, Y. Zhang, T. Guo, T. Chen, *Appl. Phys. Lett.* **2011**, *99*, 042108.

[259] O. O. Ekiz, M. Urel, H. Guner, A. K. Mizrak, A. Dana, *ACS Nano* **2011**, *5*, 2475.

[260] K. S. Vasu, S. Sampath, A. K. Sood, *Solid State Commun.* **2011**, *151*, 1084.

[261] L. H. Wang, W. Yang, Q.-Q. Sun, P. Zhou, H.-L. Lu, S.-J. Ding, D. Wei Zhang, *Appl. Phys. Lett.* **2012**, *100*, 063509.

[262] H. D. Kim, M. J. Yun, J. H. Lee, K. H. Kim, T. G. Kim, *Sci. Rep.* **2014**, *4*, 4614.

[263] C. L. He, Z. W. Shi, L. C. Zhang, W. Yang, R. Yang, D. X. Shi, G. Y. Zhang, *ACS Nano* **2012**, *6*, 4214.

[264] J. Kim, M. Siddik, J. Shin, K. P. Biju, S. Jung, H. Hwang, *Appl. Phys. Lett.* **2011**, *99*, 042101.

[265] X. Dong, W. Huang, P. Chen, *IEEE Trans. Nanotechnol.* **2010**, *99*, 1.

[266] S. K. Hong, J. E. Kim, S. O. Kim, B. J. Cho, *J. Appl. Phys.* **2011**, *110*, 044506.

[267] Z. R. Wang, V. Tjoa, L. Wu, W. J. Liu, Z. Fang, X. A. Tran, J. Wei, W. G. Zhu, H. Y. Yod, *J. Electrochem. Soc.* **2012**, *159*, 177.

[268] K. C. Chang, R. Zhang, *IEEE Electron Device Lett.* **2013**, *34*.

[269] X. L. Li, H. L. Wang, J. T. Robinson, H. Sanchez, G. Diankov, H. J. Dai, *J. Am. Chem. Soc.* **2009**, *131*, 15939.

[270] F. Zhao, J. Liu, X. Huang, X. Zou, G. Lu, P. Sun, S. Wu, W. Ai, M. D. Yi, X. Y. Qi, L. Xie, J. Wang, H. Zhang, W. Huang, *ACS Nano* **2012**, *6*, 3027.

[271] S. K. Pradhan, B. Xiao, S. Mishra, A. Killam, A. K. Pradhan, *Sci. Rep.* **2016**, *6*, 26763.

[272] N. T. Ho, V. Senthilkumar, Y. S. Kim, *Solid-State Electron.* **2014**, *94*, 61.

[273] B. L. Hu, R. Quhe, C. Chen, F. Zhuge, X. J. Zhu, S. S. Peng, X. X. Chen, L. Pan, Y. Z. Wu, W. Zheng, Q. Yan, J. Lu, R. W. Li, *J. Mater. Chem.* **2012**, *22*, 16422.

[274] J. C. Jin, J. Lee, E. Lee, F. Hwang, H. Lee, *Chem. Commun.* **2012**, *48*, 4235.

[275] G. Liu, X. D. Zhuang, Y. Chen, B. Zhang, J. H. Zhu, C. X. Zhu, K. G. Neoh, E. T. Kang, *Appl. Phys. Lett.* **2009**, *95*, 253301.

[276] X. D. Zhuang, Y. Chen, G. Liu, P. P. Li, C. X. Zhu, E. T. Kang, K. G. Noeh, B. Zhang, J. H. Zhu, Y. X. Li, *Adv. Mater.* **2010**, *22*, 15, 1731.

[277] L. Valentini, M. Cardinali, E. Fortunti, J. M. Kenny, *Appl. Phys. Lett.* **2014**, *105*, 153111.

[278] S. Pinto, R. Krishna, C. Dias, G. Pimentel, G. N. P. Oliveira, J. M. Teixeira, P. Aguiar, F. Titus, J. Gracio, J. Ventura, J. P. Araujo, *Appl. Phys. Lett.* **2012**, *101*, 063104.

[279] J. Liu, Z. Yin, X. Cao, F. Zhao, A. Lin, L. Xie, Q. Fan, F. Boey, H. Zhang, W. Huang, *ACS Nano* **2010**, *4*, 3987.

[280] CareRAMM public summary: http://emps.exeter.ac.uk/media/universityofexeter/emps/careramm/D4.4_Public_summary_of_project_results_from_the_third_year_of_the_project.pdf (accessed August 2016).

[281] C. Casiraghi, J. Robertson, A. C. Ferrari, *Mater. Today* **2007**, *10*, 44

[282] A. C. Ferrari, *Surf. Coat. Technol.* **2004**, *180*, 190.

[283] P. J. Fallon, V. S. Veerasaamy, C. A. Davis, J. Robertson, G. A. J. Amaratunga, W. I. Milne, J. Koskinen, *Phys. Rev. B* **1993**, *48*, 4777.

[284] M. C. Polo, J. L. Andujar, A. Hart, J. Robertson, W. I. Milne, *Diamond Relat. Mater.* **2000**, *9*, 663.

[285] A. Sebastian, A. Pauza, C. Rossel, R. M. Shelby, A. F. Rodriguez, H. Pozidis, E. Eleftheriou, *New J. Phys.* **2011**, *13*, 013020.

[286] F. Kreupl, R. Bruchhaus, P. Majewski, J. B. Philipp, R. Symanczyk, T. Happ, C. Arndt, M. Vogt, R. Zimmermann, A. Buerke, A. P. Graham, M. Kund, *IEDM 2008 Technical Digest* **2008**, *15*, 521.

[287] P. Peng, D. Xie, Y. Yang, C. Zhou, S. Ma, T. Feng, H. Tian, T. Ren, *J. Phys D: Appl. Phys.* **2012**, *45*, 365103.

[288] D. Fu, D. Xie, T. T. Feng, C. H. Zhang, J. B. Niu, H. Qian, L. T. Liu, *IEEE Electron Device Lett.* **2011**, *32*, 80.

[289] J. Xu, D. Xie, T. Feng, C. Zhang, X. Zhang, P. Peng, D. Fu, H. Qian, T. L. Ren, L. Liu, *Carbon* **2014**, *75*, 255.

[290] Y. Chai, Y. Wu, K. Takei, H.-Y. Chen, S. Yu, P. C. H. Chan, A. Javey, H.-S. P. Wong, *IEEE Trans. Electron. Dev.* **2011**, *58*, 3933.

[291] B. Ren, L. Wang, L. Wang, J. Huang, K. Tang, Y. Lou, D. Yuan, Z. Pan, Y. Xia, *Vacuum* **2014**, *107*, 1.

[292] K. Antonowicz, A. Jesmanowicz, J. Wieczorek, *Carbon* **1972**, *10*, 81.

[293] A. C. Ferrari, J. Robertson, *Philos. Trans. R. Soc.*, **A 2004**, *362*, 2477.

[294] C. A. Santini, A. Sebastian, C. Marchiori, V. P. Jonnalagadda, L. Dellmann, W. W. Koelmans, M. D. Rossell, C. P. Rössel, E. Eleftheriou, *Nat. Commun.* **2015**, *6*, 1.

[295] H. Chen, F. Zhuge, B. Fu, J. Li, J. Wang, W. Wang, Q. Wang, L. Li, F. Li, H. Zhang, L. Liang, H. Luo, M. Wang, J. Gao, H. Cao, H. Zhang, Z. Li, *Carbon* **2014**, *76*, 459.

Article 6

[296] V. K. Nagareddy, A. K. Ott, C. Dou, T. Tsvetkova, M. Sandulov, M. F. Craciun, A. C. Ferrari, C. D. Wright, unpublished.

[297] S. Zhang, J. Zhou, D. Zhang, B. Ren, L. Wang, J. Huang, L. Wang, *Vacuum* 2016, 125, 189.

[298] M. Pyun, H. Choi, J. B. Park, D. Lee, M. Hasan, R. Dong, S. J. Jung, J. Lee, D. J. Seong, J. Yoon, H. Hwang, *Appl. Phys. Lett.* 2008, 93, 212907.

[299] H. Choi, M. Pyun, T. W. Kim, M. Hasan, R. Dong, J. Lee, J. B. Park, J. Yoon, D. J. Seong, T. Lee, H. Hwang, *IEEE Electron Dev. Lett.* 2009, 30, 302.

[300] D. I. Kim, J. Yoon, J.-B. Park, H. Hwang, Y. M. Kim, S. H. Kwon, K. H. Kim, *Appl. Phys. Lett.* 2011, 98, 152107.

[301] W. W. Koelmans, T. Bachmann, F. Zipoli, A. K. Ott, C. Dou, A. C. Ferrari, O. Cojocaru-Mirédin, S. Zhang, M. Wuttig, V. K. Nagareddy, M. F. Craciun, A. Alexeev, C. D. Wright, V. P. Jonnalagadda, A. Curioni, A. Sebastian, E. Eleftheriou, *Carbon-based resistive memories, IEEE (IMW)*, 1–4 (2016).

[302] T. A. Bachmann, A. M. Alexeev, W. W. Koelmans, F. Zipoli, A. K. Ott, C. Dou, A. C. Ferrari, V. K. Nagareddy, M. F. Craciun, V. P. Jonnalagadda, A. Curioni, A. Sebastian, E. Eleftheriou, C. D. Wright, *IEEE NMDC* 2016.

[303] F. Kreupl, *Carbon Memory Assessment*, white paper for the ITRS meeting on emerging research devices (ERD) in Albuquerque, New Mexico, on August 25–26, 2014.

[304] X. Zhao, H. Xu, Z. Wang, L. Zhang, J. Ma, Y. Liu, *Carbon* 2015, 91, 38.

[305] W. Dai, P. Ke, A. Wang, *J. Vac. Sci. Technol. B* 2013, 31, 031207.

[306] J. Park, M. Jo, J. Lee, S. Jung, W. Lee, S. Kim, S. Park, J. Shin, H. Hwang, *Microelectron. Eng.* 2011, 88, 935.

[307] X. Gao, X. Zhang, C. Wan, J. Wang, X. Tan, D. Zeng, *Diamond Rel. Mater.* 2012, 22, 37.

[308] A. K. Ott, C. Dou, U. Sassi, I. Goykhman, A. Katsounaros, D. Yoon, X. Chen, J. Wu, A. Lombardo, A. C. Ferrari, unpublished.

[309] Y. Chai, A. Hazeghi, K. Takei, H. Y. Chen, P. C. H. Chan, A. Javey, H. S. P. Wong, *International Electron Devices Meeting - Technical Digest* 2010, 210.

[310] D. Fu, D. Xie, C.-H. Zhang, D. Zhang, J.-B. Niu, H. Qian, L.-T. Liu, *Chin. Phys. Lett.* 2010, 27, 098102.

[311] Y. He, J. Zhang, X. Guan, L. Zhao, Y. Wang, H. Qian, Z. Yu, *IEEE Trans. Electron Devices* 2010, 57, 3434.

[312] S. Qin, J. Zhang, D. Fu, D. Xie, Y. Wang, H. Qian, L. Liu, Z. Yu, *Nanoscale* 2012, 4, 6658.

[313] C. D. Wright, P. Hosseini, J. A. Vazquez Diosdado, *Adv. Funct. Mater.* 2013, 23, 2248.

[314] Y. V. Pershin, M. Di Ventra, *Adv. Phys.* 2011, 60, 145.

[315] K. Sangwan, D. Jariwala, I. S. Kim, K. S. Chen, T. J. Marks, L. J. Lauhon, M. C. Hersam, *Nat. Nanotechnol.* 2015, 10, 403.

[316] P. F. Cheng, K. Sun, Y. H. Hu, *Nano Lett.* 2016, 16, 572.

[317] X. Zhang, H. Qiao, X. Nian, Y. Huang, X. Pang, *J. Mater. Sci. Mater. Electron.* 2016, 27, 7609.

[318] C. X. Hao, F. S. Wen, J. Y. Xiang, S. J. Yuan, B. C. Yang, L. Li, W. H. Wang, Z. M. Zeng, L. M. Wang, Z. Y. Liu, Y. J. Tian, *Adv. Funct. Mater.* 2016, 26, 2016.

[319] A. Ambrosi, Z. Sofer, M. Pumera, *Chem. Commun.* 2015, 51, 8450.

[320] J. Liu, Z. Zeng, X. Cao, G. Lu, L. H. Wang, Q. L. Fan, W. Huang, H. Zhang, *Small* 2012, 8, 3517.

[321] S. Ali, J. Bae, C. H. Lee, *Proc. SPIE* 2015, 9553, 95530T.

[322] K. Ali, J. Ali, S. M. Mehdi, K. Choi, Y. J. An, *Appl. Surf. Sci.* 2015, 353, 1186.

[323] S. T. Han, Y. Zhou, B. Chen, C. Wang, L. Zhou, Y. Yan, J. Zhuang, Q. Sun, H. Zhang, V. A. L. Roy, *Small* 2015, 12, 390.

[324] S. M. Shinde, G. Kalita, M. Tanemura, *J. Appl. Phys.* 2014, 116, 214306.

[325] J. Mangalam, S. Agarwal, A. N. Resmi, M. Sundararajan, K. B. Jinesh, *Org. Electron.* 2016, 29, 33.

[326] Z. Yin, Z. Zeng, J. Liu, Q. He, P. Chen, H. Zhang, *Small* 2013, 9, 727.

[327] A. A. Bessonov, M. N. Kirikova, D. I. Petukhov, M. Allen, T. Ryhänen, M. J. a Bailey, *Nat. Mater.* 2015, 14, 199.

[328] X. Y. Xu, Z. Y. Yin, C. X. Xu, J. Dai, J. G. Hu, *Appl. Phys. Lett.* 2014, 104, 033504.

[329] K. Qian, R. Y. Tay, V. C. Nguyen, J. X. Wang, G. F. Cai, T. P. Chen, E. H. T. Teo, P. S. Lee, *Adv. Funct. Mater.* 2016, 26, 2176.

[330] N. D. Zhigadlo, *J. Cryst. Growth* 2014, 402, 308.

[331] D. Golberg, P. M. F. J. Costa, O. Lourie, M. Mitome, X. Bai, K. Kurashima, C. Zhi, C. Tang, Y. Bando, *Nano Lett.* 2007, 7, 2146.

[332] C. W. Chang, D. Okawa, A. Majumdar, A. Zettl, *Science* 2006, 314, 1121.

[333] B. Ilhan, M. Kurt, H. Ertuk, *Exp. Therm. Fluid Sci.* 2016, 77, 272.

[334] P. Y. Huang, C. S. R. Vargas, A. M. V. D. Zande, W. S. Whitney, M. P. Levendorf, J. W. Kevek, S. Garg, J. S. Alden, C. J. Hustedt, Y. Zhu, J. Park, P. L. McEuen, D. A. Muller, *Nature* 2011, 469, 389.

[335] A. L. Gibb, N. Alem, J. H. Chen, K. J. Erickson, J. Ciston, A. Gautam, M. Linck, A. Zettl, *J. Am. Chem. Soc.* 2013, 135, 6758.

[336] Y. Hattori, T. Taniguchi, K. Watanabe, K. Nagashio, *ACS Nano* 2015, 9, 916.

[337] Y. F. Ji, C. B. Pan, M. Y. Zhang, S. B. Long, X. J. Lian, F. Miao, F. Hui, Y. Y. Shi, L. Larcher, E. Wu, M. Lanza, *Appl. Phys. Lett.* 2016, 108, 012905.

[338] N. Jain, R. B. Jacobs-Gedrim, B. Yu, *Mater. Res. Soc. Symp. Proc.* 2014, 1658, DOI: 10.1557/opl.2014.503.

[339] J. D. Wood, S. A. Wells, D. Jariwala, K. S. Chen, E. Cho, V. K. Sangwan, X. L. Liu, L. J. Lauhon, T. J. Marks, M. C. Hersam, *Nano Lett.* 2014, 14, 6964.

[340] D. Hanlon, C. Backes, E. Doherty, C. S. Cucinotta, N. C. Berner, C. Boland, K. Lee, P. Lynch, Z. Gholamvand, A. Harvey, S. F. Zhang, K. P. Wang, G. Moynihan, A. Pokle, Q. M. Ramasse, N. McEvoy, W. J. Blau, J. Wang, S. Sanvito, D. D. Regan, G. S. Duesberg, V. Nicolosi, J. N. Coleman, *Nat. Commun.* 2015, 6, 8563.

[341] T. W. Hickmott, *J. Appl. Phys.* 1962, 33, 2669.

[342] J. G. Simmons, R. R. Verderber, *Proc. R. Soc. London Ser. A* 1967, 301, 77.

[343] R. Degraeve, B. Kaczer, G. Groeseneken, *Microelectron. Reliab.* 1999, 39, 1445.

[344] L. Miranda, J. Sune, R. Rodriguez, M. Nafria, X. Aymerich, *Appl. Phys. Lett.* 1998, 73, 490.

[345] M. Ahmad, S. A. Han, D. H Tien, J. Jung, Y. Seo, *J. Appl. Phys.* 2011, 110, 054307.

[346] J. J. Yang, M. X. Zhang, J. P. Strachan, F. Miao, M. D. Pickett, R. D. Kelley, G. M. Ribeiro, R. S. Williams, *Appl. Phys. Lett.* 2010, 97, 232102.

[347] Y. B. Kim, S. R. Lee, D. Lee, C. B. Lee, M. Chang, J. H. Hur, M. J. Lee, G. S. Park, C. J. Kim, U. Chung, I. K. Yoo, K. Kim, 2011, *Symposium on VLSI Technology*.

[348] C. H. Cheng, A. Chin, F. S. Yeh, *2010 Symposium on VLSI Technology Digest of Technical Papers* 2010, 85.

[349] S. Liu, N. D. Lu, X. L. Zhao, H. Xu, W. Banerjee, H. B. Lv, S. B. Long, Q. J. Li, Q. Liu, M. Liu, *Adv. Mater.* 2016, 28, 10623.

[350] X. Cao, X. M. Li, X. D. Gao, W. D. Yu, X. J. Liu, Y. W. Zhang, L. D. Chen, X. H. Cheng, *J. Appl. Phys.* 2009, 106, 073723.

[351] B. Zhang, G. Liu, Y. Chen, L. J. Zeng, C. X. Zhu, K. G. Neoh, C. Wang, E. T. Kang, *Chem. Eur. J.* 2011, 17, 13646.

[352] A. Midya, N. Gogurla, S. K. Ray, *Current Appl. Phys.* 2015, 15, 706.

[353] A. N. Aleshin, P. S. Krylova, A. S. Berestennikova, I. P. Shcherbakova, V. N. Petrova, V. V. Kondratievb, S. N. Eliseeva, *Synth. Met.* 2016, 217, 7.

[354] Y. Sun, J. Lu, C. Ai, D. Wen, *Phys. Chem. Chem. Phys.* 2016, 18, 11341.

[355] A. Thakre, H. Borkar, B. P. Singha, A. Kumar, *RSC Adv.* 2015, 5, 57406.

[356] S. Valanarasu, I. Kulandaisamy, A. Kathalingam, J. K. Rhee, T. A. Vijayan, R. Chandramohan, *J. Nanosci. Nanotechnol.* 2013, 13, 6755.

[357] O. O. Kapitanova, G. N. Panin, O. V. Kononenko, A. N. Baranov, T. W. Kang, *J. Korean Phys. Soc.* 2014, 64, 1399.

[358] G. Khurana, P. Misra, N. Kumar, S. Kooriyattil, J. F Scott, R. S Katiyar, *Nanotechnology* 2016, 27, 015702.

[359] J. R. Rani, S. I. Oh, J. M. Woo, J. H. Jang, *Carbon* 2015, 94, 367.

[360] A. D. Yu, C. L. Liu, W. C. Chen, *Chem. Commun.* 2012, 48, 383.

[361] H. Y. Wu, C. C. Lin, C. H. Lin, *Ceram. Int.* 2015, 41, S823.

[362] X. Huang, B. Zheng, Z. Liu, C. Tan, J. Liu, B. Chen, H. Li, J. Chen, X. Zhang, Z. Fan, W. Zhang, Z. Guo, F. Huo, Y. Yang, L. H. Xie, W. Huang, H. Zhang, *ACS Nano* 2014, 8, 8695.

[363] C. Tan, X. Qi, Z. Liu, F. Zhao, H. Li, X. Huang, L. Shi, B. Zheng, X. Zhang, L. Xie, Z. Tang, W. Huang, H. Zhang, *J. Am. Chem. Soc.* 2015, 137, 1565.

[364] Z. Wei, Y. Kanzawa, K. Arita, Y. Katoh, K. Kawai, S. Muraoka, S. Mitani, S. Fujii, K. Katayama, M. Iijima, T. Mikawa, T. Ninomiya, R. Miyanaga, Y. Kawashima, K. Tsuji, A. Himeno, T. Okada, R. Azuma, K. Shimakawa, H. Sugaya, T. Takagi, R. Yasuhara, K. Horiba, H. Kumigashira, M. Oshima, *IEEE IEDM*, 15–17 Dec, 2008.

- 138 -

# Chapter 5:

# Conclusions and perspectives

In conclusion, during this PhD thesis I have learned how to grow high quality and large area *h*-BN stacks with different thicknesses using CVD method. This skill is very important because *h*-BN is a very demanded material, and because CVD is a method that can be used to grow many other 2D materials. I also learned how to transfer the 2D materials on target (arbitrary) substrates using three different methods (wet, bubbling, and roll-to-roll transfer). I also learned how to analyze the properties of 2D materials using different equipment (e.g. CAFM, SEM, Optical microscope, Raman and TEM). Moreover, I also fabricated *h*-BN based capacitors and memristors using photolithography, thermal evaporation, E-beam evaporation and sputtering, and I analyzed the properties of the devices using a probestation. Theoretical modeling and fittings (carried out with the help of my collaborators) helped me to understand the functioning of the devices. Overall, the main conclusions of my work are:

- Monolayer and multilayer *h*-BN can be grown by CVD on Pt, Cu and Fe substrates. The main parameters affecting the growth of the *h*-BN are: *i)* a proper temperature determines the decomposition of the precursor. Temperatures below a threshold value produce remaining particles and more defects in *h*-BN stack. *ii)* The flow rate of precursor/$H_2$ influences the density of seeds. Excessive precursor produces multilayer *h*-BN islands. *iii)* High vacuum and low pressure help to remove impurities in the tube furnace (e.g. oxygen, carbon), and therefore it produces better quality *h*-BN, i.e. uniform thickness with less defects.

- *h*-BN sheets grown on polycrystalline Pt substrates show different thicknesses depending on the crystallographic orientation at the surface of each Pt grain. This produces an undesired fluctuation on the leakage current from one Pt grain to another. However, the leakage current across the *h*-BN on the same Pt grain is very uniform, much more than that observed across amorphous $HfO_2$ and $TiO_2$ thin films. This phenomenon doesn't take place when growing the *h*-BN on Cu substrates. For example, the leakage current across *h*-BN grown on Cu substrates display small current variability among different Cu grains.

- The dielectric breakdown behavior in multilayer *h*-BN shows surface extrusion, similar to what happens in $SiO_2$, $HfO_2$ and $Al_2O_3$. However, monolayer *h*-BN keeps unaltered its structure even for harder breakdown events. The reason may be the extremely high thermal conductivity of monolayer *h*-BN.

- Multilayer *h*-BN shows random telegraph noise signals when applying constant voltage stresses, both at the device level and at the nanoscale. This strongly indicates the trapping and de-trapping of charges during the stress. This observation has been confirmed by the detection of charges at the dielectric breakdown location. The breakdown spot shows a singular ring-like structure that contains fixed negative charges, mobile negative charges, and positive fixed charges.

- The synthesis of *h*-BN on polycrystalline Fe substrates required longer cooling down times than when using Pt and Cu substrates. The reason is that the growth of *h*-BN on Fe substrates mainly takes place by surface precipitation mechanism, while on Pt and Cu substrates the mechanism is by surface-mediated reaction.

- Memristors with Ag/*h*-BN/Fe structure show threshold resistive switching when the set is induced by applying positive voltage to the Ag electrode, and bipolar resistive switching when the set/reset processes are induced by applying negative/positive voltage to the Ag electrode. The reason should be that in threshold mode the filament is formed by $Ag^+$ ions that penetrate in the h-BN stack, while in bipolar mode $Fe^+$ ions penetrate in the h-BN stack. $Ag^+$ ions show higher diffusivity than $Fe^+$ ions and produce volatile switching.

Apart from the technical skills gained from the experiments, during my PhD I have made a huge effort on literature revision and knowledge organization. In my case this contribution is bigger than in other PhD thesis, as I have written two extensive review papers with, in total, more than 543 references. In the first one, published in Microelectronics Engineering, I analyzed the status of h-BN as dielectric in electronic devices (prior to this PhD thesis). And in the second one I analyzed the use of 2D materials in resistive switching devices. This second review paper has been written in collaboration with Prof. Andrea Ferrari from University of Cambridge, and has been highlighted as front cover in Advanced Electronic Materials. This has given me a very wide vision on the use of 2D materials as dielectric, which is a skill that I wish to exploit in the future.

Future works in this direction should conduct RS studies in smaller devices, using cross point structures and the CAFM. Statistical analyses about the dielectric breakdown voltage and time in real devices are necessary. Analyzing the leakage current across the domain boundaries of the $h$-BN would be interesting to understand potential weaknesses of the material. Several parameters related to the dielectric breakdown process, such as charge-to-breakdown should be also analyzed. The most important characterization study would be to describe the performance of $h$-BN at high temperatures, as well as to observe the relationship between thermal conductivity and the degradation. However, the biggest improvement would be to grow *h*-BN using a single seed. This method has been used in the past to grow graphene, but it has never been applied to *h*-BN. In addition, not only *h*-BN but also other 2D insulating material (like graphene oxide, black phosphorus) should be explored and studied as dielectric. The range of possibilities is very wide, and the experiments and findings that will come in the next years very exciting !

# References

[1]     F. Hui, E. Grustan-Gutierrez, S. Long, Q. Liu, A. K. Ott, A. C. Ferrari, M. Lanza, Graphene and related materials for resistive random access memories. *Advanced Electronic Materials* **2017**, 1600195.

[2]     G. I. Meijer. Who wins the nonvolatile memory race? *Science* **2008**, *319*, 1625.

[3]     R. Chau, S. Datta, M. Doczy, J. Kavalieros, M. Metz. Gate dielectric scaling for high-performance CMOS: from $SiO_2$ to high-k, in *International Workshop on Gate Insulator 2003*, **2003**, 124.

[4]     R.M. Wallace, G. Wilk. Alternative gate dielectrics for microelectronics, *MRS Bulletin* **2002**, *27*, 186.

[5]     V. Mistra, G. Lucovsky, G. Parsons. Issue in high-k gate stack interfaces, *MRS Bulletin* **2002**, *27*,212.

[6]     X. Wang, S. M. Tabakman, H. Dai. Atomic layer deposition of metal oxides on pristine and functionalized graphene, *Journal of the American Chemical Society* **2008**, *130*, 8152.

[7]     S. McDonnell, B. Brennan, A. Azcatl, N. Lu, H. Dong, C. Buie, J. Kim, C. L. Hinkle, M. J. Kim, and R. M. Wallace. HfO2 on MoS2 by Atomic Layer Deposition: Adsorption Mechanisms and Thickness Scalability, *ACS Nano* **2013**, *7*, 10354.

[8]     K. Watanabe, T. Taniguchi, H. Kanda. Direct-bandgap properties and evidence for ultraviolet lasing of hexagonal boron nitride single crystal, *Nature Materials* **2004**, *3*, 404.

[9]     Z. Liu, Y. J. Gong, W. Zhou, L. L. Ma, J. J. Yu, J. C. Idrobo, J. Jung, A. H. MacDonald, R. Vajtai, J. Lou, P. M. Ajayan. Ultrathin High-Temperature Oxidation-Resistant Coatings of Hexagonal Boron Nitride, *Nature Communications* **2013**, *4*, 2541.

[10]    L. H. Li, E. J. G. Santos, T. Xing, E. Cappelluti, R. Roldán, Y. Chen, K. Watanabe, T. Taniguchi. Dielectric Screening in Atomically Thin Boron Nitride Nanosheets, *Nano Letters* **2015**, *15*, 218.

[11]    L. Song; L. J. Ci, H. Lu, P. B. Sorokin, C. H. Jin, J. Ni, A. G. Kvashnin, D. G. Kvashnin, J. Lou, B. I. Yakobson, P. M. Ajayan. Large Scale Growth and Characterization of Atomic Hexagonal Boron Nitride Layers, *Nano Letters* **2010**, *10*, 3209.

[12]    L. Lindsay, D. A. Broido. Enhanced Thermal Conductivity and Isotope Effect in Single-Layer Hexagonal Boron Nitride, *Physical Review B* **2011**, *84*, 155421.

[13]    Z. Lin, Y. Liu, S. Raghavan, K. Moon, S. K. Sitaraman, C. Wong. Magnetic

Alignment of Hexagonal Boron Nitride Platelets in Polymer Matrix: Toward High Performance Anisotropic Polymer Composites for Electronic Encapsulation, *ACS Applied Materials & Interfaces* **2013**, *5*, 7633.

[14] R. V. Gorbachev, I. Riaz, R. R. Nair, R. Jalil, L. Britnell, B. D. Belle, E. W. Hill, K. S. Novoselov, K. Watanabe, T. Taniguchi, A. K. Geim, P. Blake. Hunting for Monolayer Boron Nitride: Optical and Raman Signatures, *Small* **2011**, *7*, 465.

[15] C. R. Dean, A. F. Young, I. Meric, C. Lee, L. Wang, S. Sorgenfrei, K. Watanabe, T. Taniguchi, P. Kim, K. L. Shepard, J. Hone. Boron nitride substrates for high-quality graphene electronic, *Nature Nanotechnology* **2010**, *5*, 722.

[16] M. Monajjemi. Metal-doped graphene layers composed with boron nitride–graphene as an insulator: a nano-capacitor, *Journal of Molecular Modeling* **2014**, *20*, 2507.

[17] A. F. Khan, D. A. C. Brownson, C. W. Foster, G. C. Smith, C. E. Banks. Surfactant exfoliated 2D hexagonal Boron Nitride (2D-hBN) explored as a potential electrochemical sensor for dopamine: surfactants significantly influence sensor capabilities, *Analyst* **2017**, *142*, 1756.

[18] C. B. Pan, Y. F. Ji, N. Xiao, F. Hui, K. C. Tang, Y. Z. Guo, X. M. Xie, F. M. Puglisi, L. Larcher, E. Miranda, L. L. Jiang, Y. Y. Shi, I. Valov, P. C. McIntyre, R. Waser, M. Lanza. Coexistence of Grain-Boundaries-Assisted Bipolar and Threshold Resistive Switching in Multilayer Hexagonal Boron Nitride, *Advanced Functional Materials* **2017**, *27*, 1604811.

[19] C.B. Pan, E. Miranda, M. A. Villena, N. Xiao, X. Jing, X. Xie, T. Wu, F. Hui, Y. Shi, M. Lanza. Model for multi-filamentary conduction in graphene/hexagonalboron-nitride/graphene based resistive switching devices, *2D Materials* **2017**, *4*, 025099.

[20] F. M. Puglisi, L. Larcher, C. Pan, N. Xiao, Y. Shi, F. Hui, M. Lanza. 2D h-BN based RRAM devices, in *2016 IEEE International Electron Devices Meeting*, **2016**, 34.8. 1.

[21] M. Lanza, M. Porti, M. Nafria, G. Benstetter, W. Frammelsberger, H. Ranzinger, E. Lodermeier, G. Jaschke. Influence of the manufacturing process on the electrical properties of thin (<4 nm) Hafnium based high-k stacks observed with CAFM, *Microelectronics Reliability* **2007**, *47*, 1424.

[22] A. Bayerl, M. Lanza, M. Porti, F. Campabadal, M. Nafría, X. Aymerich, G. Benstetter. Reliability and gate conduction variability of $HfO_2$-based MOS devices: A combined nanoscale and device level study, *Microelectronic Engineering* **2011**, *88*, 1334.

[23] A. Bayerl, M. Lanza, L. Aguilera, M. Porti, M. Nafría, X. Aymerich, S. De Gendt. Nanoscale and device level electrical behavior of annealed ALD Hf-based gate oxide stacks grown with different precursors, *Microelectronics Reliability* **2013**, *53*, 867.

[24]  K. McKenna, A. Shluger, V. Iglesias, M. Porti, M. Nafría, M. Lanza, G. Bersuker. Grain boundary mediated leakage current in polycrystalline $HfO_2$ films, *Microelectronic Engineering* **2011**, *88*, 1272.

[25]  M. Lanza, M. Porti, M. Nafria, X. Aymerich, G. Benstetter, E. Lodermeier, H. Ranzinger, G. Jaschke, S. Teichert, L. Wilde, P. Michalowski. Crystallization and silicon diffusion nanoscale effects on the electrical propertiesof $Al_2O_3$ based devices, *Microelectronic Engineering* **2009**, *86*, 1921.

[26]  O. Pirrotta, L. Larcher, M. Lanza, A. Padovani, M. Porti, M. Nafría, G. Bersuker. Leakage current through the poly-crystalline $HfO_2$: Trap densities at grains and grain boundaries, *Journal of Applied Physics* **2013**, *114*, 134503.

[27]  M. Lanza, M. Porti, M. Nafría, X. Aymerich, G. Ghidini, A. Sebastiani. Trapped charge and stress induced leakage current (SILC) in tunnel $SiO_2$ layers of de-processed MOS non-volatile memory devices observed at the nanoscale, *Microelectronics Reliability* **2009**, *49*, 1188.

[28]  M. Lanza, M. Porti, M. Nafria, X. Aymerich, A. Sebastiani, G. Ghidini, A. Vedda, M. Fasoli, Combined nanoscale and device-level degradation analysis of $SiO_2$ layers of MOS nonvolatile memory devices, *IEEE Transactions on Device and Materials Reliability* **2009**, *9*, 529.

[29]  S. M. Sze. Current transport and maximum dielectric strength of silicon nitride films. *Journal of applied physics* **1967**, *38*, 2951.

[30]  M. Lanza, M. Porti, M. Nafría, X. Aymerich, E. Whittaker, B. Hamilton. UHV CAFM characterization of high-k dielectrics: Effect of the technique resolution on the pre- and post-breakdown electrical measurements, *Microelectronics Reliability* **2010**, *50*, 1312.

[31]  V. Iglesias, M. Lanza, A. Bayerl, M. Porti, M. Nafría, X. Aymerich, L. F. Liu, J. F. Kang, G. Bersuker, K. Zhang, Z. Y. Shen. Nanoscale observations of resistive switching high and low conductivity states on $TiN/HfO_2/Pt$ structures, *Microelectronics Reliability* **2012**, *52*, 2110.

[32]  N. Xiao, M. A. Villena, B. Yuan, S. Chen, B. Wang, M. Eliáš, Y. Shi, F. Hui, X. Jing, A. Scheuermann, K. Tang, P. C McIntyre, M. Lanza. Resistive random access memory cells with a bilayer $TiO_2/SiO_X$ insulating stack for simultaneous filamentary and distributed resistive switching, *Advanced Functional Materials* **2017**, *27*, 1700384.

[33]  Y. Shi, Y. Ji, F. Hui, V. Iglesias, M. Porti, M. Nafria, E. Miranda, G. Bersuker, M. Lanza. Elucidating the origin of resistive switching in ultrathin hafnium oxides through high spatial resolution tools, *ECS Transactions* **2014**, *64*, 19.

[34]  M. Lanza, U. Celano, F. Miao. Nanoscale characterization of resistive switching using advanced conductive atomic force microscopy based setups, *Journal of Electroceramics* **2017**, *39*, 94.

[35]   F. Hui, C. B. Pan, Y. Y. Shi, Y. F. Ji, E. Grustan-Gutierrez, M. Lanza. On the Use of Two Dimensional Hexagonal Boron Nitride as Dielectric, *Microelectronic Engineering* **2016**, *163*, 119.

[36]   Y. N. Xu, W. Y. Ching. Calculation of ground-state and optical properties of boron nitrides in the hexagonal, cubic, and wurtzite structures, *Physical Review B* **1991**, *44*, 7787.

[37]   N. Ooi, A. Rairkar, L. Lindsley, J. B. Adams. Electronic structure and bonding in hexagonal boron nitride, *Journal of Physics: Condensed Matter* **2006**, *18*, 97.

[38]   G. Giovannetti, P. A. Khomyakov, G. Brocks, P.J. Kelly, J. van den Brink. Substrate-induced band gap in graphene on hexagonal boron nitride: Ab initio density functional calculations, *Physical Review B* **2007**, *76*, 73103.

[39]   D. Akinwande, N. Petrone, J. Hone. Two-dimensional flexible nanoelectronics, *Nature Communications* **2014**, 5678.

[40]   A. L. Gibb, N. Alem, J. H. Chen, K. J. Erickson, J. Ciston, A. Gautam, M. Linck, A. Zettl. Atomic Resolution Imaging of Grain Boundary Defects in Monolayer Chemical Vapor Deposition-Grown Hexagonal Boron Nitride, *Journal of the American Chemical Society* **2013**, *135*, 6758.

[41]   K. S. Novoselov, D. Jiang, F. Schedin, T. J. Booth, V. V. Khotkevich, S. V. Morozov, A. K. Geim. Two-dimensional atomic crystals, *Proceedings of the National Academy of Sciences of the United States of America* **2005**, *102*, 10451.

[42]   X. L. Li, X. P. Hao, M. W. Zhao, Y. Z. Wu, J. X. Yang, Y. P. Tian, G. D. Qian. Exfoliation of hexagonal boron nitride by molten hydroxides, *Advanced Materials* **2013**, *25*, 2200.

[43]   J. N. Coleman, M. Lotya, A. O'Neill, S. D. Bergin, P. J. King, U. Khan, K. Young, A. Gaucher, S. De, R. J. Smith, I. V. Shvets, S. K. Arora, G. Stanton, H.-Y. Kim, K. Lee, G. T. Kim, G. S. Duesberg, T. Hallam, J. J. Boland, J. J. Wang, J. F. Donegan, J. C. Grunlan, G. Moriarty, A. Shmeliov, R. J. Nicholls, J. M. Perkins, E. M. Grieveson, K. Theuwissen, D. W. McComb, P. D. Nellist, V. Nicolosi. *Science*, **2011**, *331*, 568.

[44]   Y. F. Xue, Q. Liu, G. J. He, K. B. Xu, L. Jiang, X. H. Hu, J. Q. Hu. Excellent electrical conductivity of the exfoliated and fluorinated hexagonal boron nitride nanosheets, *Nanoscale Research Letters* **2013**, *8*, 49.

[45]   P. Sutter, J. Lahiri, P. Zahl, B. Wang, E. Sutter. Scalable synthesis of uniform few-layer hexagonal boron nitride dielectric films, *Nano Letter* **2013**, *13*, 276.

[46]   S. Nakhaie, J. M. Wofford, T. Schumann, U. Jahn, M. Ramsteiner, M. Hanke, J. M. J. Lopes, H. Riechert. Synthesis of atomically thin hexagonal boron nitride films on nickel foils by molecular beam epitaxy, *Applied Physics Letter* **2015**, *106*, 213108.

[47]  S. K. Bae, H. K. Kim, Y. B. Lee, X. F. Xu, J. S. Park, Y. Zheng, J. Balakrishnan, T. Lei, H. R. Kim, Y. I. Song, Y. J. Kim, K. S. Kim, B. Oͤ zyilmaz, J. H. Ahn, B. H. Hong, S. Iijima. Roll-to-roll production of 30-inch graphene films for transparent electrodes, *Nature Nanotechnology* **2010**, *5*, 574.

[48]  B. J. Choi. Chemical vapor deposition of hexagonal boron nitride films in the reduced pressure, *Materials Research Bulletin* **1999**, *34*, 2215.

[49]  M. A. Shehzad, D. H. Tien, M. W. Iqbal, J. W. Eom, J. H. Park, C. Y. Hwang, Y. H. Seo. Nematic Liquid Crystal on a Two Dimensional Hexagonal Lattice and its Application, *Scientific Reports* **2015**, *5*, 13331.

[50]  R. A. Doganov, E. C. T. O'Farrell, S. P. Koenig, Y. T. Yeo, A. Ziletti, A. Carvalho, D. K. Campbell, D. F. Coker, K. J. Watanabe, T. Taniguchi, A. H. C. Neto, B. Özyilmaz. Transport properties of pristine few-layer black phosphorus by van der Waals passivation in an inert atmosphere, *Nature Communications* **2015**, *6*, 6647.

[51]  Y. T. Lee, W. K. Choi, D. K. Hwang. Chemical free device fabrication of two dimensional van der Waals materials based transistors by using one-off stamping, *Applied Physics Letter* **2016**, *108*, 253105.

[52]  X. L. Fu, Y. F. Hu, T. Zhang, S. F. Chen. The role of ball milled h-BN in the enhanced photocatalytic activity: A study based on the model of ZnO, *Applied Surface Science* **2013**, *280*, 828.

[53]  Y. C. Yang, P. Gao, S. Gaba, T. Chang, X. Q. Pan, W. Lu. Observation of conducting filament growth in nanoscale resistive memories, *Nature Communications* **2012**, *3*, 732.

[54]  International Technology Roadmap for Semiconductors, 2013 Edition, Process Integration, Devices, and Structures section, http://www.itrs.net (accessed February 11th 2015).

[55]  F. Hui, C. B. Pan, Y. Y. Shi, Y. F. Ji, E. Grustan-Gutierrez, M. Lanza. On the use of two dimensional hexagonal boron nitride as dielectric, *Microelectronic Engineering* **2016**, *163*, 119.

[56]  X. Li, C. H. Tung, K. L. Pey, V. L. Lo. The physical origin of random telegraph noise after dielectric breakdown, *Applied Physics Letter* **2009**, *94*, 132904.

[57]  K. K. Kim, A. Hsu, X. T. Jia, S. M. Kim, Y. M. Shi, M. Dresselhaus, T. Palacios, J. Kong. Synthesis and characterization of hexagonal boron nitride film as a dielectric layer for graphene devices, *ACS Nano* **2012**, *6*, 8583.

[58]  C. R. Dean, A. F. Young, P. Cadden-Zimansky, L. Wang, H. Ren, K. Watanabe, T. Taniguchi, P. Kim, J. Hone, K. L. Shepard. Multicomponent fractional quantum Hall effect in graphene, *Nature Physics* **2011**, *7*, 693.

[59]  T. Taychatanapat, K. Watanabe, T. Taniguchi, P. Jarillo-Herrero. Quantum Hall effect and Landau-level crossing of Dirac fermions in trilayer graphene, *Nature Physics* **2011**, *7*, 621.

[60]  M. S. Bresnehan, M. J. Hollander, M. Wetherington, M. LaBella, K. A. Trumbull, R. Cavallero, D. W. Snyder, J. A. Robinson. Integration of Hexagonal Boron Nitride with Quasi-freestanding Epitaxial Graphene: Toward Wafer-Scale, High-Performance Devices, *ACS Nano* **2012**, *6*, 5234.

[61]  E. Kim, T. H. Yu, E. S. Song, B. Yu. Chemical vapor deposition-assembled graphene field-effect transistor on hexagonal boron nitride, *Applied Physics Letters* **2011**, *98*, 109.

[62]  L. Britnell, R. V. Gorbachev, R. Jalil, B. D. Belle, F. Schedin, M. I. Katsnelson, L. Eaves, S. V. Morozov, A. S. Mayorov, N. M. R. Peres, A. H. C. Neto, J. Leist, A. K. Geim, L. A. Ponomarenko, K. S. Novoselov. Electron Tunneling through Ultrathin Boron Nitride Crystalline Barriers. *Nano Letters* **2012**, *12*, 1707.

[63]  Y. Hattori, T. Taniguchi, K. Watanabe, K. Nagashio. Layer-by-layer dielectric breakdown of hexagonal boron nitride, *ACS Nano* **2015**, *9*, 916.

[64]  Y. F. Ji, C. B. Pan, M. Y. Zhang, S. B. Long, X. J. Lian, F. Miao, F. Hui, Y. Y. Shi, L. Larcher, E. Wu, M. Lanza. Boron Nitride as Two Dimensional Dielectric: Reliability and Dielectric Breakdown, *Applied Physics Letters* **2016**, *108*, 012905.

[65]  K. Qian, R. Y. Tay, V. C. Nguyen, J. Wang, G. Cai, T. Chen, E. H. T. Teo, P. S. Lee. Hexagonal Boron Nitride Thin Film for Flexible Resistive Memory Applications, *Advanced Functional Materials* **2016**, *26*, 2176.

[66]  K. Qian, R. Y. Tay, M. F. Lin, J. Chen, H. Li, J. Lin, J. Wang, G. Cai, V. C. Nguyen, E. H. Teo, T.Chen, P. S. Lee. Direct Observation of Indium Conductive Filaments in Transparent, Flexible, and Transferable Resistive Switching Memory, *ACS nano* **2017**, *11*, 1712.

[67]  J. C. Koepke , J. D. Wood, Y. Chen, et al. Role of Pressure in the Growth of Hexagonal Boron Nitride Thin Films from Ammonia-Borane, *Chemistry of Materials* **2016**, *28*, 4169.

[68]  Y. Jin, B. S. Hu, Z. D. Wei, Z. T. Luo, D. P. Wei, Y. Xi, Y. Zhang, Y. L. Liu. Roles of H2 in annealing and growth times of graphene CVD synthesis over copper foil, *Journal of Materials Chemistry A* **2014**, *2*, 16208.

[69]  W. J. Fang. PhD thesis. Synthesis of bilayer graphene and hexagonal boron nitride by chemical vapor deposition method, *Massachusetts Institute of Technology* **2016**.

[70]  S. M. Kim, A. Hsu, M. H. Park, S. H. Chae, S. J. Yun, J. S. Lee, D. H. Cho, W. J. Fang, C. G. Lee, T. Palacios, M. Dresselhaus, K. K. Kim, Y. H. Lee, Jing Kong. Synthesis of large-area multilayer hexagonal boron nitride for high material performance, *Nature Communication* **2015**, *6*, 8662.

[71]    T. Wu, X. Zhang, Q. Yuan, J. Xue, G. Lu, Z. Liu, H.Wang, H.Wang, F. Ding, Q. Yu, X. Xie, M. Jiang. Fast growth of inch-sized single-crystalline graphene from a controlled single nucleus on Cu–Ni alloys. *Nature Materials* **2016**, *15*, 43.

[72]    A. Nagashima, N. Tejima, Y. Gamou, T. Kawai, C. Oshima, Electronic Structure of Monolayer Hexagonal Boron Nitride Physisorbed on Metal Surfaces, Physical Review letters **1995**, *51*, 4606.

[73]    S. Caneva, R. S. Weatherup, B. C. Bayer, B. Brennan, S. J. Spencer, K. Mingard, A. Cabrero-Vilatela, C. Baehtz, A. J. Pollard, S. Hofmann, Nucleation Control for Large, Single Crystalline Domains of Monolayer Hexagonal Boron Nitride via Si-Doped Fe Catalysts. Nano Lette **2015**, *15*, 1867.

[74]    S. Vongehr, X. K. Meng. The missing memristor has not been found, *Scientific Reports* **2015**, *5*, 11657.

[75]    D. B. Strukov, G. S. Snider, D. R. Stewart, R. S. Williams, The missing memristor found. *Nature*, **2008**, *453*, 80.

[76]    C. Pan, Y. Fu, J. Wang, J. Zeng, G. Su, M. Long, E. Liu, C. Wang, A. Gao, M. Wang, Y. Wang, Z. Wang, S. Liang, R. Huang, F. Miao, Analog Circuit Applications based on Ambipolar Graphene/MoTe2, *Advanced Electronic Materials* **2018**, *1700662*.

[77]    R. J. Ge, X. H. Wu, M. Kim, J. P. Shi, S. Sonde, L. Tao, Y. F. Zhang, J. C. Lee, D. Akinwande. Atomristor: Nonvolatile Resistance Switching in Atomic Sheets of Transition Metal Dichalcogenides. *Nano Letter* **2018**, *18*, 434.

[78]    R. Yang, H. Li, K. K. H. Smithe, T. R. Kim, K. Okabe, E. Pop, J. A. Fan, H. S. Philip Wong, 2D molybdenum disulfide (MoS 2) transistors driving RRAMs with 1T1R configuration, IEEE *Electron Devices Meeting* **2017**, 8268423.

[79]    Mario Lanza, H.-S. Philip Wong, Eric Pop, Daniele Ielmini, Dimitri Strukov, Brian Chris Regan, Jianghua Joshua Yang, Ludovic Goux, Attilio Belmonte, Yuchao Yang, Anthony Kenyon, Adnan Mehonic, Mark Buckwell, Luca Larcher, Blanka Magyari-Köpe, Eilam Yalon, Francesco M. Puglisi, Marco A. Villena, Alexander Shluger, Tuo-Hung Hou, Boris Hudec, Deji Akinwande, Ruijing Ge, Juan B. Roldan, Jordi Suñe, Enrique Miranda, Kin Leong Pey, Xing Wu, Ernest Wu, Wei D. Lu, Gabriele Navarro, Gabriel Molas, Weidong Zhang, Alexander Holleitner, Max Lemme, Rainer Waser, Ilia Valov, S. Ambrogio, Ming Liu, Shibing Long, Qi Liu, Hangbin Lv, Jinfeng Kang, Huaqiang Wu, Runwei Li, Haitong Li, Xu Jing, Fei Hui, Yuanyuan Shi, Unified criteria for studying resistive switching devices, *Adv. Electron. Mater.* **2018**, (in press).

[80]    J. Li, J. F. Hsu, H. Lee, S. Tripathi, Q. Guo, L. Chen, M. C. Huang, S. Dhingra, J. W. Lee, C. B. Eom, P. Irvin, J. Levy, B. D'Urso, Method for Transferring High-Mobility CVD-Grown Graphene with Perfluoropolymers, **2016**, arXiv:1606.08802.

[81]    M. Hempel, A. Y Lu, F. Hui, T. Kpulun, M. Lanza, G. Harris, T. Palacios, J. Kong. Repeated roll-to-roll transfer of two-dimensional materials by electrochemical delamination, *Nanoscale* **2018**, *10*, 5522.

[82]    L. Gao, G. X. Ni, Y. Liu, B. Liu, A. H. C. Neto, K. P. Loh. Face-to-face transfer of wafer-scale graphene films, *Naure* **2014**, *505*, 190.

[83]    A. C. Ferrari. Science and technology roadmap for graphene, related two-dimensional crystals, and hybrid systems, *Nanoscale* **2015**, *7*, 4598.

[84]    G. Vescio, A. Crespo-Yepes, D. Alonso, S. Claramunt, M. Porti, R. Rodriguez, A. Cornet, A. Cirera, M. Nafria, X. Aymerich, Inkjet Printed $HfO_2$-Based ReRAMs: First Demonstration and Performance Characterization. *IEEE Electron Device Letters*, **2017**, *38*, 457.

[85]    T. Roy, M. Tosun, J.K. Kang, A.B. Sachid, S.B. Desai, M. Hettick, C.C. Hu, A. Javery, *ACS Nano* **2004**, *8*, 6259.

[86]    H. Liu, K. Xu, X. Zhang, P. D. Ye. The Integration of High-k Dielectric on Two-Dimensional Crystals by Atomic Layer Deposition. *Appl. Phys. Lett.* **2012**, *100*, 152115.

[87]    A. Pirkle, S. McDonnell, B. Lee, J. Kim, L. Colombo, R. M. Wallace, The Effect of Graphite Surface Condition on the Composition of Al2O3 by Atomic Layer Deposition. *Appl. Phys. Lett*. **2010**, *97*, 082901.

[88]    J. Yao, J. Lin, Y. H. Dai, G. Ruan, Z. Yan, L. Li, L. Zhong, D. Natelson, J. M. Tour. Highly transparent nonvolatile resistive memory devices from silicon oxide and graphene, *Nature Communications* **2012**, *3*, 1101.

[89]    P. K. Yang, W. Y. Chang, P. Y. Teng, S. F. Jeng, S. J. Lin, P. W. Chiu, J. H. He. Fully transparent resistive memory employing graphene electrodes for eliminating undesired surface effects, *in Proceeding of the IEEE* **2013**, *101*, 1732

[90]    H. Tian, H. Y. Chen, B. Gao, S. M. Yu, J. L. Liang, Y. Yang, D. Xie, J. F. Kang, T. L. Ren, Y. G. Zhang, H. S. P. Wong. Monitoring oxygen movement by raman spectroscope of resistive random access memory with a graphene-inserted electrode, *Nano Letters* **2013**, *13*, 651.

[91]    G. Venugopal, S. J. Kim, Observations of nonvolatile resistive memory switching characteristics in Ag/Graphene-oxide/Ag devices, *Journal of Nanoscience and Nanotechnology*, **2012**, *12*, 8522.

[92]    G. N. Panin, O. O. Kapitanova, S. W. Lee, A. N. Baranov, T. W. Kang. Resistive switching in Al/Graphene/Al structure, *Japanese Journal of Applied Physics* **2011**, *50*, 070110.

[93]    B. Chakrabarti, T. Roy, E. M. Vogel. Nonlinear switching with ultralow reset power in graphene-insulator-graphene forming-free resistive memories, *IEEE Electron Device Letters* **2014**, *35*, 7.

[94]    F. Kreupl, R. Bruchhaus, P. Majewski, J. B. Philipp, R. Symanczyk, T. Happ, C. Arndt, M. Vogt, R. Zimmermann, A. Buerke, A. P. Graham, M. Kund, Carbon-based resistive memory, *in Technical Digest of International Electron Devices Meeting*, **2008**, *15*, 521.

[95]    B. L. Hu, R. Quhe, C. Chen, F. Zhuge, X. J. Zhu, S. S. Peng, X. X. Chen, L. Pan, Y. Z. Wu, W. Zheng, Q. Yan, J. Lu, R. W. Li. Electrically controlled electron transfer and resistance switching in reduced graphene oxide noncovalently functionalized with thionine, *Journal of Materials Chemistry* **2012**, *22*, 16422.

[96]    Care RAMM public summary: http://emps.exeter.ac.uk/media/universityofexeter/ emps/careramm/D4.4 Public summary of project results from the third year of the project.pdf (accessed on August 2016).

[97]    B. Zhang, G. Liu, Y. Chen, L. J. Zeng, C. X. Zhu, K. G. Neoh, C. Wang, E. T. Kang, Conjugated polymer-grafted reduced graphene oxide for nonvolatile rewritable memory, *Chemistry A European Journal* **2011**, *17*, 13646.

[98]    P. F. Cheng, K. Sun, Y. H. Hu. Memristive behavior and ideal memristor of 1T phase MoS₂ nanosheets, *Nano Letters* **2016**, *16*, 572.

[99]    S. L. Hong, J. E. Kim, S. O. Kim, S. Y. Choi, B. J. Cho. Flexible resistive switching memory device based on graphene oxide, *IEEE Electron Device Letters*, **2010**, *31*, 1005.

[100]   V. K. Nagareddy, A. K. Ott, C. Dou, T. Tsvetkova, M. Sandulov, M. F. Craciun, A. C. Ferrari, C. D. Wright, unpublished.

[101]   S. Liu, N. D. Lu, X. L. Zhao, H. Xu, W. Banerjee, H. B. Lv, S. B. Long, Q. J. Li, Q. Liu, M. Liu, Eliminating negative-SET behavior by suppressing nanofilament overgrowth in cation-based memory. *Adv. Mater.* **2016**, *28*, 10623.

[102]   Y. Park, D. Gupta, C. Lee, Y. Hong, Role of tunneling layer in graphene-oxide based organic nonvolatile memory transistors. *Org. Electron*. **2012**, *13*, 2887.

[103]   Y. C. Yang, J. Lee, S. Lee, C. H. Liu, Z. H. Zhong, W. Lu, Oxide resistive memory with functionalized graphene as built-in selector elecment. *Adv. Mater.* **2014**, *26*, 3693.

# Appendix A: Scientific vita

## Curriculum vitae

### PERSONAL INFORMATION

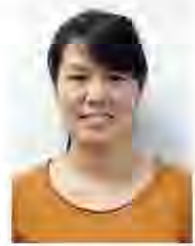

Name: Fei Hui (惠飞)
Date of birth: May 20, 1988
Address: Institute of Functional Nano and Soft Materials
Building 910, office 316, Soochow University,
199 Ren-Ai Road, Suzhou Industrial Park, 215123
Phone: +86 15370083707
Email: huifei324@126.com

### EDUCATION BACKGROUND

| | | |
|---|---|---|
| 2018-07 | PhD in Nanosciences | Universitat de Barcelona |
| | | **Supervisor: Mario Lanza** |
| 2017-05 ~ 2017-10 | Academic visiting | University of Cambridge |
| | | **Prof. Andrea Ferrari's group** |
| 2016-04 ~ 2017-03 | Academic visiting | Massachusetts Institute of Technology |
| | | **Prof. Jing Kong's group** |
| 2009-09 ~ 2013-06 | Bachelor in Chemistry | Huanghuai University |

### RESEARCH EXPERIENCE

- **Research topic 1:** Chemical vapor deposited graphene
  - Growth of graphene by chemical vapor deposition and transfer using polymer media
  - Micro and nanoscale analysis of graphene morphology with AFM, SEM and EDAX
  - Conductive AFM study (I-V curves and maps) of the local contact resistance of graphene

- **Research topic 2:** Design of commercial devices using 2D materials ink
  - Preparation of graphene inks by dispersing graphene powder in different solvents
  - Cost-effective fabrication of graphene-coated AFM tips with very long lifetime and superior reliability.
  - Preparation and nanoscale electrical characterization (using the CAFM) of piezoelectric inks using multilayer $MoS_2$ nanosheets
  - Observation of enhanced piezoelectric effect at the edges of multilayer $MoS_2$.
  - High yield fabrication of graphene coated AFM probes by spray coating (**collaborated with prof. Andrea Ferrari in Cambridge University**)

- **Research Topic 3:** Design of advanced ReRAM using h-BN as dielectric
  - CVD growth multilayer h-BN on different substrates (**collaborated with prof. Jing Kong in Massachusetts Institute of Technology**)
  - Fabrication of ReRAM non-volatile memory devices based on metal/h-BN/metal structures. To do so, I used photolithography, electron beam evaporator.
  - Fabrication of fully 2D ReRAM cells using graphene/h-BN/graphene stacks.
  - Device level characterization with probestation: cyclic voltammograms, state retention by current vs. time curves, statistical analysis of the set/reset variability using Weibull
  - Nanoscale characterization of the boron nitride using CAFM, SEM and cross sectional TEM.

### SCHOLASHIPS AND AWARDS

- 2016 International Exchange Scholarship of Soochow University
- Award the Royal Society of Chemistry mobility grant for working as a visiting scholar at University of Cambridge (UK). (Feburary 2016)
- Soochow University international travel conference award to participate in the CDE conference organized by IEEE on Feb. 11th - 13th of 2015 Madrid.
- National Scholarship during bachelor

# Appendix A: Scientific vita

## Curriculum vitae

**MEMBERSHIPS**

- Student member of the Royal Society of Chemistry (since October 2015)
- IEEE Membership (student) (October 2015)
- Member of the Electron Device Society (EDS)

**JOURNAL PAPERS**

Google citation: 314   H index: 9   i10 index: 9

1. **Fei Hui**, W. Fang, W.S. Leong, T. Kpulun, H. Wang, M. A. Villena, G. Harris, J. Kong, M. Lanza, "Electrical homogeneity of large-area chemical vapor deposited multilayer hexagonal boron nitride sheets", ACS *Applied Materials & Interfaces* 2017, 9, 39895.

2. **Fei Hui**, S. Chen, X. Liang, B. Yuan, X. Jing, Y. Shi, M. Lanza, "Graphene coated nanoprobes: a review", **Crystals** 2017, 7, 269.

3. **Fei Hui**, Pujashree Vajha, Yanfeng Ji, Chengbin Pan, Enric Grustan-Gutierrez, Huiling Duan, Peng He, Guqiao Ding, Yuanyuan Shi, Mario Lanza*, "Variability of graphene devices fabricated using graphene inks: atomic force microscope tips", *Surface & Coatings Technology* 2017, 320, 391.

4. **Fei Hui**, Enric Grustan-Gutierrez, Qi Liu, Shibing Long, Anna. L. Ott, Andrea C. Ferrari, Mario Lanza. "A review on the use of two dimensional materials in resistive random access memories". *Advanced Electronic Materials* 2017, 3, 1600195. - **Highlighted as front cover**.

5. **Fei Hui**, Chengbin Pan, Yuanyuan Shi, Yanfeng Ji, Enric Grustan-Gutierrez, Mario Lanza*, "On the use of two dimensional hexagonal boron nitride as dielectric", *Microelectronic Engineering* 2016, 163, 119.

6. **Fei Hui**, Pujashree Vajha, Yuanyuan Shi, Yanfeng Ji, Huiling Duan, Andrea Padovani, Luca Larcher, Xiao-Rong Li, Jing-Juan Xu, Mario Lanza*, "Moving graphene devices from lab to market: advanced graphene-coated nanoprobes", *Nanoscale* 8: 8466-84723 (2016) - **Highlighted as front cover.**

7. **Fei Hui**, Yuanyuan Shi, Yanfeng Ji, Mario Lanza, Huiling Duan. "Mechanical properties of locally oxidized graphene electrodes", *Archive of Applied Mechanics* 85: 339-345 (2014).

8. **Fei Hui**, M. Porti, M. Nafria, H. Duan, M. Lanza, "Fabrication of graphene MEMS by standard transfer: high performance atomic force microscope tips", 2015 10th Spanish conference on electron devices (CDE).

9. Bingru Wang, Na Xiao, Chengbin Pan, Yuanyuan Shi, **Fei Hui**, Xu Jing, Kaichen Zhu, Biyu Guo, Marco A Villena, Enrique Miranda, Mario Lanza. "Experimental Observation and Mitigation of Dielectric Screening in Hexagonal Boron Nitride Based Resistive Switching Devices", *Crystal Research and Technology* 2018, 1800006.

10. M Hempel, A-Y Lu, **Fei Hui**, T Kpulun, M Lanza, G Harris, T Palacios, J Kong. "Repeated roll-to-roll transfer of two-dimensional materials by electrochemical delamination", *Nanoscale* 2018, 10, 5522-5531.

11. Felix Palumbo, Xianhu Liang, Bin Yuan, Yuanyuan Shi, **Fei Hui**, Marco A. Villena, Mario Lanza, "Bimodal Dielectric Breakdown in Electronic Devices Using Chemical Vapor Deposited Hexagonal Boron Nitride as Dielectric", *Advanced Electronic Materials* 2018, 1700506.

# Appendix A: Scientific vita

## Curriculum vitae

12. Shosuke Fujii, Jean Anne C Incorvia, Fang Yuan, Shengjun Qin, **Fei Hui**, Yuanyuan Shi, Yang Chai, H-S Philip Wong, "Scaling the CBRAM switching layer diameter to 30 nm improves cycling endurance", *IEEE Electron Device Letters* 39, 23-26, 2018.

13. L. Jiang, Y. Shi, **Fei Hui**, K. Tang, Q Wu, C. Pan, X. Jing, H, Uppal, F. Palumbo, G. Lu, T. Wu, H. Wang, M. A. Villena, X. Xie, P. C. McIntyre, M. Lanza, "Dielectric breakdown in chemical vapor deposited hexagonal boron nitride", *ACS Applied Materials & Interfaces* 9, 39758, 2017.

14. N. Xiao, M.A. Villena, B. Yuan, S. Chen, B. Wang, M. Elias, Y. Shi, **Fei Hui**, X. Jing, A. Scheuermann, K. Tang, P.C. McIntyre, M. Lanza, "Resistive random access memory cells with a bilayer TiO2/SiOx insulating stack for simultaneous filamentary and distributed resistive switching", *Advanced Functional Materials* 27(33), 2017.

15. Xiaoxue Song, **Fei Hui**, Theresia Knobloch, Bingru Wang, Zhongchao Fan, Tibor Grasser, Xu Jing, Yuanyuan Shi, Mario Lanza, "Piezoelectricity in two dimensions: graphene vs. molybdenum disulfide" *Applied Physics Letters* 111(8), 083107, 2017.

16. Kechao Tang, Andrew C Meng, **Fei Hui**, Yuanyuan Shi, Trevor Petach, Chuck Hitzman, Ai Leen Koh, David Goldhaber-Gordon, Mario Lanza, Paul C McIntyre, "Distinguishing oxygen vacancy electromigration and conductive filament formation in TiO2 resistance switching using liquid electrolyte contacts" *Nano Letters* 17(7), 4390-4399, 2017.

17. Chengbin Pan, Yanfeng Ji, Na Xiao, **Fei Hui**, Kechao Tang, Yuzheng Guo, Xiaoming Xie, Francesco M. Puglisi, Luca LARCHR, Enrique Miranda, Lanlan Jiang, Yuanyuan Shi, Ilia Valov, Paul C. McIntyre, Rainer Waser, Mario Lanza. "Coexistence of Grain-Boundaries-Assisted Bipolar and Threshold Resistive Swithcing in Multilayer hexagonal boron nitride", *Advanced Functional Materials*, 1604811, 2017.

18. Chengbin Pan, Enrique Miranda, Marco A Villena, Na Xiao, Xu Jing, Xiaoming Xie, Tianru Wu, **Fei Hui**, Yuanyuan Shi, Mario Lanza. "Model for multi-filamentary conduction in graphene/hexagonal-boron-nitride/graphene based resistive switching devices". *2D Materials* 2017, 4, 025099.

19. Xu Jing, Emanuel Panholzer, Xiaoxue Song, Enric Grustan-Gutierrez, **Fei Hui**, Yuanyuan Shi, Guenther Benstetter, Yury Illarionov, Tibor Grasser and Mario Lanza*, "Fabrication of scalable and ultra low power photodetectors with high light/dark current ratios using polycrystalline monolayer MoS2 sheets", *Nano Energy*, 30, 494–502, 2016.

20. Lanlan Jiang, Na Xiao, Bingru Wang, Enric Grustan-Gutierrez, Xu Jing, Petr Babor, Miroslav Kolibal, Guangyuan Lu, Tianru Wu, Haomin Wang, **Fei Hui**, Yuanyuan Shi, Bo Song, Xiaoming Xie, Mario Lanza*, "High resolution characterization of Hexagonal Boron Nitride Coatings exposed to aqueous and air oxidative environments", *Nano Research*, 2016, DOI: 10.1007/s12274-016-1393-2.

21. Yanfeng Ji, Chengbin Pan, Meiyun Zhang, Shibing Long, Xiaojuan Lian, Feng Miao, **Fei Hui**, Yuanyuan Shi, Luca Larcher, Ernest Wu, Mario Lanza*, "Boron nitride as two dimensional dielectric: reliability and dielectric breakdown", *Applied Physics Letters* 108, 012905 (2016).

22. Yuanyuan Shi, Yanfeng Ji, **Fei Hui**, Montserrat Nafria, Marc Porti, Gennadi Bersuker, Mario Lanza*, "Atomic Force Microscope study links resistive switching to the local mechanical properties in HfO$_2$ films", *Advanced Electronic Materials*, (1-2): 1400058 (2015).

# Appendix A: Scientific vita

## Curriculum vitae

23. Yuanyuan Shi, Yanfeng Ji, **Fei Hui**, Hai-Hua Wu, Mario Lanza. "Ageing of graphene electronic properties", ***Nano Research*** 7(12): 1820-1831 (2014).

24. Y. Shi, Y. Ji, **Fei Hui**, V. Iglesias, M. Porti, M. Nafria, E. Miranda, G. Bersuker, M. Lanza, "Elucidating the Origin of Resistive Switching in Ultrathin Hafnium Oxides through High Spatial Resolution Tools", ***ECS Transactions*** 64(14): 19-28 (2014).

25. Jianchen Hu, Yanfeng Ji, Yuanyuan Shi, **Fei Hui**, Huiling Duan, Mario Lanza. "A review on the use of graphene as a protective coating against corrosion", ***Annals of Materials Science and Engineering***, 1(3): 16 (2014).

### PATENTS

- Chengbin Pan, Yuanyuan Shi, **Fei Hui**, Enric Grustan-Gutierrez, Mario Lanza. "History and Status of the CAFM", John Wiley & Sons, 2017.

### PATENTS

- **Fei Hui**, Yuanyuan Shi, Mario Lanza, "Cost-effective fabrication of ultra-durable atomic force microscope tips using graphene powder coatings ", *International Patent* PCT/CN2014/09368. In September 2016 this patent reeived $1M investment from the Beijing Institute of Collaborative Innovation for creating a start-up and introducing this product in the market.
- Yanfeng Ji, Chengbin Pan, **Fei Hui**, Yuanyuan Shi, Na Xiao, Mario Lanza. Hexagonal multilayer BN based RRAM device and the fabrication method. International Patent 201510968511.7.

### CONFERENCE

- **Fei Hui**, Xianhu Liang, Wenjing Fang, Wei Sun Leong, Haozhe Wang, Hui Ying Yang, Yuanyuan Shi, Marco A. Villena, Jing Kong and Mario Lanza. Uniformity of multilayer hexagonal boron nitride dielectric stacks grown by chemical vapor deposition on platinum and copper substrates. IPFA, 2018, Singapore.

- Xianhu Liang, Felix Palumbo, Yuanyuan Shi, **Fei Hui**, Bin Yuan, Xu Jing and Mario Lanza, Dielectric breakdown in hexagonal boron nitride dielectric stacks. China Semiconductor Technology International Conference. Shanghai March, 2018.

- **Fei Hui**, Wei Sun Leong, Yuanyuan Shi, Jing Kong, Mario Lanza. Scalable resistive random access memories made of multilayer h-BN grown by CVD on Iron and Platinum electrode. 2nd International Symposium on Science and Technology of 2D materials, February 3rd-4th 2017, Orlando, FL, USA.

- Kechao Tang, Andrew Meng, **Fei Hui**, Yuanyuan Shi, Trevor A. Petach, David Goldhaber-Gordon, Mario Lanza, and Paul C. McIntyre, Distinguishing Oxygen Vacancy Electromigration and Conductive Filament Formation in TiO2 Resistance Switching Using Liquid Electrolyte Contacts, 48th IEEE Semiconductor Interface Specialists Conference, December 6th-9th 2017, San Diego (USA).

- F. M. Puglisi, L. Larcher, C. Pan, N. Xiao, Y. Shi, **F. Hui,** M. Lanza. 2D h-BN based RRAM devices. 2016 IEEE International Electron Devices Meeting (IEDM), December 3rd-7th, 2016, San Francisco, CA, USA.

# **Appendix A: Scientific vita**

## Curriculum vitae

- **Fei Hui**, Yuanyuan Shi, Mario Lanza. 2016 IEEE International Integrated Reliability Workshop (IIRW), October 9th-13th 2016, Stanford Serria Conference Center, CA, USA.

- Yanfeng Ji, **Fei Hui**, Tingting Han, Xiaoxue Song, Yuanyuan Shi, Mario Lanza. Reliability of Boron Nitride as thin dielectric. 2015 Fall Materials research society, November 29th – December 4th 2015, Boston, USA.

- **Fei Hui**, Marc Porti, Montsrrat Nafria, Huiling Duan, Mario Lanza. Fabrication of graphene MEMS by standard transfer: high performance atomic force microscope tips, 10th Spain Conference on Electron Devices, February 11th – 13th, 2015. Aranjuez, Spain.

- Yuanyuan Shi, Yanfeng Ji, **Fei Hui**, Mario Lanza, Ageing mechanisms and reliability of graphene-based electrodes, *2nd International Workshop of Soochow University-Western University Centre for Synchrotron Radiation Research*, May 6th-8th 2014, Suzhou, China.

### REVIEW EXPERIENCE

I have the experience on reviewing papers submitted to Scientific reports and IET Nanobiotechnology. Until now, I have reviewed four papers in these two journals.

### SCIENTIFIC SEMINARS

- **Fei Hui**, Mario Lanza, Zhu Tong, Huiling Duan, Seminar about the research of air pollutants - PM2.5, College of Engineering, Peking University, December 21th 2014, Beijing, China.
- **Fei Hui**, Enric Grustan, Haiyi Liang, Huiling Duan, Research and application of advanced printing materials and devices based on 2D materials. College of Engineering, Peking University, hosted by professor Huiling Duan, July 30 th 2015, Beijing, China.

### INTERN EXPERIENCE

- 2011-10 - 2012-11 Internship: Vice secretary of the league in Chemistry department
- 2011-07 - 2011-09 Internship: Assistant in Pro. Miao Yu's labortary
- 2010-07 - 2010-08 Teaching-Assistance in country as volunteer

### OTHER SKILLS

- Able to efficiently communicate in English
  1. Used to daily research under the supervision of a European Professor using English.
  2. Fluent communication in both spoken and written.
  3. College English Test (CET) Level 6 (482 points).

# Appendix B: Summary in official language

En resumen, durante esta tesis doctoral he aprendido a crecer capas de h-BN de alta calidad y con diferentes grosores utilizando el método CVD. Esta habilidad es muy importante porque h-BN es un material muy demandado, y porque CVD es un método que puede ser utilizado para crecer muchos otros materiales bidimensionales. También he aprendido a transferir materiales bidimensionales sobre cualquier otro substrato utilizando tres métodos diferentes. También he aprendido a analizar las propiedades de los materiales bidimensionales utilizando múltiples equipos (como por ejemplo CAFM, SEM, microscopio óptico, Raman y TEM). Además he fabricado condensadores y memristores basados en h-BN (utilizando fotolitografía, evaporación de metal térmica, evaporación de metal por haz de electrones, y sputtering) y he analizado las propiedades de los dispositivos utilizando una tabla de puntas. El uso de modelos teóricos y ajustes (realizados con ayuda de mis coautores) me ha ayudado a entender el funcionamiento de los dispositivos. Las principales conclusiones de mi trabajo son:

- h-BN monocapa y multicapa pueden ser crecidos mediante CVD sobre sustratos de platino, cobre o hierro. Los principales parámetros durante el crecimiento son: *i)* una temperatura adecuada para la decomposición del precursor. Bajas temperaturas producen la acumulación de partículas y más defectos en la capa h-BN. *ii)* El ratio precursor/nitrógeno influencia la densidad de semillas. Una cantidad excesiva de precursor producirá la formación de islas multicapas. *iii)* Un alto vacío y una presión baja ayudan a eliminar las impurezas dentro del tubo CVD (por ejemplo carbón, oxígeno) y por lo tanto mejora la calidad de la capa h-BN (es decir, produce un grosor más homogeneo y con menos defectos).

- Las capas h-BN crecidas sobre sustratos de platino policristalino muestran diferentes grosores dependiendo de la orientación cristalográfica de cada cristal de platino. Esto produce una (indeseada) fluctuación de la corriente de fugas a través del h-BN. Sin embargo, la corriente de fugas a través de la capa h-BN dentro de un mismo cristal de platino es muy uniforme, mucho más que a través de capas amorfas de $HfO_2$ y $TiO_2$. Este fenómeno no se observa si el h-BN se crece sobre sustratos de cobre o hierro. Por ejemplo, la corriente de fugas a través de h-BN crecido sobre sustratos policristalinos de cobre muestran una baja variabilidad de un cristal de cobre a otro.

- La ruptura dieléctrica de h-BN multicapa muestra una extrusión de la superficie, muy similar a lo que sucede en $SiO_2$, $HfO_2$ y $Al_2O_3$. Sin embargo, las monocapas de h-BN mantienen su estructura incluso cuando la ruptura dieléctrica es mucho más brusca. La razón podría ser la elevada conductividad térmica de las monocapas de h-BN.

- Las multicapas de h-BN muestran fluctuaciones de corriente entre dos estados al aplicar una tensión constante, tanto a escala nanométrica como a nivel de dispositivo. Esta observación indica que existe atrapamiento y desatrapamiento de carga. Este fenómeno ha sido confirmado mediante la detección de cargas atrapadas en el punto de ruptura, el cual muestra una singular estructura de anillo con cargas fijas negativas, cargas móviles negativas, y cargas fijas positivas.

- La síntesis de h-BN sobre sustratos de hierro policristalinos requiere un tiempo de enfriamiento (durante el proceso CVD) mucho más elevados que sobre sustratos de platino o cobre. La razón principal es que el mecanismo de crecimiento es distinto, sobre hierro la capa h-BN crece por precipitación, mientras que sobre platino o cobre crece por reacción con la superficie.

- Los memristores con estructura Ag/h-BN/Fe muestran modulación de la resistividad de tipo volátil cuando la ruptura dieléctrica es generada aplicando tensión positiva en el electrodo de plata, y de tipo bipolar cuando la ruptura dieléctrica y la recuperación son generadas aplicando tensiones negativa y positiva (respectivamente) en el electrodo de plata. La razón es que en modo volátil el filamento está formado por iones de plata que penetran en la capa h-BN, los cuales tienen una alta difusividad y pueden retroceder a su estado de reposo cuando la tensión es desactivada. En el caso del modo bipolar, los átomos que forman el filamento son de hierro, que tienen una menor difusividad, y por lo tanto requieren la aplicación de una tensión extra para romper el filamento.

A parte de las habilidades técnicas adquiridas durante los experimentos, durante el desarrollo de esta tesis doctoral he hecho un esfuerzo muy importante en revisar la literatura relacionada y organizar la información. En mi caso, esta contribución es mayor que en otras tesis doctorales, ya que he escrito dos artículos de revisión, y en total he estudiado más de 543 artículos. En el primer artículo, publicado en la revista Microelectronics Engineering, he analizado el uso de h-BN como dieléctrico en dispositivos electrónicos (estado previo a esta tesis). Y en el segundo he analizado el uso de materiales bidimensionales para memristores. Este segundo artículo de revisión ha sido escrito en colaboración con el profesor Andrea Ferrari de la Universidad de Cambridge, y ha sido seleccionado como portada en la revista Advanced Electronic Materials. Esto me ha dado una visión muy amplia sobre el uso de materiales bidimensionales como dieléctrico, que es una habilidad que espero explotar en el futuro.

Futuros trabajos en esta dirección deberían concentrarse en el estudio del fenómeno de modulación de la resistividad a escala nanométrica, utilizando dispositivos más pequeños y el CAFM. Se deberían realizar análisis estadísticos de la tensión y tiempo de ruptura en dispositivos reales. Sería también interesante estudiar la corriente de fugas a través de las fronteras de grano en el h-BN, y así poder comprender sus potenciales puntos débiles. Algunos parámetros relacionados con la ruptura dieléctrica, como la carga de ruptura deberían ser analizados. El parámetro más importante a analizar es la influencia de la elevada constante térmica del h-BN en la ruptura dieléctrica, y también el comportamiento de este material a altas temperaturas. Sin embargo el avance más significativo sería poder crecer capas de h-BN con una semilla única. Este método ha sido utilizado anteriormente en el crecimiento de grafeno, pero nunca antes en el crecimiento de h-BN. Además, otros materiales bidimensionales (como el óxido de grafeno y el fosforeno) deberían ser estudiados desde el punto de vista dieléctrico. El rango de posibilidades es muy amplio, y los experimentos y hallazgos que vendrán en un futuro serán muy excitantes !

# Appendix C: List of acronyms

| | |
|---|---|
| 2D | Two dimensional |
| 3D | Three dimensional |
| BD | Dielectric breakdown |
| CAFM | Conductive atomic force microscopy |
| $c$-BN | Cubic boron nitride |
| CVD | Chemical vapor deposition |
| F1 | Tube line 1 |
| F2 | Tube line 2 |
| FETs | Field effect transistors |
| FIB | Focused ion beam |
| $h$-BN | Hexagonal boron nitride |
| HRS | High resistive state |
| I-V | Current vs. voltage |
| LPE | Liquid phase exfoliation |
| LRS | Low resistive state |
| MBE | Molecular beam epitaxy |
| MIM | Metal/insulator/metal |
| NVMs | Non-volatile memories |
| P | Pressure |
| PVD | Physical vapor deposition |
| RS | Resistive switching |
| RT | Room temperature |
| RTN | Random telegraph noise |
| SEM | Scanning electron microscopy |
| SILC | Stress induced leakage current |
| $SiO_2$ | Silicon dioxide |
| $T_A$ | Annealing temperature |
| $t_A$ | Annealing time |
| $t_C$ | Cooling down time |
| $t_G$ | Growth time |
| $T_G$ | Growth temperature |
| TEM | Transmission electron microscopy |
| TMOs | Transition metal oxides |
| TMDs | Transition metal dichalcogenides |
| UHV | Ultra high vacuum |
| $w$-BN | wurtzite boron nitride |